\documentclass[12pt]{report} 

\usepackage{utdiss2}    

\usepackage{amsmath,amsthm,amsfonts,amscd}
\usepackage{verbatim}       
\usepackage[numbers,sort&compress]{natbib}
\usepackage{url}  
\usepackage{tikz}
\usetikzlibrary{arrows}
\usetikzlibrary{decorations.pathreplacing}
\usepackage{xspace}
\usepackage{multirow}
\usepackage{subfig}
\usepackage{epigraph}
\usepackage{paralist}
\usepackage{dcolumn}
\usepackage{tabularx}
\usepackage{booktabs}
\usepackage{threeparttable}
\usepackage{upgreek}
\usepackage{afterpage}
\usepackage{pdflscape}
\usepackage[plainpages=false,colorlinks=true]{hyperref}
\usepackage{hypernat}
\usepackage{algorithm}
\usepackage{algorithmic}
\newcommand{\ignore}[1]{}

\hypersetup{colorlinks=false}

\let\origfootnote\footnote
\renewcommand{\footnote}[1]{\kern.06em\origfootnote{#1}}
\newcommand{\punctfootnote}[1]{\kern-.06em\origfootnote{#1}}

\author{Milad Olia Hashemi} 

\address{Unlisted} 

\title{On-Chip Mechanisms \\ to Reduce Effective Memory Access Latency} 

\supervisor
 {Yale N. Patt}

\committeemembers
 [Douglas M. Carmean]
 [Derek Chiou]
 [Mattan Erez]
 {Donald S. Fussell}

\previousdegrees{B.S.E.E.; M.S.E.}

\graduationmonth{August}

\oneandonehalfspacing

\singlespacequote

\begin{document}

\titlepage              



\begin{preface}  

Only a fraction of the work that has allowed me to write this dissertation is my own. I can't imagine the strength that it must've taken my parents to immigrate to a new and unfamiliar country with no resources and then raise two kids. They always prioritized our education over any of their own needs. I was only able to write this dissertation because of their sacrifice. I thank my parents: Homa and Mohammad, and my sister Misha for their unwavering love.

I would never have pursued a Ph.D. or arrived at UT without my wife, Kelley. Well before I had any idea, she knew that I wouldn't be happy leaving graduate school without a doctorate. I thank her for her advice, clairvoyance, and patience throughout these last seven years.

My time in graduate school has allowed me to meet and work with amazing people who have taught me far more than I could list here. This starts with my advisor, Professor Yale N. Patt. Despite his accomplishments, Professor Patt maintains a contagious passion for both teaching and research. He's taught me how to learn, how to ask questions, how to attack problems, and tried to teach me how to share knowledge with others. I'm still not sure why he agreed to let me join HPS, but it's one of the pivotal moments of my life. It's an honor to be counted as a member of his research group, I thank him for his faith in me.

I've had the incredible opportunity of learning from Doug Carmean for the past five years. Technically, Doug has taught me how to pay attention to details and more importantly, how to listen to everybody and not allow preconceptions to color your opinion of what they're saying. Beyond work, Doug is one of the kindest people that I know and he has impacted my life in more ways than I can count. I thank him for putting up with my constant pestering and being so open with me when he had no reason to be.

I'd like to thank Professor Derek Chiou, Professor Mattan Erez, and Professor Don Fussell for serving on my committee. Professor Erez and Professor Chiou are instrumental to my success at UT. The wealth of knowledge that they've shared with me has given me the foundation to work in computer architecture and motivated me to want to work in this field. The university is lucky to have such amiable and brilliant individuals. 

Many of the research directions that I've worked on have come as a result of discussions with Professor Onur Mutlu. Professor Mutlu is an incredibly motivational, hardworking, and intelligent person. I'd like to thank him for teaching me how to never be satisfied with the work that I've done, how to continuously strive for more, and for pushing me to not give up when things didn't go my way. I'd like to thank him and Professor Moinuddin Qureshi for their advice, research discussions, and for always treating me like one of their own family.

I wouldn't have joined HPS without Eiman Ebrahimi. Eiman was the first person to teach me how to do research and how to strive towards writing high-quality papers. I'd like to thank him for his advice and support throughout my time at UT. I'd like to thank Carlos Villavieja for putting up with an obstinate young graduate student and showing him how to grow both as a person and a researcher. I'd also like to thank him for proof-reading this entire dissertation. I'd like to thank the entire HPS research group while I've been at UT, and in particular Khubaib for always being eager to talk about research, Jos\'{e} Joao for his guidance and maintaining our computing systems, Rustam Miftakhutdinov for the insane amount of work that he put into our simulation infrastructure, and Faruk Guvenilir for maintaining our computing systems after Jos\'{e} and for completing countless miscellaneous tasks without complaint.

Finally, I'd like to thank my friends: Will Diel, Curtis Hickmott, Zack Smith, Trevor Kilgannon, and David Cate for keeping me sane for over a decade now.

\vspace{2em}

\noindent%
Milad Hashemi\\
August 2016, Austin, TX

\end{preface}


\utabstract
\index{Abstract}%
\indent

This dissertation develops hardware that automatically reduces the effective latency of accessing memory in both single-core and multi-core systems. To accomplish this, the dissertation shows that all last level cache misses can be separated into two categories: dependent cache misses and independent cache misses. Independent cache misses have all of the source data that is required to generate the address of the memory access available on-chip, while dependent cache misses depend on data that is located off-chip. This dissertation proposes that dependent cache misses are accelerated by migrating the dependence chain that generates the address of the memory access to the memory controller for execution. Independent cache misses are accelerated using a new mode for runahead execution that only executes filtered dependence chains. With these mechanisms, this dissertation demonstrates a 62\% increase in performance and a 19\% decrease in effective memory access latency for a quad-core processor on a set of high memory intensity workloads.

\tableofcontents   

\listoftables      
\listoffigures     



\chapter{Introduction}
\label{chap:intro}
\setlength{\epigraphwidth}{0.41\textwidth}

\section{The Problem}
\label{sec:intro:problem}

The large latency disparity between performing computation at the core and accessing data from off-chip memory is a key impediment to system performance. This problem is known as the ``memory wall" \cite {wulf:1995, wilkes:2001} and is due to two factors. First, raw main memory access latency has remained roughly constant historically \cite{kim:2013}, with the row activation time ($t_{rc}$) decreasing by only 26\% from SDR-200 DRAM to DDR3-1333. Second, increasing levels of on-chip shared-resource contention in the multi-core era have further caused the \textit{effective} latency of accessing memory from on-chip to increase. Examples of this contention include: on-chip interconnect, shared cache, DRAM queue, and DRAM bank contention.  Due to these two factors, main memory accesses are a performance bottleneck, particularly for single threaded applications where the reorder buffer (ROB) of a core cannot hide long-latency operations with thread-level parallelism.

\begin{figure}
	\centering
	\includegraphics[width=5.5in]{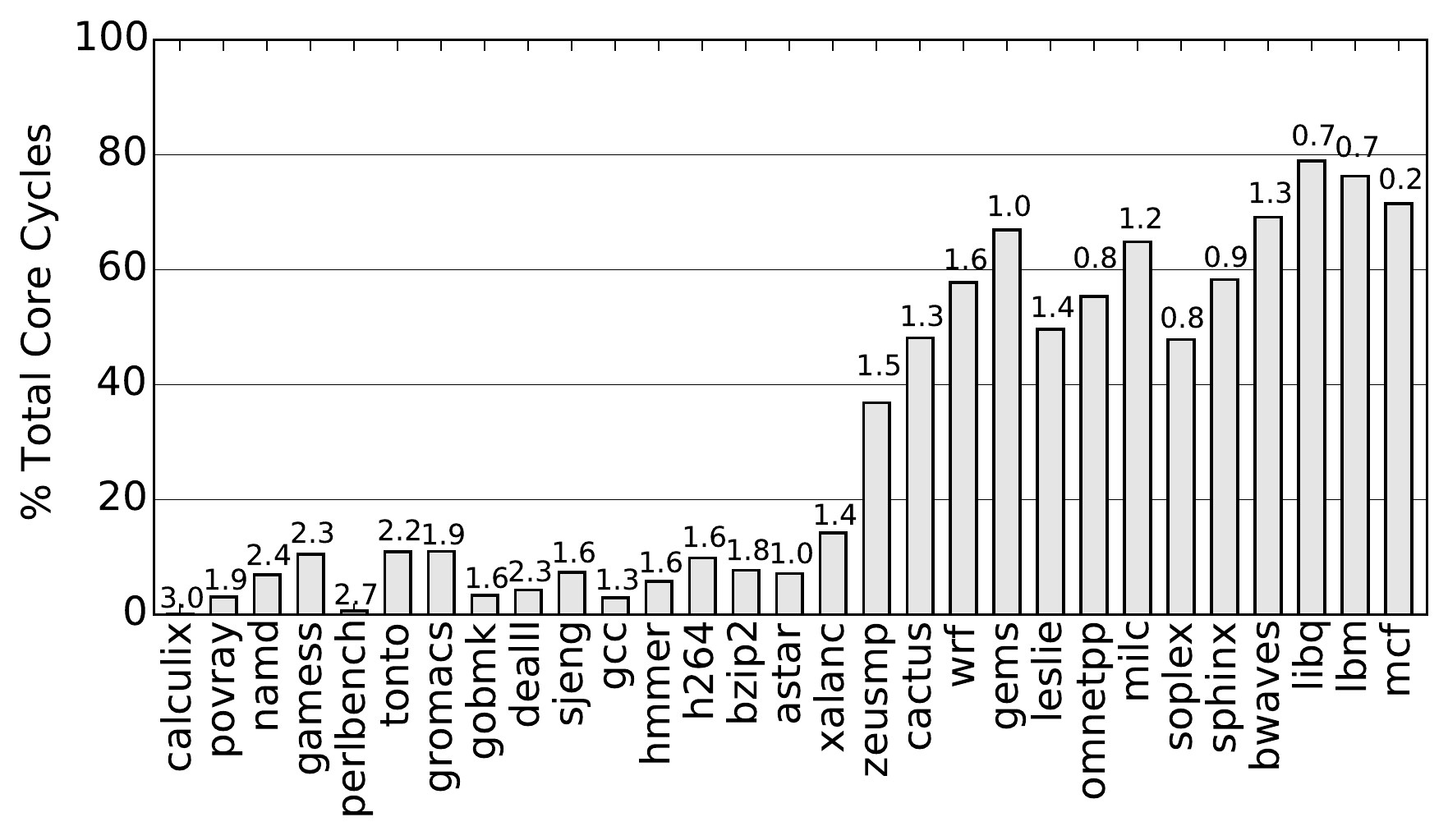}
	\caption{SPEC CPU2006 Stall Cycles}
	\label{fig:intro:stall}
\end{figure}

Figure \ref{fig:intro:stall} shows the percentage of total cycles that a 4-wide superscalar out-of-order processor with a 256-operation reorder buffer and 1MB of last level cache (LLC) is stalled and waiting for data from main memory across the \textit{SPEC CPU2006} benchmark suite. The applications are sorted from lowest to highest memory intensity and the average instructions per cycle (IPC) of each application is overlaid on top of each bar. Even with an out-of-order processor, the memory intensive applications to the right of \textit{zeusmp} in Figure \ref{fig:intro:stall} all have low IPC (generally under 1 instruction/cycle) and all spend over half of their total cycles executing the benchmark stalled waiting for data from main memory. In contrast, the non-memory intensive applications to the left of \textit{zeusmp} all spend under 20\% of total execution time stalled waiting for data from memory and have higher average IPC.  This dissertation focuses on accelerating memory intensive applications and the loads that lead to LLC misses which cause the ROB to fill.

\section{Independent vs. Dependent Cache Misses}

\begin{figure}
	\centering
	\includegraphics[width=5.0in]{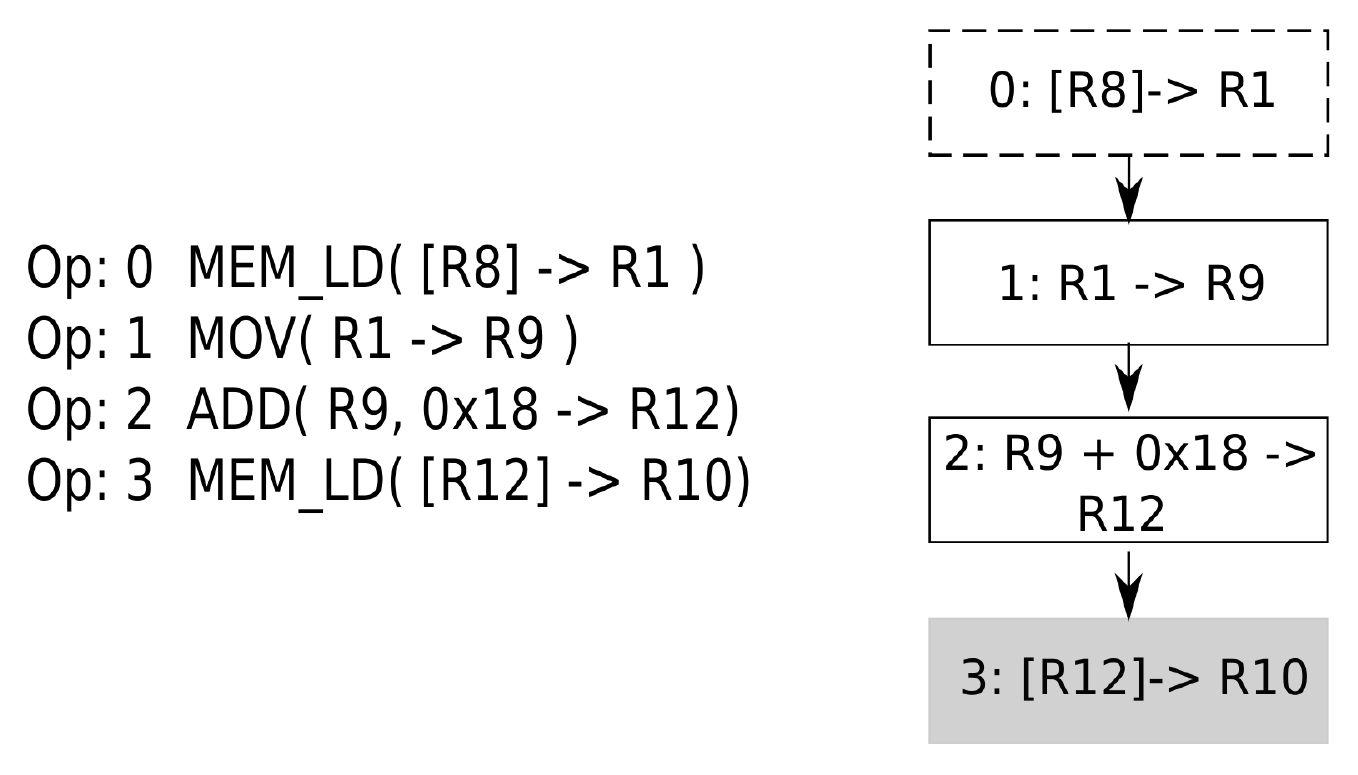}
	\caption{Dependence Chain}
	\label{fig:intro:mcfCode}
\end{figure}

Before any load instruction can access memory, it requires a memory address. This memory address is generated by a chain of earlier instructions or micro-operations (micro-ops) in the program. One example of an address generation chain is shown in Figure \ref{fig:intro:mcfCode}. A sequence of operations is shown on the left while the dataflow graph of the operations is shown on the right. Operation 0 is a load that uses the value in R8 to access memory and places the result in R1. Operation 1 moves the value in R1 to R9. Operation 2 adds 0x18 to R9 and places the result in R12. Finally, operation 3 uses R12 to access memory and places the result in R10. As R12 is the address that is used to access memory, the only operations that are required to complete before operation 3 can be executed are operations 0, 1, and 2. Therefore, I define the \textit{dependence chain} for operation 3 as consisting of operations 0, 1, and 2. 

If operation 3 results in an LLC miss, then operations 0, 1, and 2 are the dependence chain of a cache miss. I observe that all LLC misses can be split into two categories based on the source data required by their dependence chain:

\textit{Dependent Cache Misses:} Memory accesses that depend on source data that is not available on-chip. These operations cannot be executed by an out-of-order processor until source data from a prior, outstanding cache-miss returns to the core from main memory.

\textit{Independent Cache Misses:} Memory accesses that depend on source data that is available on-chip. The effective memory access latency of these operations cannot be hidden by an out-of-order processor because of the limited size of the processor's reorder buffer.

For Figure \ref{fig:intro:mcfCode}, if operation 0 is a cache hit, then operation 3 is a independent cache miss. All of the source-data that is required to generate R12 is available on-chip. However, if operation 0 is a cache miss, then operation 3 must wait to execute until operation 0 returns from memory and operations 1 and 2 execute. In this case operation 3 is a dependent cache miss.

\begin{figure}
	\centering
	\includegraphics[width=5.5in]{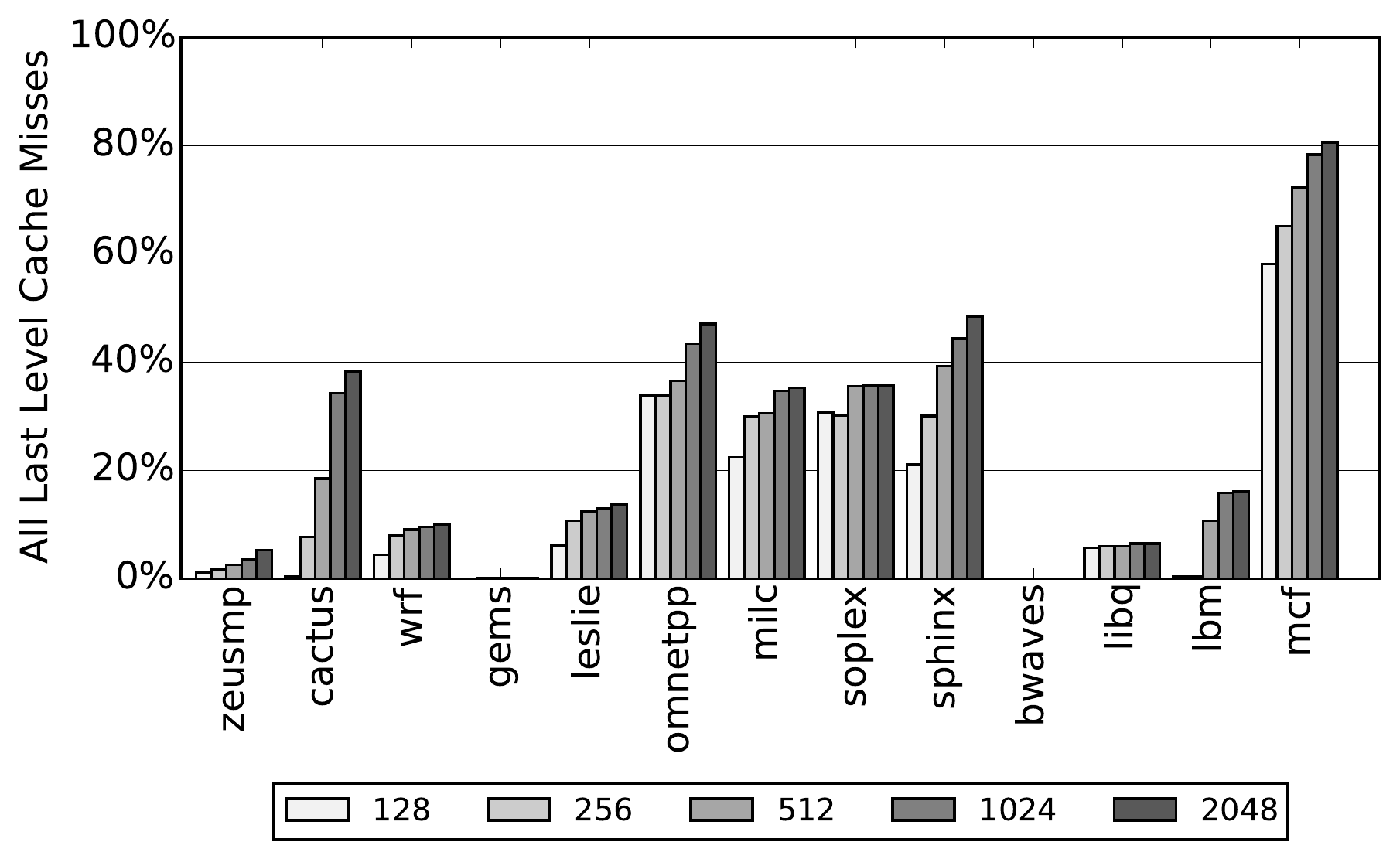}
	\caption{Fraction of all Cache Misses that are Dependent Cache Misses}
	\label{fig:intro:depMiss}
\end{figure}

Figure \ref{fig:intro:depMiss} shows the percent of all cache misses that are dependent cache misses for the memory intensive \textit{SPEC06} benchmarks. Since the number dependent cache misses is a function of the number of operations that are in-flight, ROB size is varied from 128 entries to 2048 entries, scaling support for the number of outstanding memory operations and memory bandwidth accordingly. The benchmarks with high dependent cache miss rates such as \textit{omnetpp}, \textit{milc}, \textit{soplex}, \textit{sphinx}, and \textit{mcf} all exhibit a high rate of dependent cache misses at even the smallest ROB size of 128 entries. This indicates that dependent cache misses are a property of application code, not hardware constraints. Figure \ref{fig:intro:depMiss} also shows that the fraction of all dependent cache misses grows as ROB size increases. Over the memory intensive benchmarks, \textit{mcf} has the highest rate of dependent cache misses. From Figure \ref{fig:intro:stall}, \textit{mcf} also is the most memory intensive application and has the lowest IPC across the entire benchmark suite. This highlights the negative impact that dependent cache misses have on processor performance. However, Figure \ref{fig:intro:depMiss} shows that for all applications besides \textit{mcf}, the majority of LLC misses are independent cache misses, not dependent cache misses. Accelerating both of these categories of LLC misses is critical to improving performance.

\section{Reducing Effective Memory Access Latency}

In this dissertation, I design specialized hardware to automatically reduce memory access latency for each of these two types of cache-misses in both single-core and multi-core systems. As dependent cache misses cannot be executed until data returns from main-memory, I propose dynamically identifying the dependence chain of a dependent cache miss at the core and migrating it closer to memory for execution at a compute capable, enhanced memory controller (EMC). I demonstrate that these dependence chains are short and show that this migration reduces the effective memory access latency of the subsequent dependent cache miss. 

Independent cache misses have all source data available on chip but are limited from issue by ROB size. Therefore, I revisit a prior technique for expanding the instruction window of out-of-order processors: runahead execution \cite{mut:sta03}. I identify that many of the operations that are executed in runahead are not relevant to producing the memory address of the cache miss. I propose a new hardware structure, the Runahead Buffer, that runs-ahead using only the filtered dependence chain that is required to generate cache misses. By executing fewer operations, this dissertation shows that the Runahead Buffer generates more cache misses per runahead interval when compared to traditional runahead and is more energy efficient.

Yet, while the Runahead Buffer is more effective than traditional runahead execution, I demonstrate that it is limited by the runahead paradigm. This dissertation shows that while runahead requests have very high accuracy, the Runahead Buffer is only active for a fraction of total execution time. This limits the impact that the Runahead Buffer has on reducing effective memory access latency. In this dissertation, I explore migrating the dependence chains that are used in the Runahead Buffer to the enhanced memory controller. This allows the dependence chain to execute far ahead of the program, creating a continuous prefetching effect. The result is a large reduction in effective memory access latency. I evaluate new co-ordinated dynamic throttling policies that increase performance when traditional prefetchers are added to the system. The final implementation of the EMC is a lightweight memory accelerator that reduces effective memory access latency for both independent and dependent cache misses.

\section{Thesis Statement} 

\begin{quote}
Processors can dynamically identify and accelerate the short code \linebreak segments that generate cache misses, decreasing effective memory access latency and thereby increasing single-thread performance.
\end{quote}

\section{Contributions}

This dissertation makes the following contributions:

\begin{itemize}
	\item This dissertation shows that there are two different kinds of cache misses: independent cache misses and dependent cache misses. This distinction is made on the basis of whether all source data for the cache miss is available on-chip or off-chip. By differentiating between independent and dependent cache misses, this thesis proposes dynamic hardware acceleration mechanisms for reducing effective memory access latency for each of these two types of cache misses.
	\item This dissertation observes that the dependence chains for independent cache misses are stable. That is, if a dependence chain has generated an independent cache miss, it is likely to generate more independent cache misses in the near future. In Chapter \ref{chap:raBuf}, this observation is exploited by the Runahead Buffer, a new low-overhead mode for runahead execution. The Runahead Buffer generates 57\% more memory level parallelism on average as traditional runahead execution. I show that a hybrid policy using both the Runahead Buffer and traditional runahead further increases performance, generating 82\% more memory level parallelism than traditional runahead execution alone.
	\item While the original Runahead Buffer algorithm has low complexity, this dissertation shows that it is not the optimal algorithm for picking a dependence chain to use during runahead. Chapter \ref{chap:scRaEMC} evaluates several different algorithms for Runahead Buffer chain generation and demonstrates that a more intelligent algorithm increases the performance gain of the Runahead Buffer from 11\% to 23\%.
	\item This dissertation identifies that a large component of the total effective memory access latency for dependent cache misses is a result of multi-core on-chip contention. I develop the hardware that is required to transparently migrate the dependent cache miss to a new compute capable memory controller in Chapter \ref{chap:EMC}. This enhanced memory controller (EMC) executes the dependence chain immediately when source data arrives from main memory. This is shown to result in a 20\% average reduction in effective memory access latency for dependent cache misses.
	\item This dissertation argues that runahead execution is limited by the length of each runahead interval. To solve this problem, mechanisms are developed in Chapter \ref{chap:scRaEMC} that offload Runahead Buffer dependence chains to the EMC for continuous runahead execution. This results in a 32\% average reduction in effective memory access latency and a 37\% performance increase.
	\item This dissertation shows that the final hardware mechanism, runahead at the EMC with dependent miss acceleration (RA-EMC+Dep) reduces effective memory access latency in a multi-core system by 19\% while increasing performance on a set of ten high-memory intensity workloads by 62\%. I demonstrate that this is a greater performance increase and effective memory access latency reduction than three state-of-the-art on-chip prefetchers. RA-EMC+Dep is the first combined mechanism that uses dependence chains to automatically accelerate both independent and dependent cache misses in a multi-core system. 
	
\end{itemize}

\section{Dissertation Organization}

Chapter \ref{chap:background} discusses prior work that is related to this dissertation. Chapter \ref{chap:raBuf} introduces the Runahead Buffer and explores the properties of independent cache misses in a single-core setting. Chapter \ref{chap:EMC} explores dependent cache misses and demonstrates the performance implications of migrating these operations to the EMC. In Chapter \ref{chap:scRaEMC}, I explore the optimal dependence chain to use during runahead at the EMC while Chapter \ref{chap:mcRaEMC} considers the multi-core policies that optimize runahead performance at the EMC. I conclude with Chapter \ref{chap:conclusion}.

\chapter{Related Work}
\label{chap:background}
\setlength{\epigraphwidth}{0.41\textwidth}

This dissertation is related to previous work on hardware mechanisms to reduce memory access latency in four general areas: on-chip prefetching, code pre-execution, computation near memory, and memory scheduling. This chapter describes the prior studies that this dissertation builds on.

\section{Research in Reducing Data Access Latency via Predicting Memory Access Addresses (Prefetching)}

Hardware prefetching can be generally divided into two categories: prefetchers that predict future addresses based on memory access patterns, and prefetching effects that are based on pre-execution of code-segments provided by (or dynamically generated for) the application. I discuss the first category here and the second in Section \ref{sec:spec}.

Prefetchers that uncover stream or stride patterns \cite{gin77, jou90, pal:kes94} require a small amount of hardware overhead and are commonly implemented in modern processors today \cite{haswell}. These prefetchers can significant reduce data access latency for predictable data access patterns, but suffer when requests are issued too early or too late. Additionally, stream/stride prefetchers do not handle complex access patterns well, leading to inaccurate prefetch requests that waste memory bandwidth and pollute the cache.

More advanced hardware prefetching techniques such as correlation prefetching \cite{cha:ree95,joseph:isca97,lai:fid01,somogyi:2006} aim to reduce average memory access latency for more unpredictable cache misses. These prefetchers work by maintaining large on-chip tables that correlate past cache miss addresses to future cache misses. The global-history buffer (GHB) \cite{nes:smith04} is a form of correlation prefetching that uses a two-level indexing scheme to reduce the need for large correlation tables. Some prefetching proposals use large off-chip storage to reduce the need for on-chip storage \cite{jain:2013, wenisch:2009}. These proposals incur the additional cost of transmitting meta-data over the memory bus. This dissertation focuses on evaluating on-chip mechanisms to reduce memory access latency.

Other hardware prefetching mechanisms specifically target the pointers that lead to cache misses. Roth and Sohi \cite{rot:soh99} use jump-pointers during the traversal of linked-data structures to create memory level parallelism. Roth et al. \cite{rot:mos98} identify stable dependence patterns between pointers, and store this information in a correlation table. Content-directed prefetching \cite{coo:jou02} does not require additional state to store pointers, but greedily prefetches by dereferencing values that could be memory addresses. This results in a large number of useless prefetches. Ebrahimi et al. \cite{ecdp09} developed mechanisms to throttle inaccurate content-directed prefetchers.

I show that not all cache miss addresses are easily predicted by prefetching (Chapter \ref{chap:EMC}), and the work on accelerating dependent cache misses in this dissertation targets addresses that are difficult to prefetch. My research on accelerating independent cache misses dynamically uses portions of the application's own code to prefetch. This is demonstrated to result in more accurate memory requests (Chapter \ref{chap:raBuf}). The proposed mechanisms for both independent and dependent cache miss acceleration are compared to three state-of-the-art on-chip prefetchers in the evaluation: a stream prefetcher, GHB prefetcher, and Markov correlation prefetcher.

\section{Research in Reducing Data Access Latency via Pre-Execution}
\label{sec:spec}

\textbf{Pre-Execution via Runahead Execution:} In runahead execution \cite{dun:mud97, mut:sta03, sri:raj04}, once the back-end of a processor is stalled due to a full reorder buffer, the state of the processor is checkpointed and the front-end continues to fetch operations. These operations are executed if source data is ready. Some implementations do not store runahead results \cite{mut:sta03}, while other similar proposals do \cite{sri:raj04}. The main goal is to generate additional memory-level parallelism and prefetch future cache misses.

The research in this dissertation on independent cache misses is an extension to runahead execution. Traditional runahead execution requires the front-end to always be on to fetch/decode instructions. I find that this is inefficient. Furthermore, traditional runahead issues all of these fetched instructions to the back-end of the processor for execution. I find that many of these operations are not relevant to the dependence chain of a cache miss (Chapter \ref{chap:raBuf}). I show that the core can generate more memory level parallelism by issuing only the \textit{filtered} dependence chain required to generate the cache miss to the back-end. This idea is expanded upon (Chapters \ref{chap:scRaEMC} and \ref{chap:mcRaEMC}) to allow the EMC to \textit{continuously} runahead at all times, not just when the core is stalled. To my knowledge this is the first proposal that dynamically allows runahead execution to continue when the main thread is active.

\noindent \textbf{Pre-Execution via Compiler/Hand Generated Code-Segments:} Many papers attempt to prefetch by using compiler/hand-tuned portions of code to execute ahead of the demand access stream \cite{cha:sta99, balasubramonian2001, luk01, yan:leb00}. These helper threads can execute on special hardware or on a different core of a multi-core processor. Collins et al. \cite{col:tul01-spec} generate helper-threads with compiler analysis and require free hardware thread-contexts to execute them. Other work also constructs helper threads manually \cite{zil:soh01}. Kim and Yeung \cite{Kim:2002} discuss techniques for the static compiler to generate helper threads. Similar concepts are proposed in Inspector-Executor schemes \cite{saltz1991multiprocessors}, where the computation loop is preceded by an ``inspector" loop, which prefetches data. Dynamic compilation techniques have also been pursued \cite{Zhang:2007, Lu:2005}. Hand-generated helper threads have also been proposed to run on idle-cores of a multi-core processor \cite{Kamruzzaman:2011, Brown01}. These statically generated pre-execution proposals all are based on the idea of decoupling the memory access stream in an application from the execution stream. This high-level idea was initially proposed by Pleszkun \cite{pleszkun1983structured} and Smith \cite{Smith:1984}.

In contrast to these methods, I propose mechanisms that allow dynamic generation of dependence chains in this dissertation. These chains do not require resources like free hardware cores or free thread-contexts. I tailor the memory controller to contain the specialized functionality required to execute these dependence chains (Chapter \ref{chap:EMC}).

\noindent \textbf{Speculation via automatically generated ``Helper Threads":} Research towards automatically generated helper threads is limited. For a helper-thread to be effective it needs to execute ahead of the main-thread. In prior work, this is done by using a filtered version of the main-thread (so the helper-thread can run faster than the main-thread) where unimportant instructions have been removed. Three main works are related to this thesis. 

First, in Slipstream \cite{sundaramoorthy:2000} two processors are used to execute an application. The A-stream runs a filtered version of the application ahead of the R-stream. The A-stream can then communicate performance hints such as branch-directions or memory addresses for prefetching back to the R-stream, although a main focus for Slipstream is fault-tolerance. However, the a instructions that are removed in Slipstream are generally simple. Slipstream only removes “ineffectual writes” (stores that are never referenced, stores that do not modify the state of a location) and highly biased branches. Other work uses a similar two-processor architecture, but does not allow the A-stream to stall on cache misses \cite{Zhou:2005}.

Second, Collins et al. \cite{col:tul01} propose a dynamic scheme to automatically extract helper-threads from the back-end of a processor. To do so, they require large additional hardware structures, including a buffer that is twice the size of their reorder buffer. All retired operations are filtered through this buffer. Once the helper threads are generated, they must run on full SMT thread contexts. This requires the front-end to fetch and decode operations and the SMT thread contends with the main thread for resources. An 8-way SMT core is used in their evaluation.

Third, Annavaram et al. \cite{ann:pat01} add hardware to extract a dependent chain of operations that are likely to result in a cache miss from the front-end during decode. These operations are prioritized and execute on a separate back-end. This reduces the effects of pipeline contention on these operations, but limits runahead distance to operations that the processor has already fetched.

I propose a lightweight solution to dynamically create a dependence chain (Chapter \ref{chap:raBuf}) that does not require free hardware thread contexts and filters the program down to only the dependence chain required to create a cache miss. Unlike prior work, this dependence chain is speculatively executed as if it was in a loop with minimal control overhead. Chapter \ref{chap:scRaEMC} demonstrates that this technique is limited by the length of each runahead interval and proposes using the EMC to speculatively execute dependence chains. To my knowledge this is the first work to study general dynamically generated ``helper threads" in a multi-core setting.

\section{Research in Reducing Data Access Latency via Computation Near Memory}

\textbf{Logic and memory fabricated on the same process:} Prior work has proposed performing computation inside the logic layer of 3D-stacked DRAM \cite{AhnPIM:2015, Zhang:2014}, but none has specifically targeted accelerating dependent cache misses. Both EXECUBE \cite{kogge:1994} and iRAM \cite{patterson:1997} recognize that placing compute next to memory would maximize the available memory bandwidth for computation. This proposal has been recently revisited with Micron's 3D-stacked Hybrid Memory Cube (HMC) \cite{pawlowski2011hybrid, dlugosch2014efficient}. Ahn et al. \cite{AhnGraph:2015} propose performing graph processing in an interconnected network of HMCs by changing the programming model and architecture, forfeiting cache coherence and virtual memory mechanisms. Alexander et al. \cite{alexander:hpca96} and Solihin et al. \cite{sol:lee02} propose co-locating large correlation prefetching tables at memory and using memory-side logic to decide which data elements to prefetch on-chip.

These proposals generally do not split computation between on-chip and off-chip compute engines due to the cost of data-coherence across the DRAM bus. I argue that the latency constraints of the memory bus are relatively small compared to DRAM access latency. Therefore, locating computation at the first point where data enters the chip, the memory controller, is an attractive and unexplored research direction. 

\noindent \textbf{Migrating computation closer to data:} Prior work has proposed atomically combining arithmetic with loads to shared data \cite{Gottlieb:1982} as well as migrating general purpose computation closer to the on-chip caches where data is resident \cite{michaud:hpca10, khan:csailtr10}. I use migration to reduce main-memory access latency, not cache access latency.

\section{Research in Reducing Data Access Latency via Memory Scheduling}

The order in which memory requests are serviced has a large impact on the latency of a memory request, due to DRAM row-buffer/bank contention. Prior work has researched algorithms to optimize row-buffer hit rate and data to bank mappings \cite{Carter:1999, mut:mos08, kim:2012, lee:2008}. This dissertation is orthogonal to memory scheduling. I use an advanced memory scheduler \cite{mut:mos08} throughout this dissertation as the baseline. 

\chapter{The Runahead Buffer}
\label{chap:raBuf}
\setlength{\epigraphwidth}{0.41\textwidth}

\section{Introduction}
\label{sec:raBuf:Intro}

Figure \ref{fig:intro:depMiss} showed that most last level cache (LLC) misses in an application have all of the source data that is necessary to generate the address that results in the LLC miss available on chip. I define this category of LLC-misses as independent cache misses. In this chapter, I propose an energy efficient mechanism to reduce effective memory access latency for independent cache misses. This mechanism, the runahead buffer, is based on runahead execution for out-of-order processors \cite{mut:sta03} \footnote{An earlier version of this chapter was published as: Milad Hashemi and Yale Patt. Filtered Runahead Execution with a Runahead Buffer. In \textit{MICRO}, 2015. I developed the initial idea and conducted the simulator design and evaluation for this work.}.

In runahead, once a core is stalled and waiting for memory, the processor's architectural state is checkpointed and the front-end continues to fetch and execute instructions. This creates a prefetching effect by pre-executing future load instructions. The processor is able to use the application's own code to uncover additional cache misses when it would otherwise be stalled, thereby reducing the effective memory access latency of the subsequent demand request. Runahead targets generating cache misses that have source data available on-chip but cannot be issued by the core due to limitations on the size of the reorder buffer. However, runahead execution requires the front-end to remain on when the core would be otherwise stalled. As front-end power consumption can reach 40\% of total core power \cite{tegra4}, this can result in a significant energy overhead. 

In this Chapter, I show that most of the dependence chains that lead to cache misses in runahead execution are repetitive (Section \ref{sec:raBuf:Back}). I then propose dynamically identifying these chains and using them to run ahead with a new structure called a runahead buffer (Section \ref{sec:raBuf:mechanism}). This results in two benefits. First, by targeting only the filtered dependence chain, the runahead buffer frequently generates more MLP than traditional runahead by running further ahead. Second, by clock-gating the front-end during runahead, the runahead buffer incurs a much lower energy cost than traditional runahead \cite{mut:kim05}.

\section{Background}
\label{sec:raBuf:Back}

The majority of all cache misses are independent cache misses that have all source data available on-chip. Yet, two main factors prevent an out-of-order processor from issuing these cache misses early enough to hide the effective memory access latency of the operation. The first factor is the limited resources of an out-of-order processor. An out-of-order core can only issue operations up to the size of its reorder buffer. Once this buffer is full, generally due to a long-latency memory access, the core can not issue additional operations that may result in a cache miss. The second factor is branch prediction.  Assuming that limited resources are not an issue, the out-of-order processor would have to speculate on the sequence of instructions that generates the cache misses. However, prior work has shown that even wrong-path memory requests are generally beneficial for performance \cite{mut:kim04}.

Runahead execution for out-of-order processors \cite{mut:sta03} is one solution to the first factor, the limited resources of an out-of-order processor. Runahead is a dynamic hardware mechanism that effectively expands the reorder buffer. Once the retirement of instructions is stalled by a long-latency memory access, the processor takes several steps. 

First, architectural state, along with the branch history register and return address stack, are checkpointed. Second, the result of the memory operation that caused the stall is marked as poisoned in the physical register file. Once this has occurred, the processor begins the runahead interval and continues fetching and executing instructions with the goal of generating additional cache misses.

Any operation that uses poisoned source data propagates the poison flag to its destination register. Store operations cannot allow data to become globally observable, as runahead execution is speculative. Therefore, a special runahead cache is maintained to hold the results of stores and forward this data to runahead loads. While runahead execution allows the core to generate additional MLP, it has the downside of requiring the front-end to be on and remain active when the core would be otherwise stalled, using energy. This trade-off is examined in Section \ref{sec:raBuf:Obs}.

\section{Runahead Observations}
\label{sec:raBuf:Obs}

\begin{figure}
	\centering
	\includegraphics[width=\columnwidth]{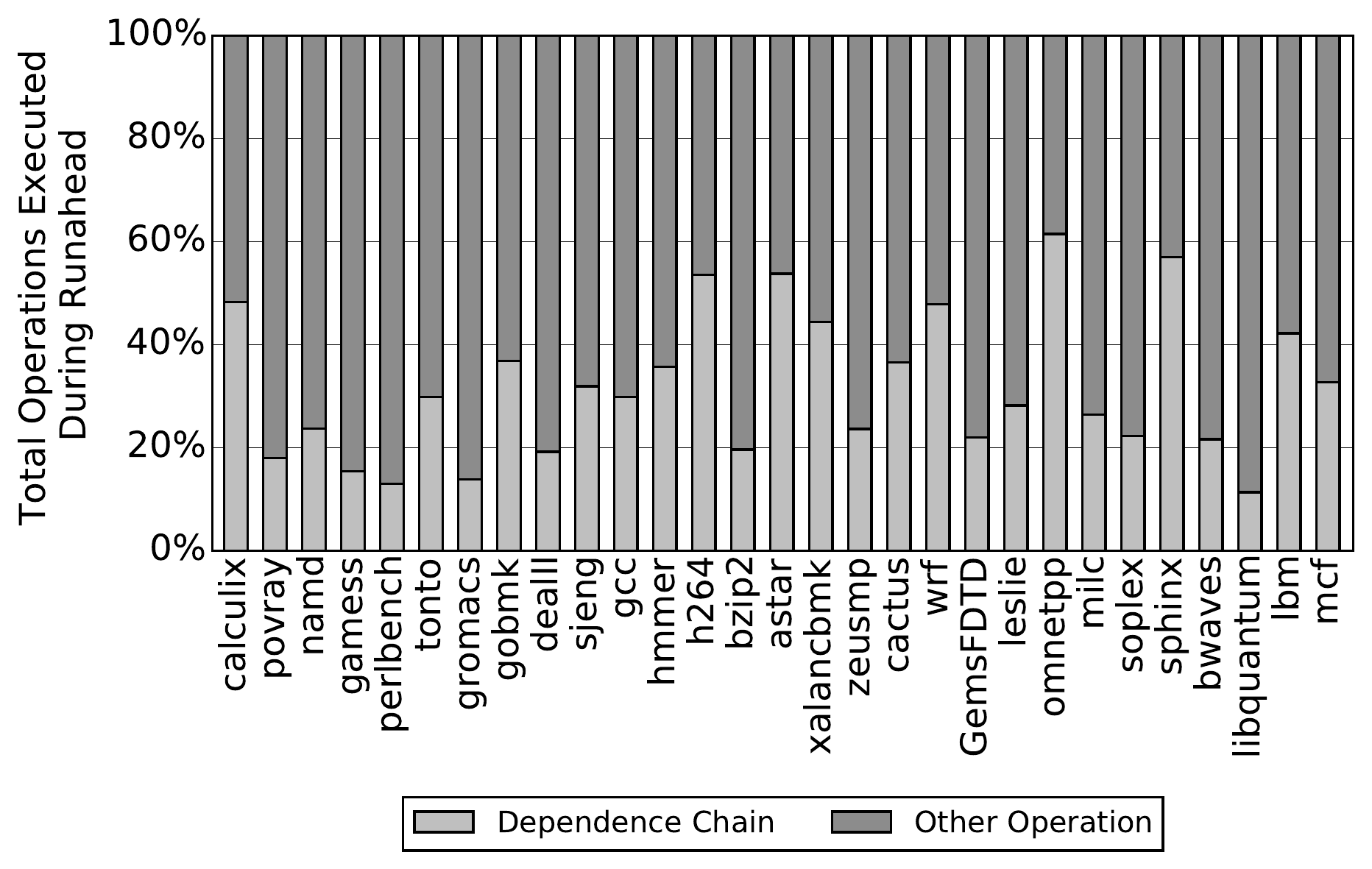}
	\caption{Breakdown of Operations Executed during Traditional Runahead}
	\label{fig:RAB:raRat}
	\vspace{.1in}
	\centering
	\includegraphics[width=\columnwidth]{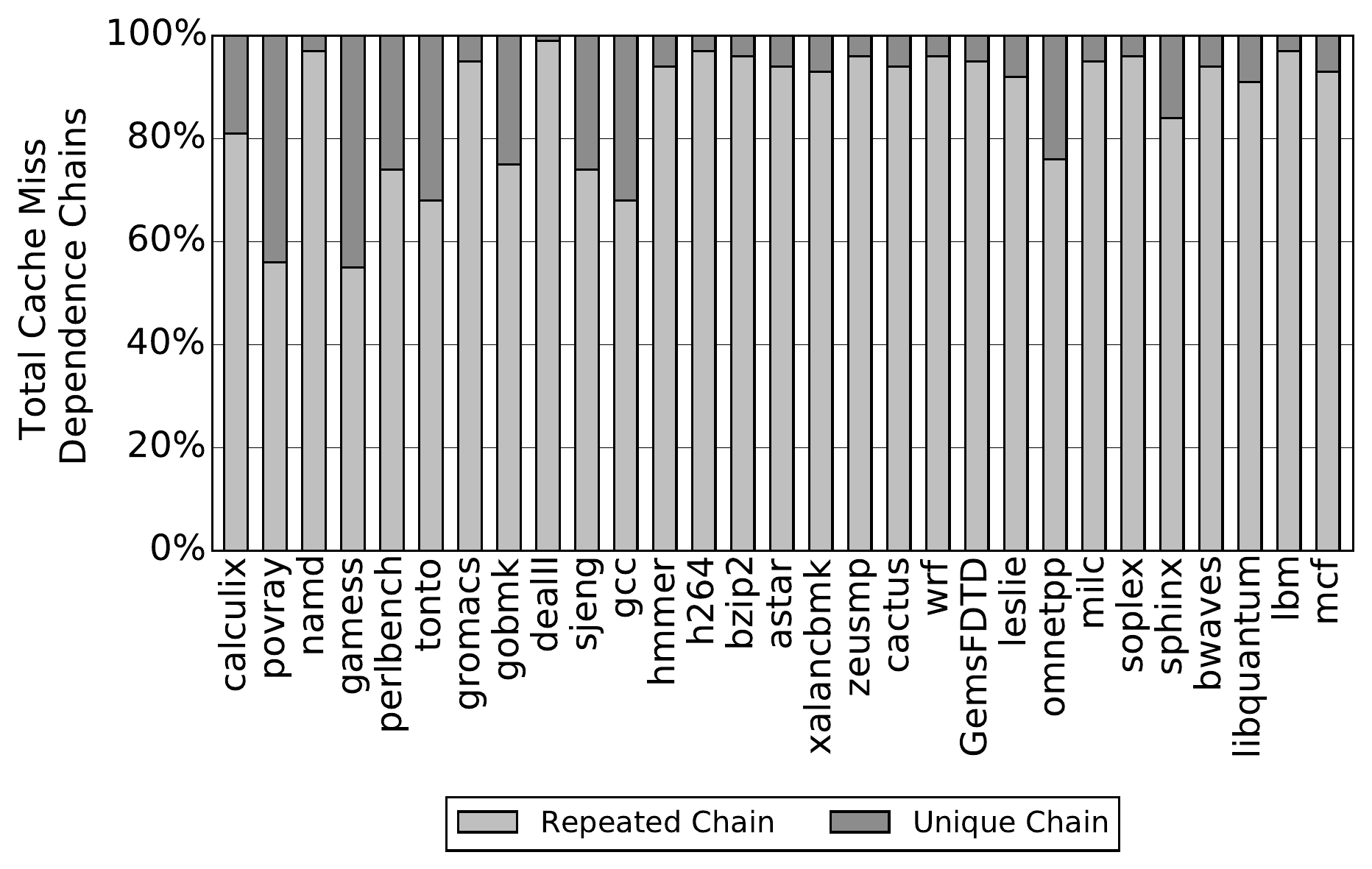}
	\caption{Unique vs. Repeated Dependence Chains}
	\label{fig:RAB:raChain}
\end{figure}

To uncover new cache misses, traditional runahead issues all of the operations that are fetched by the front-end to the back-end of the processor. Many of these operations are not relevant to calculating the address necessary for a subsequent cache miss. The operations required to execute a cache miss are encapsulated in the dependence chain of the miss, as shown in Figure \ref{fig:intro:mcfCode}. These are the only operations that are necessary to generate the memory address that causes the cache miss. Figure \ref{fig:RAB:raRat} compares the total number of operations executed in runahead to the number of operations that are actually in a dependence chain that is required to generate a cache miss. The \textit{SPEC06} benchmarks are sorted from lowest to highest memory intensity.

As Figure \ref{fig:RAB:raRat} shows, in most applications only a small fraction of the executed instructions are necessary to uncover an LLC miss. For example, in \textit{mcf} only 36 \% of the instructions executed in runahead are necessary to cause a new cache miss. Ideally, runahead would only fetch and execute these required instructions, executing other operations is a waste of energy. 

To observe how often these dynamic dependence chains vary, during each runahead interval, I trace the dependence chain for each generated cache miss. This chain is compared to all of the other dependence chains for cache misses generated during that particular runahead interval. Figure \ref{fig:RAB:raChain} shows how often each dependence chain is unique, i.e. how often a dependence chain has not been seen before in the current runahead interval.

As Figure \ref{fig:RAB:raChain} demonstrates, most dependence chains are repeated, not unique, in a given runahead interval. This means that if an operation with a given dependence chain generates a cache miss it is highly likely that a different dynamic instance of that instruction with the same dependence chain will generate another cache miss in the same interval. This is particularly true for the memory intensive applications on the right side of Figure \ref{fig:RAB:raChain}.

\begin{figure}
	\centering
	\includegraphics[width=\columnwidth]{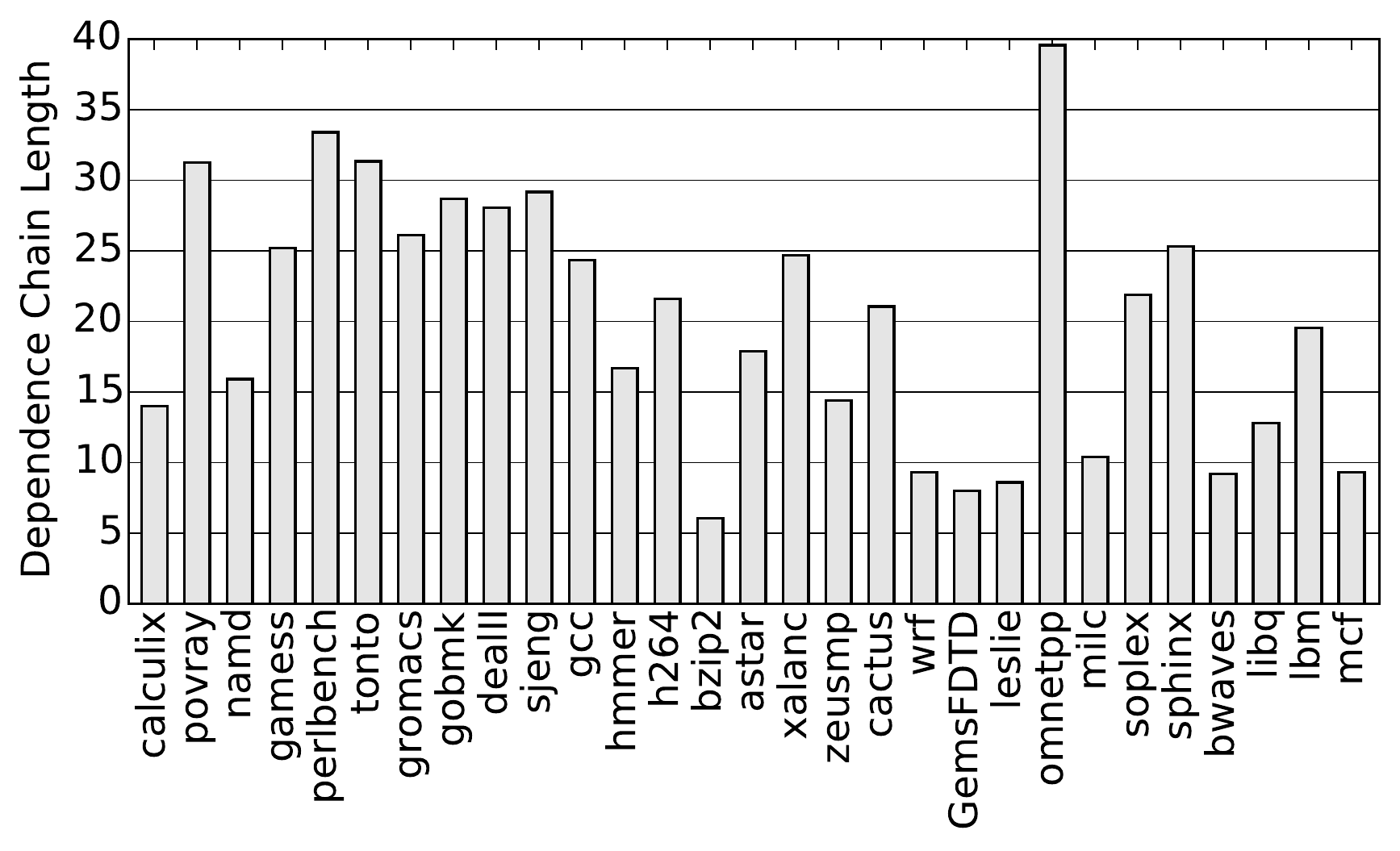}
	\caption{Average Length of a Runahead Miss Dependence Chain }
	\label{fig:RAB:raLength}
\end{figure}

Each of these dependence chains are on average reasonably short. Figure \ref{fig:RAB:raLength} lists the average length of the dependence chains for the cache misses generated during runahead in micro-operations (uops).

With the exception of \textit{omnetpp}, all of the memory intensive applications in Figure \ref{fig:RAB:raLength} have an average dependence chain length of under 32 uops. Several benchmarks including \textit{mcf, libquantum, bwaves,} and \textit{soplex}, have average dependence chain length of under 20 operations. Considering that the dependence chains that lead to cache misses during runahead are short and repetitive, I propose dynamically identifying these chains from the reorder buffer when the core is stalled. Once the chain is determined, the core can runahead by executing operations from this dependence chain. To accomplish this, the chain is placed in a runahead buffer, similar to a loop buffer \cite{cray1}. As the dependence chain is made up of decoded uops, the runahead buffer is able to feed these decoded ops directly into the back-end. Section \ref{sec:raBuf:mechanism} discusses how the chains are identified and the hardware structures required to support the runahead buffer.

\section{Mechanism}
\label{sec:raBuf:mechanism}

\subsection{Hardware Modifications} 
\label{sec:raBuf:pipe}

\begin{figure}[h]
	\centering
	\includegraphics[width=\columnwidth]{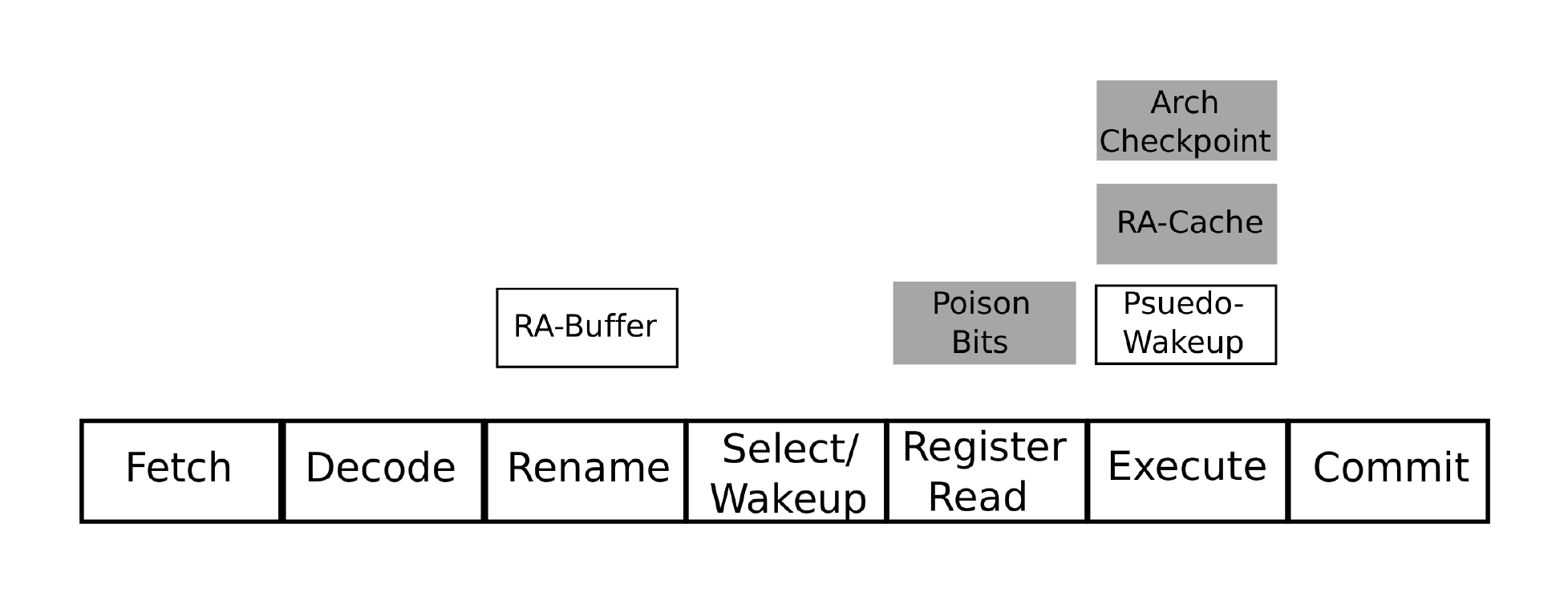}
	\caption{The Runahead Buffer Pipeline}
	\label{fig:RAB:raPipe}
\end{figure}

To support the runahead buffer, small modifications are required to the traditional runahead scheme. A high-level view of a traditional out-of-order processor is shown in Figure \ref{fig:RAB:raPipe}. The front-end includes the fetch and decode stages of the pipeline. The back-end consists of the rename, select/wakeup, register read, execute and commit stages. To support traditional runahead execution, the shaded modifications are required. The physical register file must include poison bits so that poisoned source and destination operands can be marked. This is denoted in the register read stage. Additionally, the pipeline must support new hardware paths to checkpoint architectural state, so that normal execution can recommence when the blocking operation returns from memory, and a runahead cache (RA-Cache) for forwarding store data as in \cite{mut:sta03}. These two changes are listed in the execute stage. 

The runahead buffer requires two further modifications to the pipeline: the ability to dynamically generate dependence chains in the back-end and the runahead buffer, which holds the dependence chain itself. Additionally, a small dependence chain cache (Section \ref{sec:raBuf:raCache}) reduces how often chains are generated.

To generate and read filtered dependence chains out of the ROB, the runahead buffer uses a pseudo-wakeup process. This requires every decoded uop, PC, and destination register to be available in the ROB. Both the PC and destination register are already part of the ROB entry of an out-of-order processor. Destination register IDs are necessary to reclaim physical registers at retirement. Program counters are stored to support rolling back mispredicted branches and exceptions \cite{smith:95}. However, decoded uop information can be discarded upon instruction issue. We add 4-bytes per ROB entry to maintain this information until retirement.  

The runahead buffer itself is placed in the rename stage, as operations issued from the buffer are decoded but need to be renamed for out-of-order execution. Both architectural register IDs and physical register ids are used during the psuedo-wakeup process and runahead buffer execution. Physical register ids are used during the dependence chain generation process.  Architectural register ids are used by the renamer once the operations are issued from the runahead buffer into the back-end of the processor. 

The hardware required to conduct the backwards data-flow walk to generate a dependence chain depends on ROB implementation. There are two primary techniques for reorder buffer organization in modern out-of-order processors. The first technique, used in Intel's P6 microarchitecture, allocates destination registers and ROB entries together in a circular buffer \cite{hinton2001}. This ROB implementation allows for simple lookups as destination register IDs also point to ROB entries. The second technique is used in Intel's NetBurst microarchitecture: ROB entries and destination registers are allocated and maintained separately \cite{hinton2001}. This means that destination registers are not allocated sequentially in the ROB. This second implementation is what is modeled in the performance evaluation of this dissertation. Therefore, searching for a destination register in the ROB requires additional hardware.  I modify the ROB to include a content addressable memory (CAM) for the PC and destination register ID field. This hardware is used during the pseudo-wakeup process for generating dependence chains (Section \ref{sec:raBuf:wakeup}).

\subsection{Dependence Chain Generation} 
\label{sec:raBuf:wakeup}

\begin{figure}
	\centering
	\includegraphics[width=\columnwidth, height=2.2in]{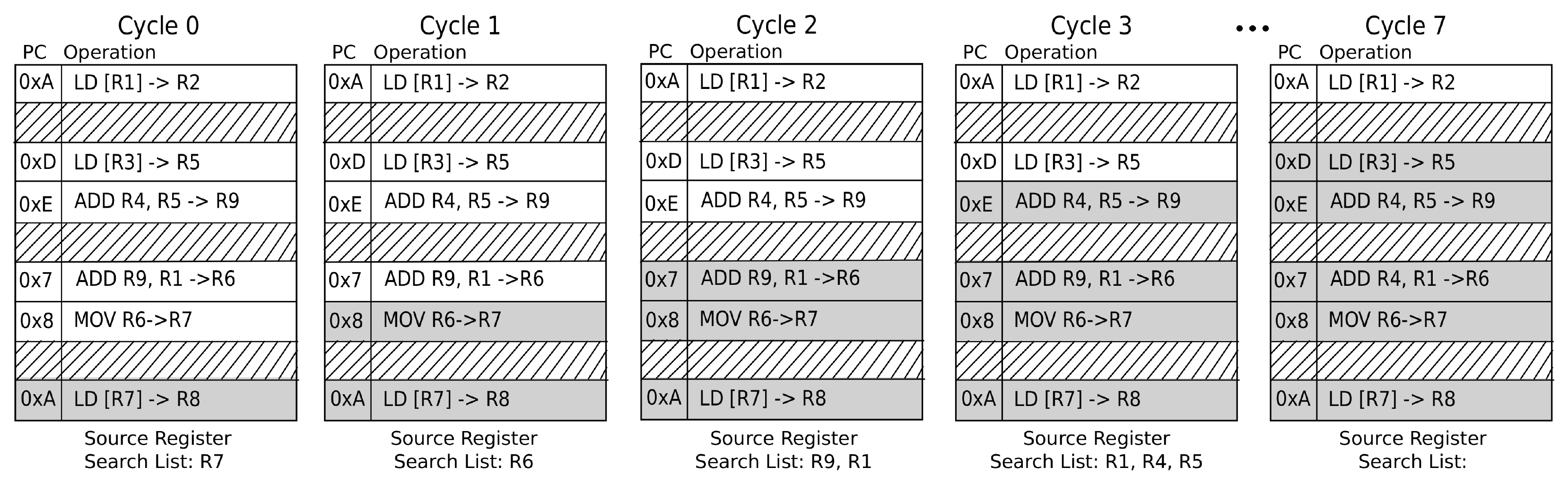}
	\caption{Dependence Chain Generation Process}
	\label{fig:RAB:traceGen}
\end{figure}

Once a miss has propagated to the top of the reorder buffer, as in the traditional runahead scheme, runahead execution begins and the state of the architectural register file is checkpointed. This also triggers creation of the dependence chain for the runahead buffer. Figure  \ref{fig:RAB:traceGen} shows an example of this process with code from \textit{mcf}. Control instructions are omitted in Figure \ref{fig:RAB:traceGen} and not included in the chain, as the ROB contains a branch-predicted stream of operations. The dependence chain does not need to be contiguous in the ROB, only relevant operations are shown and other operations are hashed out.

In Figure \ref{fig:RAB:traceGen}, the load stalling the ROB is at PC:0xA. This load cannot be used for dependence chain generation as its source operations have likely retired. Instead, I speculate that a different dynamic instance of that same load is present in the ROB. This is based on the data from Figure \ref{fig:RAB:raChain} that showed that if a dependence chain generates a cache miss, it is very likely to generate additional cache misses.

Therefore, in cycle 0, the ROB is searched for a different load with the same PC. If the operation is found with the CAM, it is included in the dependence chain (denoted by shading in Figure \ref{fig:RAB:traceGen}). Micro-ops that are included in the dependence chain are tracked using a bit-vector that includes one bit for every operation in the ROB. The source physical registers for the included operation (in this case R7) are maintained in a source register search list. These registers are used to generate the dependence chain.

During the next cycle, the destination registers in the ROB are searched using a CAM to find the uop that produces the source register for the miss. In this case, R7 is generated by a move from R6. In cycle 1, this is identified. R6 is added to the source register search list while the move operation is added to the dependence chain. 

This process continues in cycle 2. The operation that produces R6 is located in the reorder buffer, in this case an ADD, and its source registers are added to the search list (R9 and R1). Assuming that only one source register can be searched for per cycle, in cycle 3 R4 and R5 are added to the search list and the second ADD is included in the dependence chain. This process is continued until the source register search list is empty, or the maximum dependence chain length (32 uops, based on Figure \ref{fig:RAB:raLength}) is met. In Figure \ref{fig:RAB:traceGen}, this process takes 7 cycles to complete. In cycle 4 R1 finds no producers and in cycle 5 R4 finds no producing operations. In cycle 6, the load at address 0xD is included in the dependence chain, and in cycle 7 R3 finds no producers. 

As register spills and fills are common in x86, loads additionally check the store queue to see if the load value is dependent on a prior store. If so, the store is included in the dependence chain and its source registers are added to the source register search list. Note that as runahead is speculative, the dependence chains are not required to be exact. The goal is to generate a prefetching effect. While using the entire dependence chain is ideal, given the data from Figure \ref{fig:RAB:raLength}, I find that capping the chain at 32 uops is sufficient for most applications. This dependence chain generation algorithm is summarized in Algorithm \ref{alg:RAB:chainGen}.

Once the chain is generated, the operations are read out of the ROB with the superscalar width of the back-end (4 uops in our evaluation) and placed in the runahead buffer. Runahead execution then commences as in the traditional runahead policy.

\begin{algorithm}
	\caption{Runahead Buffer dependence chain \newline generation. \newline SRSL: Source Register Search List \newline ROB: Reorder Buffer \newline DC: Dependence Chain \newline MAXLENGTH: 32}
	\label{alg:RAB:chainGen}
	\begin{algorithmic}
		\IF{ROB Full}
		\STATE Get PC of op causing stall.
		\STATE Search ROB for another op with same PC.
		\IF{Matching PC found}
		\STATE Add oldest matching op to DC.
		\STATE Enqueue all source registers to SRSL.
		\WHILE {SRSL != EMPTY \AND \\ DC \textless MAXLENGTH}
		\STATE Dequeue register from SRSL.
		\STATE Search ROB for op that produces register.
		\IF {Matching op found}
		\STATE Add matching op to DC.
		\STATE Enqueue all source registers to SRSL.
		\IF{Matching op is load}
		\STATE Search store buffer for load address.
		\IF{Store buffer match}
		\STATE Add matching store to DC.
		\STATE Enqueue all source registers to SRSL.
		\ENDIF
		\ENDIF
		\ENDIF
		\ENDWHILE
		\STATE Fill runahead buffer with DC from ROB.
		\STATE Start runahead execution.
		\ENDIF
		\ENDIF
	\end{algorithmic}
\end{algorithm}

\subsection{Runahead Buffer Execution}
\label{sec:raBuf:raExec}

Execution with the runahead buffer is similar to traditional runahead execution except operations are read from the runahead buffer as opposed to the front-end. The runahead buffer is placed in the rename stage. Since the dependence chain is read out of the ROB, operations issued from the runahead buffer are pre-decoded but must be renamed to physical registers to support out-of-order execution. Operations are renamed from the runahead buffer at up to the superscalar width of the processor. Dependence chains in the buffer are treated as loops; once one iteration of the dependence chain is completed the buffer starts issuing from the beginning of the dependence chain once again. As in traditional runahead, stores write their data into a runahead cache (Table \ref{tab:RAB:systemConfig}) so that data may be forwarded to runahead loads. The runahead buffer continues issuing operations until the data of the load that is blocking the ROB returns. The core then exits runahead, as in \cite{mut:sta03}, and regular execution commences. 

\subsection{Dependence Chain Cache}
\label{sec:raBuf:raCache}

A cache to hold generated dependence chains can significantly reduce how often chains need to be generated prior to using the runahead buffer. I use a 2-entry cache that is indexed by the PC of the operation that is blocking the ROB. Dependence chains are inserted into this cache after they are filtered out of the ROB. The chain cache is checked for a hit before beginning the construction of a new dependence chain. Path-associativity is disallowed, so only one dependence chain may exist in the cache for every PC. As dependence chains can vary between dynamic instances of a given static load, I find that it is important for this cache to remain small. This allows old dependence chains to age out of the cache. Note that chain cache hits do not necessarily match the exact dependence chains that would be generated from the reorder buffer, this is explored further in Section \ref{sec:raBuf:results}

\subsection{Runahead Buffer Hybrid Policies}
\label{sec:raBuf:raHybrid}

Algorithm \ref{alg:RAB:chainGen} describes the steps that are necessary to generate a dependence chain for the runahead buffer. In addition to this algorithm, I propose a hybrid policy that uses traditional runahead when it is best and the runahead buffer with the chain cache otherwise. For this policy, if one of two events occur during the chain generation process, the core begins traditional runahead execution instead of using the runahead buffer. These two events are: an operation with the same PC as the operation that is blocking the ROB is not found in the ROB, or the generated dependence chain is too long (more than 32 operations). 

\begin{figure}[h]
	\centering
	\includegraphics[width=3in, height=4.25in]{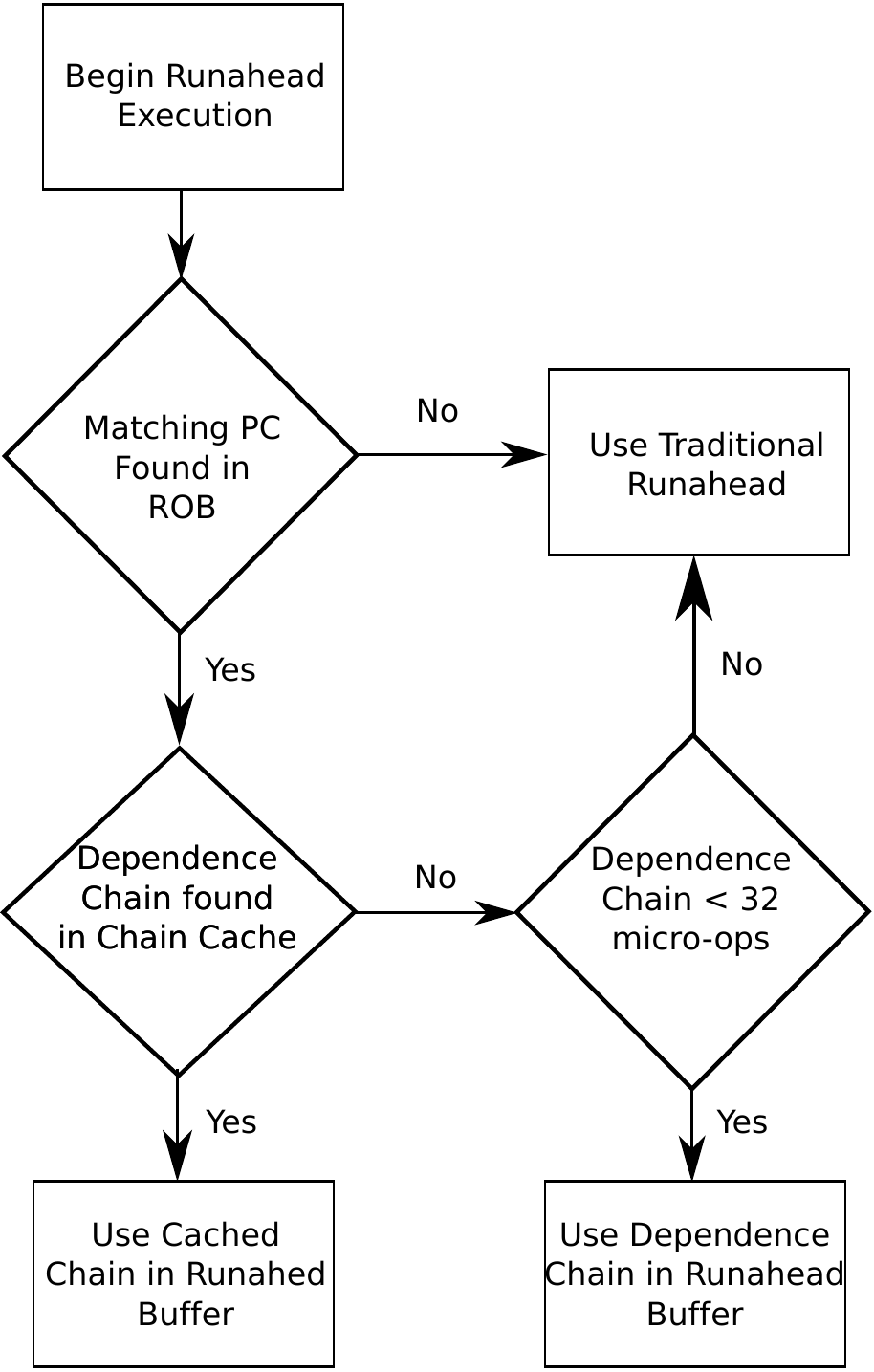}
	\caption{Flow Chart of the Hybrid Policy}
	\label{fig:RAB:hybridFlow}
\end{figure}

If an operation with the same PC is not found in the ROB, the policy predicts that the current PC will not generate additional cache misses in the near future. Therefore, traditional runahead will likely be more effective than the runahead buffer. Similarly, if the dependence chain is longer than 32 operations, the policy predicts that the dynamic instruction stream leading to the next cache miss is likely to differ from the dependence chain that will be obtained from the ROB (due to a large number of branches). Once again, this means that traditional runahead is preferable to the runahead buffer, as traditional runahead can dynamically predict the instruction stream using the core's branch predictor, while the runahead buffer executes a simple loop. This hybrid policy is summarized in Figure \ref{fig:RAB:hybridFlow} and evaluated in Section \ref{sec:raBuf:results}.

\subsection{Runahead Enhancements}
\label{sec:raBuf:raEnhance}

I find that the traditional runahead execution policy significantly increases the total dynamic instruction count. This is due to repetitive and unnecessary runahead intervals as discussed in \cite{mut:kim05}. Therefore, I implement the two hardware controlled policies from that paper. These policies limit how often the core can enter runahead mode. The first policy states that the core does not enter runahead mode unless the operation blocking the ROB was issued to memory less than a threshold number of instructions ago (250 instructions). The goal of this optimization is to ensure that the runahead interval is not too short. It is important for there to be enough time to enter runahead mode and generate MLP. The second policy states that the core does not enter runahead unless it has executed further than the last runahead interval. The goal of this optimization is to eliminate overlapping runahead intervals. This policy helps ensure that runahead does not waste energy uncovering the same cache miss over and over again. 

These policies are implemented in the runahead enhancements policy (evaluated in Section \ref{sec:raBuf:energy}) and the Hybrid policy (Section \ref{sec:raBuf:raHybrid}). As the runahead buffer does not use the front-end during runahead, I find that these enhancements do not noticeably effect energy consumption for the runahead buffer policies.

\section{Methodology}
\label{sec:raBuf:method}

The simulations for this dissertation are conducted with an execution driven, x86 cycle-level accurate simulator. The front-end of the simulator is based on Multi2Sim \cite{multi2sim}. The simulator faithfully models core microarchitectural details, the cache hierarchy, wrong-path execution, and includes a detailed non-uniform access latency DDR3 memory system \cite{rustam}.

The proposed mechanisms are evaluated on the \textit{SPEC06} benchmark suite. However, since the focus of this dissertation is accelerating memory intensive applications, my evaluation targets the medium and high memory intensive applications in the benchmark suite (Table \ref{tab:RAB:workloadClass}). The words application and benchmark are used interchangeably throughout the evaluation and refer to a single program in Table \ref{tab:RAB:workloadClass}. Workloads are collections of applications/benchmarks and are used in the multi-core evaluations.

\begin{table}
	\small
	\centering
	\begin{tabular}{|p{1.5in}|p{3.75in}|}
		\hline High Intensity \newline (MPKI \textgreater= 10) &  omnetpp, milc, soplex, sphinx3, bwaves, libquantum, lbm, mcf\\ 
		\hline Medium Intensity \newline (MPKI \textgreater=5) & zeusmp, cactusADM, wrf, GemsFDTD, leslie3d \\
		\hline Low Intensity \newline (MPKI \textless 5) & calculix, povray, namd, gamess, perlbench, tonto, gromacs, gobmk, dealII, sjeng, gcc, hmmer, h264ref, bzip2, astar, xalancbmk \\ 
		\hline
	\end{tabular} 
	\caption{SPEC06 Classification by Memory Intensity}
	\label{tab:RAB:workloadClass}
\end{table}

Each application is simulated from a representative SimPoint \cite{simpoint}. The simulation has two stages. First, the cache hierarchy and branch predictor warm up with a 50 million instruction warmup period. Second, the simulator conducts a 50 million instruction detailed cycle accurate simulation. Table \ref{tab:RAB:comp} lists the raw IPC and MPKI for this two-phase technique and a 100 million instruction detailed simulation for the \textit{SPEC06} benchmarks. There is an average IPC error of 1\% and an average MPKI error of 3\% between the 100 million instruction simulation and the 2-phase scheme used in this dissertation. 

Chip energy is modeled using McPAT 1.3 \cite{li:micro09} and computed using total execution time, ``runtime dynamic'' power, and ``total leakage power''. McPAT models clock-gating the front-end during idle cycles for all simulated systems.  The average runtime power for the memory intensive single core applications is listed in Table \ref{tab:RAB:power}. DRAM power is modeled using CACTI 6.5 \cite{Muralimanohar_cacti6.0}. A comparison between the CACTI DRAM power values and a MICRON 2Gb DDR3 module \cite{dram:micronPower} is shown in Table \ref{tab:RAB:micron} and raw average DRAM bandwidth consumption values are listed in Table \ref{tab:RAB:BW}.

\ignore{
		& calculix & povray & namd & gamess & perlbench & tonto  \\
		& gromacs & gobmk & dealII & sjeng & gcc & hmmer  \\
		& h264ref & bzip2 & astar & xalancbmk & zeusmp & cactus \\
		& wrf & gems & leslie & omnetpp & milc & soplex \\
		& sphinx & bwaves & libq & lbm & mcf & \\
}

\begin{table}
	\centering
	\begin{tabular}{|c|cccccc|}\hline
		\centering
		\multirow{10}{1.0in}{\centering 100M IPC} 
		& calculix & povray & namd & gamess & perlbench & tonto  \\
		& 2.98 & 1.85 &	2.37 &	2.29 &	2.67 &	2.22 \\
		& gromacs & gobmk & dealII & sjeng & gcc & hmmer  \\
		& 1.94 & 1.61 &	2.34 &	1.62 &	1.32 &	1.57 \\	
		& h264ref & bzip2 & astar & xalancbmk & zeusmp & cactus \\
		& 1.56 & 1.79 &	0.97 &	1.42 &	1.51 &	1.28 \\	
		& wrf & gems & leslie & omnetpp & milc & soplex \\
		& 1.62 & 1.01 &	1.35 &	0.77 &	1.22 &	0.83 \\	
		& sphinx & bwaves & libq & lbm & mcf & \\
		& 0.89 & 1.27 &	0.71 &	0.73 &	0.21 & \\
		\hline
		\multirow{10}{1.0in}{\centering 2-Phase IPC} 
		& calculix & povray & namd & gamess & perlbench & tonto  \\
		& 2.97 &	1.82 & 2.37	& 2.17	&    2.68 &	2.33 \\
		& gromacs & gobmk & dealII & sjeng & gcc & hmmer  \\
		& 2.01  &	1.56 &	2.32 &	1.62 &	1.23 &	1.50 \\
		& h264ref & bzip2 & astar & xalancbmk & zeusmp & cactus \\
		& 1.76 &	1.53 &	0.99 &	1.40 &	1.46 &	1.25 \\
		& wrf & gems & leslie & omnetpp & milc & soplex \\
		& 1.55 &	1.04 &	1.35 &	0.76 &	1.16 &	0.74\\
		& sphinx & bwaves & libq & lbm & mcf & \\
		& 0.88 &	1.50 &	0.80 &	0.73 &	0.19 & \\
		\hline
		\hline
		\multirow{10}{1.0in}{\centering 100M MPKI} 
		& calculix & povray & namd & gamess & perlbench & tonto  \\
		& 0.02 &	0.04 &	0.05 &	0.06 &	0.06 &	0.11 \\	
		& gromacs & gobmk & dealII & sjeng & gcc & hmmer  \\
		& 0.36 &	0.39 &	0.43 &	0.49 &	0.80 &	1.01\\
		& h264ref & bzip2 & astar & xalancbmk & zeusmp & cactus \\
		& 1.64 &	1.97 &	2.59 &	3.46 &	5.30 &	5.40\\
		& wrf & gems & leslie & omnetpp & milc & soplex \\
		& 5.54 &	10.42 &	10.96 &	12.29 &	13.81 &	16.05\\
		& sphinx & bwaves & libq & lbm & mcf & \\
		& 17.07 &	18.26 &	19.85 & 23.68 &	28.57 &\\
		\hline
		\multirow{10}{1.0in}{\centering 2-Phase MPKI} 
		& calculix & povray & namd & gamess & perlbench & tonto  \\
		& 0.03 &	0.04 &	0.03 &	0.04 &	0.05 &	0.09 \\		
		& gromacs & gobmk & dealII & sjeng & gcc & hmmer  \\
		& 0.31 &	0.42 &	0.40 &	0.53 &	0.70 &	0.94 \\		
		& h264ref & bzip2 & astar & xalancbmk & zeusmp & cactus \\
		& 2.06 &	2.09 &	2.69 &	3.89 &	5.34 &	5.53\\		
		& wrf & gems & leslie & omnetpp & milc & soplex \\
		& 5.68 &	10.29 &	12.02 &	12.24 &	13.81 &	14.21\\
		& sphinx & bwaves & libq & lbm & mcf & \\
		& 15.92 &	18.51 &	21.77 &	23.69 &	29.52 & \\			
		\hline
	\end{tabular}
	\caption{Simulation Comparison}
	\label{tab:RAB:comp}
\end{table}

\begin{table*}[h]
	\centering
	\begin{tabular}{|c|ccccccc|}\hline
		\centering
		\multirow{4}{1.0in}{\centering Runtime Power (W)} & zeusmp & cactus & wrf & gems & leslie & omnetpp & milc \\
		& 62.0 &	60.1 &	63.2 &	55.9 &	63.0 &	53.3 &	55.2  \\\cline{2-8}
		& soplex & sphinx & bwaves & libq & lbm & mcf & \\
		& 59.0 &	54.0 &	52.2 &	47.9 &	50.9 &	48.4 & \\\cline{2-8}
		\hline
	\end{tabular}
	\caption{Runtime Power Consumption (W)}
	\label{tab:RAB:power}
\end{table*}

\begin{table*}[h]
	\centering
	\begin{tabular}{|c|cccc|}\hline
		\centering
		\multirow{2}{1.5in}{CACTI Power (mW)} & Activate & Read & Write & Static \\
		& 122.9 &	31.2 &	32.5 &	110.1 \\\hline
		\multirow{2}{1.5in}{Micron Power (mW)} & Activate & Read & Write & Static \\
		& 249.4 &	77.1 &	44.6 &	66.8 \\
		\hline
	\end{tabular}
	\caption{DRAM Power Comparison}
	\label{tab:RAB:micron}
\end{table*}

\begin{table*}[h]
	\centering
	\begin{tabular}{|c|ccccccc|}\hline
		\centering
		\multirow{4}{1.0in}{\centering Average DRAM Bandwidth (GB/S)} & zeusmp & cactus & wrf & gems & leslie & omnetpp & milc \\
		& 1.6 &	1.5 &	1.9 &	2.9 &	3.5 &	2.8 &	4.3 \\\cline{2-8}
		& soplex & sphinx & bwaves & libq & lbm & mcf & \\
		&	3.8 &	3.6 &	4.8 &	4.9 &	6.1 &	3.0 & \\\cline{2-8}
		\hline
	\end{tabular}
	\caption{Average DRAM Bandwidth Consumption (GB/S)}
	\label{tab:RAB:BW}
\end{table*}

System details are listed in Table \ref{tab:RAB:systemConfig}. The core uses a 256 entry reorder buffer. The cache hierarchy contains a 32KB instruction cache and a 32KB data cache with 1MB of last level cache. Three different on-chip prefetchers are used in the evaluation. A stream prefetcher (based on the stream prefetcher in the IBM POWER4 \cite{ten:dod01}), a Markov prefetcher \cite{joseph:isca97}, and a global-history-buffer (GHB) based global delta correlation (G/DC) prefetcher \cite{nes:smith04}. Prior work has shown a GHB prefetcher to outperform a large number of other prefetchers \cite{microlib}. I find that the Markov prefetcher alone has a small impact on performance for most applications and therefore always use it with a stream prefetcher. This configuration always has higher performance than using just the Markov prefetcher.

The runahead buffer used in the evaluation can hold up to 32 micro-ops, this number was determined as best through sensitivity analysis (Section \ref{sec:raBuf:sens}). The dependence chain cache for the runahead buffer consists of two 32 micro-op entries, sensitivity to this number is also shown in Section \ref{sec:raBuf:sens}. Additional hardware requirements include a 32 byte bit vector to mark the operations in the ROB that are included in the dependence chain during chain generation, an eight element source register search list, and 4-bytes per ROB entry to store micro-ops. The total storage overhead for the runahead buffer system is listed in Table \ref{tab:RAB:rabHW}.

\begin{table}
	\small
	\centering
	\begin{tabular}{|p{2.2in}|p{3.0in}|}
		\hline
		\textbf{Component} & \textbf{Bytes} \\
		\hline
		\hline
		Runahead Buffer & 8 Bytes * 32 Entries = 256 Bytes \\
		\hline
		ROB Bit-Vector & 1 Bit * 256 Entries = 32 Bytes \\
		\hline
		New ROB Storage & 4 Bytes * 256 Entries = 1024 Bytes \\
		\hline
		Chain Cache & 8 Bytes * 64 Entries = 512 Bytes \\
		\hline
		Source Register Search List & 4 Bytes * 8 Entries = 64 Bytes \\
		\hline
		Total New Storage & 1888 Bytes \\
		\hline
	\end{tabular} 
	\caption{Additional Runahead Buffer Hardware Overhead}
	\label{tab:RAB:rabHW}
\end{table}

To enter runahead, both traditional runahead and the runahead buffer require checkpointing the current architectural state. This is modeled by copying the physical registers pointed to by the register alias table (RAT) to a checkpoint register file. This process occurs concurrently with dependence chain generation for the runahead buffer or before runahead can begin in the baseline runahead scheme. For dependence chain generation, a CAM is modeled for the destination register id field where up to two registers can be matched every cycle. 

Runahead buffer dependence chain generation is modeled by the following additional energy events. Before entering runahead, a single CAM on PCs of operations in the ROB is required to locate a matching load for dependence chain generation. Each source register included in the source register search list requires a CAM on the destination register ids of operations in the ROB to locate producing operations. Each load instruction included in the chain requires an additional CAM on the store queue to search for source data from prior stores. Each operation in the chain requires an additional ROB read when it is sent to the runahead buffer. The energy events corresponding to entering runahead are: register alias table (RAT) and physical register reads for each architectural register and a write into a checkpoint register file.

\begin{table}
	\small
	\centering
	\begin{tabular}{|p{1.2in}|p{4.5in}|}
		\hline Core & 4-Wide Issue, 192 Entry ROB, 92 Entry Reservation Station, Hybrid Branch Predictor, 3.2 GHz Clock Rate. \\ 
		\hline Runahead Buffer & 32-entry. Micro-op size: 8 Bytes. 256 Total Bytes.\\
		\hline Runahead Cache & 512 Byte, 4-way Set Associative, 8 Byte Cache Lines. \\
		\hline Chain Cache & 2-entries, Fully Associative, 512 Total Bytes. \\
		\hline L1 Caches & 32 KB I-Cache, 32 KB D-Cache, 64 Byte Cache Lines, 2 Ports, 3 Cycle Latency, 8-way Set Associative, Write-through. \\ 
		\hline \hline Last Level Cache &  1MB, 8-way Set Associative, 64 Byte Cache Lines, 18-cycle Latency, Write-back, Inclusive. \\ 
		\hline Memory \linebreak Controller & 64 Entry Memory Queue, Priority scheduling. Priority order: row hit, demand (instruction fetch or data load), oldest.  \\ 
		\hline Prefetcher & Stream \cite{ten:dod01}: 32 Streams, Distance 32. Markov: 1MB Correlation Table, 4 addresses per entry. GHB G/DC: 1k Entry Buffer, 12KB total size. All configurations: FDP \cite{fdp07}, Dynamic Degree: 1-32, prefetch into Last Level Cache. \\  
		\hline DRAM & DDR3\cite{dram:micron}, 1 Rank of 8 Banks/Channel, 2 Channels, 8KB Row-Size, CAS 13.75ns.  CAS = $t_{RP}$ = $t_{RCD}$ = CL. Other modeled DDR3 constraints: BL, CWL, $t_{RC, RAS, RTP, CCD, RRD, FAW, WTR, WR}$. 800 MHz Bus, Width: 8 B. \\
		\hline
	\end{tabular} 
	\caption{System Configuration}
	\label{tab:RAB:systemConfig}
\end{table}

\section{Results}
\label{sec:raBuf:results}

Instructions per cycle (IPC) is used as the performance metric for the single core evaluation. During the performance evaluation I compare the runahead buffer to performance optimized runahead (without the enhancements discussed in Section \ref{sec:raBuf:raEnhance}) as these enhancements negatively impact performance. During the energy evaluation, the runahead buffer is compared to energy optimized runahead which uses these enhancements. I begin by evaluating the runahead buffer without prefetching in Section \ref{sec:raBuf:perf} and then with prefetching in Section \ref{sec:raBuf:Pref}. 

\subsection{Performance Results}
\label{sec:raBuf:perf}

\begin{figure}
	\centering
	\includegraphics[width=\columnwidth]{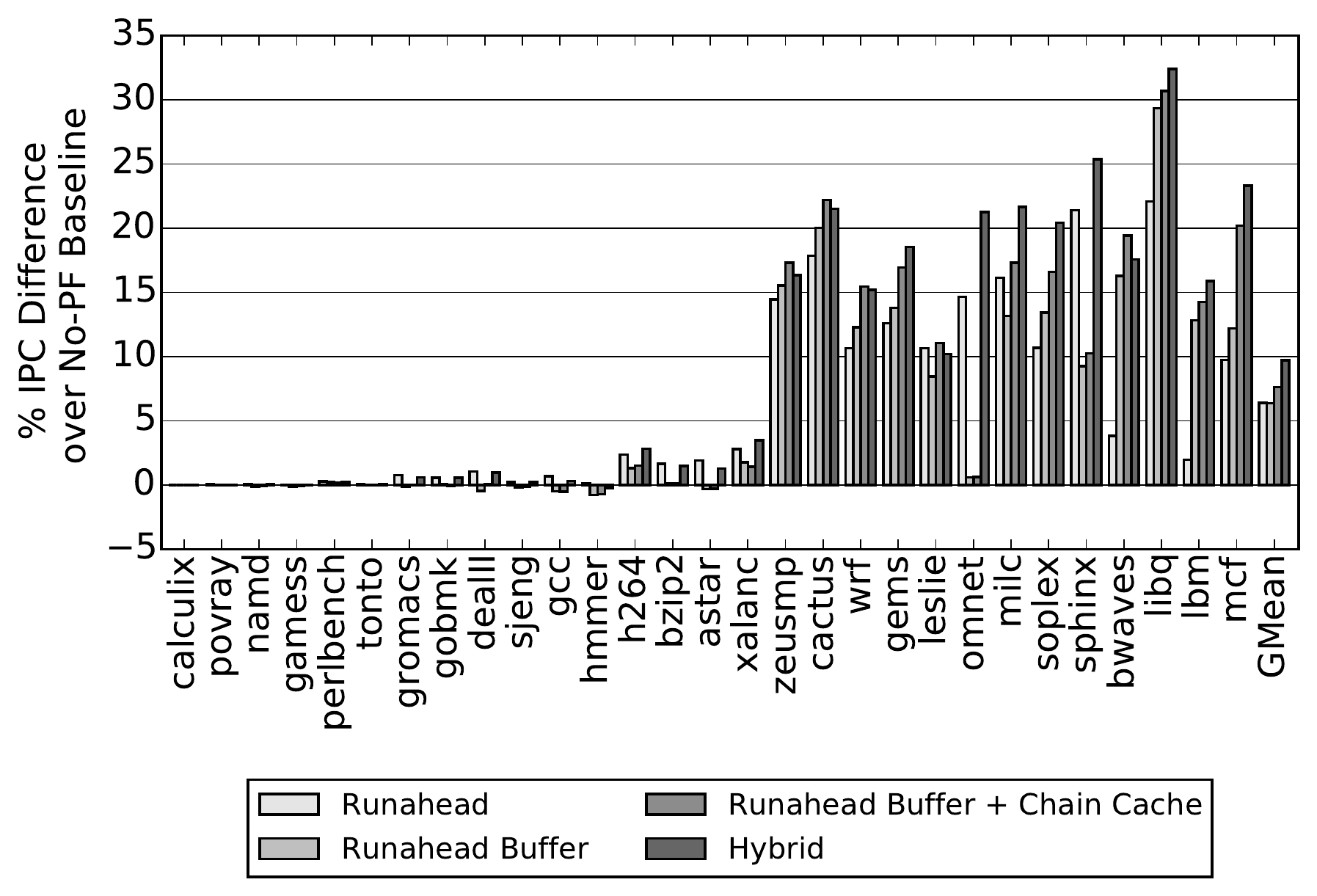}
	\caption{Runahead Performance Normalized to a No-Prefetching System}
	\label{fig:RAB:perf}
\end{figure}

Figure \ref{fig:RAB:perf} shows the results of our experiments on the \textit{SPEC06} benchmark suite. Considering only the low memory intensity applications in Table \ref{tab:RAB:workloadClass}, we observe an average 0.8\% speedup with traditional runahead. These benchmarks are not memory-limited and the techniques that I am evaluating have little to no-effect on performance. I therefore concentrate and the medium and high memory intensity benchmarks for this evaluation.

Figure \ref{fig:RAB:perf} shows the results of evaluating four different systems against a no-prefetching baseline. The ``Runahead'' system utilizes traditional out-of-order runahead execution. The ``Runahead Buffer'' system utilizes our proposed mechanism and does not have the ability to traditionally runahead. The ``Runahead Buffer + Chain Cache'' is the Runahead Buffer system but with an added cache that stores up to two, 32-operation dependence chains. The final system uses a ``Hybrid'' policy that combines the Runahead Buffer + Chain Cache system with traditional Runahead.

Considering only the medium and high memory intensity benchmarks, runahead results in performance improvements of 10.9\%, 11.0\%, 13.1\% and 19.3\% with traditional Runahead, the Runahead Buffer, Runahead Buffer + Chain Cache and Hybrid policy systems respectively. Traditional runahead performs well on \textit{omnetpp} and \textit{sphinx}, two benchmarks with longer average dependence chain lengths in Figure \ref{fig:RAB:raLength}. The runahead buffer does particularly well on \textit{mcf}, an application with short dependence chains, as well as \textit{lbm} and \textit{soplex}, which have longer average dependence chains but a large number of unnecessary operations executed during traditional runahead (Figure \ref{fig:RAB:raRat}). 

\begin{figure}
	\centering
	\includegraphics[width=\columnwidth]{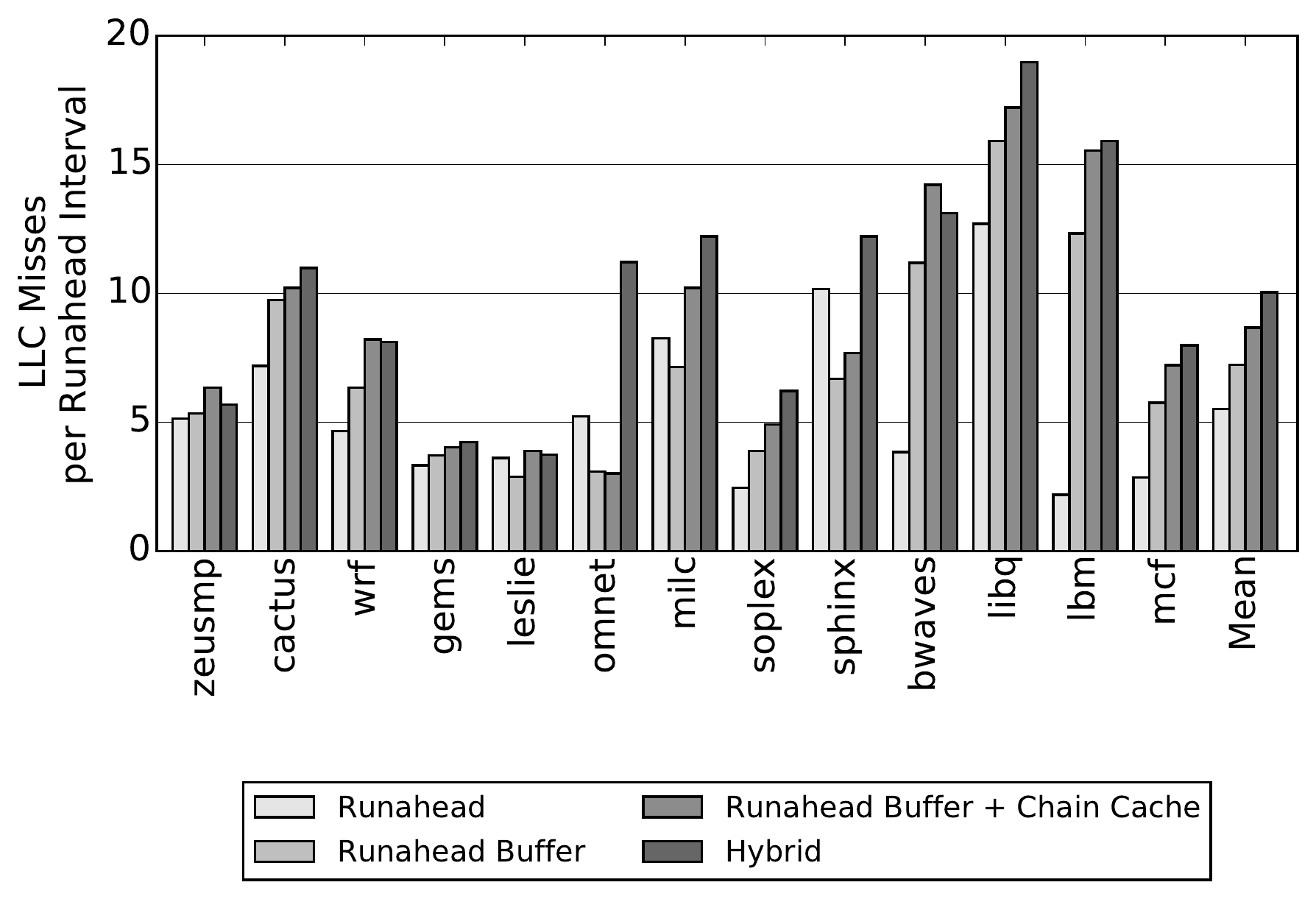}
	\caption{Average Number of Memory Accesses per Runahead Interval}
	\label{fig:RAB:mlp}
\end{figure}

By not executing these excess operations the runahead buffer is able to generate more MLP than traditional runahead. Figure \ref{fig:RAB:mlp} shows the average number of cache-misses that are generated by runahead execution and the runahead buffer for the medium and high memory intensity \textit{SPEC06} benchmarks.

The runahead buffer alone generates 32\% more cache misses when compared to traditional runahead execution. Since the chain cache eliminates chain generation delay on a hit, the runahead buffer + chain cache system is able to generate 57\% more cache misses than traditional runahead. The hybrid policy generates 83\% more cache misses than traditional runahead. Benchmarks where the runahead buffer shows performance gains over traditional runahead such as \textit{cactus, bwaves, lbm} and \textit{mcf} all show large increases in the number of cache misses produced by the runahead buffer. 

\begin{figure}
	\centering
	\includegraphics[width=\columnwidth]{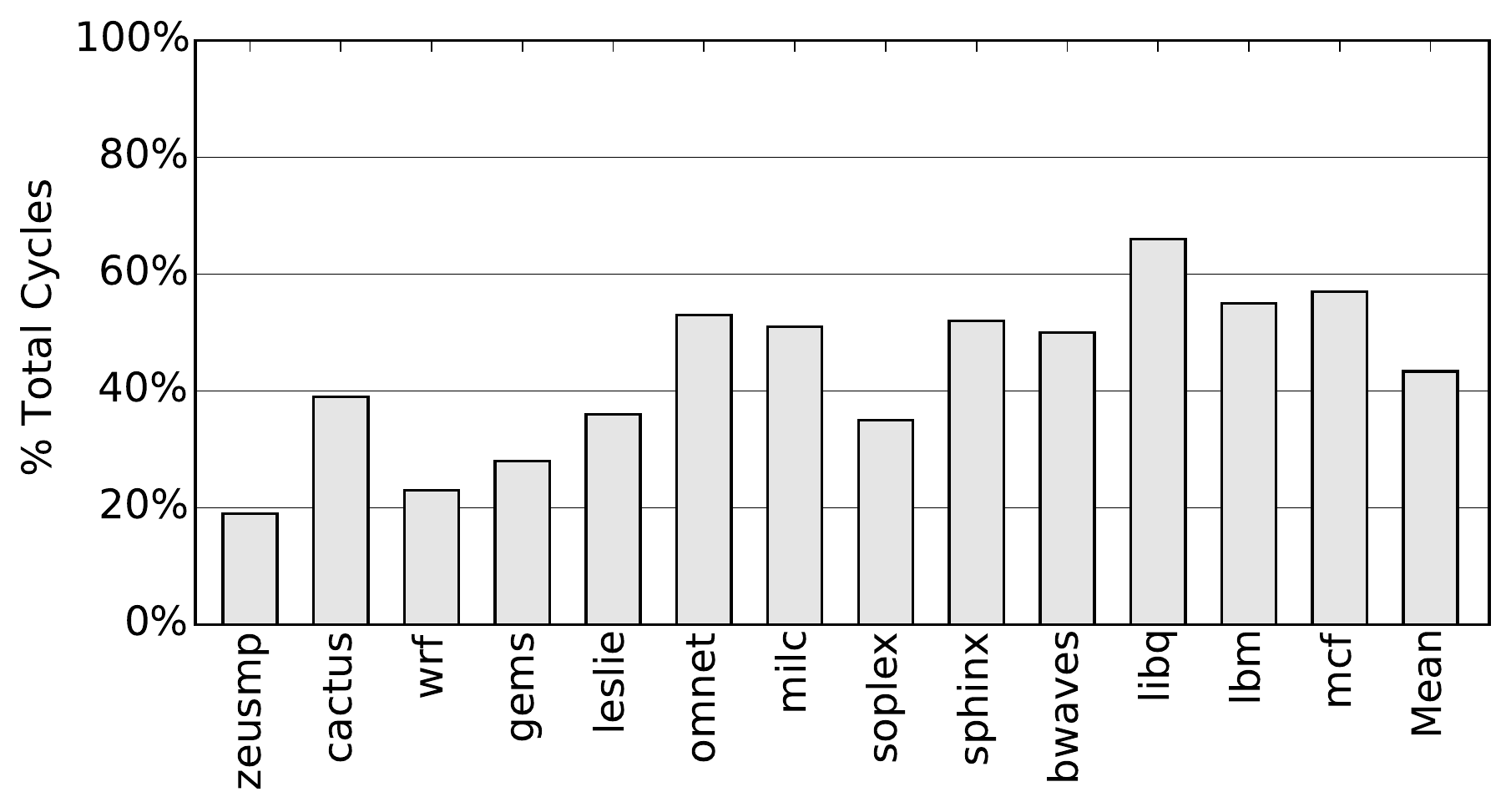}
	\caption{Percent of Time the Core is in Runahead Buffer Mode}
	\label{fig:RAB:cycles}
\end{figure}

In addition to generating more MLP than traditional runahead on average, the runahead buffer also has the advantage of not using the front-end during runahead. The percent of total cycles that the front-end is idle and can be clock-gated with the runahead buffer are shown in Figure \ref{fig:RAB:cycles} for the medium and high memory intensity \textit{SPEC06} benchmarks.

On average, 43\% of total execution cycles are spent in runahead buffer mode. By not using the front-end during this time, we reduce dynamic energy consumption vs. traditional runahead execution on average, as discussed in the energy evaluation (Section \ref{sec:raBuf:energy}).

\textbf{Dependence Chain Cache:} Looking beyond the simple runahead buffer policy, Figure \ref{fig:RAB:perf} also shows the result of adding a small dependence chain cache to the runahead buffer system. This chain cache generally improves performance, particularly for \textit{mcf}, \textit{soplex}, and \textit{GemsFDTD}. Table \ref{tab:RAB:trace} shows the hit rate for the medium and high memory intensity applications in the chain cache.

The applications that show the highest performance improvements with a chain cache show very high hit rates in Figure \ref{tab:RAB:trace}, generally above 95\%. The chain cache broadly improves performance over using a runahead buffer alone.  

The dependence chains in the chain cache do not necessarily match the exact dependence chains that would be generated from the reorder buffer. A chain cache hit is speculation that it is better to runahead with a previously generated chain than it is to take the time to generate a new chain, this is generally an acceptable trade-off. In Table \ref{tab:RAB:trace}, all chain cache hits are analyzed to determine if the stored dependence chain matches the dependence chain that would be generated from the ROB.
 
 \begin{table}
 	\centering
 	\begin{tabular}{|c|ccccccc|}\hline
 		\centering
 		\multirow{4}{1.0in}{\centering Chain Cache Hit Rate} & zeusmp & cactus & wrf & gems & leslie & omnetpp & milc \\
 		& 0.79 &	0.82 &	0.76 &	0.97 &	0.88 &	0.32 &	0.41 \\\cline{2-8}
 		& soplex & sphinx & bwaves & libq & lbm & mcf & Mean \\
 		& .0.99 &	0.76 &	0.44 &	0.99 &	0.45 &	0.97 & 	0.73 \\\cline{2-8}
 		\hline
 		\multirow{4}{1.0in}{\centering Chain Cache Exact Match} & zeusmp & cactus & wrf & gems & leslie & omnetpp & milc \\
 		& 0.76 &	0.62 &	0.82 &	0.59 &	0.64 &	0.27 &	0.48 \\\cline{2-8}
 		& soplex & sphinx & bwaves & libq & lbm & mcf & Mean \\
 		& 0.56 &	0.39 &	0.75 &	0.38 &	0.56 &	0.58 &	0.56 \\\cline{2-8}
 		\hline
 	\end{tabular}
 	\caption{Chain Cache Statistics}
 	\label{tab:RAB:trace}
 \end{table}

On average, the chain cache is reasonably accurate, with 60\% of all dependence chains matching exactly. The two applications with longer dependence chains, \textit{omnetpp} and \textit{sphinx}, show significantly less accurate chain cache hits than the other benchmarks. 

\begin{figure}
	\centering
	\includegraphics[width=\columnwidth]{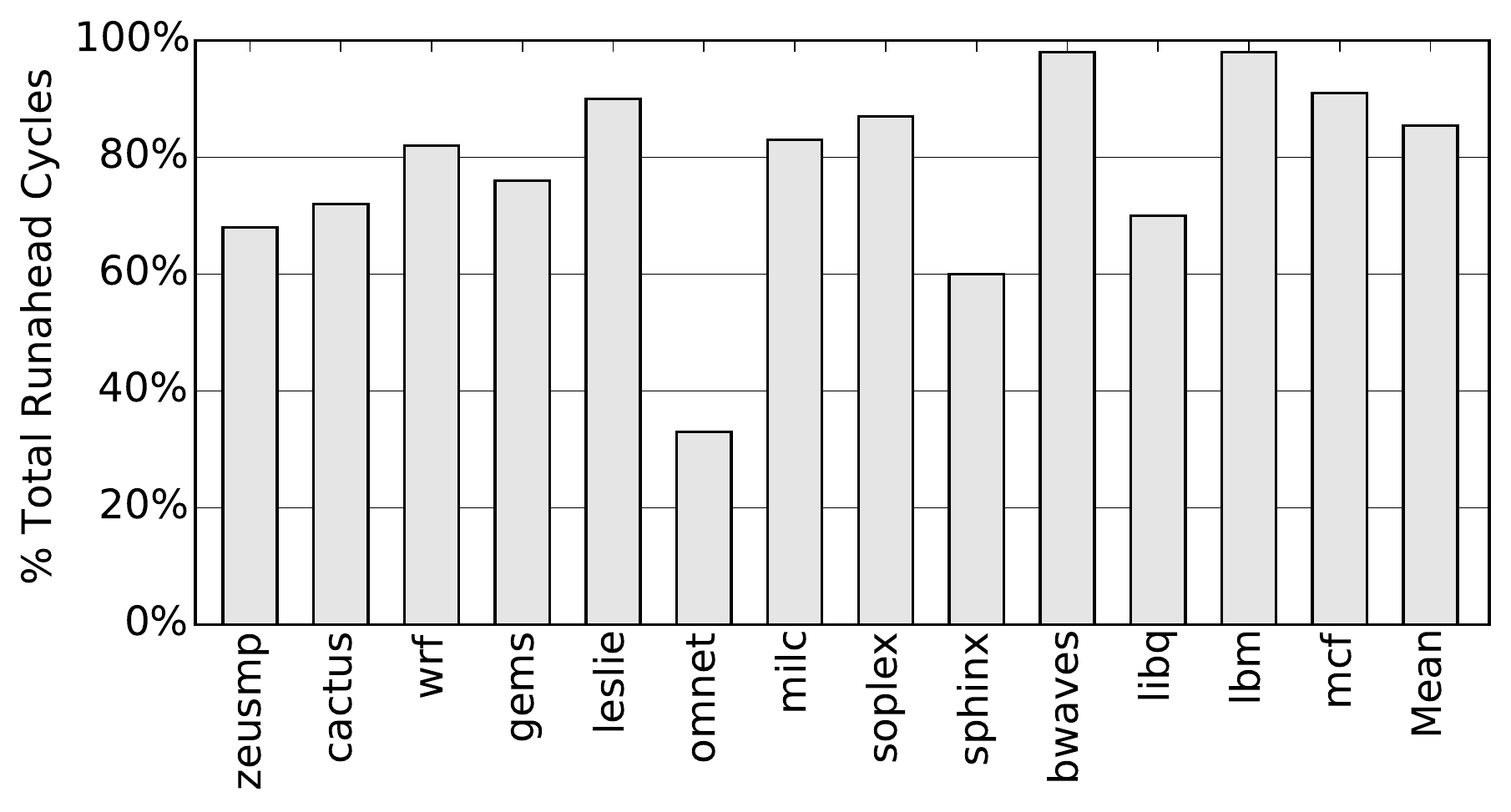}
	\caption{Cycles Spent in Runahead Buffer Mode during the Hybrid Policy}
	\label{fig:RAB:hybrid-state}
\end{figure}

\textbf{Hybrid Policy:} Lastly, the hybrid policy results in an average performance gain of 19.3\% over the baseline. Figure \ref{fig:RAB:hybrid-state} displays the fraction of time spent using the runahead buffer during the hybrid policy.

As Figure \ref{fig:RAB:hybrid-state} shows, the hybrid policy favors the runahead buffer, spending 85\% of the time on average in runahead buffer mode. The remainder is spent in traditional runahead. Applications that do not do well with the runahead buffer spend either the majority of the time (\textit{omnetpp}), or a large fraction of the time (\textit{sphinx}), in traditional runahead.  We conclude that the hybrid policy improves performance over the other schemes by using traditional runahead when it is best to do so (as in \textit{omnetpp}) and leveraging the runahead buffer otherwise (as in \textit{mcf}).

\subsection{Sensitivity to Runahead Buffer Parameters}
\label{sec:raBuf:sens}

In Section \ref{sec:raBuf:method} the two main parameters of the runahead buffer were stated to have been chosen via sensitivity analysis. Table \ref{tab:RAB:rabSens} shows sensitivity to these parameters: runahead buffer size and the number of chain cache entries. The sizes used in the evaluation (32-Entries/2-Entries) are bolded. The runahead buffer size experiments are conducted without a chain cache and the chain cache experiments are conducted assuming a 32-entry runahead buffer.

\begin{table}
	\begin{minipage}[bht*]{1.00\columnwidth}
		\centering
		\footnotesize
		\begin{tabular}{|c||c||c||c||c||c|} \hline
			\multicolumn{6}{|c|}{{\bf Runahead Buffer Size (Operations)} } \\ \hline 
			\multicolumn{1}{|c||}{2} & \multicolumn{1}{c||}{4} & \multicolumn{1}{c||}{8} & \multicolumn{1}{c||}{16} & \multicolumn{1}{c||}{\textbf{32}} & \multicolumn{1}{c|}{64}\\ \hline
			$\Delta$ Perf & $\Delta$ Perf & $\Delta$ Perf & $\Delta$ Perf & $\Delta$ Perf & $\Delta$ Perf \\ \hline
			1.11\% & 2.5\% & 4.9\% & 5.5\% & 11.0\% & 8.6\% \\ 
			\hline
		\end{tabular}
	\end{minipage}
	
	\begin{minipage}[bht*]{1.00\columnwidth}
		\centering
		\footnotesize
		\begin{tabular}{|c||c||c||c||c||c|} \hline
			\multicolumn{6}{|c|}{{\bf Chain Cache Entries} } \\ \hline 
			\multicolumn{1}{|c||}{\textbf{2}} & \multicolumn{1}{c||}{4} & \multicolumn{1}{c||}{8} & \multicolumn{1}{c||}{16} & \multicolumn{1}{c||}{32} & \multicolumn{1}{c|}{64}\\ \hline
			$\Delta$ Perf & $\Delta$ Perf & $\Delta$ Perf & $\Delta$ Perf & $\Delta$ Perf & $\Delta$ Perf \\ \hline
			13.1\% & 11.6\% & 11.4\% & 11.1\% & 10.8\% & 10.8\% \\ 
			\hline
		\end{tabular}
	\end{minipage}
	\begin{small}
		\caption{Performance Sensitivity to Runahead Buffer Parameters}
		\label{tab:RAB:rabSens}
	\end{small}
\end{table}

Overall, the runahead buffer performance gains suffer significantly as runahead buffer size is reduced, as the runahead buffer is not able to hold full dependence chains. Increasing the size of the runahead buffer increases dependence chain generation time, thereby decreasing performance. The chain cache shows the highest performance when the size is small, as recent history is important for high dependence chain accuracy, but shows little sensitivity to increasing storage capacity. This is explored further in Section \ref{sec:scRaEMC:depChain}.

\subsection{Performance with Prefetching}
\label{sec:raBuf:Pref}

\begin{figure}
	\centering
	\includegraphics[width=\columnwidth]{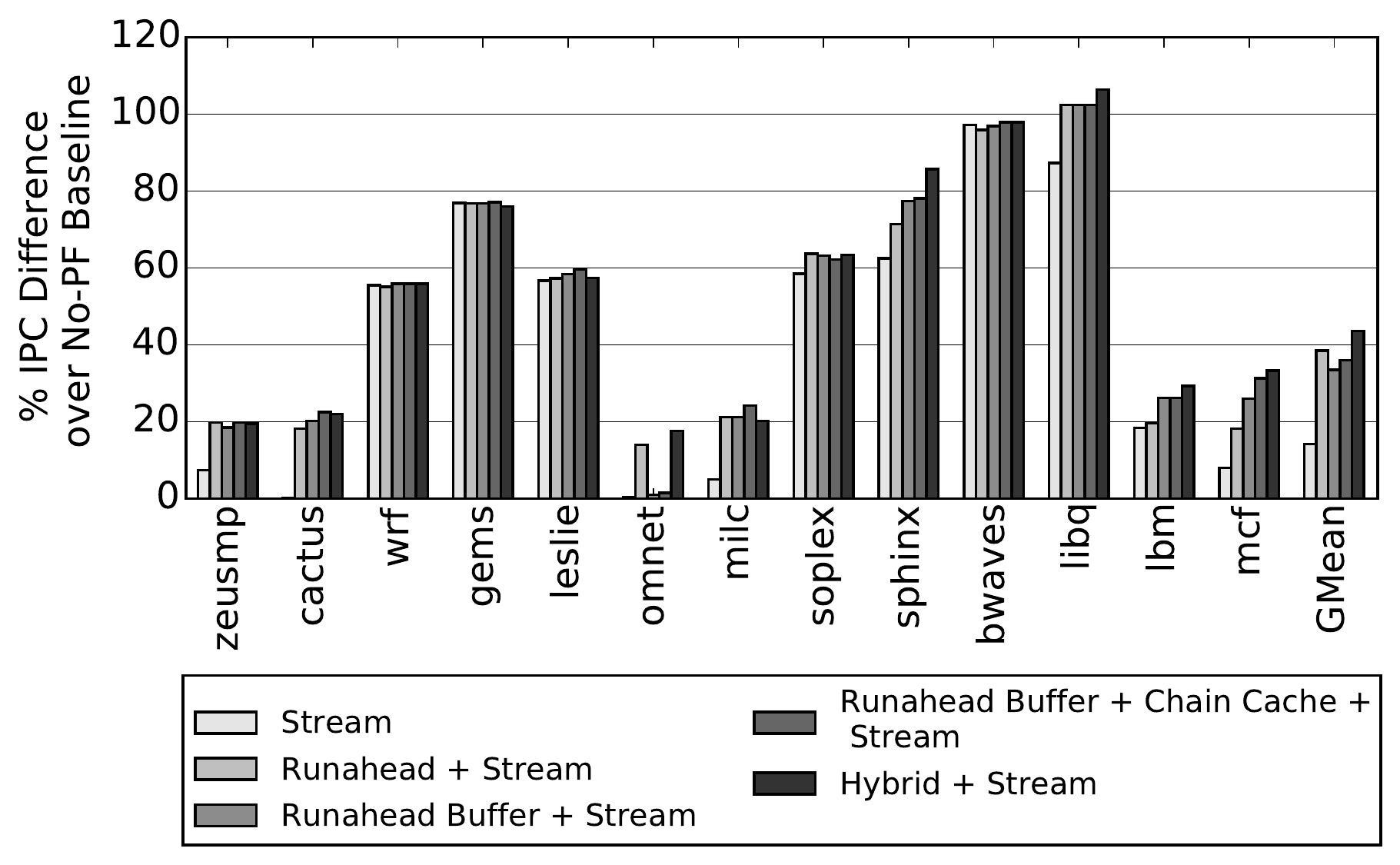}
	\caption{System Performance with Stream Prefetching}
	\label{fig:RAB:perfSt}
\end{figure}

\begin{figure}
	\centering
	\includegraphics[width=\columnwidth]{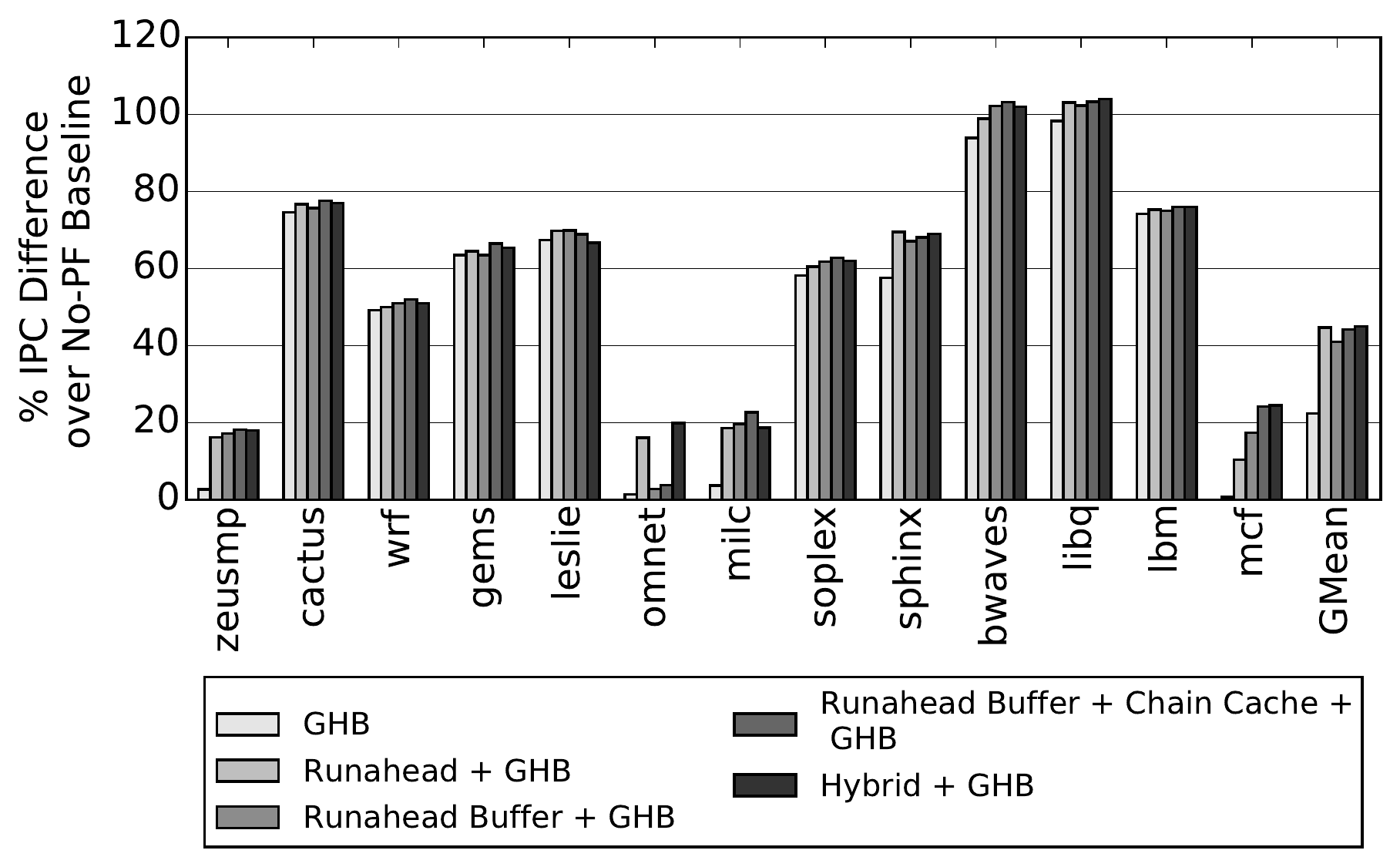}
	\caption{System Performance with GHB Prefetching}
	\label{fig:RAB:perfGHB}
	\centering
	\includegraphics[width=\columnwidth]{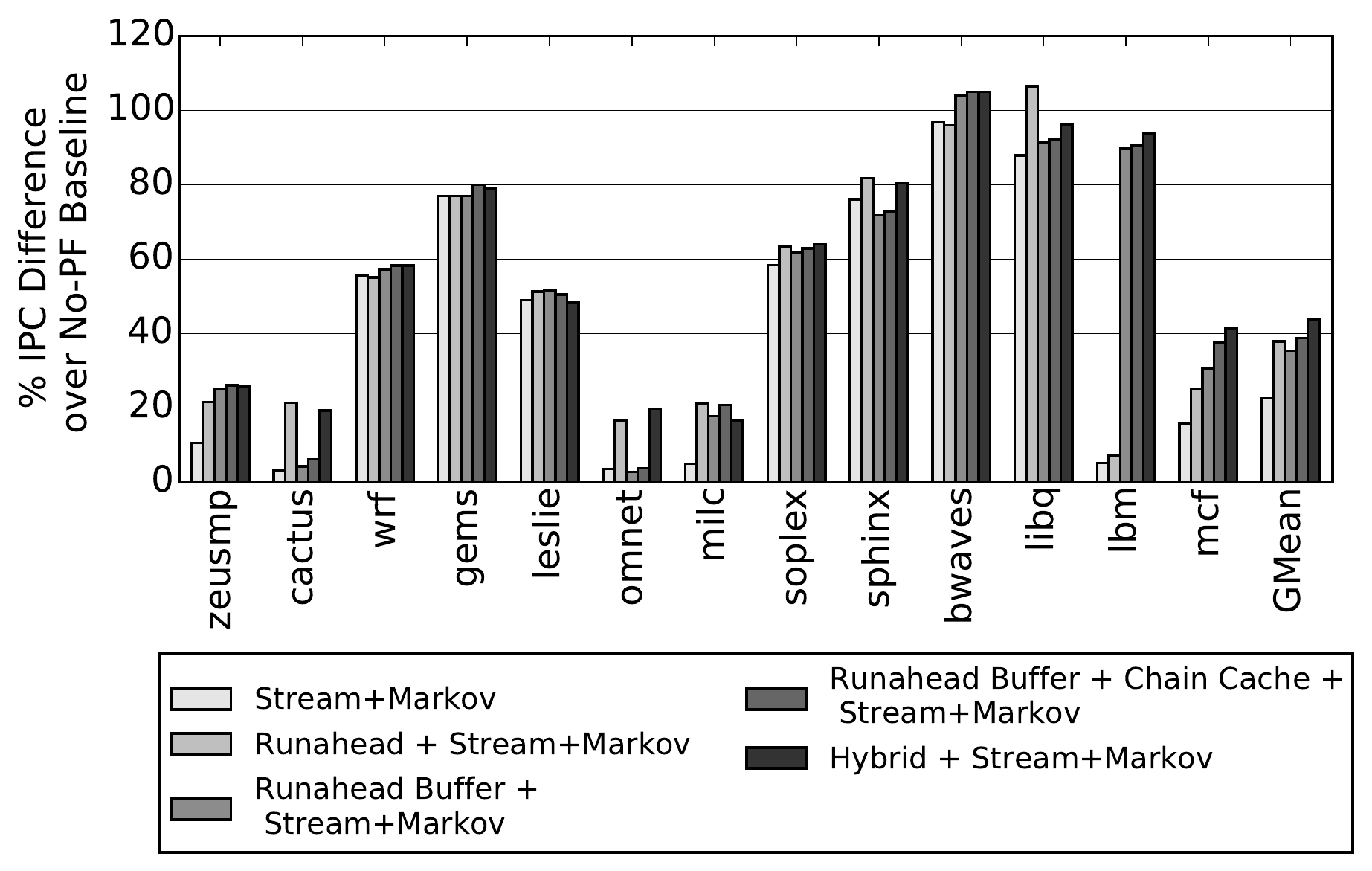}
	\caption{System Performance with Stream+Markov Prefetching}
	\label{fig:RAB:perfMarks}
\end{figure}

Figures \ref{fig:RAB:perfSt}/\ref{fig:RAB:perfGHB}/\ref{fig:RAB:perfMarks} shows the effect of adding a Stream/GHB/Markov+Stream prefetcher to the system respectively. 

The GHB prefetcher results in the largest performance gain across these three prefetchers. Combining the GHB prefetcher with runahead results in the highest performing system on average with a 45\% performance gain over the no-prefetching baseline. Applications that do not significantly improve in performance with GHB prefetching such as \textit{zeusmp, omnetpp, milc,} and \textit{mcf} all result in performance improvements when combined with runahead execution. Similar cases occur in Figures \ref{fig:RAB:perfSt}/\ref{fig:RAB:perfMarks} where applications like \textit{cactus} or \textit{lbm} do not improve in performance with prefetching but result in large performance gains when runahead is added to the system. Overall, traditional runahead and the runahead buffer both result in similar performance gains when prefetching is added to the system while the hybrid policy is the highest performing policy on average.

\begin{figure}
	\centering
	\includegraphics[width=\columnwidth]{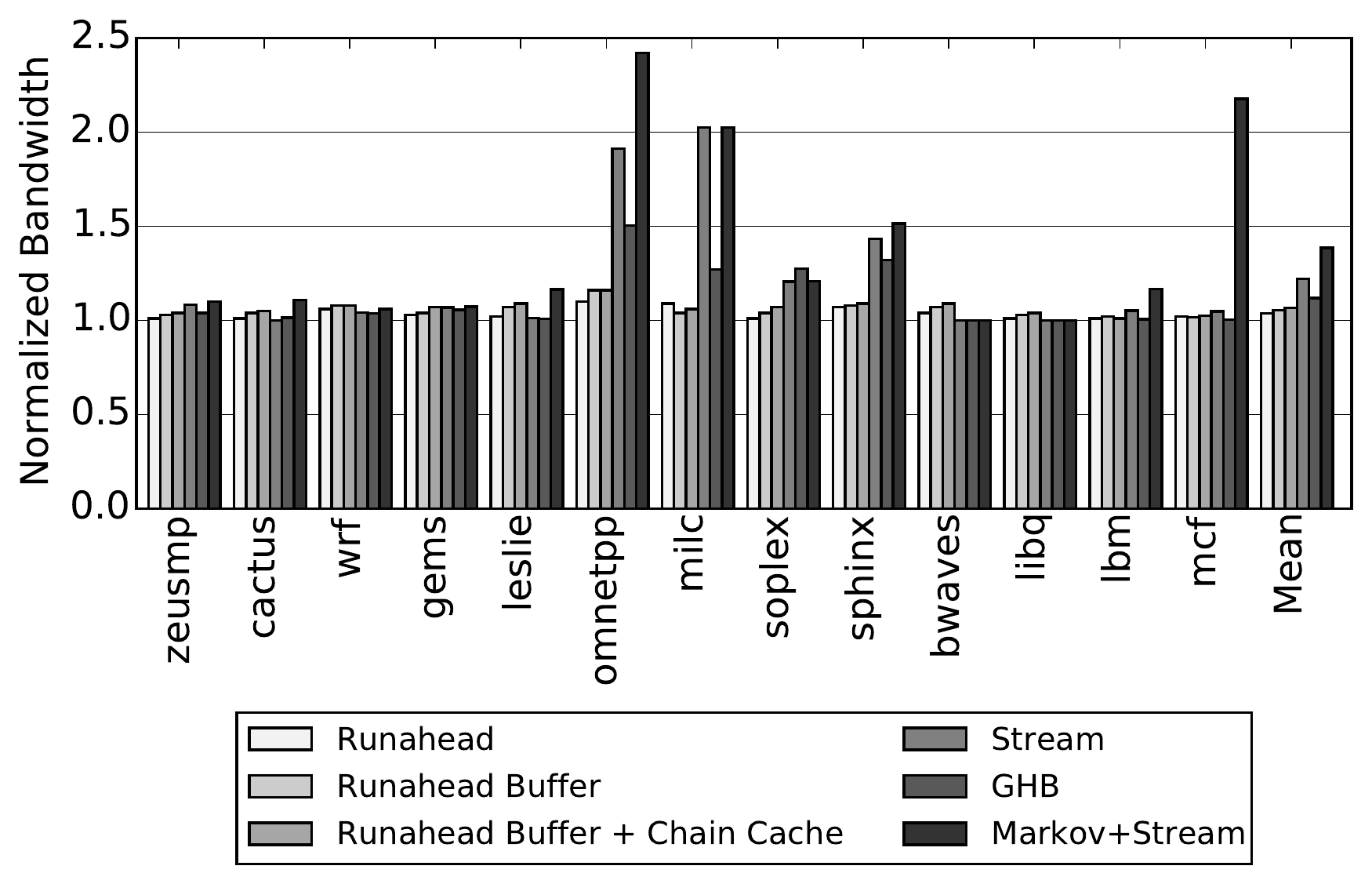}
	\caption{Normalized Bandwidth Consumption}
	\label{fig:RAB:mem-traffic}
\end{figure}

However, in addition to performance, the effect of prefetching on memory bandwidth is an important design consideration as prefetching requests are not always accurate. Figure \ref{fig:RAB:mem-traffic} quantifies the memory system overhead for prefetching and runahead. 

On average, the memory bandwidth requirements of runahead execution are small, especially when compared to the other prefetchers. Traditional runahead has a very small impact on memory traffic, increasing the total number of DRAM requests by 4\%. This highlights the accuracy benefit of using fragments of the application's own code to prefetch. Using the runahead buffer alone increases memory traffic by 6\% and the chain cache increases bandwidth overhead to 7\%. The runahead buffer consumes the most additional bandwidth on \textit{omnetpp}. Even with prefetcher throttling, the prefetchers all result in a larger bandwidth overhead than the runahead schemes. The Markov+Stream prefetcher has the largest overhead, at 38\%, while the GHB prefetcher is the most accurate with a 12\% bandwidth overhead. I conclude that while prefetching can significant increase performance, it also significantly increases memory traffic.

\subsection{Energy Evaluation}
\label{sec:raBuf:energy}

\begin{figure}[h]
	\centering
	\includegraphics[width=\columnwidth]{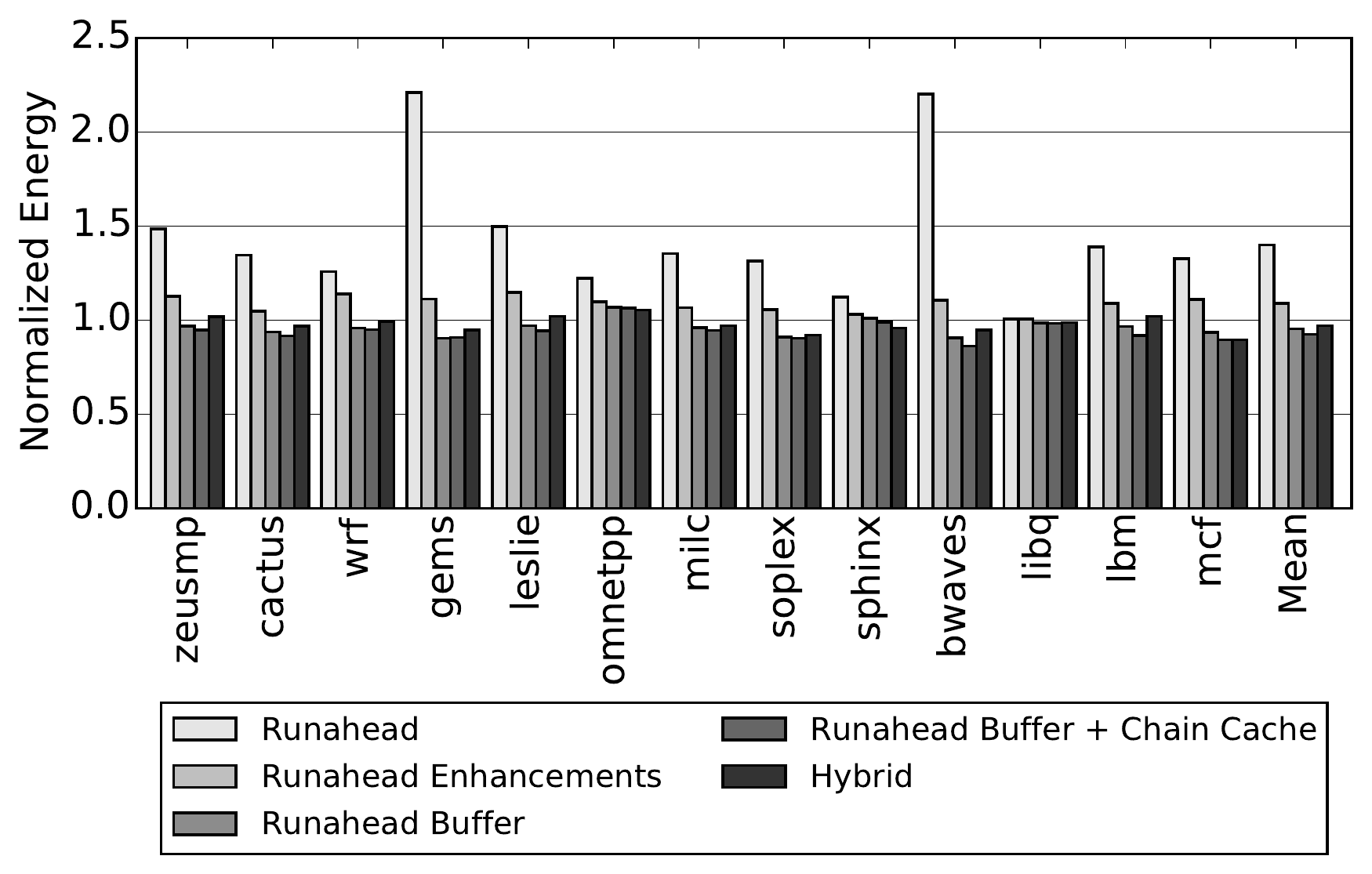}
	\caption{Normalized Energy Consumption}
	\label{fig:RAB:energy}	
\end{figure}

\begin{figure}
	\centering
	\includegraphics[width=\columnwidth]{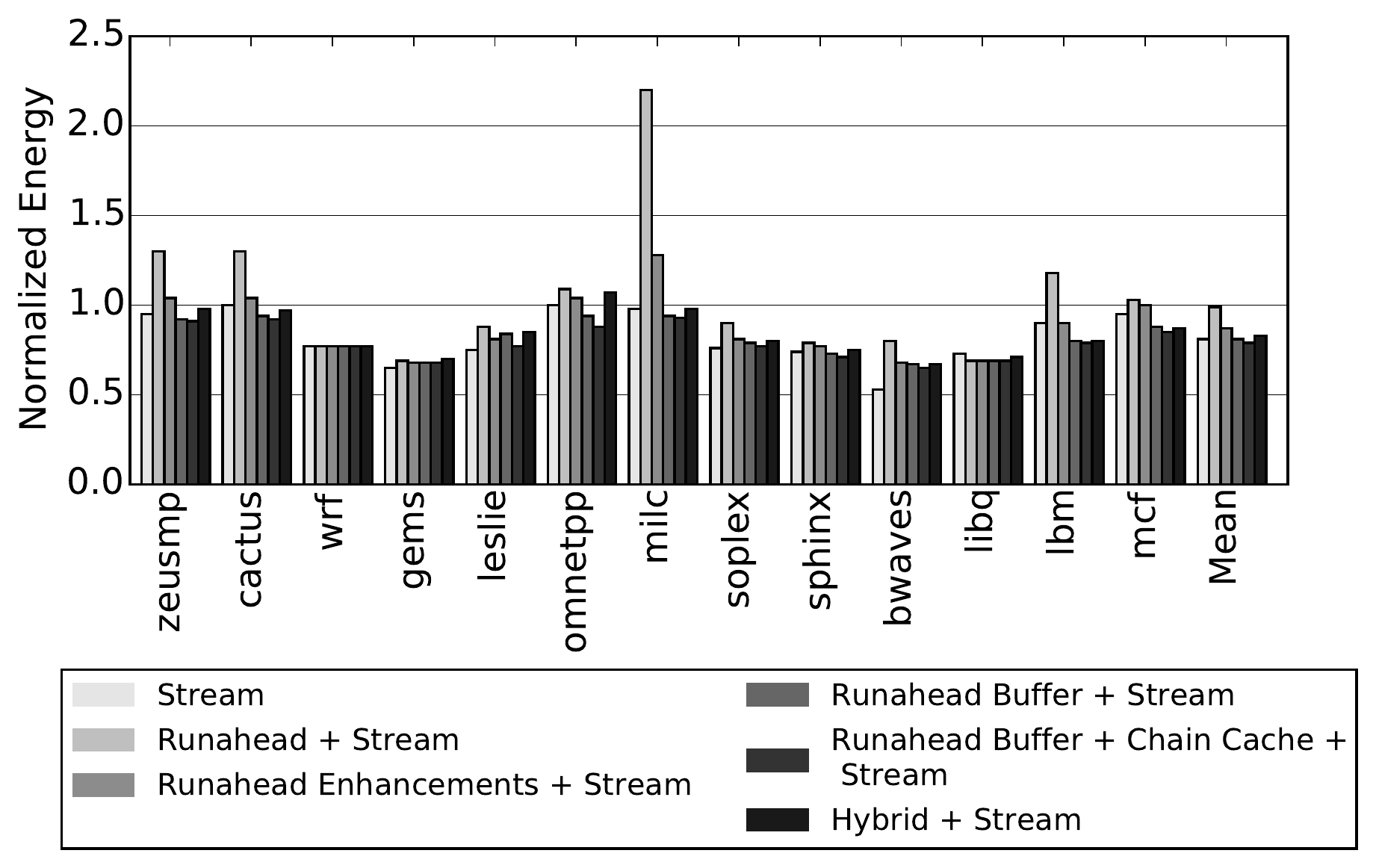}
	\caption{Normalized Energy Consumption with Stream Prefetching}
	\label{fig:RAB:energyStream}	
\end{figure}

\begin{figure}
	\centering
	\includegraphics[width=\columnwidth]{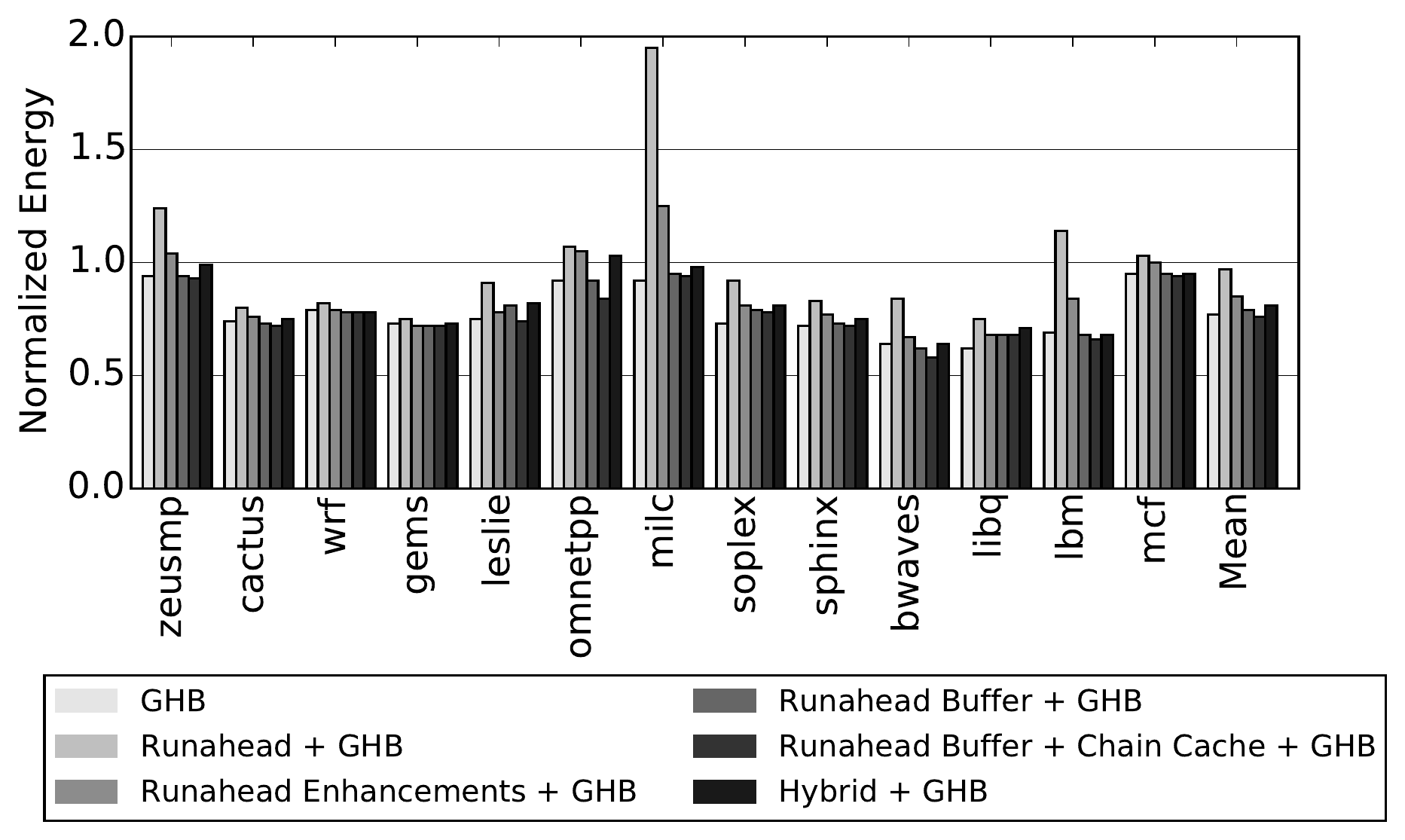}
	\caption{Normalized Energy Consumption with GHB Prefetching}
	\label{fig:RAB:energyGHB}	
\end{figure}

\begin{figure}
	\centering
	\includegraphics[width=\columnwidth]{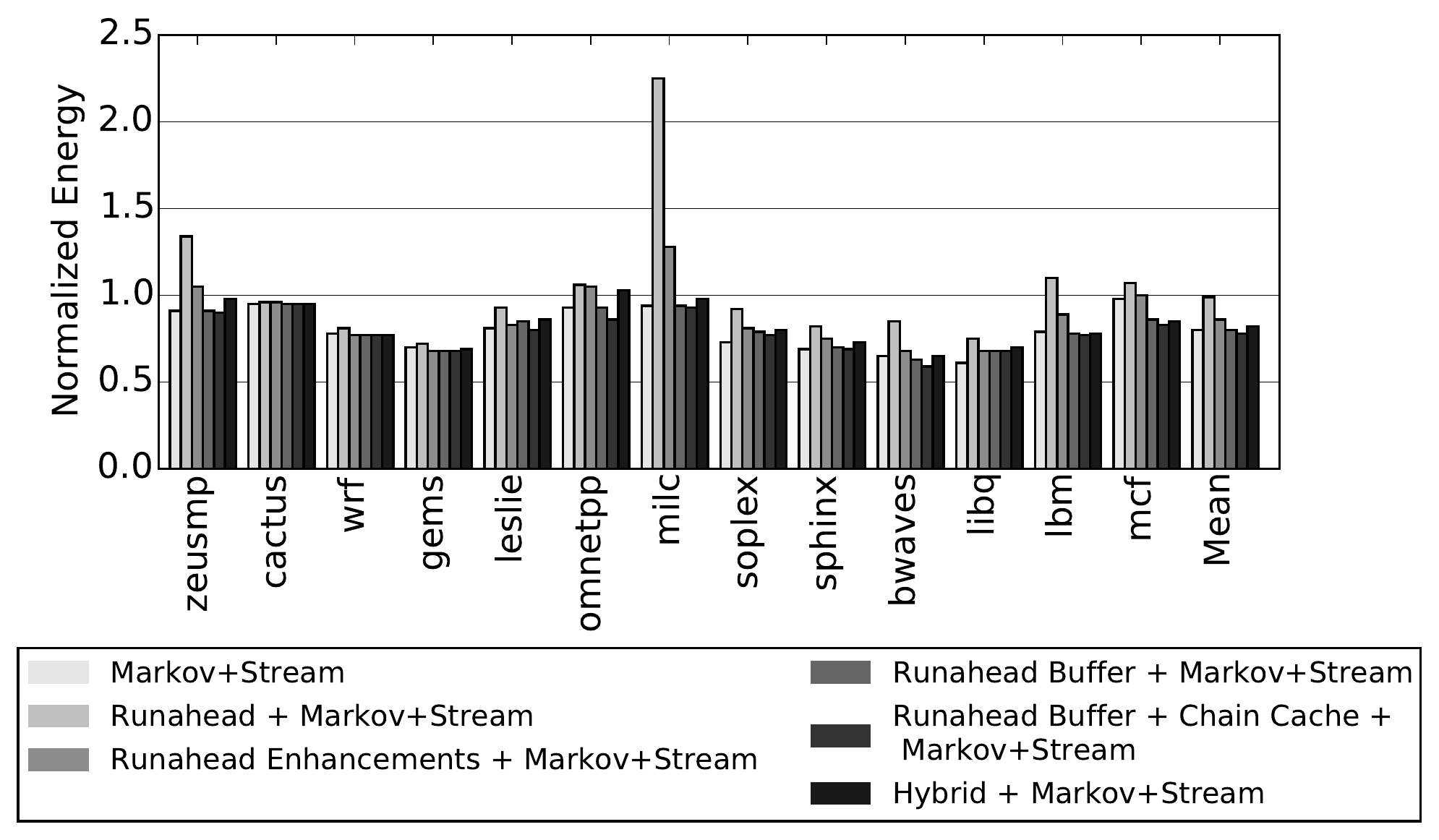}
	\caption{Normalized Energy Consumption with Markov+Stream Prefetching}
	\label{fig:RAB:energyMarks}	
\end{figure}

The normalized results for the system without prefetching are shown in Figure \ref{fig:RAB:energy}. Normalized energy consumption with prefetching is shown in Figures \ref{fig:RAB:energyStream}-\ref{fig:RAB:energyMarks}.

Runahead alone drastically increases energy consumption due to very high dynamic instruction counts, as the front-end fetches and decodes instructions during periods where it would be otherwise idle. This observation has been made before \cite{mut:kim05}, and several mechanisms have been proposed to reduce the dynamic instruction count (Section \ref{sec:raBuf:raEnhance}). From this work, I implement the two hardware-based mechanisms that reduce the dynamic instruction count the most in ``Runahead Enhancements''. These mechanisms seek to eliminate short and overlapping runahead intervals. 

With these enhancements, traditional runahead results in drastically lower energy consumption with a 2.1\% average degradation of runahead performance vs. the baseline (2.6\% with prefetching). Traditional runahead increases system energy consumption by 40\% and the system with the runahead enhancements increases energy consumption by 9\% on average. 

The runahead buffer reduces dynamic energy consumption by leaving the front-end idle during runahead periods. This allows the runahead buffer system to decrease average energy consumption by 4.5\% without a chain cache and 7.5\% with a chain cache. The hybrid policy decreases energy consumption by 2.3\% on average. The runahead buffer decreases energy consumption more than the hybrid policy because the hybrid policy spends time in the more inefficient traditional runahead mode to maximize performance. 

When prefetching is added to the system, similar trends hold. Traditional runahead execution increases energy consumption over the prefetching baseline in all three cases. Runahead enhancements cause 6\%/8\%/6\% over the Stream/GHB/ Markov+Stream prefetching baseline. The Runahead Buffer and Runahead Buffer + Chain Cache schemes both reduce energy consumption. The Runahead Buffer + Chain Cache + GHB prefetcher system is the most energy efficient, resulting in a 24\% energy reduction over the no-prefetching baseline. I conclude that this system is both the most energy efficient and highest performing system in this evaluation.

\subsection{Sensitivity to System Parameters}
\label{sec:raBuf:sysSens}

Table \ref{tab:RAB:sysSens} shows performance and energy sensitivity of the runahead buffer to three system parameters: LLC capacity, the number of memory banks per channel, and ROB size. Performance and energy are shown as average numbers relative to a baseline system with no-prefetching of that identical configuration. For example, the 2MB LLC data point shows that the runahead buffer results in a performance gain of 12.1\% and an energy reduction of 4.9\% over a system with a 2MB LLC and no-prefetching.

The runahead buffer shows some sensitivity to LLC size. As LLC size decreases, the performance and energy gains also decrease as the system has less LLC capacity to devote to prefetching effects. As LLC sizes increase, particularly at the 4MB data point, runahead buffer gains once again decrease as the application working set begins to fit in the LLC. Table \ref{tab:RAB:sysSens} also shows that increasing the number of memory banks per channel to very large numbers and increasing ROB size generally have negative effects on runahead buffer performance and energy consumption.

\begin{table}[htb*]
	\begin{minipage}[bht*]{1.00\columnwidth}
		\centering
		\footnotesize
		\begin{tabular}{|c|c||c|c||c|c||c|c|} \hline
			\multicolumn{8}{|c|}{{\bf LLC Cache Size} } \\ \hline 
			\multicolumn{2}{|c||}{512 KB} & \multicolumn{2}{c||}{\textbf{1 MB}} & \multicolumn{2}{c||}{2 MB} & \multicolumn{2}{c|}{4 MB} \\ \hline
			$\Delta$ Perf & $\Delta$ Energy & $\Delta$ Perf & $\Delta$ Energy & $\Delta$ Perf & $\Delta$ Energy & $\Delta$ Perf & $\Delta$ Energy \\ \hline
			8.2\% & -3.2\% & 11.0\% & -4.5\% & 12.1\% & -4.9\% & 9.1\% & -4.6\% \\ \hline
		\end{tabular}
	\end{minipage}
	
	\begin{minipage}[bht*]{1.00\columnwidth}
		\centering
		\footnotesize
		\begin{tabular}{|c|c||c|c||c|c||c|c|} \hline
			\multicolumn{8}{|c|}{{\bf Number of Memory Banks} } \\ \hline 
			\multicolumn{2}{|c||}{\textbf{8}} & \multicolumn{2}{c||}{16} & \multicolumn{2}{c||}{32} & \multicolumn{2}{c|}{64} \\ \hline
			$\Delta$ Perf & $\Delta$ Energy & $\Delta$ Perf & $\Delta$ Energy & $\Delta$ Perf & $\Delta$ Energy & $\Delta$ Perf & $\Delta$ Energy \\ \hline
			11.0\% & -4.5\% & 7.4\% & -2.9\% & 6.6\% & -2.5\% & 4.7\% & -2.2\% \\ \hline
		\end{tabular}
	\end{minipage}
	
	\begin{minipage}[bht*]{1.00\columnwidth}
		\centering
		\footnotesize
		\begin{tabular}{|c|c||c|c||c|c||c|c|} \hline
			\multicolumn{8}{|c|}{{\bf ROB Size} } \\ \hline 
			\multicolumn{2}{|c||}{192} & \multicolumn{2}{c||}{\textbf{256}} & \multicolumn{2}{c||}{384} & \multicolumn{2}{c|}{512} \\ \hline
			$\Delta$ Perf & $\Delta$ Energy & $\Delta$ Perf & $\Delta$ Energy & $\Delta$ Perf & $\Delta$ Energy & $\Delta$ Perf & $\Delta$ Energy \\ \hline
			14.4\% & -4.4\% & 11.0\% & -4.5\% & 10.5\% & -4.1\% & 6.5\% & -2.1\% \\ \hline
		\end{tabular}
	\end{minipage}
	\begin{small}
		\caption{Runahead Buffer Performance and Energy Sensitivity}
		\label{tab:RAB:sysSens}
	\end{small}
\end{table}

\section{Conclusion}
\label{sec:raBuf:conclusion}

In this chapter, I presented an approach to increase the effectiveness of runahead execution for out-of-order processors. I identify that many of the operations that are executed in traditional runahead execution are unnecessary to generate cache-misses. Using this insight, I enable the core to dynamically generate filtered dependence chains that only contain the operations that are required for a cache-miss. These chains are generally short. The operations in a dependence chain are read into a buffer and speculatively executed as if they were in a loop when the core would be otherwise idle. This allows the front-end to be idle for 44\% of the total execution cycles of the medium and high memory intensity \textit{SPEC06} benchmarks on average.

The runahead buffer generates 57\% more MLP on average as traditional runahead execution. This leads to a 13.1\% performance increase and 7.5\% decrease in energy consumption over a system with no-prefetching. Traditional runahead execution results in a 10.9\% performance increase and 9\% energy increase, assuming additional optimizations. Overall, the runahead buffer is a small, easy to implement structure (requiring 1.9 kB of additional total storage) that increases performance for memory latency-bound, single-threaded applications. Chapters \ref{chap:scRaEMC} and \ref{chap:mcRaEMC} further develop the mechanisms from the runahead buffer to accelerate independent cache misses in a multi-core system. However, in the next chapter I shift focus from independent cache misses to reducing effective memory access latency for critical dependent cache misses.

\chapter{The Enhanced Memory Controller}
\label{chap:EMC}
\setlength{\epigraphwidth}{0.41\textwidth}

\section{Introduction}
\label{sec:EMC:Intro}

The impact of effective memory access latency on processor performance is magnified when a last level cache (LLC) miss has dependent operations that also result in an LLC miss. These dependent cache misses form chains of long-latency operations that fill the reorder buffer (ROB) and prevent the core from making forward progress. This is highlighted by the \textit{SPEC06} benchmark \textit{mcf} which has the lowest IPC in Figure \ref{fig:intro:stall} and the largest fraction of dependent cache misses in Figure \ref{fig:intro:depMiss}. The result of changing all of these dependent cache misses into LLC hits is a performance gain of 95\%. This chapter shows that dependent cache misses are difficult to prefetch as they generally have data-dependent addresses (Section \ref{sec:EMC:Background}). I then propose a new hardware mechanism to minimize effective memory access latency for all dependent cache-misses (Section \ref{sec:EMC:Mechanism}) \footnote{An earlier version of this chapter was published as: Milad Hashemi, Khubaib, Eiman Ebrahimi, Onur Mutlu, and Yale Patt. Accelerating Dependent Cache Misses with an Enhanced Memory Controller. In \textit{ISCA}, 2016. I developed the initial idea in collaboration with Professor Onur Mutlu and conducted the performance simulator design and evaluation for this work.}.

\section{Background}
\label{sec:EMC:Background}

\begin{figure}
	\centering
	\includegraphics[width=\columnwidth]{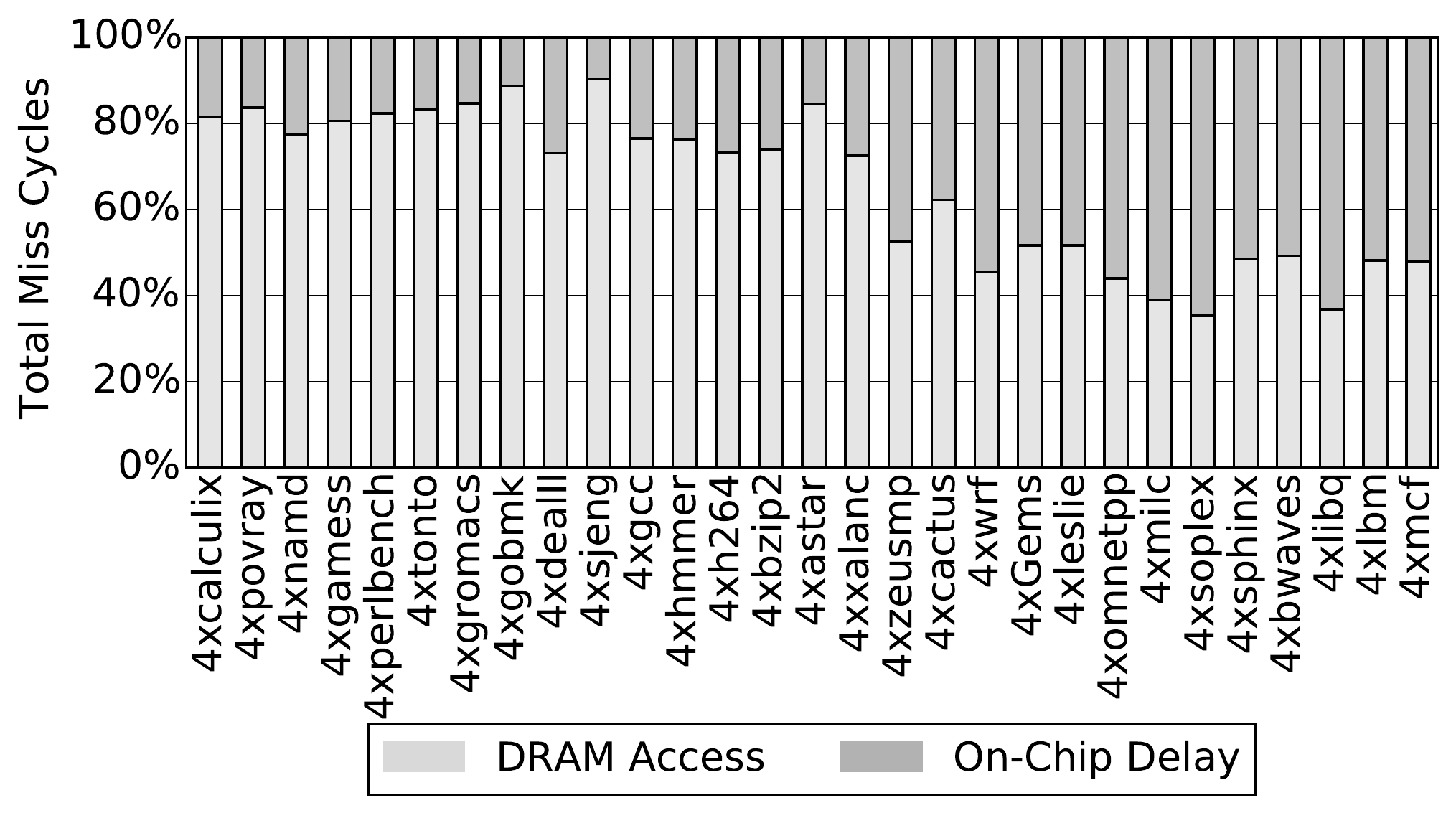}
	\caption{Breakdown of the Cycles to Satisfy a Memory Request}
	\label{fig:EMC:contention}
\end{figure}

While single-core systems have a significant latency disparity between performing computation at the core and accessing data from off-chip DRAM, the effective memory access latency problem (Section \ref{sec:intro:problem}) worsens in a multi-core system. This is because multi-core systems have high levels of on-chip delay that result from the different cores contending for on-chip shared resources. To demonstrate this, Figure \ref{fig:EMC:contention} shows the effect of on-chip latency on DRAM accesses for \textit{SPEC06}. A quad-core processor is simulated where each of the cores is identical to the single core described in Section \ref{sec:raBuf:method}. The delay incurred by a DRAM access is separated into two categories:  the average cycles that the request takes to access DRAM and return data to the chip and all other on-chip delays that the access incurs after missing in the LLC. 

In Figure \ref{fig:EMC:contention}, benchmarks are sorted in ascending memory intensity.  For the memory intensive applications to the right of \textit{gems}, defined as having an MPKI (misses per thousand instructions) of over 10, the actual DRAM access is less than half the total latency of the memory request.  Roughly half the cycles spent satisfying a memory access are attributed to on-chip delays. This extra on-chip latency is due to shared resource contention among the multiple cores on the chip. This contention happens in the shared on-chip interconnect, cache, and DRAM buses, row-buffers, and banks. Others \cite{mut:mos07, mut:mos08, lee:2008} have pointed out the large effect of on-chip contention on performance, and Awasthi et al. \cite{Awasthi:2010} have noted that this effect will increase as the number of cores on the chip grows. 

\begin{figure}
	\centering
	\includegraphics[width=\columnwidth]{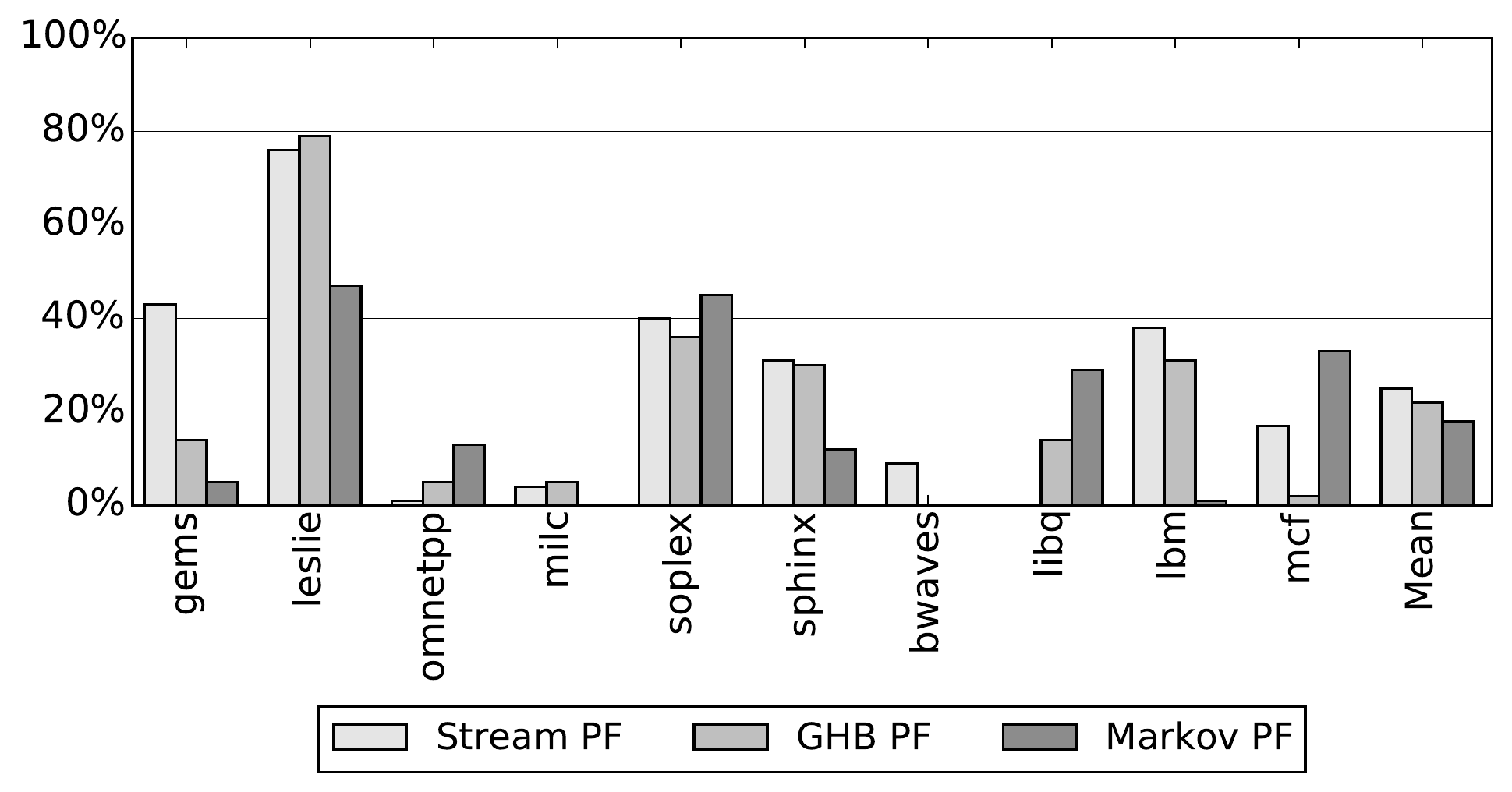}
	\caption{Percent of Dependent Cache Misses Prefetched with a GHB, Stream, and Markov prefetcher}
	\label{fig:EMC:depPer}
\end{figure}

Several techniques have attempted to reduce the effect of dependent cache misses on performance.  The most common is prefetching.  Figure \ref{fig:EMC:depPer} shows the percent of all dependent cache misses that are prefetched by three different on-chip prefetchers: a stream prefetcher, a Markov prefetcher \cite{joseph:isca97}, and a global history buffer (GHB) \cite{nes:smith04} for the memory intensive \textit{SPEC06} benchmarks. The average percentage of all dependent cache misses that are prefetched is small, under 25\% on average.

Prefetchers have difficulty with dependent cache misses because their addresses are reliant on the source data that originally resulted in a cache miss. This leads to data dependent patterns that are hard to capture.  Moreover, inaccurate and untimely prefetch requests lead to a large increase in bandwidth consumption, a significant drawback in a bandwidth constrained multi-core system.  The stream, GHB, and Markov prefetchers increase bandwidth consumption by 22\%, 12\%, and 39\% respectively, even with prefetcher throttling enabled \cite{fdp07}.

Note that pre-execution techniques such as traditional Runahead Execution \cite{dun:mud97, mut:sta03}, the Runahead Buffer, and Continual Flow Pipelines \cite{sri:raj04} target prefetching \textit{independent} cache misses. Unlike dependent cache misses, independent misses only require source data that is available on chip. These operations can be issued and executed by an out-of-order processor as long as the ROB is not full. Runahead and CFP discard slices of operations that are dependent on a miss (including any dependent cache misses) in order to generate memory level parallelism with new independent cache misses.

Since dependent cache misses are difficult to prefetch and are delayed by on-chip contention, I propose a different mechanism to accelerate these dependent accesses. I observe that the number of operations between a cache miss and its dependent cache miss is usually small (Section \ref{sec:EMC:Mechanism}). If this is the case, I propose migrating these dependent operations to the memory controller where they can be executed immediately after the original cache miss data arrives from DRAM. This new enhanced memory controller (EMC) generates cache misses faster than the core since it is able to bypass on-chip contention, thereby reducing the on-chip delay observed by the critical dependent cache miss.

\section{Mechanism}
\label{sec:EMC:Mechanism}

\begin{figure}
	\centering
	\includegraphics[width=3.5in, height = 2.5in]{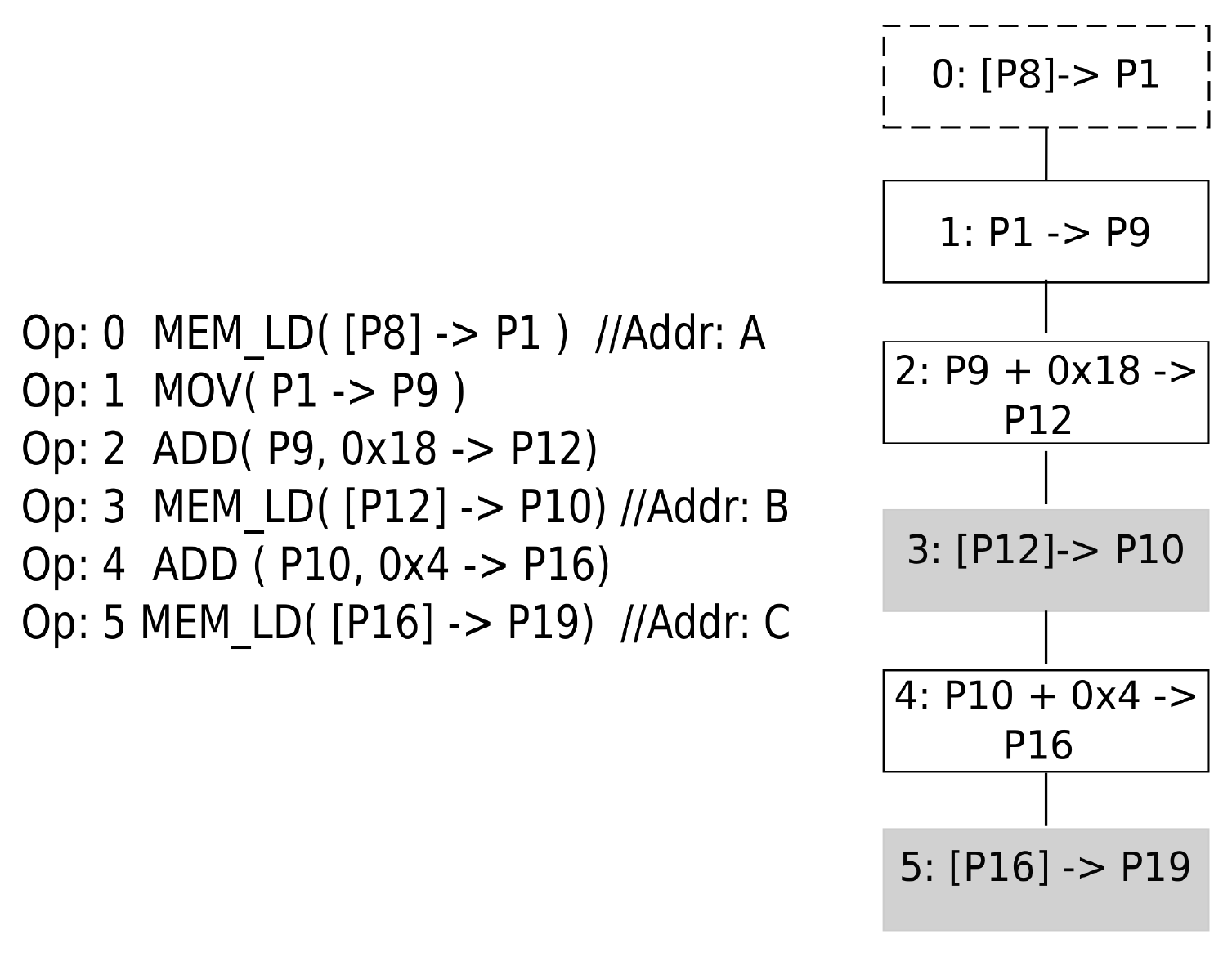}
	\caption{Dynamic Sequence of Micro-ops Based on \textit{mcf}}
	\label{fig:EMC:mcfCode}
\end{figure}

Figure \ref{fig:EMC:mcfCode} presents one example of the dependent cache miss problem. A dynamic sequence of micro-operations (uops) has been adapted from \textit{mcf}. The uops are shown on the left and the data dependencies, omitting control uops, are illustrated on the right. A, B, C represent cache line addresses. Core physical registers are denoted by a `P'. Assume a scenario where Operation 0 is an outstanding cache miss. I call this uop a source miss and denote it with a dashed box. Operations 3 and 5 will result in cache misses when issued, shaded gray. However, their issue is blocked as Operation 1 has a data dependence on the result of the source miss, Operation 0. Operations 3 and 5 are delayed from execution until the data from Operation 0 returns to the chip and flows back to the core through the interconnect and cache hierarchy. Yet, there are a small number of relatively simple uops between Operation 0 and Operations 3/5.

I propose that these operations that are dependent on a cache miss can be executed as soon as the source data enters the chip, at the memory controller. This avoids on-chip interference and reduces the overall latency to issue the dependent memory requests.

\begin{figure}
	\centering
	\includegraphics[width=\columnwidth]{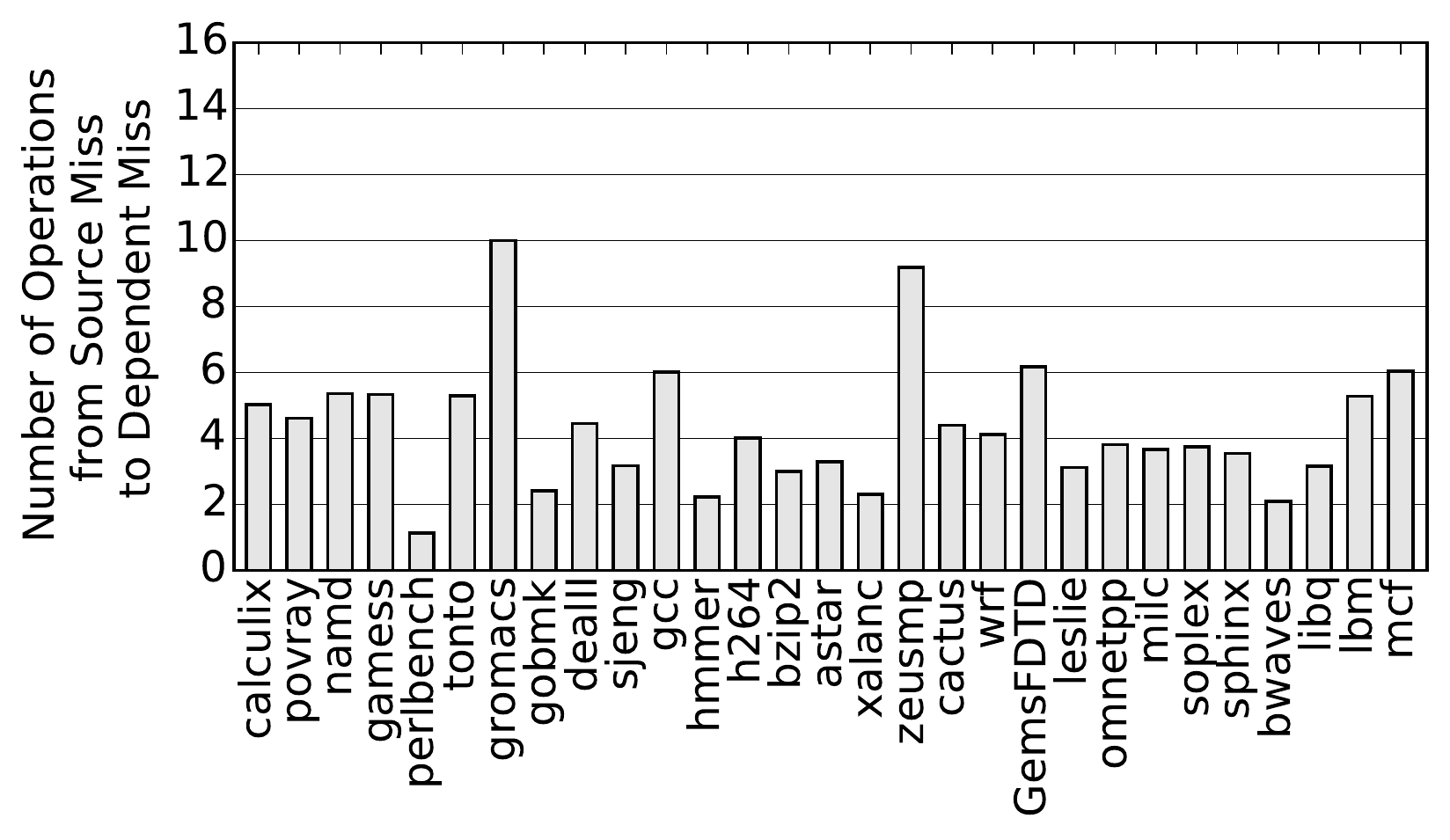}
	\caption{Average Number of Dependent Operations between a Source Miss and a Dependent Miss}
	\label{fig:EMC:depLen}
	\vspace{.5in}
	\centering
	\includegraphics[width=4.3in, height=2.3in]{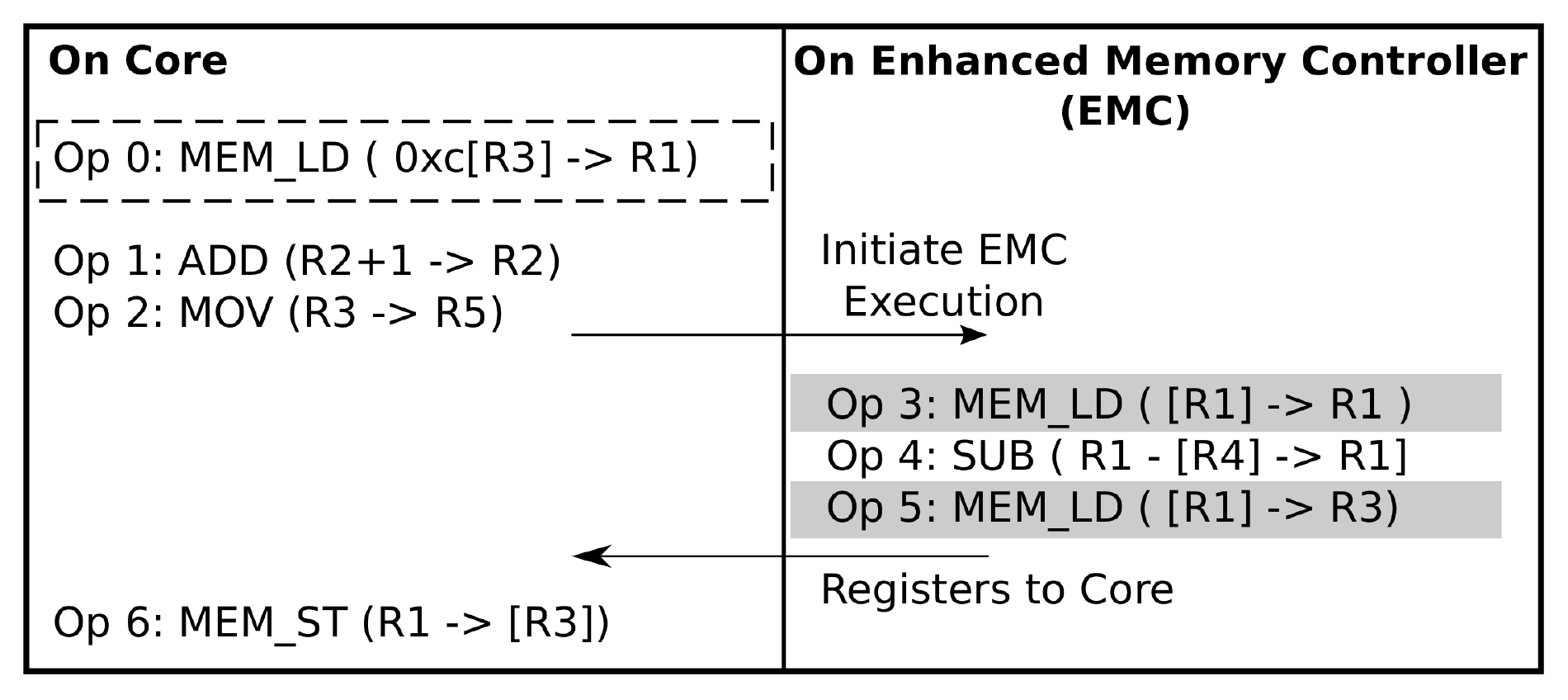}
	\caption{Partitioning the Instruction Stream between the EMC and the Core}
	\label{fig:EMC:milcSplit}
\end{figure}

Figure \ref{fig:EMC:mcfCode} shows one dynamic instance where there are a small number of simple integer operations between the source and dependent miss. I find that this trend holds over the memory intensive applications of \textit{SPEC06}. Figure \ref{fig:EMC:depLen} shows the average number of operations in the dependence chain between a source and dependent miss, if a dependent miss exists. A small number of operations between a source and dependent miss means that the enhanced memory controller (EMC) does not have to do very much work to uncover a cache miss and that it requires a small amount of input data to do so.

I therefore tailor the memory controller to execute dependent chains of operations such as those listed in Figure \ref{fig:EMC:mcfCode}. The added compute capability is described in detail in Section \ref{sec:EMC:trace_uarch}. Since the instructions have already been fetched and decoded at the core and are sitting in the reorder buffer, the core can automatically determine the uops to include in the dependence chain of a cache miss by leveraging the existing out-of-order execution hardware (Section \ref{sec:EMC:trace_gen}). The chain of decoded uops is then sent to the EMC. I refer to the core that generated the dependence chain as the home core.

With this mechanism, a slice of the operations in the ROB are executed at the home core, while others are executed remotely at the EMC. Figure \ref{fig:EMC:milcSplit} provides a high level view of partitioning a sequence of seven instructions from  \textit{mcf} between the EMC and the core. 

In Figure \ref{fig:EMC:milcSplit}, instruction 0 is the first cache miss. Instructions 1 and 2 are independent of instruction 0 and therefore execute on the core while instruction 0 is waiting for memory data. Instructions 3, 4, and 5 are dependent on instruction 0. The core recognizes that instructions 3 and 5 will likely miss in the LLC, i.e., they are dependent cache misses, and so will transmit instructions 3, 4, and 5 to execute at the EMC. When EMC execution completes, R1 and R3 are returned to the core so that execution can continue. To maintain the sequential execution model, operations sent to the EMC are not retired at the EMC, only executed. Retirement state is maintained at the ROB of the home core and physical register data is transmitted back to the core for in-order retirement. 

Once the cache line arrives from DRAM for the original source miss, the chain of dependent uops are executed by the EMC. The details of execution at the EMC are discussed in Section \ref{sec:EMC:trace_exec}. 

\subsection{EMC Compute Microarchitecture}
\label{sec:EMC:trace_uarch}

\begin{figure}
	\centering
	\includegraphics[width=4.0in, height=3.0in]{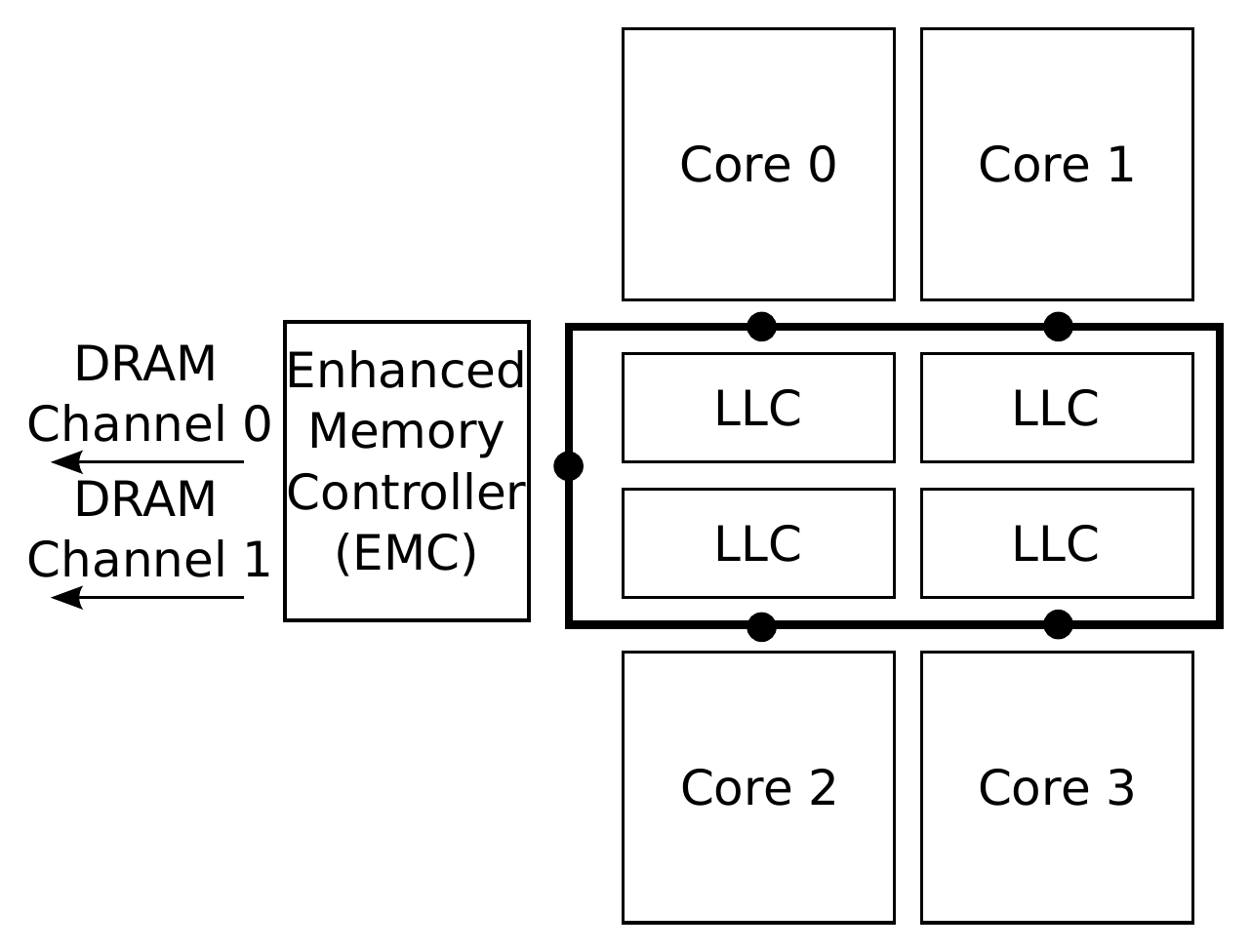}
	\caption{High Level View of a Quad-Core processor with an Enhanced Memory Controller}
	\label{fig:EMC:highLevel}
	\centering
	\includegraphics[width=5.5in, height=4.0in]{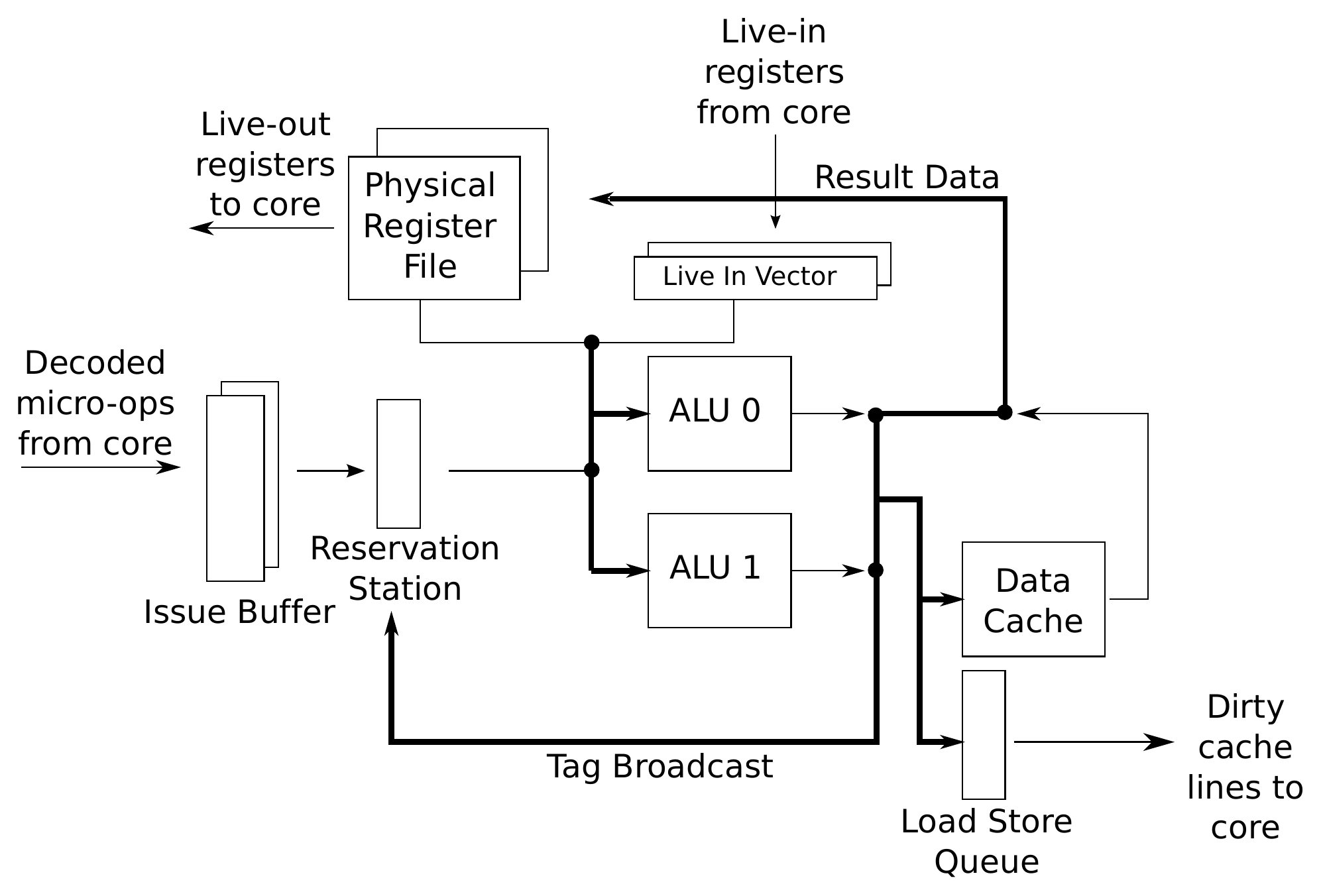}
	\caption{Microarchitecture of the EMC}
	\label{fig:EMC:engine}
\end{figure}

A quad-core multiprocessor that uses the proposed enhanced memory controller is shown in Figure \ref{fig:EMC:highLevel}. The four cores are connected with a bi-directional ring. The memory controller is located at a single ring-stop, along with both memory channels, similar to Intel's Haswell microarchitecture \cite{haswell}. The EMC adds two pieces of hardware to the processor: limited compute capability at the memory controller (described first in Section \ref{sec:EMC:trace_uarch}) and a dependence chain-generation unit at each of the cores (Section \ref{sec:EMC:trace_gen}).

The EMC is designed to have the minimum functionality required to execute the pointer-arithmetic that generates dependent cache misses. Instead of a front-end, the EMC utilizes small uop buffers (Section \ref{sec:EMC:EMC_FE}). For the back-end, the EMC uses 2 ALUs and provide a minimal set of caching and virtual address translation capabilities (Section \ref{sec:EMC:EMC_BE}). Figure \ref{fig:EMC:engine} provides a high level view of the compute microarchitecture that we add to the memory controller.

\subsubsection{Front-End}
\label{sec:EMC:EMC_FE}

The front-end of the EMC consists of two small uop buffers that can each hold a single dependence chain of up to 16 uops. With multiple buffers, the EMC can be shared between the cores of a multi-core processor. The front-end of the EMC consists only of this buffer, it does not contain any fetch, decode, or register rename hardware. Chains of dependent operations are renamed for the EMC using the out-of-order capabilities of the core (Section \ref{sec:EMC:trace_gen}).

\subsubsection{Back-End}
\label{sec:EMC:EMC_BE}

As the EMC is targeting the pointer-arithmetic that generates dependent cache misses, it is limited to executing a subset of the total uops that the core is able to execute. Only integer operations are allowed (Table \ref{tab:EMC:systemConfig}). Floating point and vector operations are not allowed.  This simplifies the microarchitecture of the EMC, and enables the EMC to potentially execute fewer operations to get to the dependent cache miss. The core is creating a \textit{filtered} chain of operations for the EMC to execute, only the operations that are required to generate the address for the dependent cache miss are included in the uop chain.

These filtered dependence chains are issued from the uop buffers to the 2-wide EMC back-end. For maximum performance it is important to exploit the memory level parallelism present in the dependence chains. Therefore, the EMC has the capability to issue non-blocking memory accesses. This requires a small load/store queue along with out-of-order issue and wakeup using a small reservation station and common data bus (CDB). In Figure \ref{fig:EMC:engine} the CDB is denoted by the result and tag broadcast buses. Stores are only executed at the EMC if they are a register spill that is filled later in the dependence chain. Sensitivity to these EMC parameters is shown in Section \ref{sec:EMC:emcSens}.

Each of the issue buffers in the front-end is allocated a private physical register file (PRF) that is 16 registers large and a private live-in source vector. As the out-of-order core has a much larger physical register file than the EMC (256 vs. 16 registers), operations are renamed by the core (Section \ref{sec:EMC:trace_gen}) to be able to execute using the EMC physical register file.

\subsubsection{Caches}

The EMC contains no instruction cache, but it does contain a small data cache that holds the most recent lines that have been transmitted from DRAM to the chip to exploit temporal locality. This requires minimal modifications to the existing cache coherence scheme, as we are simply adding an additional logical first level cache to the system. We add an extra bit to each directory entry for every line at the inclusive LLC to track the cache lines that the EMC holds.

\subsubsection{Virtual Address Translation}

Virtual memory translation at the EMC occurs through a small 32 entry TLB for each core. The TLBs act as a circular buffer and cache the page table entries (PTE) of the last pages accessed by the EMC for each core. The PTEs of the home core add a bit to each TLB entry to track if a page translation is resident in the TLB at the EMC. This bit is used to invalidate TLB entries resident at the EMC during the TLB shootdown process \cite{villavieja:2011}. Before a chain is executed, the core sends the EMC the PTE for the source miss if it is determined not to be resident at the EMC TLB. The EMC does not handle page-faults, if the PTE is not available at the EMC, the EMC halts execution and signals the core to re-execute the entire chain. 

\subsection{Generating Chains of Dependent Micro-Operations}
\label{sec:EMC:trace_gen}

The EMC leverages the out-of-order execution capability of the core to generate the short chains of operations that the EMC executes. This allows the EMC to have no fetch, decode, or rename hardware, as shown in Figure \ref{fig:EMC:engine}, significantly reducing its area and energy consumption. 

The core can generate dependence chains to execute at the EMC once there is a full-window stall due to a LLC miss blocking retirement. If this is the case, a 3-bit saturating counter is used to  determine if a dependent cache miss is likely. This counter is incremented if any LLC miss has a dependent cache miss and decremented if any LLC miss has no dependent cache misses. Dependent cache misses are tracked by poisoning the destination register of a cache miss. Poison values are propagated for 16 operations. When an LLC miss retires with poisoned source registers the counter is incremented, otherwise it is decremented.  If either of the top 2-bits of the saturating counter are set, the core begins the following process of generating a dependence chain for the EMC to accelerate.

\begin{figure*}[t]
	\centering
	\includegraphics[width=\columnwidth, height =2.2in]{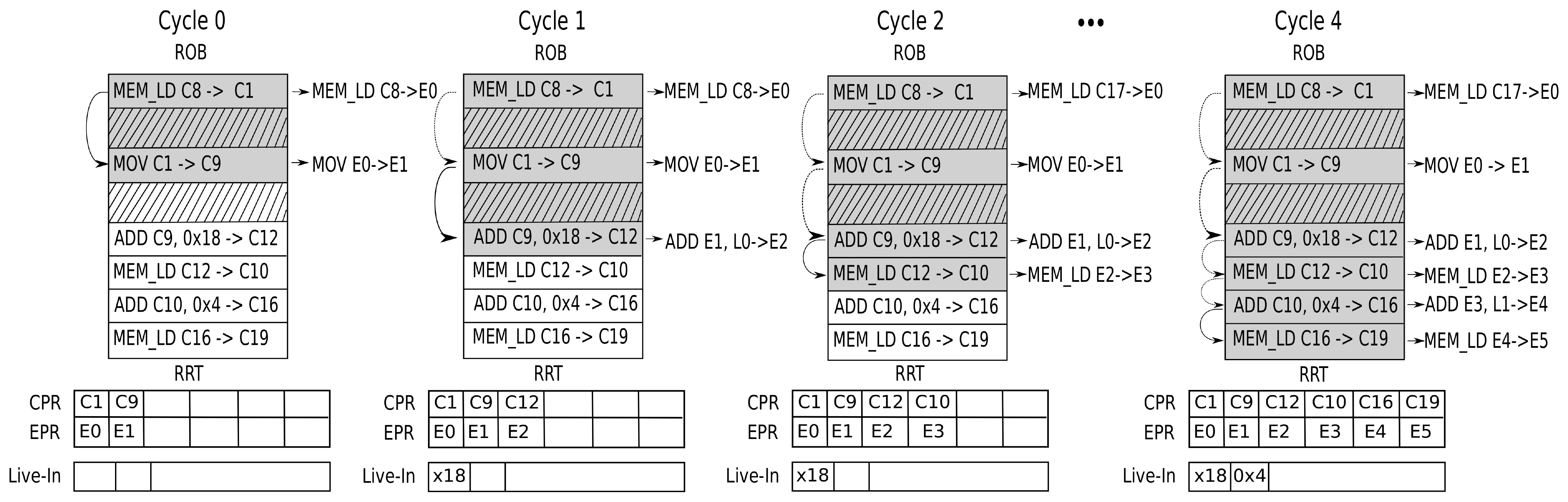}
	\caption{EMC Dependence Chain Generation Example}
	\label{fig:EMC:traceCreation}
\end{figure*}

The dynamic micro-op sequence from Figure \ref{fig:EMC:mcfCode} is used to demonstrate the chain generation process, illustrated by Figure \ref{fig:EMC:traceCreation}. This process takes a variable number of cycles based on dynamic chain length (5 cycles for Figure \ref{fig:EMC:traceCreation}). As the uops are included in the chain, they are stored in a buffer maintained at the core until the entire chain has been assembled. At this point the chain is transmitted to the EMC. 
 
For each cycle three structures are shown in Figure \ref{fig:EMC:traceCreation}. The reorder buffer of the home core (ROB), the register remapping table (RRT), and a live-in source vector. The RRT is functionally similar to a register alias table and maps core physical registers to EMC physical registers. The operations in the chain have to be remapped to a smaller set of physical registers so that the memory controller can execute them. The live-in source vector is a shift register that holds the input data necessary to execute the chain of operations. Only the relevant portion of the ROB is shown. Irrelevant operations are denoted by stripes. Processed operations are shaded after every cycle.

In Figure \ref{fig:EMC:traceCreation} the cycle 0 frame shows the source miss at the top of the ROB. It has been allocated core physical register number 1 (C1) to use as a destination register. This register is remapped to an EMC register using the RRT. EMC physical registers are assigned using a counter that starts at 0 and saturates at the maximum number of physical registers that the EMC contains (16). In the example, C1 is renamed to use the first physical register of the EMC (E0) in the RRT.

Once the source miss has been remapped to EMC physical registers, chains of decoded uops are created using a forward dataflow walk that tracks dependencies through renamed physical registers. The goal is to mark uops that would be ready to execute when the load has completed. Therefore, the load that has caused the cache miss is pseudo ``woken up'' by broadcasting the tag of the destination physical register onto the common data bus (CDB) of the home core. A uop wakes up when the physical register tag of one of its source operands matches the tag that is broadcast on the CDB, and all other source operands are ready. By pseudo waking up the uop it does not execute or commit the uop, it simply broadcasts its destination tag on the CDB. A variable number of uops are broadcast every cycle, up to the back-end width of the home core.

In the example, there is only a single ready uop to broadcast in Cycle 0. The destination register of the source load (C1) is broadcast on the CDB. This wakes up the second operation in the chain, which is a MOV instruction that uses C1 as a source register. It reads the remapped register id from the RRT for C1, and uses E0 as its source register at the EMC. The destination register (C9) is renamed to E1. 

Operations continue to ``wake-up'' dependent operations until either the maximum number of operations in a chain is reached or there are no more operations to awaken. Thus, in the next cycle, the core broadcasts C9 on the CDB. The result of this operation is shown in Cycle 1, an ADD operation is woken up. This operation has two sources, C9 and an immediate value, 0x18. The immediate is shifted into a live-in source vector which will be sent to the EMC along with the chain. The destination register C12 is renamed to E2 and written into the RRT.

In the example, the entire process takes five cycles to complete. In cycle 4, once the final load is added to the chain, a filtered portion of the execution window has been assembled for the EMC to execute. These uops are read out of the instruction window and sent to the EMC for execution along with the live-in vector. Algorithm \ref{alg:EMC:chainGen} describes our mechanism for dynamically generating a filtered chain of dependent uops.

\begin{algorithm}
	\caption{EMC Dependence Chain Generation}
	\label{alg:EMC:chainGen}
	\begin{algorithmic}
		\STATE \textit{Process the source uop at ROB full stall}
		\STATE Allocate EPR for destination CPR of source uop in RRT. 
		\STATE Broadcast destination CPR tag on CDB.
		\FOR {each dependent uop}
		\IF {uop Allowed \textbf{and} (source CPR ready \textbf{or} source CPR in RRT)}
		\STATE \textit{Prepare the dependent uop to be sent to the EMC}
		\FOR {each source operand}
		\IF {CPR ready}
		\STATE Read data from physical register file into live-in vector.
		\ELSE
		\STATE EPR = RRT[CPR]
		\ENDIF
		\ENDFOR
		\STATE Allocate EPR for destination CPR of dependent uop in RRT. 
		\STATE Broadcast destination CPR tag of dependent uop on CDB.
		\ENDIF
		\IF {Total uops in Chain == MAXLENGTH}
		\STATE{break}
		\ENDIF
		\ENDFOR
		\STATE Send filtered chain of uops and live-in vector to the EMC
	\end{algorithmic}
\end{algorithm}

The proposals in this thesis require data from the back-end of an out-of-order processor to be used in unconventional ways and therefore require hardware paths into/out-of back-end structures that do not currently exist. This thesis does not explore the ramifications of these micro-architectural changes on physical layout. For example, while the instruction window may be used to provide the uops for EMC dependence chain generation, it may be simpler to augment to ROB to store all uops until retirement and instead obtain the uops from the ROB. Generally, the proposals in this thesis require new paths into/out-of the ROB (for uops and branch conditions), the PRF (for live-in/live-out data), the LSQ (for EMC memory data) and the TLB (for page translations).

\subsection{EMC Execution}
\label{sec:EMC:trace_exec}

To start execution, the enhanced memory controller (EMC) takes two inputs: the source vector of live-in registers and the executable chain of operations. The EMC does not commit any architectural state, it executes the chain of uops speculatively and sends the destination physical registers back to the core. Two special cases arise with respect to control operations and memory operations. First, I discuss control operations.

The EMC does not fetch instructions and is sent the branch predicted stream that has been fetched in the ROB. We send branch directions along with computation to the EMC so that the EMC does not generate wrong path memory requests if it is on the wrong path. The EMC evaluates each condition and determines if the chain that it was sent to execute contains the correct path of execution. If the EMC realizes it is on the wrong-path, execution is stopped and the core is notified of the mis-predicted branch. 

For memory operations, a load first queries the EMC data cache, if it misses in the data cache it generates an LLC request. The EMC has the ability to predict if any given load is going to result in a cache miss. This enables the EMC to directly issue the request to memory if it is predicted to miss in the cache, thus saving the latency to access the on-chip cache hierarchy. To enable this capability we keep an array of 3-bit counters for each core, similar to \cite{qureshi:2012, Yoon:2012}. The PC of the miss causing instruction is used to hash into the array. On a miss the corresponding counter is incremented, a hit decrements the counter. If the counter is above a threshold the request is sent directly to memory. 

Stores are only included in the dependence chain by the home core if the store is a register spill. This is determined by searching the home core LSQ for a corresponding load with the same address (fill) during dependence chain generation. A store executed at the EMC writes its value into the EMC LSQ. 

Loads and stores are retired in program order back at the home core. Every load or store executed at the EMC sends a message on the address ring back to the core. The core snoops this request and populates the relevant entry in the LSQ. This serves two purposes. First, if a memory disambiguation problem arises, for example if there is a store to the same address as a load executed at the EMC in program order at the core, execution of the chain can be canceled. Second, for consistency reasons, stores executed at the EMC are not made globally observable until the store has been drained from the home core store-queue in program order. In our evaluation, we only allow the EMC to execute a dependence chain while the core is already stalled. This prevents these disambiguation scenarios from occurring when the EMC is executing a dependence chain. While this simplifies the execution model, it is not required for the EMC to correctly function.

Executing chains of instructions remotely requires these modifications to the core. However, transactional memory implementations that are built into current hardware \cite{hsw:tsx} provide many similar guarantees for memory ordering. Remote execution at the EMC is simpler than a transaction, as there is no chance for a conflict or rollback due to simultaneous execution. As the chains that are executed at the EMC are very short, the overhead of sending messages back to the core for all loads/stores is smaller than a transaction signature \cite{Ceze:2006}.  

Once each dependence chain has completed execution, the live-outs, including the store data from the LSQ, are sent back to the core. Physical register tags are broadcast on the CDB, and execution on the main core continues. As the home core maintains all instruction state for in-order retirement, any bad-event (branch misprediction, EMC TLB-miss, EMC exception) causes the home core to re-issue and execute the entire chain normally.

\section{Methodology}
\label{sec:EMC:method}

The multi-core chip used to evaluate the EMC consists of four cores that are each identical to those used in the single core evaluation of Chapter \ref{chap:raBuf}. The details of the system configuration are listed in Table \ref{tab:EMC:systemConfig}. The cache hierarchy of each core contains a 32KB instruction cache and a 32KB data cache. The LLC is divided into 1MB cache slices per core. The interconnect is composed of two bi-directional rings, a control ring and a data ring. Each core has a ring-stop that is shared with the LLC slice. Each core can can access the LLC slice at its own ring stop without getting onto the ring (using a bypass) to not overstate ring contention. 
 
\begin{table}
 	\small
 	\centering
 	\begin{tabular}{|p{.75in}|p{4.5in}|}
 		\hline Core & 4-Wide Issue, 256 Entry ROB, 92 Entry Reservation Station, Hybrid Branch Predictor, 3.2 GHz Clock Rate \\ 
 		\hline L1 Caches & 32 KB I-Cache, 32 KB D-Cache, 64 Byte Lines, 2 Ports, 3 Cycle Latency, 8-way, Write-Through. \\ 
 		\hline \hline L2 Cache &  1MB 8-way, 18-cycle latency, Write-Back. \\ 
 		\hline \hline EMC \newline Compute & 2-wide issue. 8 Entry Reservation Stations. 4KB Cache 4-way, 2-cycle access, 1-port. 1 Runahead dependence chain context with 32 entry uop buffer, 32 entry physical register file. 1 Dependent cache miss context with 16 entry uop buffer, 16 entry physical register file. Micro-op size: 8 bytes in addition to any live-in source data. \\
 		\hline EMC \newline Instructions & Integer: add/subtract/move/load/store. \newline \hbox{Logical: and/or/xor/not/shift/sign-extend.} \\
 		\hline Memory Controller & Batch Scheduling \cite{mut:mos08}. 128 Entry Memory Queue.\\ 
 		\hline Prefetchers & Stream: 32 Streams, Distance 32. Markov: 1MB Correlation Table, 4 addresses per entry. GHB G/DC: 1k Entry Buffer, 12KB total size. All configurations: FDP \cite{fdp07}, Dynamic Degree: 1-32, prefetch into Last Level Cache. \\  
 		\hline DRAM & DDR3\cite{dram:micron}, 1 Rank of 8 Banks/Channel, 2 Channels, 8KB Row-Size, CAS 13.75ns.  CAS = $t_{RP}$ = $t_{RCD}$ = CL. Other modeled DDR3 constraints: BL, CWL, $t_{RC, RAS, RTP, CCD, RRD, FAW, WTR, WR}$. 800 MHz Bus, Width: 8 B.  \\
 		\hline
 	\end{tabular} 
 	\caption{Multi-core System Configuration}
 	\label{tab:EMC:systemConfig}
 \end{table}

The baseline memory controller uses a sophisticated multi-core scheduling algorithm, batch scheduling \cite{mut:mos08}, and Feedback Directed Prefetching (FDP) \cite{fdp07} to throttle prefetchers. The parameters for the EMC listed in Table \ref{tab:EMC:systemConfig} (TLB size, cache size, number/size of contexts) have been chosen via sensitivity analysis. This analysis is shown in Section \ref{sec:EMC:emcSens}. 

I use the \textit{SPEC06} application classification from Table \ref{tab:RAB:workloadClass} to randomly generate two sets of ten quad-core workloads (Table \ref{tab:EMC:workloadChoices}). Each benchmark only appears once in every workload combination. As the EMC is primarily intended to accelerate memory intensive applications, the focus is on high memory intensity workloads in this evaluation. The first set of workloads is numbered H1-H10 and consists of four high memory intensity applications. M11-M15 consist of 2 high intensity applications and 2 medium intensity applications. L16-L20 consist of 2 high intensity applications and 2 low intensity application. In addition to these workloads, the evaluation also shows results for a set of workloads that consist of four copies of each of the high and medium memory intensity benchmarks in Table \ref{tab:RAB:workloadClass}. These workloads are referred to as the Copy workloads.
  
\begin{table}
 	\small
 	\centering
 	\begin{tabular}{|p{.25in}|p{2.25in}||p{.25in}|p{2.25in}|}
 		\hline H1 & bwaves+lbm+milc+omnetpp & M11 & soplex+gems+wrf+mcf \\ 
 		\hline H2 & soplex+omnetpp+bwaves+libq & M12 &  milc+zeusmp+lbm+cactus  \\ 
 		\hline H3 & sphinx3+mcf+omnetpp+milc & M13 & gems+wrf+mcf+omnetpp \\
 		\hline H4 & mcf+sphinx3+soplex+libq  & M14 &  cactus+gems+soplex+sphinx3  \\ 
 		\hline H5 & lbm+mcf+libq+bwaves & M15 &  libq+leslie3d+wrf+lbm \\ 
 		\hline H6 & lbm+soplex+mcf+milc& L16 & h264ref+lbm+omnetpp+povray \\  
 		\hline H7 & bwaves+libq+sphinx3+omnetpp & L17 & tonto+sphinx3+sjeng+mcf \\ 
 		\hline H8 & omnetpp+soplex+mcf+bwaves & L18 & bzip2+namd+mcf+sphinx3 \\ 
 		\hline H9 & lbm+mcf+libq+soplex & L19 & omnetpp+soplex+namd+xalanc \\
 		\hline H10 & libq+bwaves+soplex+omentpp & L20 & soplex+bwaves+bzip2+perlbench \\ 
 		\hline
 	\end{tabular} 
 	\caption{Multi-Core Workloads}
 	\label{tab:EMC:workloadChoices}
\end{table}

Chip energy is modeled using McPAT 1.3 \cite{li:micro09} and DRAM power is modeled using CACTI \cite{Muralimanohar_cacti6.0}. Static power of shared structures is dissipated until the completion of the entire workload. Dynamic counters stop updating upon each benchmark's completion. The EMC is modeled as a stripped down core and does not contain structures like an instruction cache, decode stage, register renaming hardware, or a floating point pipeline.  

The chain generation unit is modeled by adding the following additional energy events corresponding to the chain generation process at each home core. Each of the uops included in the chain requires an extra CDB access (tag broadcast) due to the pseudo wake-up process. Each of the source operations in every uop require a Register Remapping table (RRT) lookup, and each destination register requires a RRT write since the chain is renamed to the set of physical registers at the EMC. Each operation in the chain requires an additional ROB read when it is transmitted to the EMC. Data and instruction transfer overhead to/from the EMC is taken into account via additional messages sent on the ring.

Using this simulation methodology, Table \ref{tab:EMC:IPC} lists the raw baseline IPCs for the simulated high intensity workloads. Table \ref{tab:EMC:BW} lists the raw DRAM bandwidth consumption (GB/S) and average chip power consumption (W).

\begin{table}
	\small
	\centering
	\begin{tabular}{|p{.3in}|c|c|c|c||p{.6in}|c|c|c|c|}
		\hline & Core0 & Core1 & Core2 & Core3 & & Core0 & Core1 & Core2 & Core3 \\
		\hline
		\hline H1 & 0.43 &	0.52 &	0.64 &	0.44 & 4xzeus & 1.37 &	1.37 &	1.36 &	1.36 \\ 
		\hline H2 & 0.4 &	0.48 &	0.41 &	0.25 & 4xcactus &  0.7 &	0.65 &	0.65 &	0.65  \\ 
		\hline H3 & 0.5 &	0.22 &	0.53 &	0.76 & 4xwrf & 1.37 &	1.33 &	1.25 &	1.24 \\
		\hline H4 & 0.21 &	0.4 &	0.37 &	0.26  & 4xgems &  0.65 &	0.59 &	0.58 &	0.56  \\ 
		\hline H5 & 0.45 &	0.17 &	0.18 &	0.32 & 4xleslie &  0.9 &	0.84 &	0.8 &	0.79 \\ 
		\hline H6 & 0.48 &	0.35 &	0.18 &	0.58 & 4xoment & 0.52 &	0.49 &	0.48 &	0.47 \\  
		\hline H7 & 0.43 &	0.25 &	0.41 &	0.48 & 4xmilc & 0.87 &	0.84 &	0.84 &	0.84 \\ 
		\hline H8 & 0.47 &	0.52 &	0.48 &	0.21 & 4xsoplex & 0.52 &	0.52 &	0.51 &	0.51 \\ 
		\hline H9 & 0.42 &	0.17 &	0.18 &	0.31 & 4xsphinx & 0.6 &	0.57 &	0.56 &	0.56 \\
		\hline H10 & 0.27 &	0.4 &	0.38 &	0.44 & 4xbwaves & 0.7 &	0.68 &	0.66 &	0.66 \\ 
		\hline  & & & & & 4xlibq & 0.31 & 0.31 &	0.31 &	0.31 \\ 
		\hline  & & & & & 4xlbm & 0.33 &	0.33 &	0.32 &	0.32\\ 
		\hline  & & & & & 4xmcf & 0.19 &	0.19 &	0.19 &	0.18  \\ 
		\hline
	\end{tabular} 
	\caption{Multi-Core Workload IPC}
	\label{tab:EMC:IPC}
\end{table}

\begin{table}
	\small
	\centering
	\begin{tabular}{|p{.3in}|c|c||p{.6in}|c|c|}
		\hline & BW (GB/S) & Power (W) & & BW (GB/S) & Power (W) \\
		\hline
		\hline H1 & 10.3 & 118.5 & 4xzeus & 4.7 & 130.6 \\ 
		\hline H2 & 5.6 & 118.8 & 4xcactus & 2.5 & 122.4  \\ 
		\hline H3 & 3.7 & 121.4 & 4xwrf & 4.5 & 128.2 \\
		\hline H4 & 8.4 & 116.3 & 4xgems & 5.9 & 122.4 \\ 
		\hline H5 & 3.6 & 120.6 & 4xleslie & 7.4 & 127.3 \\ 
		\hline H6 & 3.5 & 121.2 & 4xoment & 6.6 & 121.1 \\  
		\hline H7 & 5.1 & 119.0 & 4xmilc & 10.8 & 123.0 \\ 
		\hline H8 & 3.9 & 121.3 & 4xsoplex &  9.4 & 122.3 \\ 
		\hline H9 & 3.8 & 120.6 & 4xsphinx & 8.3 & 121.6 \\
		\hline H10 & 8.9 & 116.5 & 4xbwaves & 13.2 & 122.2 \\ 
		\hline  & & & 4xlibq & 15.6 & 118.5 \\ 
		\hline  & & & 4xlbm & 10.5 & 117.3 \\ 
		\hline  & & & 4xmcf &  6.9 & 117.8 \\ 
		\hline
	\end{tabular} 
	\caption{Multi-Core Workload Memory Bandwidth (GB/S) and Power (W)}
	\label{tab:EMC:BW}
\end{table}

\ignore{
\begin{table}
	\small
	\centering
	\begin{tabular}{|p{.5in}|c||p{.6in}|c|}
		\hline H1 & 118.5 & 4xzeus & 130.6 \\ 
		\hline H2 & 118.8 & 4xcactus & 122.4  \\ 
		\hline H3 & 121.4 & 4xwrf & 128.2 \\
		\hline H4 & 116.3 & 4xgems & 122.4 \\ 
		\hline H5 & 120.6 & 4xleslie & 127.3 \\ 
		\hline H6 & 121.2 & 4xoment & 121.1 \\  
		\hline H7 & 119.0 & 4xmilc & 123.0 \\ 
		\hline H8 & 121.3 & 4xsoplex &  122.3 \\ 
		\hline H9 & 120.6 & 4xsphinx & 121.6 \\
		\hline H10 & 116.5 & 4xbwaves & 122.2 \\ 
		\hline  & & 4xlibq & 118.5 \\ 
		\hline  & & 4xlbm & 117.3 \\ 
		\hline  & & 4xmcf &  117.8 \\ 
		\hline
	\end{tabular} 
	\caption{Multi-Core Workload Runtime Power Consumption (W)}
	\label{tab:EMC:Power}
\end{table}
}

\newpage

\section{Results}
\label{sec:EMC:result}

Instead of using IPC to evaluate the performance of a multi-core system \cite{eyerman2008system}, I use a system level metric, weighted speedup \cite{sna:tul00}: 

\begin{equation}
\displaystyle Wspeedup = \sum\limits_{i=0}^{n-1} \frac{IPC^{shared}_{i}}{IPC^{alone}_{i}}
\label{eqn:EMC:wspeedup}
\end{equation}

\begin{figure}
	\centering
	\includegraphics[width=\columnwidth]{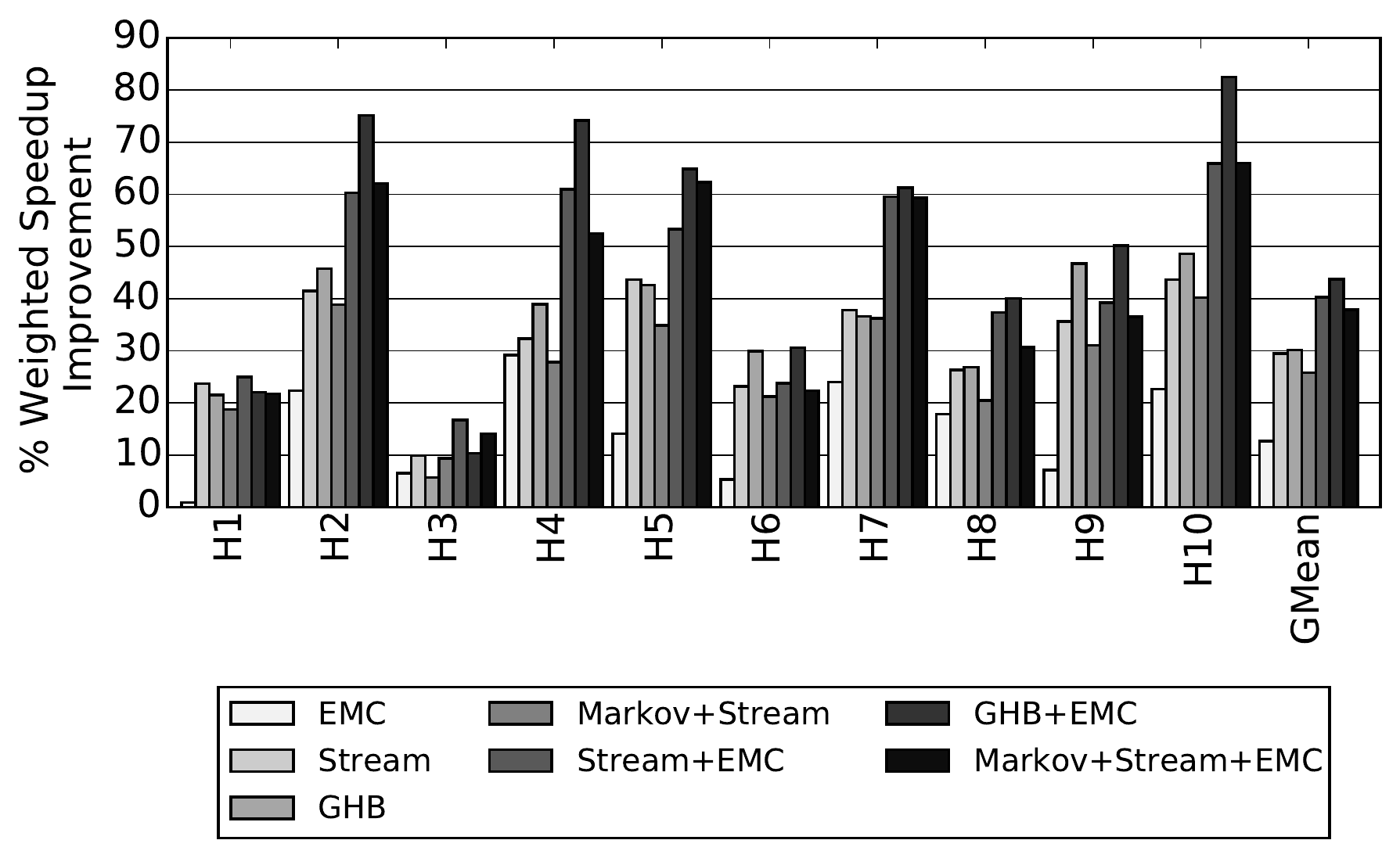}
	\caption{Quad-Core Performance for H1-H10}
	\label{fig:EMC:highPerf}
	\centering
	\includegraphics[width=\columnwidth]{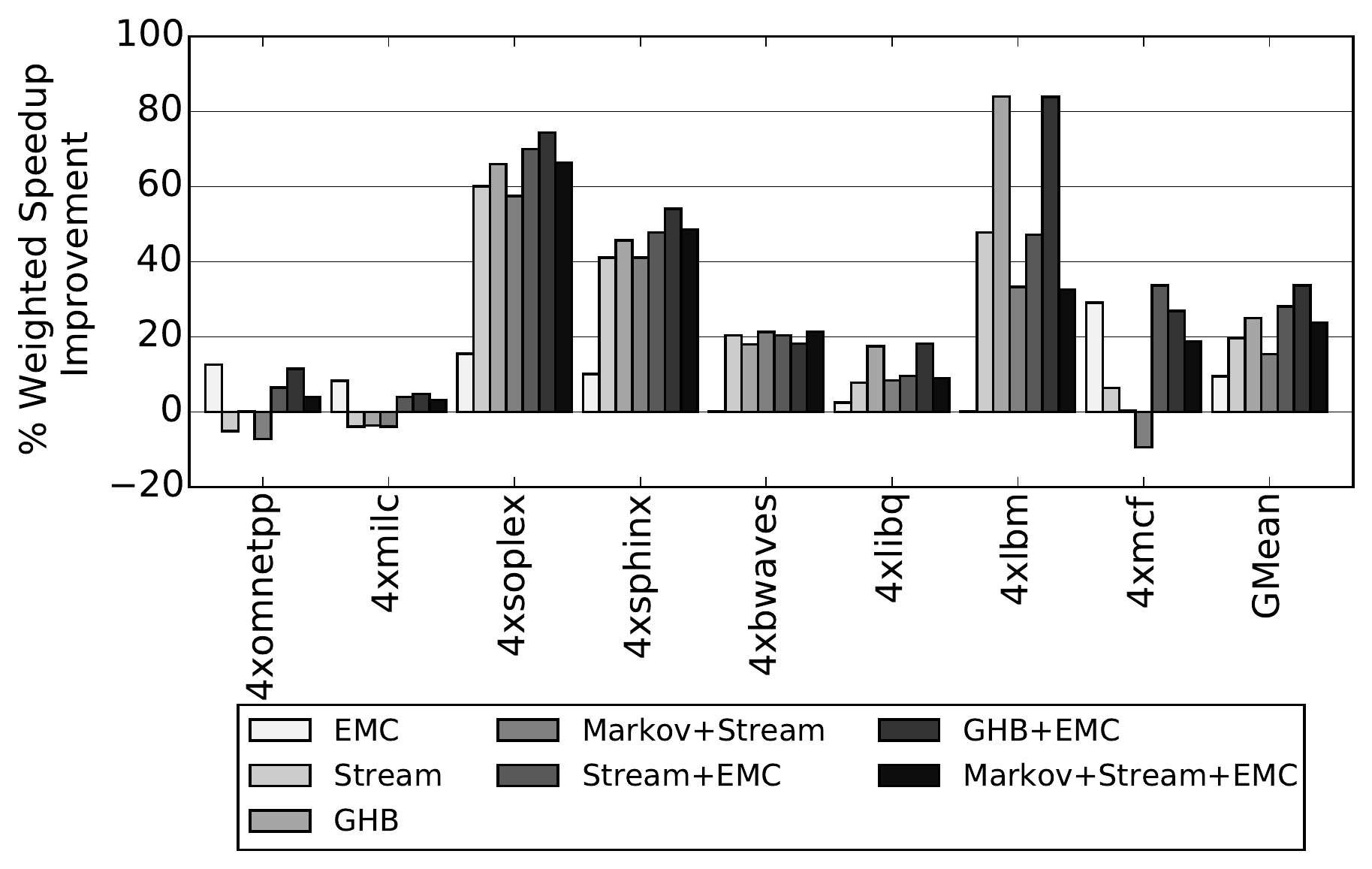}
	\caption{Quad-Core Performance for the Copy Workloads}
	\label{fig:EMC:copyPerf}
\end{figure}

The performance of the quad-core system for the workloads listed in Table \ref{tab:EMC:workloadChoices} is shown in Figure \ref{fig:EMC:highPerf}. 

On the memory intensive workloads, the EMC improves performance on average by 15\% over a no-prefetching baseline, by 10\% over a baseline with stream prefetching, 13\% over a baseline with a GHB prefetcher and by 11\% over a baseline with both a stream and Markov prefetcher. Workloads that include a SPEC2006 benchmark with a high rate of dependent cache misses (Figure \ref{fig:intro:depMiss}) such as \textit{mcf} or \textit{omnetpp} tend to perform well, especially when paired with other highly memory intensive workloads like \textit{libquantum} or \textit{bwaves}. Workloads with \textit{lbm} tend not to perform well: \textit{lbm} contains essentially zero dependent cache misses and has a regular access pattern that utilizes most of the available bandwidth, particularly with prefetching enabled, making it difficult for the EMC to satisfy latency-critical requests. 

To isolate the performance implications of the EMC, Figure \ref{fig:EMC:copyPerf} shows a system running four copies of each high memory intensity \textit{SPEC06} benchmark.

The workloads in Figure \ref{fig:EMC:copyPerf} are sorted from lowest to highest memory intensity. Overall, the EMC results in a 9.5\% performance advantage over a no-prefetching baseline and roughly 8\% over each prefetcher. The highest performance gain is on \textit{mcf}, at 30\% over a no-prefetching baseline. All of the benchmarks with a high rate of dependent cache misses show performance improvements with an EMC. These applications also generally observe overall performance degradations when prefetching is employed.

\begin{figure}
	\centering
	\includegraphics[width=\columnwidth]{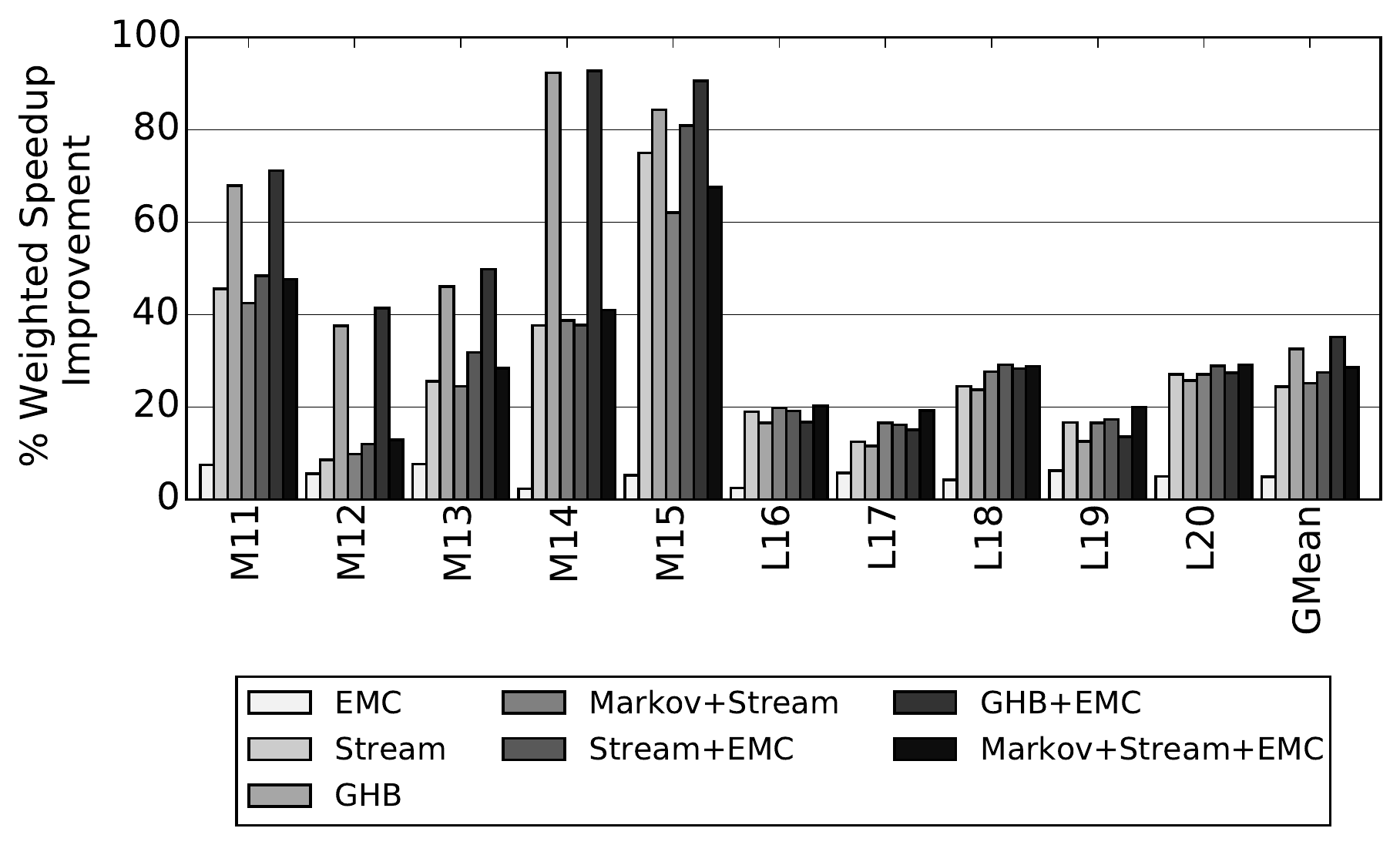}
	\caption{Quad-Core Performance for M11-L20}
	\label{fig:EMC:mixPerf}
\end{figure}

Lastly Figure \ref{fig:EMC:mixPerf} shows the results of the EMC on the M11-L20 workload suite. As this workload suite has lower memory intensity, smaller performance gains are expected. The largest EMC gains are 7.5\% on workloads M11 and M13, which both contain \textit{mcf}. Overall, the EMC results in a 4.9\% performance gain over the no-prefetching baseline and 3.0\% when combined with each of the three prefetchers.

\subsection{Performance Analysis}
\label{sec:EMC:benefitEval}

\begin{figure}
	\centering
	\includegraphics[width=\columnwidth]{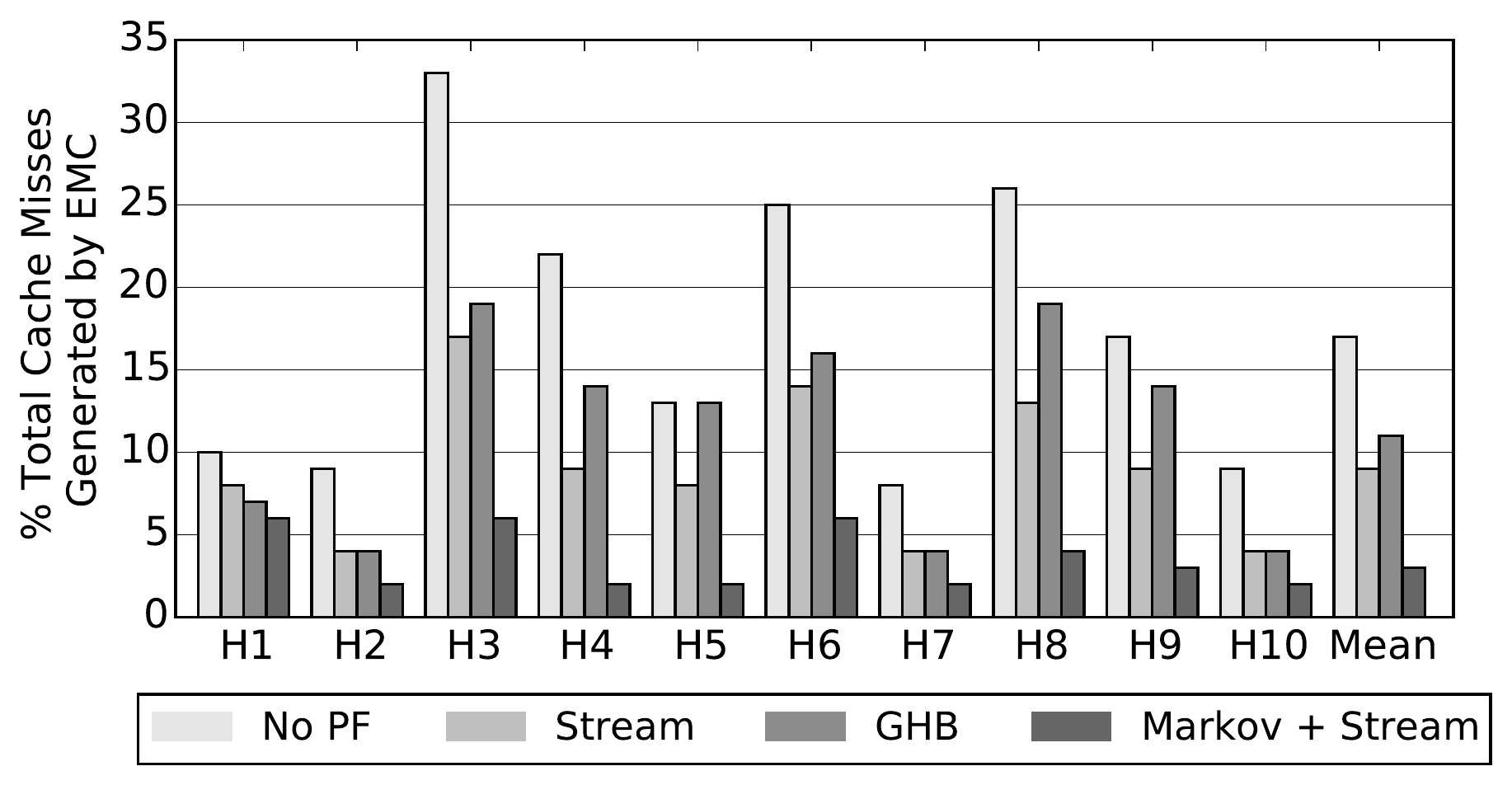}
	\caption{Fraction of Total Cache Misses Generated by the EMC for H1 - H10}
	\label{fig:EMC:tot-GenReq}
\end{figure}

To examine the reasons behind the performance benefit of the EMC, I contrast workload H1 (1\% performance gain) and workload H4 (33\% performance gain). While there is no single indicator for the performance improvement that the EMC provides, I identify three statistics that correlate to increased performance. First, Figure \ref{fig:EMC:tot-GenReq} shows the percentage of total cache misses that the EMC generates. As workloads H1 and H4 are both memory intensive workloads, the EMC generating a larger percentage of the total cache misses indicates that its latency reduction features result in a larger impact on workload performance. The EMC generates about 10\% of all of the cache misses in H1 and 22\% of the misses in H4. The Markov + Stream PF configuration generates 25\% more memory requests than any other configuration on average, diminishing the impact of the EMC in Figure \ref{fig:EMC:tot-GenReq} and one reason for lower relative performance.  

Second, the EMC should produce a reduction in DRAM contention for requests issued by the EMC. As requests are generated and issued to memory faster than in the baseline, a request can reach an open DRAM row before the row can be closed by a competing request from a different core. This results in a reduction in row-buffer conflicts. There are two different scenarios where this occurs. First, the EMC can issue a dependent request that hits in the same row-buffer as the original request. Second, multiple dependent requests to the same row-buffer are issued together and can coalesce into a batch. I observe that the first scenario occurs about 15\% of the time while the second scenario is more common, occurring about 85\% of the time on average.

\begin{figure}
	\centering
	\includegraphics[width=\columnwidth]{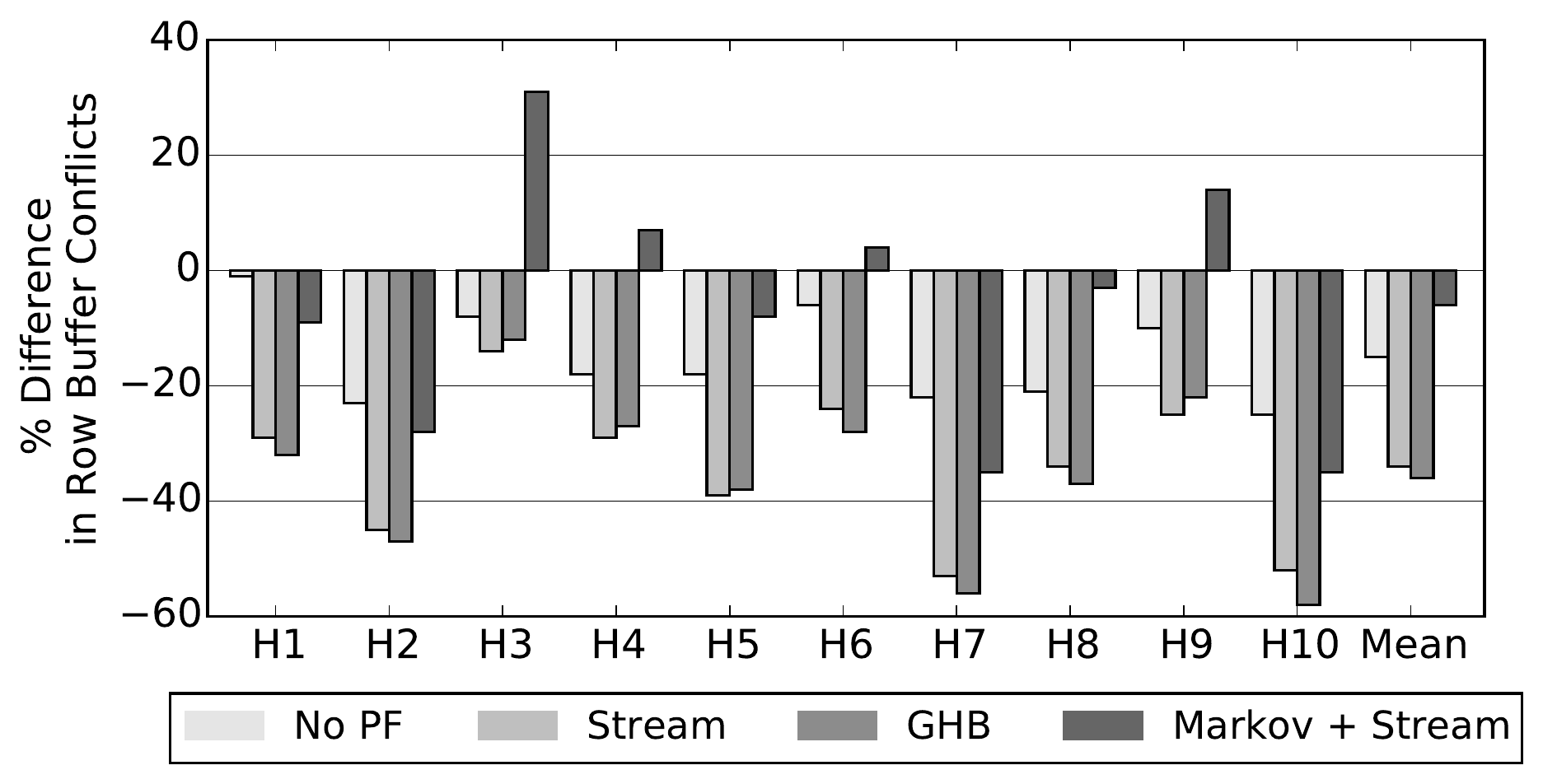}
	\caption{Difference in Row-Buffer Conflict Rate for H1-H10}
	\label{fig:EMC:tot-RowBuffer}
\end{figure}

Figure \ref{fig:EMC:tot-RowBuffer} shows the difference in row-buffer conflict reduction. This statistic strongly correlates to how much latency reduction the EMC achieves, as the latency for a row-buffer conflict is much higher than the latency of a row-buffer hit. For example, the reduction in H1 is less than 1\%. This is much smaller than the 19\% reduction exhibited by H4 (and the 23\% reduction in H2, a workload that has other indicators that are very similar to H1).

\begin{figure}
	\centering
	\includegraphics[width=\columnwidth]{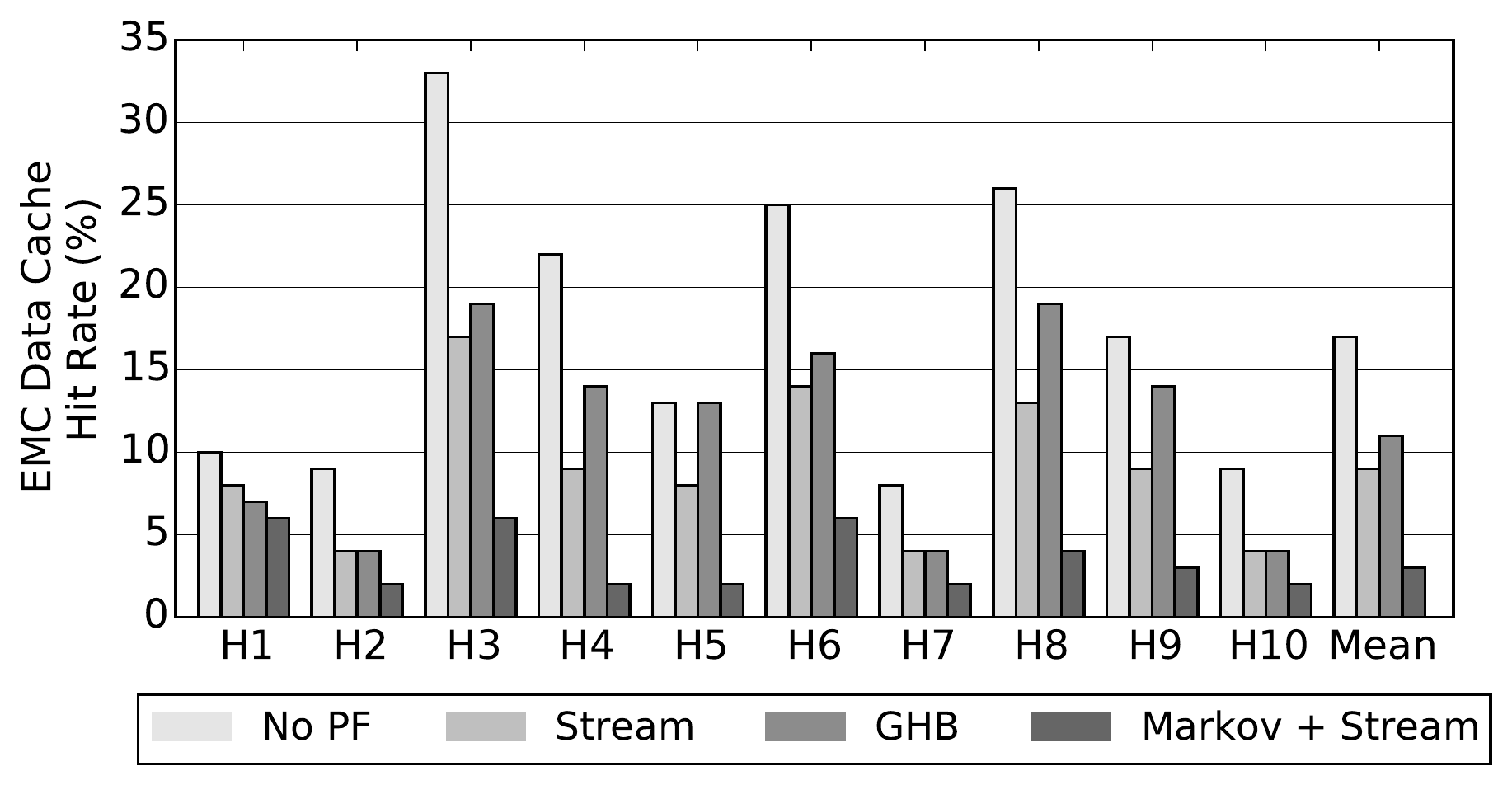}
	\caption{Data Cache Hit Rate at the EMC.}
	\label{fig:EMC:tot-HitRate}
\end{figure}

Between these two factors, the percent of total cache misses generated by the EMC and the reduction in row-buffer conflicts, it is clear that the EMC has a much smaller impact on performance in workload H1 than workload H4. One other factor is also important to note. The EMC exploits temporal locality in the memory access stream with a small data cache. If the dependence chain executing at the EMC contains a load to data that has recently entered the chip, this will result in a very short-latency cache hit instead of an LLC lookup. Figure \ref{fig:EMC:tot-HitRate} shows that Workload H1 has a much smaller hit rate in the EMC data cache than Workload H4.

These three statistics: the fraction of total cache misses generated by the EMC, the reduction in row-buffer conflict rate, and the EMC data cache hit rate are indicators that demonstrate why the performance gain in Workload H4 is much more significant than the performance gain in Workload H1.  

\begin{figure}
	\centering
	\includegraphics[width=\columnwidth]{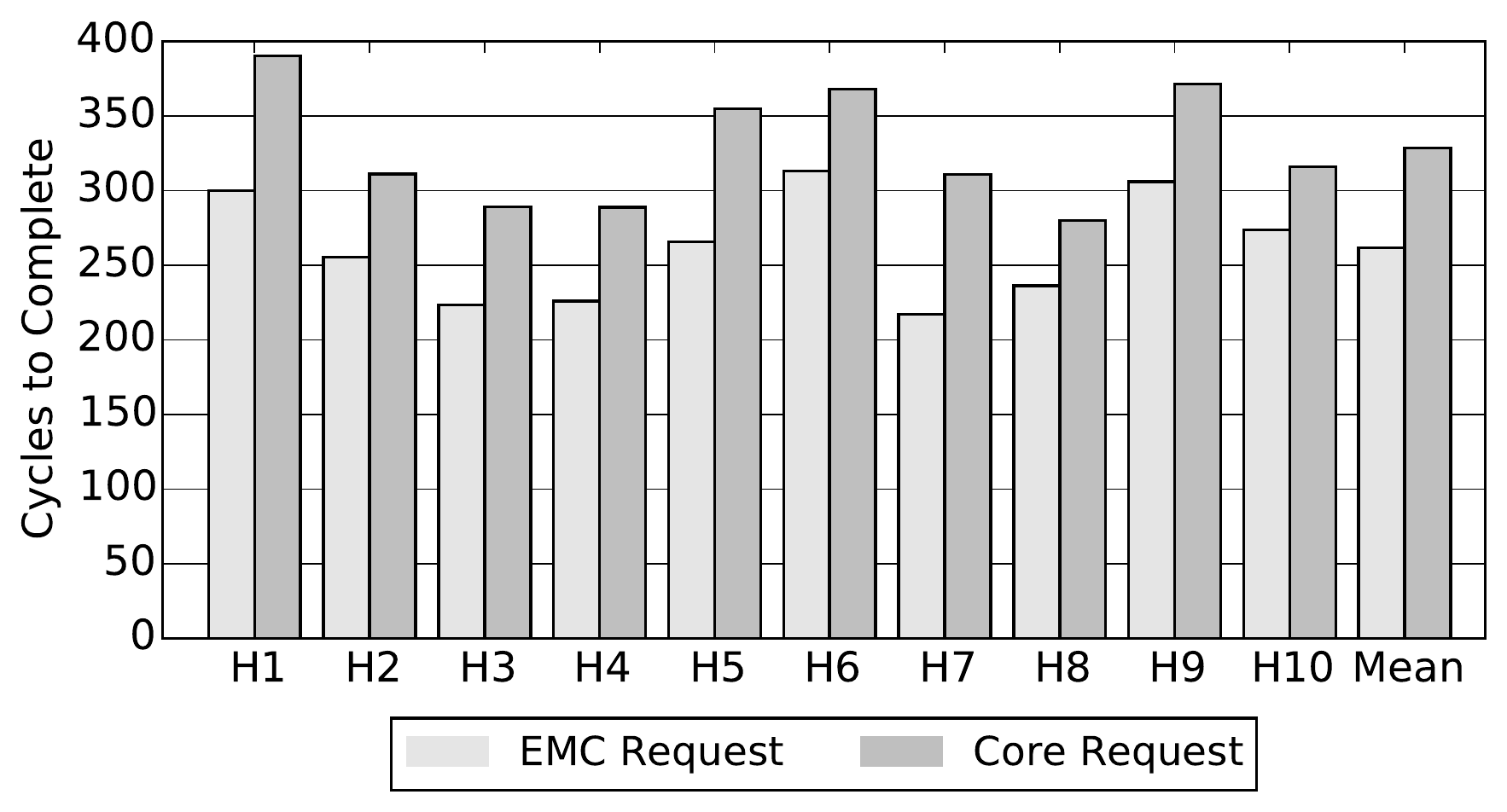}
	\caption{EMC Cache Miss Latency vs Core Cache Miss Latency}
	\label{fig:EMC:latRed}
\end{figure}

The net result of the EMC is a raw latency difference for cache misses that are generated by the EMC and cache misses that are generated by the core. This is shown in Figure \ref{fig:EMC:latRed}. Latency is given in cycles observed by the miss before dependent operations can be executed and is inclusive of accessing the LLC, interconnect, and DRAM. On average, a cache miss generated by the EMC observes a 20\% lower latency than a cache miss generated by the core. 

The critical path to executing a dependent cache miss includes three areas where the EMC saves latency. First, in the baseline, the source cache miss is required to go through the fill path back to the core before dependent operations can be woken up and executed. Second, the dependent cache miss must go through the on-chip cache hierarchy and interconnect before it can be sent to the memory controller. Third, the request must be selected by the memory controller to be issued to DRAM.

\begin{figure}
	\centering
	\includegraphics[width=\columnwidth]{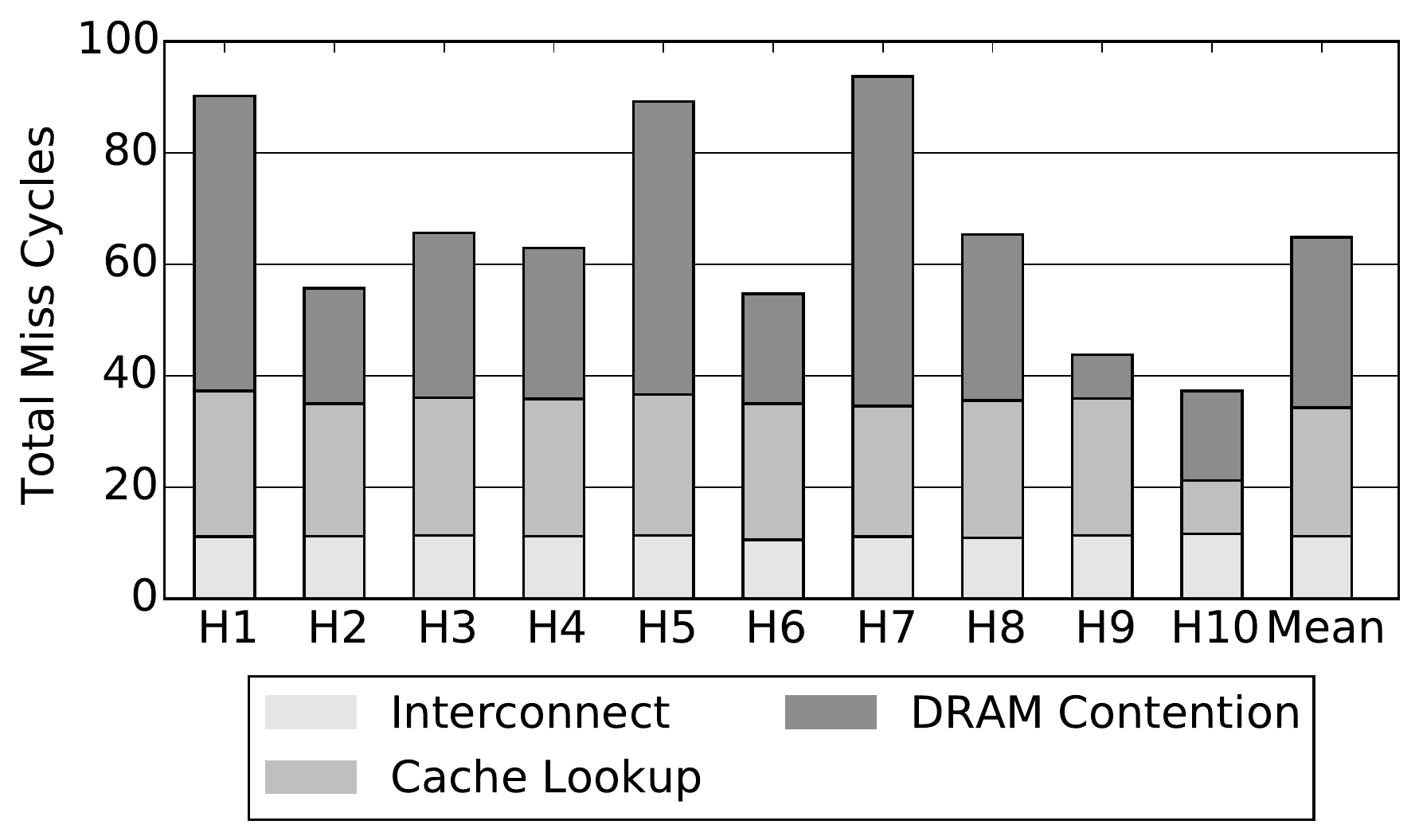}
	\caption{Average Number of Cycles Saved by the EMC on each Memory Request}
	\label{fig:EMC:perfGain}
\end{figure}

I attribute the latency reduction of requests issued by the EMC in Figure \ref{fig:EMC:latRed} to these three sources: bypassing the interconnect back to the core, bypassing cache accesses, and reduced contention at the memory controller. The average number of cycles saved by each of these factors are shown in Figure \ref{fig:EMC:perfGain}.

Overall, the interconnect savings for requests issued by the EMC is about 11 cycles on average. I observe a 20 cycle average reduction in cache access latency and a 30 cycle reduction in DRAM contention. The reduction in average cache access latency is due to issuing predicted cache misses directly to the memory controller. While we utilize a miss predictor to decide when to send misses to the memory controller, this miss predictor acts as a bandwidth filter to reduce the off-chip bandwidth cost of the EMC. Removing the miss predictor and issuing all EMC loads to main memory results in a 5\% average increase in system bandwidth consumption. 

\subsection{Prefetching and the EMC}
\label{sec:EMC:prefEval}

This section discusses the interaction between the EMC and prefetching when they are employed together. Figure \ref{fig:EMC:tot-GenReq} shows that the fraction of total cache misses that are generated by the EMC with prefetching is, on average, about 2/3 of the fraction of total cache misses generated without prefetching. However, the total number of memory requests is different between the prefetching and the non-prefetching case. This is because the prefetcher generates many memory requests, some requests are useful while others are useless. Thus, the impact of prefetching on the EMC is more accurately illustrated by considering how many fewer cache misses the EMC generates when prefetching is on versus when prefetching is off. This fraction is shown below in Figure \ref{fig:EMC:pf-loss}.

\begin{figure}
	\centering
	\includegraphics[width=\columnwidth]{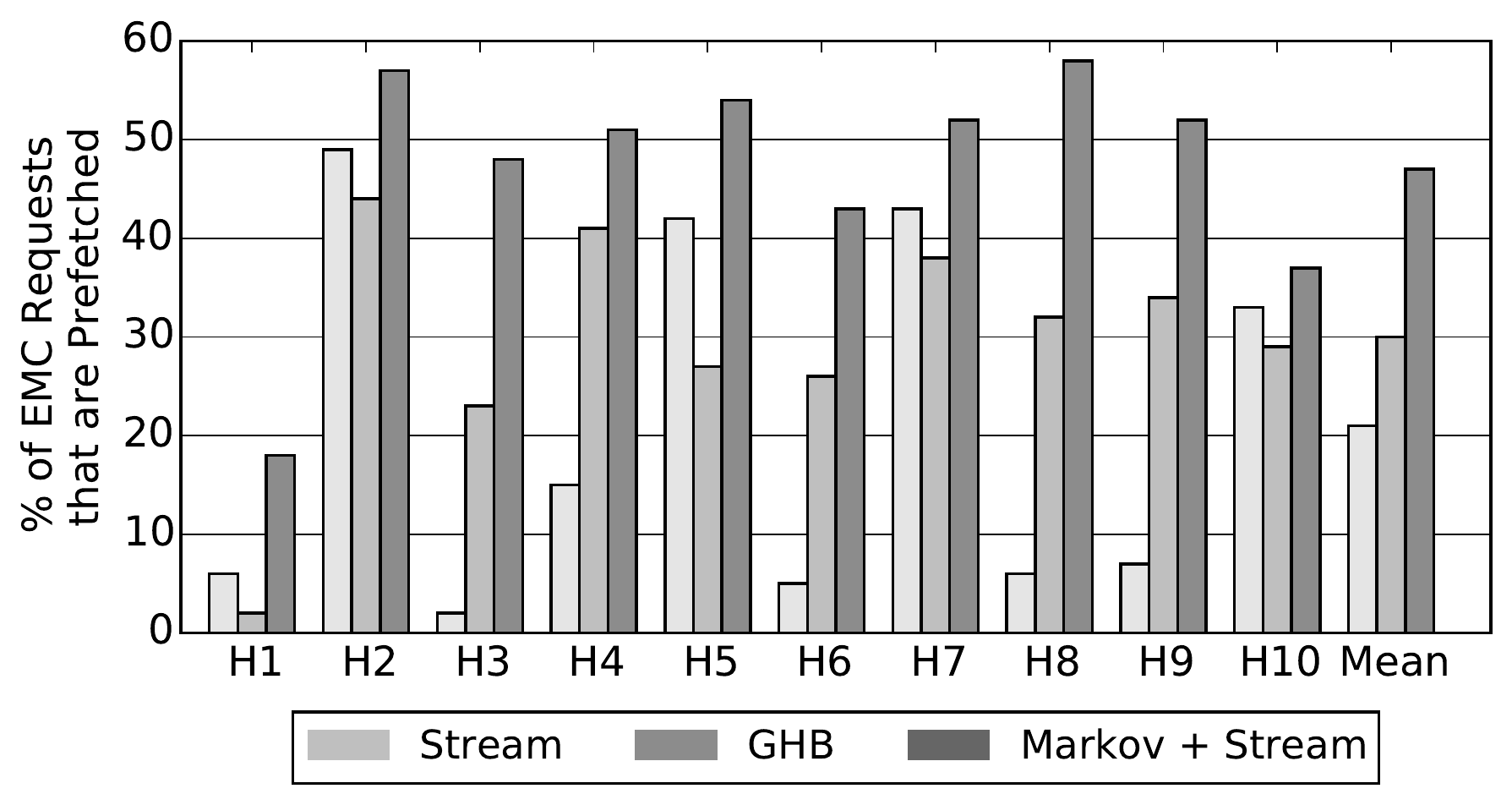}
	\caption{Effect of Prefetching on EMC Memory Requests}
	\label{fig:EMC:pf-loss}
\end{figure}

On average, the Stream/GHB/Markov+Stream prefetchers can prefetch about 21\%, 30\%, 48\% of the requests that the EMC issued in the non-prefetching case respectively. This shows that prefetching does diminish the benefit of the EMC to some extent, but the EMC also supplements the prefetcher by reducing the latency to access memory addresses that the prefetcher can not predict ahead of time. 

\subsection{Sensitivity to EMC Parameters}
\label{sec:EMC:emcSens}

The EMC is tailored to have the minimum functionality that is required to execute short chains of dependent operations. This requires making many design decisions as to the specific parameters listed in Table \ref{tab:EMC:systemConfig}. The sensitivity analysis used to make these decisions is discussed in this section.

First, Table \ref{tab:EMC:emcSens} shows sensitivity to the main micro-architectural parameters: issue width, data cache size, dependence chain buffer length, number of dependence chains contexts, and EMC TLB size. Performance is given as the average geometric mean weighted speedup of workloads H1-H10.

\begin{table}
	\begin{minipage}[bht*]{1.00\columnwidth}
		\centering
		\begin{tabular}{|c||c||c||c|} \hline
			\multicolumn{4}{|c|}{{\bf Issue Width} } \\ \hline 
			\multicolumn{1}{|c||}{1} & \multicolumn{1}{c||}{\textbf{2}} & \multicolumn{1}{c||}{4} & \multicolumn{1}{c|}{8} \\ \hline
			$\Delta$ Perf & $\Delta$ Perf & $\Delta$ Perf & $\Delta$ Perf \\ \hline
			10.7\% & 12.7\% & 13.6\% & 14.0\% \\ 
			\hline
		\end{tabular}
	\end{minipage}
	
	\begin{minipage}[bht*]{1.00\columnwidth}
		\centering
		\begin{tabular}{|c||c||c||c|} \hline
			\multicolumn{4}{|c|}{{\bf Data Cache Size} } \\ \hline 
			\multicolumn{1}{|c||}{2KB} & \multicolumn{1}{c||}{\textbf{4KB}} & \multicolumn{1}{c||}{8KB} & \multicolumn{1}{c|}{16KB} \\ \hline
			$\Delta$ Perf & $\Delta$ Perf & $\Delta$ Perf & $\Delta$ Perf \\ \hline
			9.91\% & 12.7\% & 15.4\% & 16.6\% \\ 
			\hline
		\end{tabular}
	\end{minipage}
	
	\begin{minipage}[bht*]{1.00\columnwidth}
		\centering
		\begin{tabular}{|c||c||c||c|} \hline
			\multicolumn{4}{|c|}{{\bf Dependence Chain Length} } \\ \hline 
			\multicolumn{1}{|c||}{4} & \multicolumn{1}{c||}{8} & \multicolumn{1}{c||}{\textbf{16}} & \multicolumn{1}{c|}{32} \\ \hline
			$\Delta$ Perf & $\Delta$ Perf & $\Delta$ Perf & $\Delta$ Perf \\ \hline
			3.5\% & 8.5\% & 12.7\% & 11.9\% \\ 
			\hline
		\end{tabular}
	\end{minipage}
	
	\begin{minipage}[bht*]{1.00\columnwidth}
		\centering
		\begin{tabular}{|c||c||c||c|} \hline
			\multicolumn{4}{|c|}{{\bf Number of Contexts} } \\ \hline 
			\multicolumn{1}{|c||}{1} & \multicolumn{1}{c||}{\bf 2} & \multicolumn{1}{c||}{4} & \multicolumn{1}{c|}{8} \\ \hline
			$\Delta$ Perf & $\Delta$ Perf & $\Delta$ Perf & $\Delta$ Perf \\ \hline
			8.5\% & 12.7\% & 16.1\% & 16.3\% \\ 
			\hline
		\end{tabular}
	\end{minipage}
		
	\begin{minipage}[bht*]{1.00\columnwidth}
		\centering
		\begin{tabular}{|c||c||c||c|} \hline
			\multicolumn{4}{|c|}{{\bf TLB Entries} } \\ \hline 
			\multicolumn{1}{|c||}{8} & \multicolumn{1}{c||}{16} & \multicolumn{1}{c||}{\bf 32} & \multicolumn{1}{c|}{64} \\ \hline
			$\Delta$ Perf & $\Delta$ Perf & $\Delta$ Perf & $\Delta$ Perf \\ \hline
			8.1\% & 9.8\% & 12.7\% & 15.6\% \\ 
			\hline
		\end{tabular}
	\end{minipage}
	\begin{small}
		\caption{Performance Sensitivity to EMC Parameters}
		\label{tab:EMC:emcSens}
	\end{small}
\end{table}

Increasing the data cache size, number of dependence chain contexts, and the number of TLB entries all result in performance gains. Increasing issue width leads to marginal benefits. The largest performance sensitivity is to an increased data cache, going from a 4kB structure to a 16kB structure increases the EMC performance gain by 4.2\%. The EMC also shows some sensitivity from increasing the number of dependence chain contexts from 2 to 4, resulting in a 3.4\% performance gain. Overall, the parameters picked for the EMC are the smallest parameters that allow the EMC to achieve a performance gain of over 10\%. 

The results of varying the maximum dependence chain length parameter differs from the other parameters listed in Table \ref{tab:EMC:emcSens}. Increasing dependence chain length both increases the communication overhead with the EMC and increases the amount of work that the EMC must complete before the core can resume execution. Therefore, a long dependence chain can result in performance degradation. Table \ref{tab:EMC:emcSens} shows that the 16-uop performance chain is the optimal length for these workloads.

Two other high-level parameters also influenced the design of the EMC. First, the x86 instruction set has only eight architectural registers. This means that register spills/fills (push and pop instructions) are common. While executing stores at the EMC complicates the memory consistency model, eliminating all stores from dependence chains that are executed at the EMC results in a performance gain of only 3.9\%. By including register spills/fills in dependence chains, this performance gain increases to the 12.7\% in Figure \ref{fig:EMC:highPerf}. An EMC for a processor with an instruction set that has a larger set of architectural registers would not need to perform stores. Second, the EMC is allowed to issue operations out-of-order. This is necessary because the dependence chains sent to the EMC contain load operations. The latency of these loads is variable, some may hit in the EMC data cache or the LLC while others may result in LLC misses. To minimize the latency impact of dynamic loads on dependent cache misses, out-of-order issue is required. A strictly in-order EMC only results in a 2.1\% performance improvement on workloads H1-H10.

\subsection{Single-Core Results}
\label{sec:EMC:singleCore}

While the EMC is designed to accelerate memory intensive applications in a multi-core system, it provides some small utility in a single core setting as well. Performance results for using the EMC in a single core system is shown in Table \ref{tab:EMC:singleC} for the medium and high memory intensity benchmarks. While all applications with a large fraction of dependent cache misses show some performance gain, the only significant performance gain occurs for \textit{mcf}.

\begin{table*}[h]
	\centering
	\begin{tabular}{|c|ccccccc|}\hline
		\centering
		\multirow{4}{1.0in}{\centering Performance Gain} & zeusmp & cactus & wrf & gems & leslie & omnetpp & milc \\
		& 0\% & 0\% & 0\% & 0\% & 0\% & 6.9\% & 4.8\% \\\cline{2-8}
		& soplex & sphinx & bwaves & libq & lbm & mcf & GMean \\
		& 7.8\% & 4.6\% & 0\% & 0\% & 0\% & 10.8\% & 2.6\%\\\cline{2-8}
		\hline
	\end{tabular}
	\caption{EMC Single Core Performance}
	\label{tab:EMC:singleC}
\end{table*}

\subsection{Multiple Memory Controllers}
\label{sec:EMC:multiple}

As this evaluation is aimed at accelerating single threaded applications, the multi-core system primarily centers around a common quad-core processor design, where one memory controller has access to all memory channels from a single location on the ring (Figure \ref{fig:EMC:highLevel}). However, with large core counts multiple memory controllers can be distributed across the interconnect. In this case, with our mechanism, each memory controller would be compute capable. On cross-channel dependencies (where one EMC has generated a request to a channel located at a different enhanced memory controller) the EMC directly issues the request to the new memory controller without migrating execution of the chain. This cuts the core, a middle-man, out of the process (in the baseline the original request would have to travel back to the core and then on to the second memory controller).

\begin{figure}	
	\centering
	\includegraphics[height=2.4in, width=4.5in]{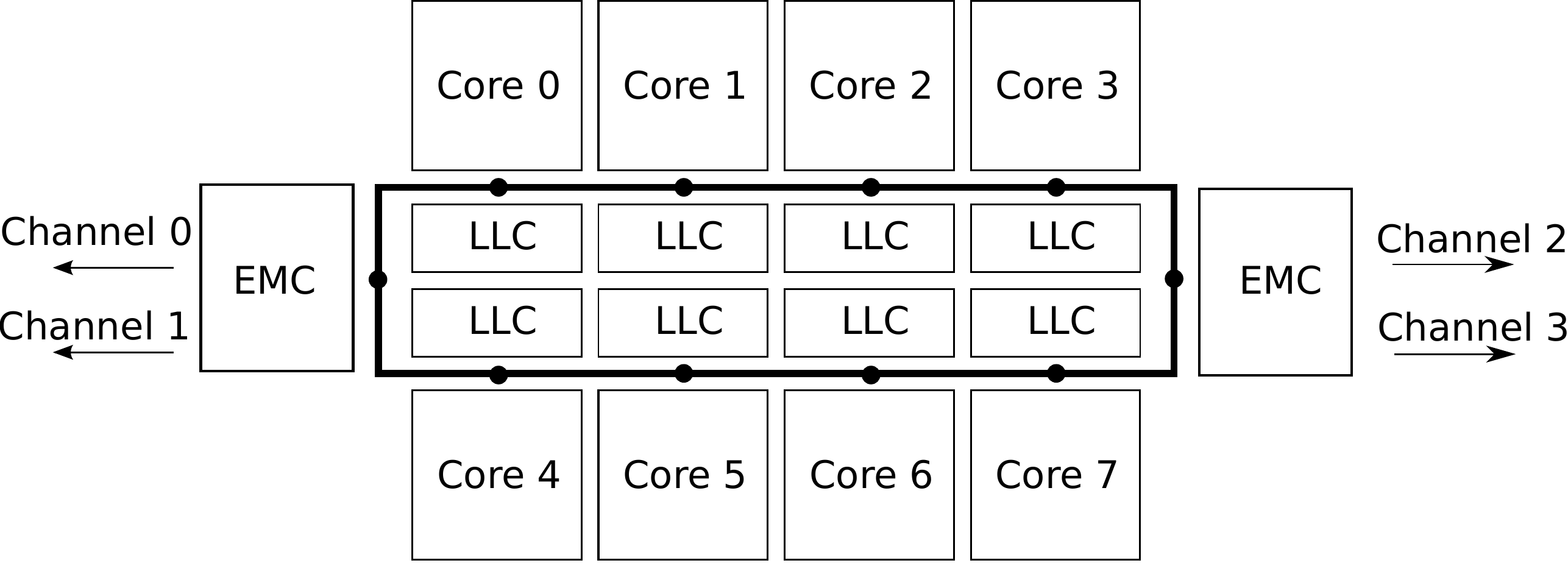}
	\caption{Eight-Core Spread Configuration}
	\label{fig:EMC:eightSpread}
\end{figure}

This scenario is evaluated with an eight-core processor as shown in Figure \ref{fig:EMC:eightSpread}. The results are compared to an eight-core processor with a single memory controller that co-locates all four memory channels at one ring stop. The eight core workloads consists of two copies of workloads H1-H10. Average performance results are shown below in Table \ref{tab:EMC:mMem}. 

\begin{table}[h]
	\centering
	\begin{tabular}{|c||c|c|}\hline
		\centering
		& Single & Spread \\
		\hline
		Baseline & 0\% & -.8\% \\
		\hline
		EMC & 16.9\% & 15.8\% \\
		\hline
		Stream & 24.5\% & 24.3\% \\
		\hline
		Stream + EMC & 41.3\% & 39.6\% \\
		\hline
		GHB & 36.4\% & 34.3\% \\
		\hline
		GHB + EMC & 51.0\% & 49.1 \% \\
		\hline
		Markov + Stream & 22.1\% & 21.2\% \\
		\hline
		Markov + Stream + EMC & 36.9\% & 35.07\% \\
		\hline
	\end{tabular}
	\caption{EMC and Multiple Memory Controllers}
	\label{tab:EMC:mMem}
\end{table}

Overall, the performance benefit of the EMC is slightly larger in the eight-core case than the quad-core case, due to a more heavily contested memory system. The single memory controller configuration gains 17\%, 14\%, 13\%, and 13\% over the no-prefetching, stream, GHB and stream+Markov prefetchers respectively. The dual memory controller baseline system shows a slight (-.8\%) performance degradation over the single memory controller system, and gains slightly less on average over each baseline (16\%, 14\%, 11\%, 12\% respectively) than the single memory controller, due to the overhead of communication between the EMCs. I conclude that there is not a significant performance degradation when using two enhanced memory controllers in the system.

\subsection{EMC Overhead}
\label{sec:EMC:overhead}

The data traffic overhead of the EMC consists of three main components: sending both dependent operation chains and the source registers (live-ins) that these chains require to the EMC, and sending destination registers (live-outs) back to the core from the EMC.

Table \ref{tab:EMC:depChain} shows the average chain length in terms of uops for the chains that are sent to the EMC along with the number of live-ins per dependence chain. The chain length defines both the number of uops which must be sent to the EMC and the number of registers that must be shipped back to the core. This is because all physical registers are sent back to the core (Section \ref{sec:EMC:trace_exec}) and each uop produces a live-out/physical register.

\begin{table*}
	\centering
	\begin{tabular}{|c|cccccc|}\hline
		\centering
		\multirow{4}{1.0in}{\centering Live-Ins} & H1 & H2 & H3 & H4 & H5 & H6 \\
		& 3.2 & 8.8 & 9.0 & 7.8 & 7.3 & 1.9 \\ \cline{2-7}
		& H7 & H8 & H9 & H10 & Mean & \\
		& 8.8 &	9 &	7.4 & 8.8 & 7.2 & \\\cline{2-7}
		\hline
		\multirow{4}{1.0in}{\centering Dependence Chain Length} & H1 & H2 & H3 & H4 & H5 & H6 \\
		& 4.5 &	11.7 &	11.0 &	8.8 &	8.4 &	3.8  \\\cline{2-7}
		& H7 & H8 & H9 & H10 & Mean & \\
		& 11.2 &	11.6 &	8.5 &	11.7 &	9.1 &  \\\cline{2-7}
		\hline
	\end{tabular}
	\caption{EMC Dependence Chain Length}
	\label{tab:EMC:depChain}
\end{table*}

On average, the dependence chains executed at the EMC for H1-H10 are short, under 10 uops on average. These chains require 7 live-ins on average. The destination registers that are shipped back to the home core result in roughly a cache line of data per chain. Transmitting the uops to the EMC results in a transfer of 1-2 cache lines on average. This relatively small amount of data transfer motivates why we do not see a performance loss due to the EMC.  The interconnect overhead of the EMC for each executed chain is small and we accelerate the issue and execution of integer dependent operations only if they exist. As shown in Table \ref{tab:EMC:icOverhead}, these messages result in a 34\% average increase in data ring activity across Workloads H1-H10 while using the EMC and a 7\% increase in control ring activity. 

\begin{table*}
	\centering
	\begin{tabular}{|c|cccccc|}\hline
		\centering
		\multirow{4}{1.0in}{\centering Data Ring Overhead} & H1 & H2 & H3 & H4 & H5 & H6 \\
		& 26.8 & 	48.2 & 	31.9 & 	21.3 & 	16.1 & 	26.1 \\ \cline{2-7}
		& H7 & H8 & H9 & H10 & Mean & \\
		& 44.7 & 56.3 & 34.6 & 	35.9 & 	34.2 &  \\\cline{2-7}
		\hline
		\multirow{4}{1.0in}{\centering Control Ring Overhead} & H1 & H2 & H3 & H4 & H5 & H6 \\
		& 3.4 & 	10.2 & 	11.8 & 	6.4 & 	9.4 & 	9.8 \\\cline{2-7}
		& H7 & H8 & H9 & H10 & Mean & \\
		& 	7.5 & 	2.9 & 	7.6 & 	4.9 & 	7.4 & \\\cline{2-7}
		\hline
	\end{tabular}
	\caption{EMC Interconnect Overhead}
	\label{tab:EMC:icOverhead}
\end{table*}

\subsection{Energy and Area}
\label{sec:EMC:energy}

\begin{figure}
	\centering
	\includegraphics[width=\columnwidth]{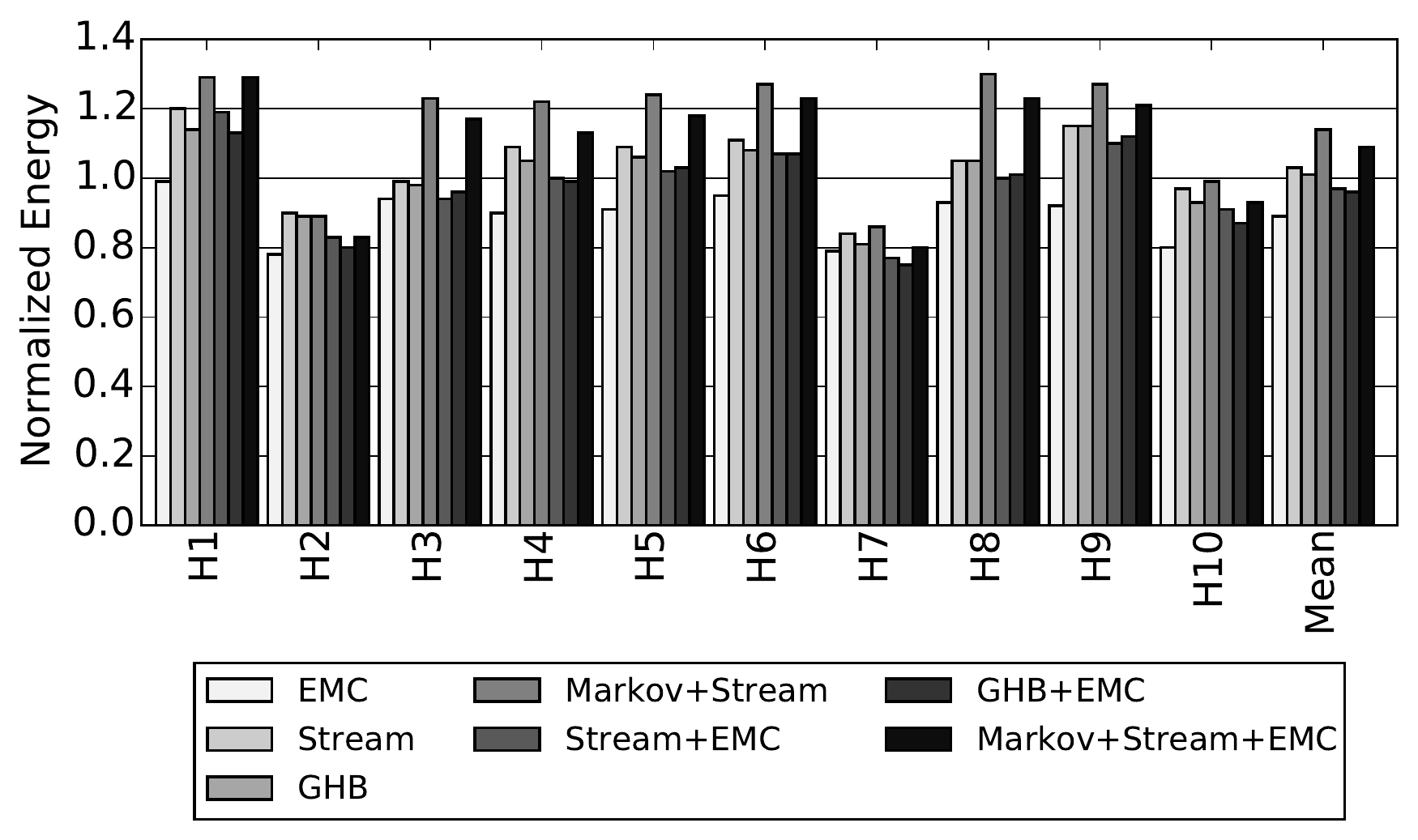}
	\caption{Normalized Energy Consumption for H1-H10}
	\label{fig:EMC:energyHigh}
\end{figure}

\begin{figure}
	\centering
	\includegraphics[width=\columnwidth]{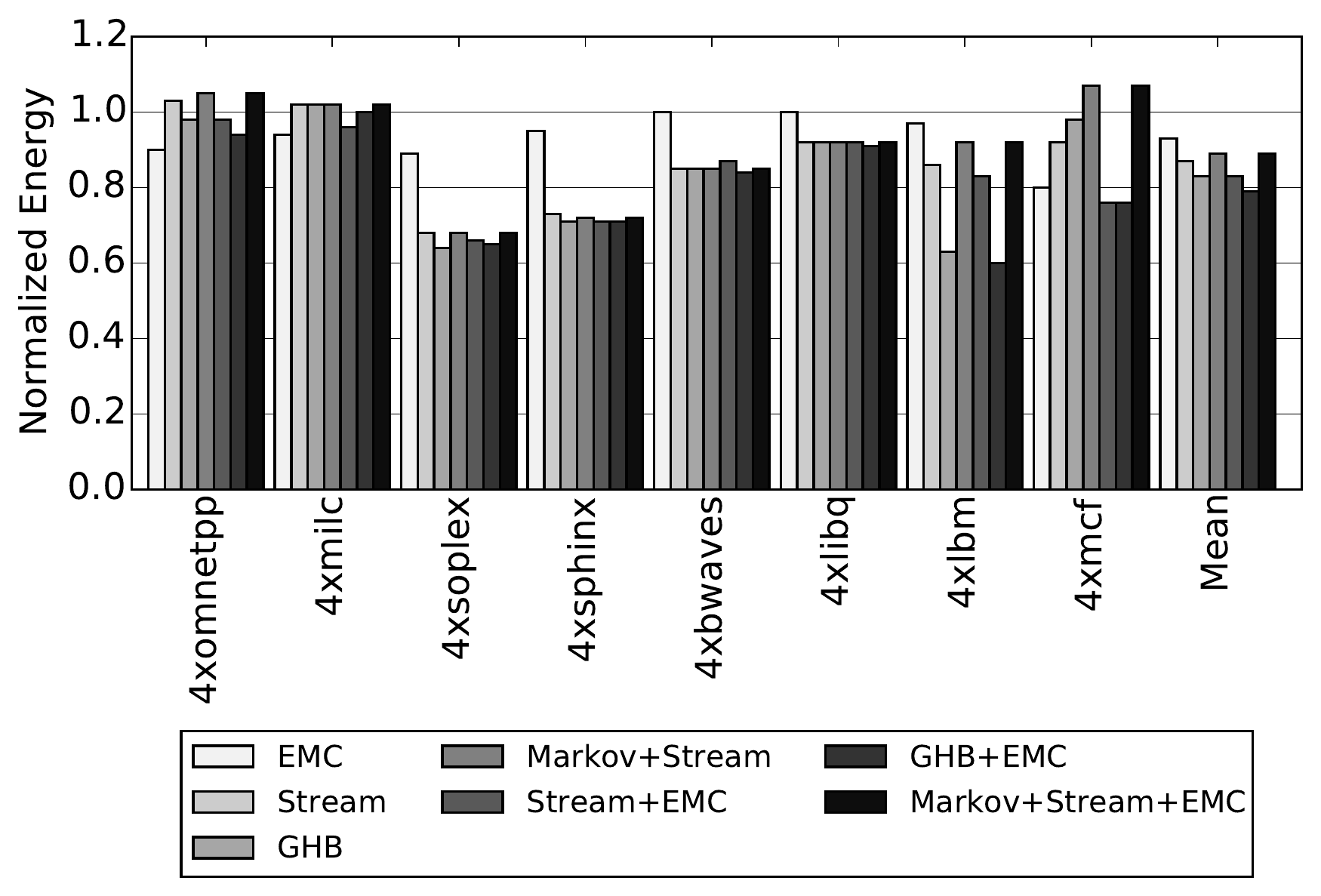}
	\caption{Normalized Energy Consumption for Copy Workloads}
	\label{fig:EMC:energyCopy}
	\centering
	\includegraphics[width=\columnwidth]{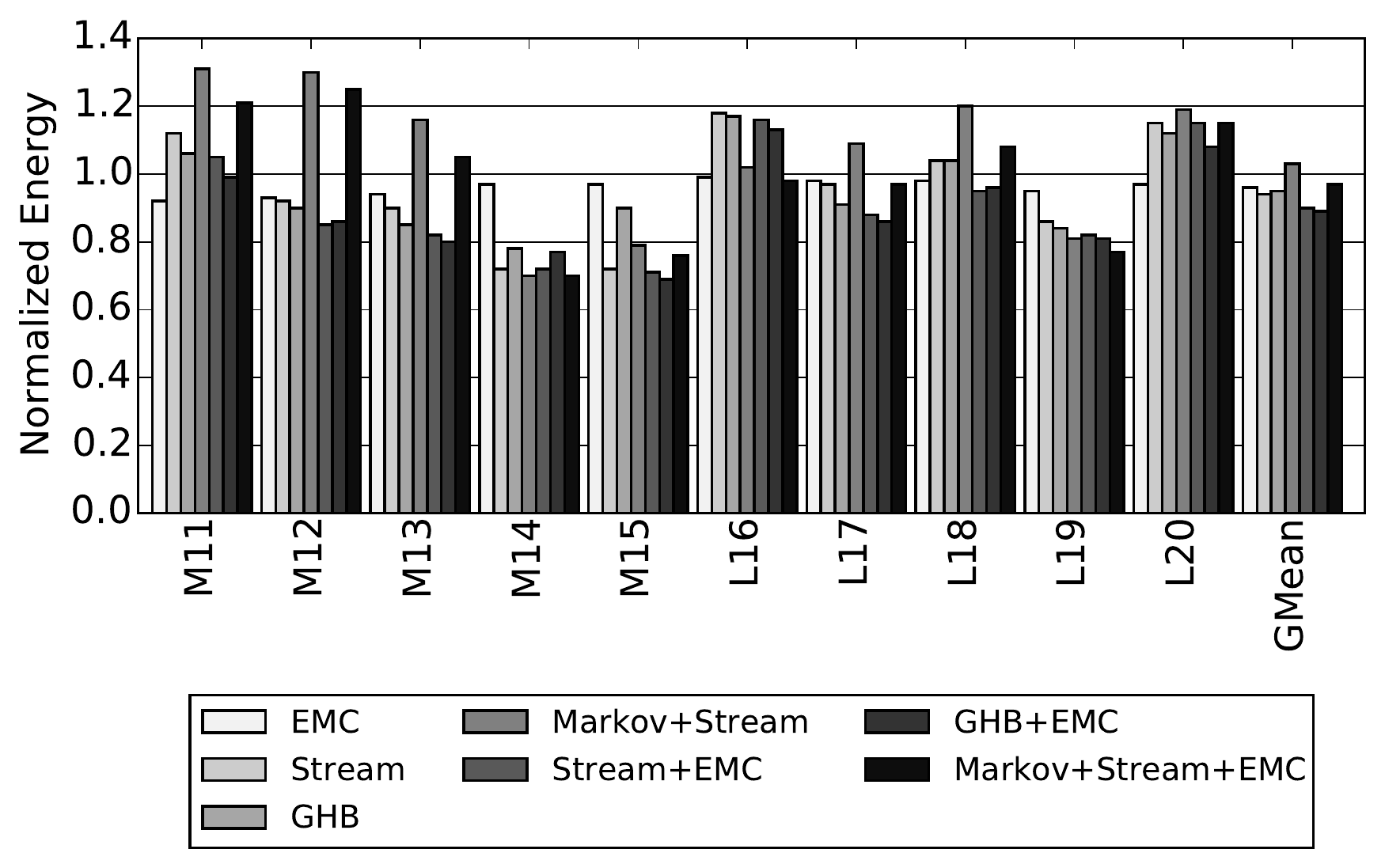}
	\caption{Normalized Energy Consumption for M11-L20}
	\label{fig:EMC:energyMix}
\end{figure}

The energy results for the quad-core workloads are shown in Figures \ref{fig:EMC:energyHigh}, \ref{fig:EMC:energyCopy}, and \ref{fig:EMC:energyMix}. All charts present the cumulative results for the energy consumption of the chip and DRAM as a percentage difference in energy consumption from the no-EMC, no-prefetching baseline.

Overall, the EMC is able to reduce energy consumption (Chip+DRAM) on average by about 11\% for H1-H10, 7\% for the copy workloads and by 5\% for M11-L20. This reduction is predominantly due to a reduction in static energy consumption (as the performance improvement caused by the EMC decreases the total execution time of a workload). 

\begin{figure}
	\centering
	\includegraphics[width=\columnwidth]{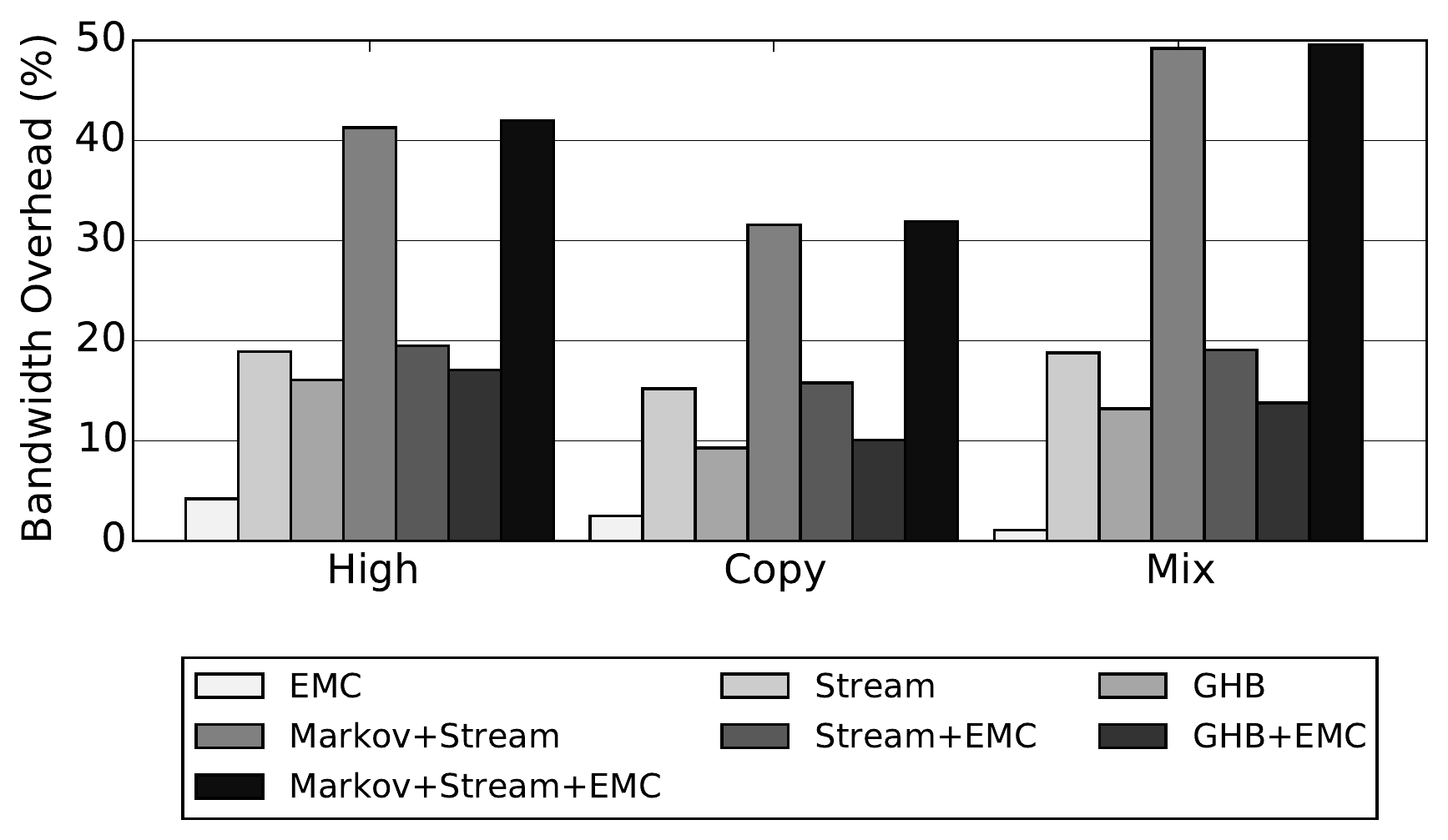}
	\caption{System Bandwidth Overhead with Prefetching}
	\label{fig:EMC:bw}
\end{figure}

In the prefetching cases, the energy consumption charts illustrate the cost of prefetching in a multi-core system. As in the performance results, combining prefetching and the EMC result in better energy efficiency than just using a prefetcher. The GHB+EMC system has the lowest average energy consumption across all three workloads.  All three of the evaluated prefetchers cause an increase in energy consumption, particularly the Markov+Stream prefetcher. This is due to inaccurate prefetch requests, which occur despite prefetcher throttling in the baseline. In Figure \ref{fig:EMC:energyHigh}, the GHB, Stream, Markov+Stream systems increase memory traffic by 19\%, 16\% and 41\% respectively while the EMC increases traffic by 4\%. A similar trend holds in Figure \ref{fig:EMC:energyCopy} where the prefetchers increase traffic by 15\%, 9\% and 32\% respectively while the EMC increases traffic by 3\%. The average bandwidth increase over the no-prefetching baseline for the EMC and prefetching is shown in Figure \ref{fig:EMC:bw} for all workloads.

\ignore{
\begin{table}[h]
	\small
	\centering
	\begin{tabular}{|p{1.75in}|p{1in}|}
		\hline
		& Bandwidth \newline Overhead (\%) \\
		\hline
		\hline
		High EMC & 4.2\% \\
		\hline
		High Stream & 18.9\% \\
		\hline
		High GHB & 16.1\% \\
		\hline
		High Markov+Stream & 41.3\%  \\
		\hline
		High Stream+EMC & 19.5\%  \\
		\hline
		High GHB+EMC & 17.1\% \\
		\hline
		High Markov+Stream+EMC & 42.0\%  \\
		\hline				
		\hline
		Copy EMC & 2.5\%  \\
		\hline
		Copy Stream & 15.2\%  \\
		\hline
		Copy GHB & 9.3\%  \\
		\hline
		Copy Markov+Stream & 31.6\% \\
		\hline
		Copy Stream+EMC & 15.8\%  \\
		\hline
		Copy GHB+EMC & 10.1\%  \\
		\hline
		Copy Markov+Stream+EMC & 31.9\% \\
		\hline
		\hline
		Mix EMC & 1.1\% \\
		\hline
		Mix Stream & 18.8\%  \\
		\hline
		Mix GHB & 13.2\% \\
		\hline
		Mix Markov+Stream & 49.2\%  \\
		\hline
		Mix \newline Stream+EMC & 19.1\%  \\
		\hline
		Mix \newline GHB+EMC & 13.8\% \\
		\hline
		Mix Markov+Stream+EMC & 49.6\% \\
		\hline
	\end{tabular} 
	\caption{System Bandwidth Overhead with Prefetching}
	\label{tab:EMC:bw}
\end{table}
}

The components of the storage overhead that are required to implement the EMC are listed in Table \ref{tab:EMC:storage}. The total additional storage required is 6kB. Most of this storage is due to the data cache at the EMC. Based on McPAT, this overhead translates to an area overhead of $2.2mm^2$, roughly 2\% of total quad-core chip area. The area overhead is listed in Table \ref{tab:EMC:area}. Over half of this additional area is due to the 4kB cache located at the EMC. The small out-of-order engine constitutes 8\% of the additional area, while the two integer ALUs make up 5\%. As McPAT estimates the area of a full out-of-order core as $21.2mm^2$, the EMC is 10.4\% of a full core and is shared by all of the cores on a multi-core processor.

\begin{table*}[h]
	\small
	\centering
	\begin{tabular}{|p{2.2in}|p{2.8in}|}
		\hline
		\textbf{Component} & \textbf{Bytes} \\
		\hline
		\hline
		\textbf{Core} & \\
		\hline
		RRT & 32 Entries * 1 Byte = 64 Bytes \\
		\hline
		Dependence Chain Buffer & 8 Bytes * 16 Entries = 128 Bytes \\
		\hline
		Live-In Vector & 4 Bytes * 16 Entries = 64 Bytes \\
		\hline
		Total New Core Storage & 256 Bytes \\
		\hline
		\hline
		\textbf{EMC} & \\
		\hline
		Instruction Buffers & 8 Bytes * 16 Entries * 2 Contexts = \newline 256 Bytes \\
		\hline
		Register Files & 4 Bytes * 16 Entries * 2 Contexts = \newline 128 Bytes\\
		\hline
		Live-In Vectors & 4 Bytes * 16 Entries * 2 Contexts = \newline 128 Bytes \\
		\hline
		Reservation Station & 8 Bytes * 8 Entries = \newline 64 Bytes \\
		\hline
		TLB & 8 Bytes * 32 Entries * 4 Cores = \newline 1024 Bytes\\
		\hline
		Data Cache & 4096 Bytes\\
		\hline
		Load Store Queue & 4 Bytes * 16 Entries = 64 Bytes\\
		\hline
		Miss Predictor & 384 Bytes\\		
		\hline				
		Total EMC Storage & 6096 Bytes\\
		\hline
	\end{tabular} 
	\caption{Additional EMC Storage Overhead}
	\label{tab:EMC:storage}
\end{table*}

\begin{table*}[h]
	\small
	\centering
	\begin{tabular}{|p{2.2in}|p{1.2in}|}
		\hline
		\textbf{Component} & \textbf{Size} \\
		\hline
		\hline
		Instruction Buffers & $.07mm^2$ \\
		\hline
		Register Files & $.12mm^2$ \\
		\hline
		Execute Logic & $.17mm^2$ \\
		\hline
		TLB & $.20mm^2$ \\
		\hline
		Data Cache & $1.52mm^2$\\
		\hline
		Load Store Queue & $.07mm^2$\\
		\hline
		Miss Predictor & $.05mm^2$\\		
		\hline				
		Total EMC Area & $2.2mm^2$\\
		\hline
	\end{tabular} 
	\caption{Additional EMC Area Overhead}
	\label{tab:EMC:area}
\end{table*}

\subsection{Sensitivity to System Parameters}
\label{sec:EMC:systemSens}

In this section, the performance sensitivity of the EMC to three key system parameters is measured. Table \ref{tab:EMC:sysSens} shows performance and energy sensitivity of the EMC to LLC capacity, the number of memory banks per channel, and ROB capacity. First, unlike the Runahead Buffer, the EMC does not show significant performance sensitivity to LLC capacity. The irregular dependent cache misses that the EMC targets do not become cache hits even with a 16MB LLC. However, the EMC does show performance sensitivity to very large numbers of banks per memory channel. One of the main performance gains of the EMC is reducing row-buffer miss rate. If a memory system has 64 banks/channel, this row-buffer contention lessens and the performance gain of the EMC is degraded. Lastly, I find that the EMC does not show significant performance sensitivity to ROB capacity. Hiding dependent cache miss latency requires a ROB to tolerate two serialized memory accesses. A 512-entry ROB is unable to generally hide the latency of even a single memory access without stalling the pipeline, so the lack of performance sensitivity is unsurprising.

\begin{table}[h]
	\begin{minipage}[bht*]{1.00\columnwidth}
		\centering
		\footnotesize
		\begin{tabular}{|c|c||c|c||c|c||c|c|} \hline
			\multicolumn{8}{|c|}{{\bf LLC Cache Size} } \\ \hline 
			\multicolumn{2}{|c||}{2 MB} & \multicolumn{2}{c||}{\textbf{4 MB}} & \multicolumn{2}{c||}{8 MB} & \multicolumn{2}{c|}{16 MB} \\ \hline
			$\Delta$ Perf & $\Delta$ Energy & $\Delta$ Perf & $\Delta$ Energy & $\Delta$ Perf & $\Delta$ Energy & $\Delta$ Perf & $\Delta$ Energy \\ \hline
			14.2\% & -10.1\% & 12.7\% & -11.0\% & 13.8\% & -11.6\% & 11.7\% & -10.1\% \\ \hline
		\end{tabular}
	\end{minipage}
	
	\begin{minipage}[bht*]{1.00\columnwidth}
		\centering
		\footnotesize
		\begin{tabular}{|c|c||c|c||c|c||c|c|} \hline
			\multicolumn{8}{|c|}{{\bf Number of Memory Banks} } \\ \hline 
			\multicolumn{2}{|c||}{\textbf{8}} & \multicolumn{2}{c||}{16} & \multicolumn{2}{c||}{32} & \multicolumn{2}{c|}{64} \\ \hline
			$\Delta$ Perf & $\Delta$ Energy & $\Delta$ Perf & $\Delta$ Energy & $\Delta$ Perf & $\Delta$ Energy & $\Delta$ Perf & $\Delta$ Energy \\ \hline
			12.7\% & -11.0 & 14.7\% & -11.9\% & 11.6\% & -10.1\% & 9.9\% & -7.9\% \\ \hline
		\end{tabular}
	\end{minipage}
	
	\begin{minipage}[bht*]{1.00\columnwidth}
		\centering
		\footnotesize
		\begin{tabular}{|c|c||c|c||c|c||c|c|} \hline
			\multicolumn{8}{|c|}{{\bf ROB Size} } \\ \hline 
			\multicolumn{2}{|c||}{192} & \multicolumn{2}{c||}{\textbf{256}} & \multicolumn{2}{c||}{384} & \multicolumn{2}{c|}{512} \\ \hline
			$\Delta$ Perf & $\Delta$ Energy & $\Delta$ Perf & $\Delta$ Energy & $\Delta$ Perf & $\Delta$ Energy & $\Delta$ Perf & $\Delta$ Energy \\ \hline
			12.9\% & -11.1\% & 12.7\% & -11.0 & 14.5\% & -11.9\% & 11.7\% & -9.8\% \\ \hline
		\end{tabular}
	\end{minipage}
	\begin{small}
		\caption{EMC Performance and Energy Sensitivity}
		\label{tab:EMC:sysSens}
	\end{small}
\end{table}

\newpage

\section{Conclusion}
\label{sec:EMC:conclusion}

This chapter identifies dependent cache misses as a critical impediment to processor performance for memory intensive applications. A mechanism is proposed for minimizing the latency of a dependent cache miss by performing computation where the data first enters the chip, at the memory controller. By migrating the dependent cache miss to the memory controller, I show that the EMC reduces effective memory access latency by 20\% for dependent cache misses. This results in a 13\% performance improvement and a 11\% energy reduction on a set of ten high memory intensity quad-core workloads. In the next chapter, I examine how to use the compute capability of the EMC to accelerate independent cache misses in addition to dependent cache misses. The analysis starts in Chapter \ref{chap:scRaEMC} for a single core system and continues in Chapter \ref{chap:mcRaEMC} for a multi-core system.

\chapter{Runahead at the Enhanced Memory Controller}
\label{chap:scRaEMC}
\setlength{\epigraphwidth}{0.41\textwidth}

\section{Introduction}
\label{sec:scRaEMC:Intro}

In Chapter \ref{chap:raBuf} I develop a mechanism that identifies the micro-operations (micro-ops) that are required to generate the address of a memory access. These micro-ops constitute the dependence chain of the memory operation. I propose pre-executing these dependence chains using runahead execution \cite{mut:sta03} with the goal of generating new independent cache misses. Section \ref{sec:raBuf:perf} demonstrates that pre-executing a dependence chain generates more independent cache misses than traditional runahead execution as traditional runahead fetches and executes many irrelevant operations.

\begin{figure}
	\centering
	\includegraphics[width=\columnwidth]{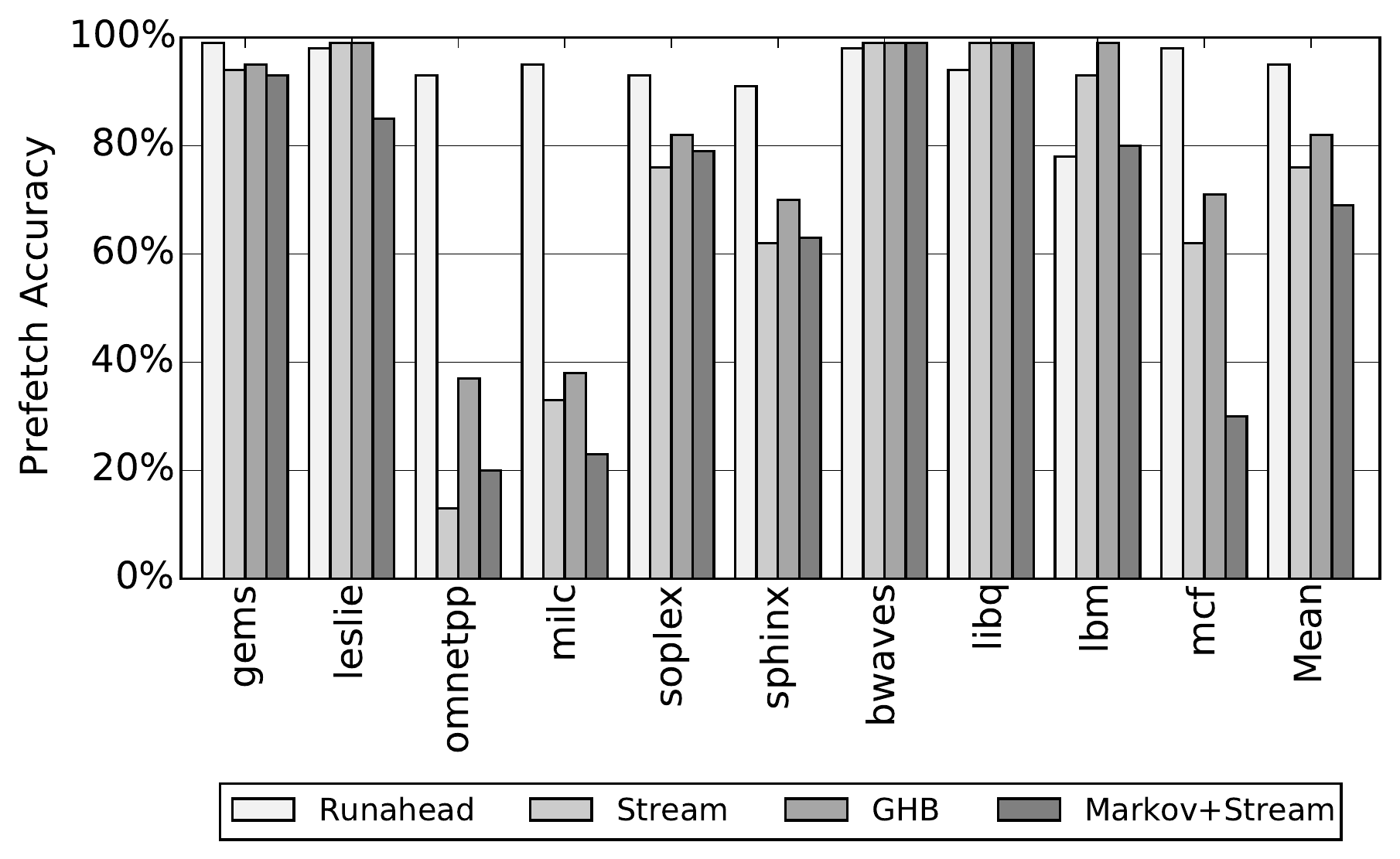}
	\caption{Percent of Useful Runahead Requests}
	\label{fig:raTouch}
\end{figure}

Based on Chapter \ref{chap:raBuf}, I make three observations that motivate this chapter: runahead requests are overwhelmingly accurate, the core spends only a fraction of total execution time in runahead mode, and runahead interval length is generally short. First, Figure \ref{fig:RAB:mem-traffic} illustrates that runahead has very low memory-bandwidth overhead, particularly when compared to traditional prefetching. This highlights the benefit of using the application's code to predict future memory accesses. To further explore this point, Figure \ref{fig:raTouch} displays the average percentage of useful runahead requests (defined as the number of cache lines prefetched by a runahead request that are accessed by the core before eviction from the LLC) for the high memory intensity \textit{SPEC06} benchmarks. On average, runahead requests are very accurate, with 95\% of all runahead accesses prefetching useful data. This is 13\% more accurate than a GHB prefetcher that uses dynamic throttling \footnote{An earlier version of this chapter was published as: Milad Hashemi, Onur Mutlu, and Yale Patt. Continuous Runahead: Transparent Hardware Acceleration for Memory Intensive Workloads. In \textit{MICRO}, 2016. I developed the initial idea and conducted the performance simulator design and evaluation for this work.}.

\begin{figure}
 	\centering
 	\includegraphics[width=\columnwidth]{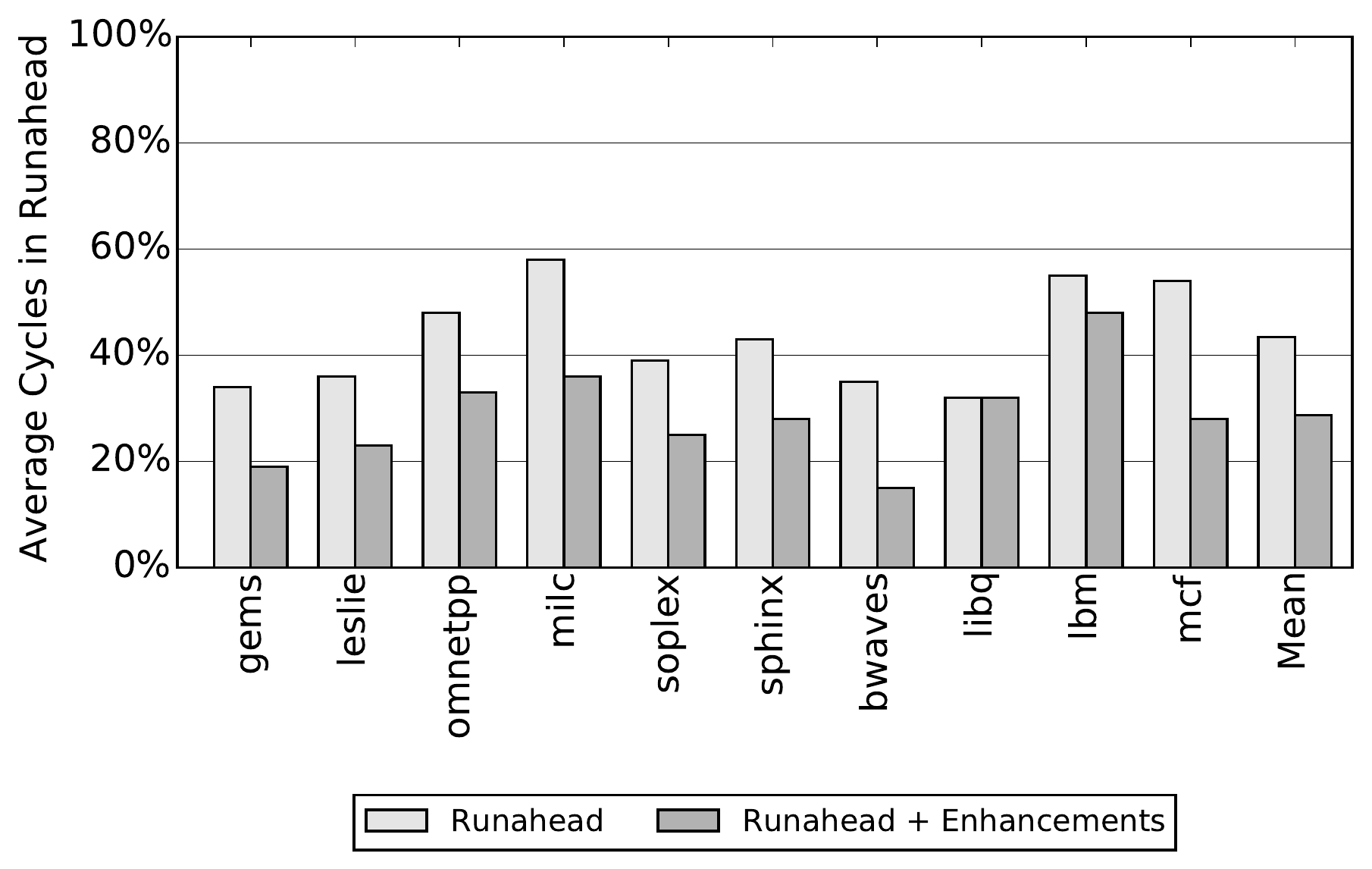}
 	\caption{Percent of Execution Time in Runahead}
 	\label{fig:raTime}
 \end{figure}

Second, given this high accuracy, it could be advantageous to spend large periods of time in runahead, prefetching independent cache misses. However, I find that this is not generally the case. Figure \ref{fig:raTime} displays the percentage of total execution cycles that the core spends in runahead. Two data points are shown: traditional runahead and traditional runahead plus energy optimizations (Section \ref{sec:raBuf:raEnhance}). Since these optimizations are intended to eliminate short or wasteful runahead intervals, this data more accurately demonstrates the number of useful runahead cycles. From Figure \ref{fig:raTime}, the core spends less than half of all execution time in traditional runahead on average. Runahead + Enhancements further reduces the cycles spent in runahead mode as repetitive runahead cycles are eliminated. 

\begin{figure}
 	\centering
 	\includegraphics[width=\columnwidth]{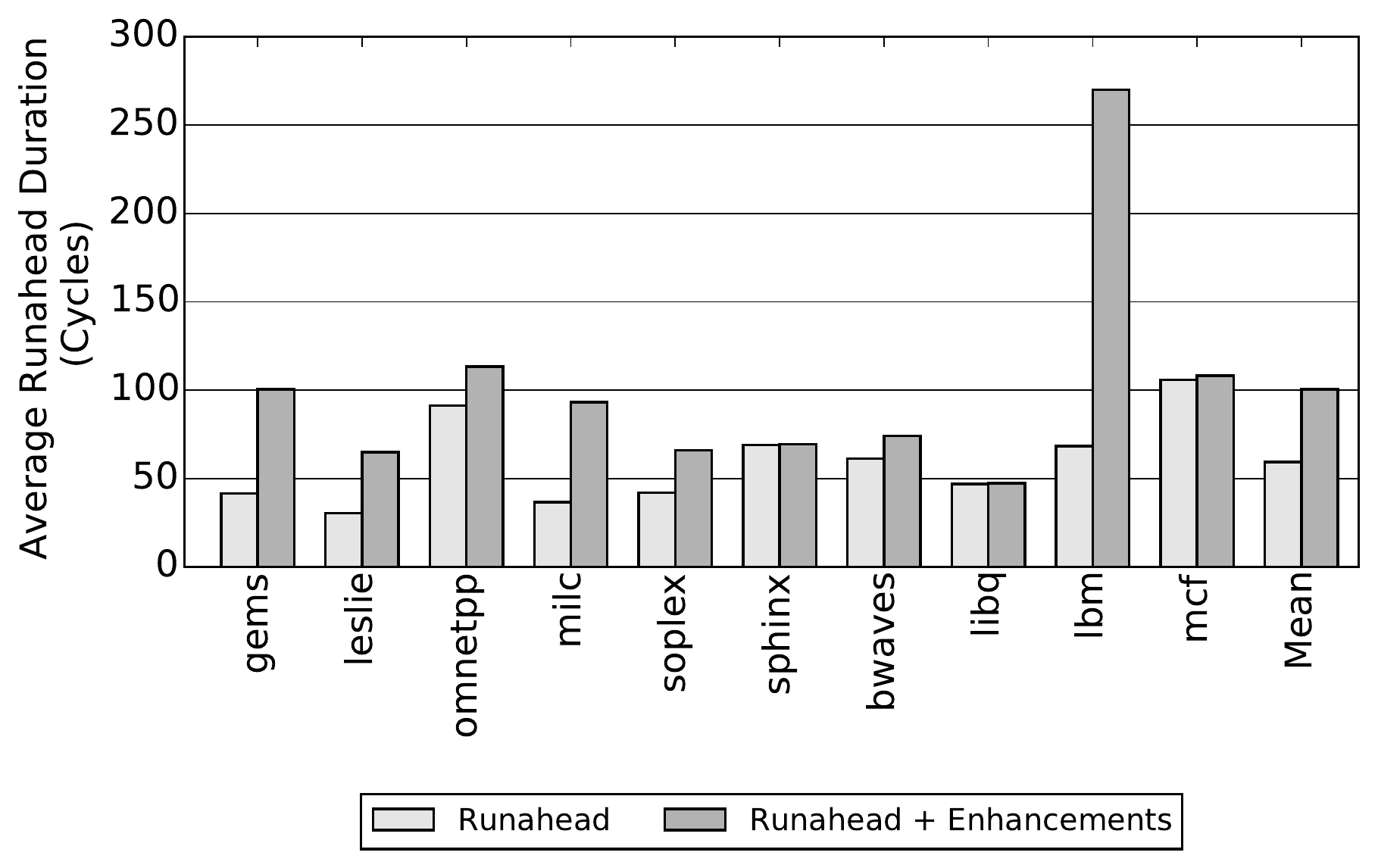}
 	\caption{Average Number of Cycles in Each Runahead Interval}
 	\label{fig:raCycles}
 \end{figure}

Third, in addition to the percentage of total cycles spent in runahead, the length of each runahead interval is also important. The average duration of each runahead interval is the number of cycles from when runahead begins to when runahead terminates. A large interval length indicates that runahead is able to get ahead of the demand execution stream and generate more cache misses. A short interval length is more likely to result in few to no new independent cache misses. Figure \ref{fig:raCycles} shows that each runahead interval is short. On average, each runahead interval is 55 cycles long without the efficiency enhancements and 101 cycles with additional enhancements. This is significantly less than the amount of time to access DRAM.

From this data, I conclude that current runahead techniques are not active for a large portion of execution time. This is because runahead imposes constraints on how often and how long the core is allowed to speculatively execute operations. First, the core is required to completely fill its reorder buffer before runahead begins. Memory level parallelism causes DRAM access latency to overlap, reducing the amount of time that the core is able to runahead as ROB size grows. This limits how often the core can enter runahead mode, particularly as ROB sizes increase.  Second, the runahead interval terminates when the operation that is blocking the pipeline retires. This limits the duration of each runahead interval. Since runahead shares pipeline resources with the main thread of execution, these constraints are necessary to maintain maximum performance when the main thread is not stalled. However, despite the high accuracy of runahead, these constraints force current runahead policies to remain active for a small fraction of total execution time. 

In this chapter, I explore removing the constraints that lead to these short intervals. Traditional runahead is a reactive mechanism that requires the pipeline to be stalled before pre-execution begins. Instead, I will explore using the additional hardware resources of the Enhanced Memory Controller (EMC) to pre-execute code arbitrarily ahead of the demand access stream. The goal is to develop a proactive policy that uses runahead to prefetch independent cache misses so that the core stalls less often.

There are two major challenges to remote runahead at the EMC. First, the core must decide which operations to execute remotely. In the case of dependent cache misses (Chapter \ref{chap:EMC}) this problem is straightforward. If a core is predicted to have dependent cache misses, the dependence chain is migrated to the EMC for execution. In the case of independent cache misses, the answer is not as clear. The independent cache miss chains can execute arbitrarily far ahead of the demand access stream. For high accuracy, it is important to make the correct choice of which dependence chain to migrate to the EMC. This is the first question that I examine in Section \ref{sec:scRaEMC:depChain}. Second, EMC memory accesses need to be timely. If runahead requests are too early they can harm cache locality by evicting useful data. If they are too late they will not reduce effective memory access latency. I examine this trade-off in Section \ref{sec:scRaEMC:dist}. As these two questions are both predominantly single-thread decisions, I focus on a single core setting in this chapter and explore multi-core policies for runahead at the EMC in Chapter \ref{chap:mcRaEMC}.

\section{Mechanism}
\label{sec:scRaEMC:mechanism}

The goal of this section is to determine a policy that decides: 1) which dependence chain to use during runahead at the EMC and 2) how long that dependence chain should execute remotely. To answer the first question, Section \ref{sec:scRaEMC:depChain} explores three different policies with unbounded storage constraints while \ref{sec:scRaEMC:hwDepPolicy} translates the highest performing policy into hardware at the EMC. To answer the second question, Section \ref{sec:scRaEMC:dist} examines how often a dependence chain needs to be sent to the EMC to maximize prefetch accuracy.

\subsection{Runahead Oracle Policies}
\label{sec:scRaEMC:depChain}

The runahead buffer uses a simple and greedy mechanism to generate dependence chains at a full-window stall. Figure \ref{fig:RAB:raChain} demonstrates that for many benchmarks if a operation is blocking retirement it is likely that a different dynamic instance of the same static load is also present in the reorder buffer. This second dynamic operation is then used during a backwards dataflow walk to generate the dependence chain for runahead. While Section \ref{sec:raBuf:perf} shows that the chain uncovered this way is useful to increase performance, it is not clear that this policy is ideal. In this section, I relax the constraints that the runahead buffer uses to choose a dependence chain and explore three new policies. While I call these policies oracles, each policy only uses unlimited storage. The policies do not have oracle knowledge of which dependence chain is optimal to run.

\begin{figure}
	\centering
	\includegraphics[width=\columnwidth]{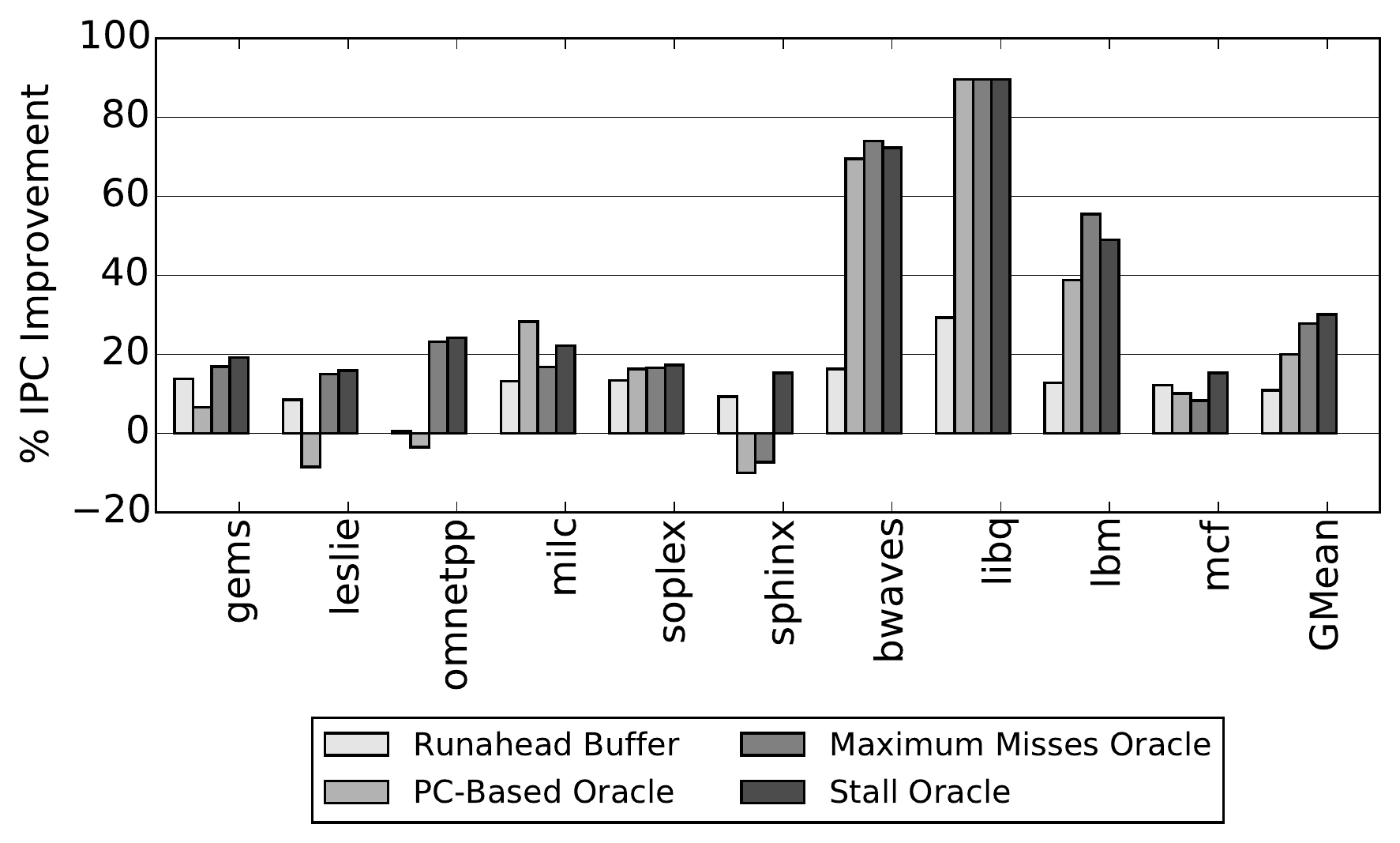}
	\caption{Performance Impact of Dependence Chain Selection Policies}
	\label{fig:threadPolicy}
	\vspace{.2in}
	\centering
	\includegraphics[width=\columnwidth]{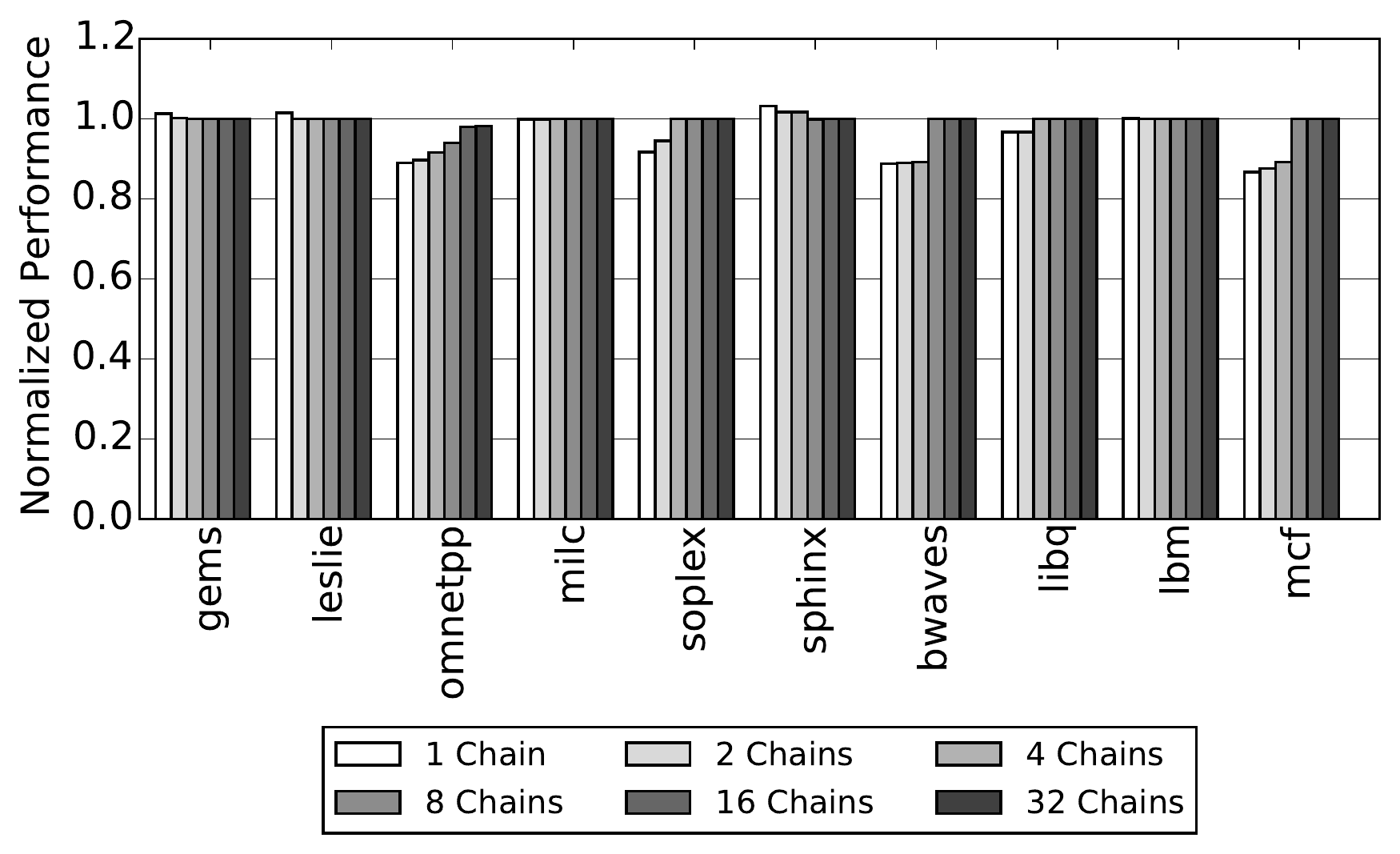}
	\caption{Varying the Number of Dependence Chains Stored per Miss PC}
	\label{fig:depTransSens}
\end{figure}

\noindent \textbf{PC-Based Oracle:} In order to generate a dependence chain on demand, the runahead buffer policy restricts itself to using a dependence chain that is available in the reorder buffer. For the first policy, this restriction is relaxed. The simulator maintains a table of all PCs that cause last level cache (LLC) misses. For each PC, the simulator also maintains a list of all of the unique dependence chains that have led to an LLC miss in the past. When the pipeline stalls due to a full reorder buffer, the runahead buffer uses the PC miss table to identify the dependence chain that has generated the most LLC misses for the PC that is blocking retirement. The performance results of the PC-based oracle are shown in Figure \ref{fig:threadPolicy}.

On average the PC-based oracle improves performance over the runahead buffer policy. However, the policy also causes performance degradations on \textit{leslie}, \textit{sphinx}, and \textit{gems}. One reason for this performance reduction is evident in Figure \ref{fig:depTransSens} where the number dependence chains that are stored for each miss PC are varied from 1 to 32. Performance is normalized to the system that stores all dependence chains for each miss PC. All three of these applications have maximum performance with only one stored dependence chain. This suggests that recent path history is more important than historical miss data for these two applications. The greedy runahead buffer algorithm is already optimized for this case (since the runahead buffer uses the last dependence chain present in the ROB during runahead). Moreover, storing 16 or 32 dependence chains is on average only marginally higher performing than storing one chain. With this data, I conclude that storing large numbers of dependence chains per miss PC is not required for high runahead buffer performance.
 
\begin{figure}
	\centering
	\includegraphics[width=\columnwidth]{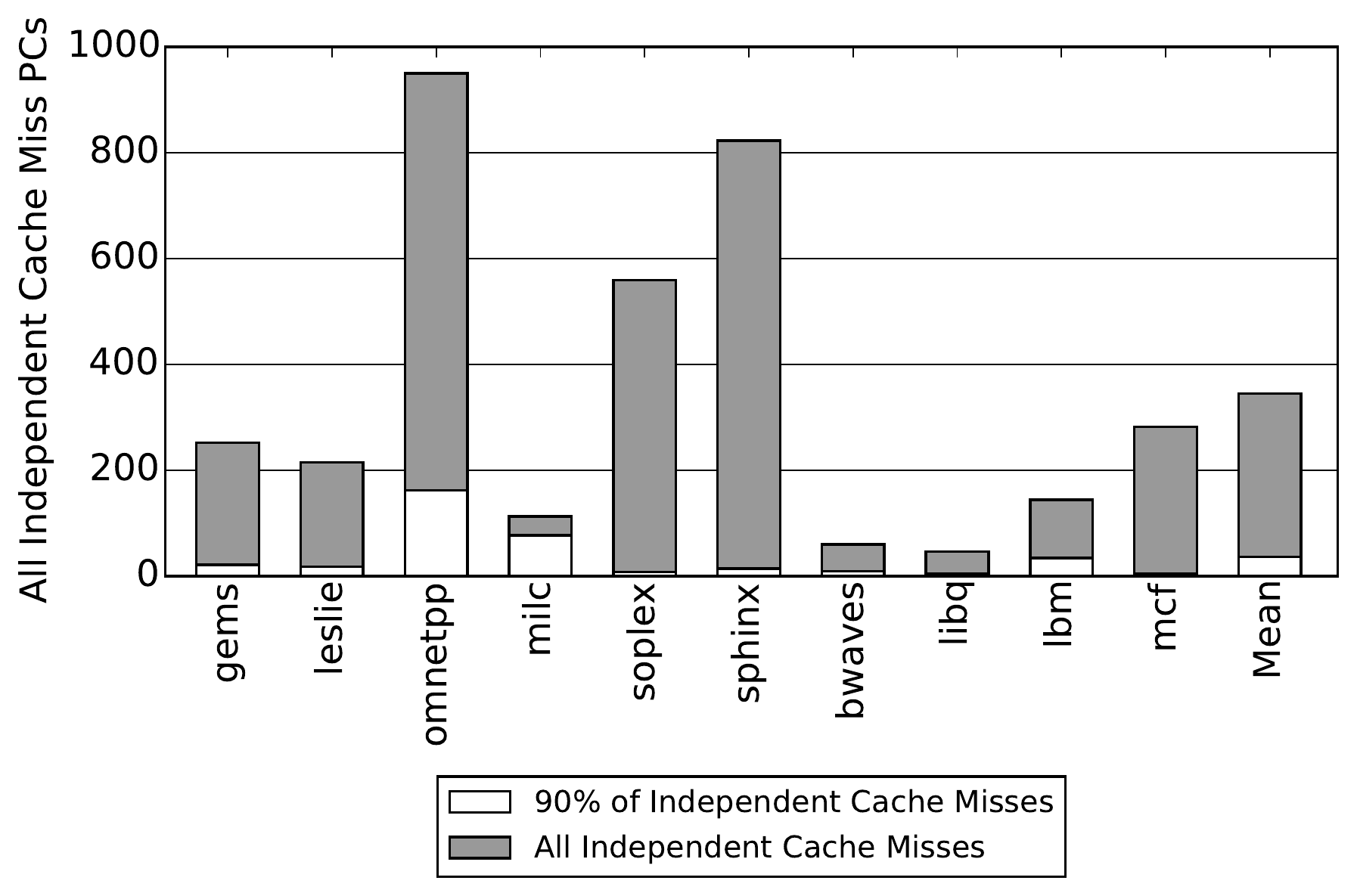}
	\caption{Number of Different PCs Generating Cache Misses}
	\label{fig:missPCStack}
\end{figure}

Beyond the performance benefits of the PC-based oracle policy, by tracking all miss PCs and all dependence chains I find two additional pieces of data that provide insight into the nature of dependence chains that generate LLC misses. First I observe that a small number of static PCs cause all LLC misses. Figure \ref{fig:missPCStack} shows the total number of static PCs that cause LLC misses and the number of static PC's that cause 90\% of all LLC misses in each of the memory intensive \textit{SPEC06} benchmarks. On average, there are 345 PCs per benchmark that cause LLC misses: \textit{omnetpp} has the most instructions causing cache misses at 950 while \textit{bwaves} has the fewest at 46. I find that on average the number of instructions that cause 90\% of all cache misses is very small across the high memory intensity \textit{SPEC06} benchmarks. For example, in \textit{libquantum}, only 4 static instructions account for 90\% of all cache misses.  With numbers this small, I conclude that it is practical for hardware to dynamically track the exact instructions that often lead to an LLC miss.

\begin{figure}	
	\centering
	\includegraphics[width=\columnwidth]{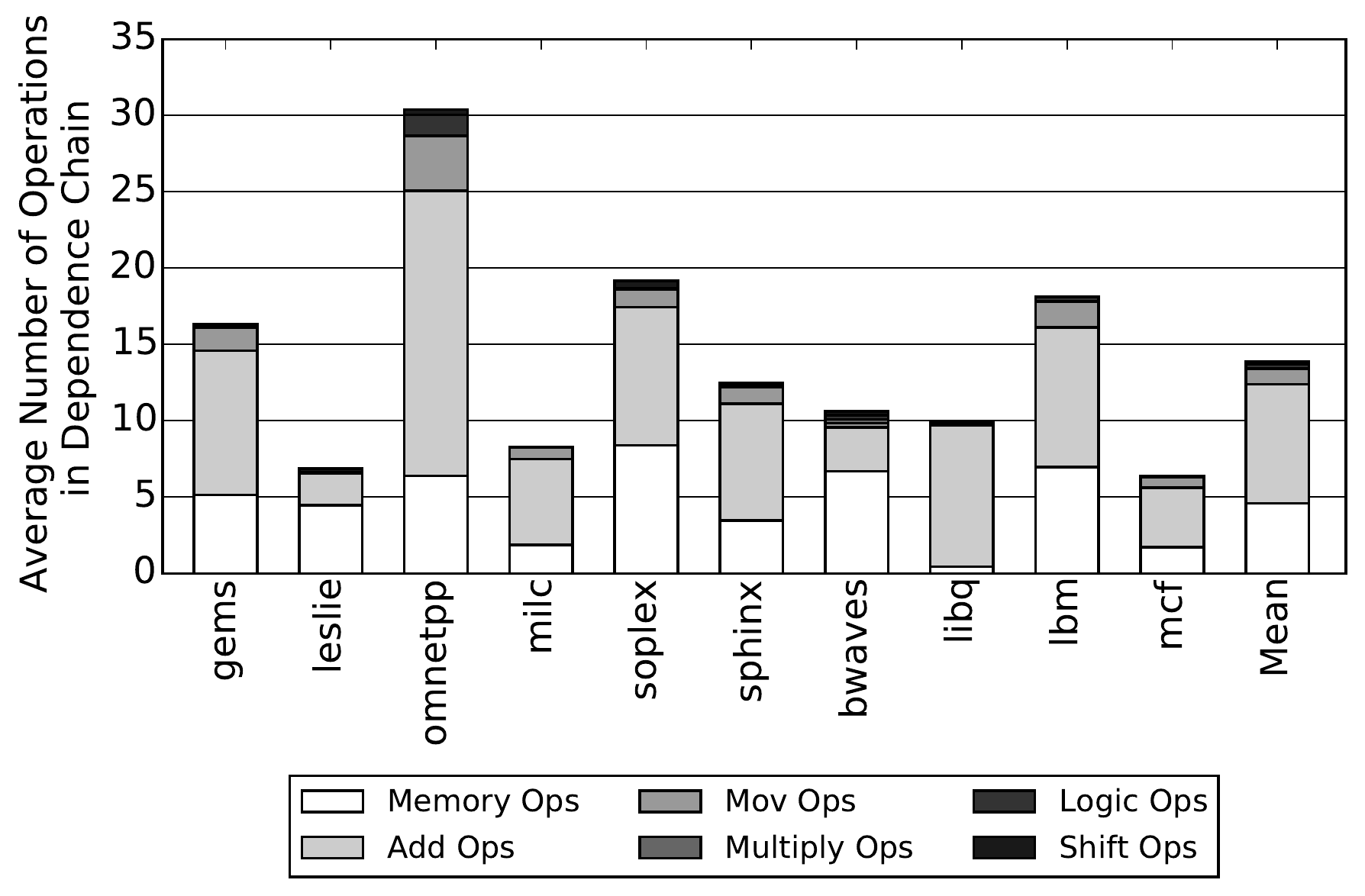}
	\caption{Length and Breakdown of Dependence Chains}
	\label{fig:depChainStack}
	\vspace{-.15in}
\end{figure}

Second, when tracking all independent cache miss chains, I verify the observation from Section \ref{sec:raBuf:Obs} that the average dependence chain length is short. Figure \ref{fig:depChainStack} shows that chains are 14 operations on average and consist of mainly memory, add, and move operations. Multiply, logical, and shift operations all add less than one operation on average to dependence chain length. This also suggests that it is practical for hardware to dynamically uncover the dependence chains for all independent cache misses, not just those that fit the constraints of the runahead buffer. Note that control operations do not propagate a register dependency and therefore do not appear in the backwards data-flow walks that generate these dependence chains. The operations that appear in the dependence chain comprise the speculated path through the program that the branch-predictor has predicted.

\noindent \textbf{Maximum Misses Oracle:} For hardware simplicity, the runahead buffer constrains itself to using a dependence chain based on the PC that is blocking retirement. However, hardware can take advantage of the observation that the total number of PCs that cause the majority of all independent cache misses is small. In the second oracle policy, instead of using the PC of the operation blocking retirement to index the PC miss table, the simulator searches the entire table for the PC that has resulted in the most LLC misses over the history of the application so far. This assigns criticality to the loads that miss most often. 

The maximum misses oracle policy maintains all PCs that have generated LLC misses and every LLC miss dependence chain for each PC. At a full window stall, the dependence chain that has caused the most cache misses for the chosen PC is loaded into the runahead buffer and runahead execution begins. The performance of this policy is shown in Figure \ref{fig:threadPolicy}. Choosing a dependence chain based on the PC that has generated the most LLC misses improves performance by 8\% on average over the PC-Based oracle.

\begin{figure}
	\centering
	\includegraphics[width=\columnwidth]{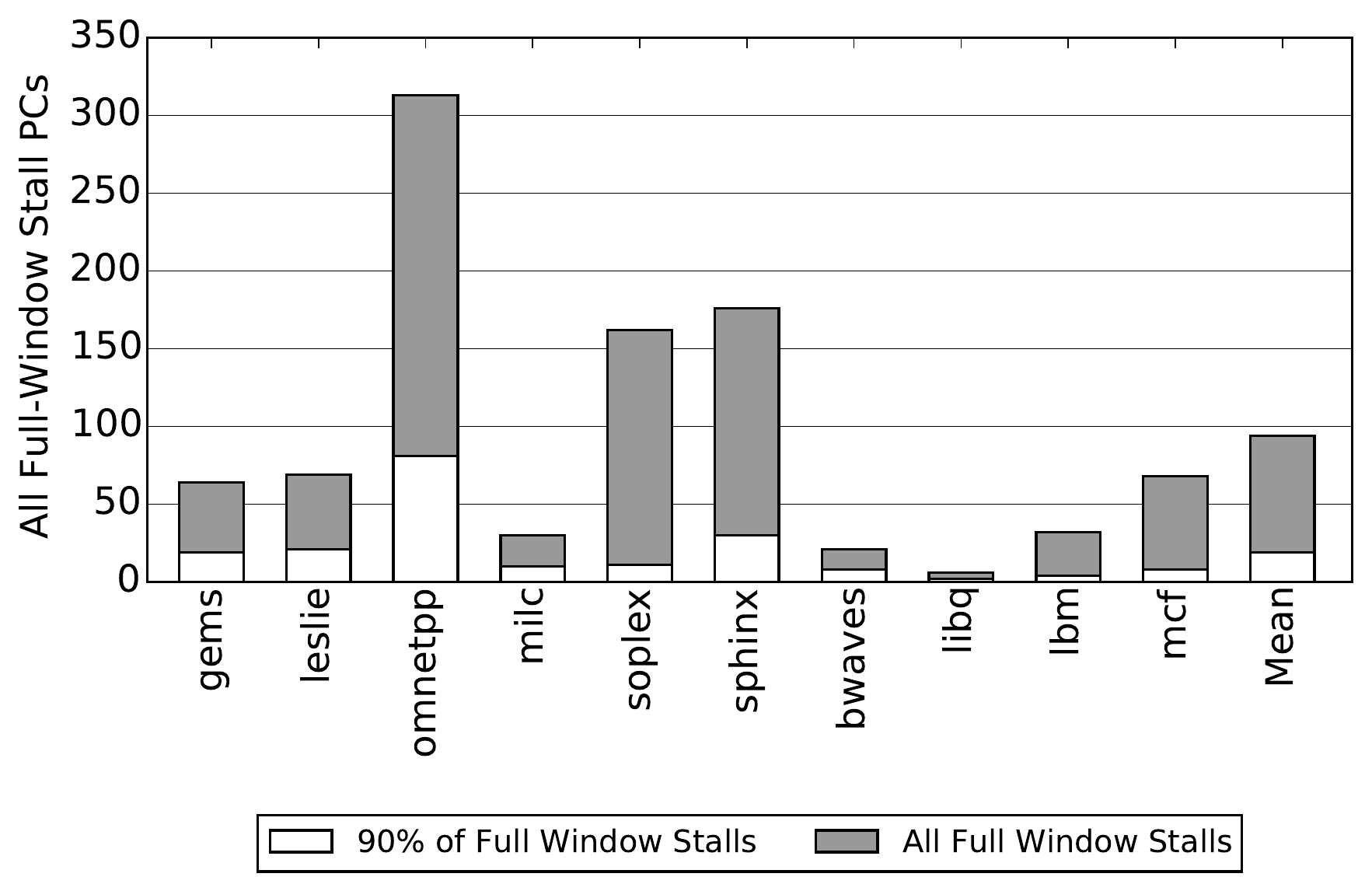}
	\caption{Number of Different Instructions that cause Full-Window Stalls}
	\label{fig:missStallStack}
\end{figure}

\noindent \textbf{Stall Oracle:} Due to overlapping memory access latencies in an out-of-order processor, the load with the highest number of total cache misses is not necessarily the most critical load operation. Instead, the most important memory operations to accelerate are those that cause the pipeline to stall due to a full reorder buffer. Using this insight, Figure \ref{fig:missStallStack} displays both the total number of different instructions that cause full-window stalls in the high memory intensity \textit{SPEC06} applications and the number of operations that cause 90\% of all full-window stalls. On average, 94 operations cause full-window stalls per benchmark and 19 operations cause 90\% of all full-window stalls. This is much smaller than the number of operations that cause cache misses in Figure \ref{fig:missPCStack}, particularly for \textit{omnetpp} and \textit{sphinx} and provides a filter for identifying the most critical loads.

For the final policy, the simulator tracks each PC that has caused a full-window stall and every dependence chain that has caused a full-window stall for each PC. Each PC has a counter that is incremented when a load operation blocks retirement. At a full-window stall, the simulator searches the table for the PC that has caused the most full-window stalls. The dependence chain that has caused the most stalls for the chosen PC is then loaded into the runahead buffer. Figure \ref{fig:threadPolicy} shows that this is the highest performing oracle policy on average.

In conclusion, I find that the highest performing policy is the stall oracle. While the runahead buffer uses the operation blocking retirement to generate a dependence chain, it is ideal to use the operation that has caused the most full-window stalls. However, from Figure \ref{fig:depTransSens}, it is reasonable to only track the last dependence chain for the chosen PC. It is not necessary to maintain a large cache of dependence chains. These observations are used to turn the stall oracle into a realizable hardware policy in Section \ref{sec:scRaEMC:hwDepPolicy}.

\subsection{Hardware Stall Policy}
\label{sec:scRaEMC:hwDepPolicy}

In this Section I develop a hardware algorithm based on the stall oracle to identify a dependence chain for use during runahead execution at the EMC. The stall oracle uses unbounded space to track all PCs that cause full-window stalls. However, from Figure \ref{fig:missStallStack}, the average number of operations that cause full-window stalls per benchmark is only 94, an unbounded amount of space is not necessary. 

\begin{figure}
	\centering
	\includegraphics[width=\columnwidth]{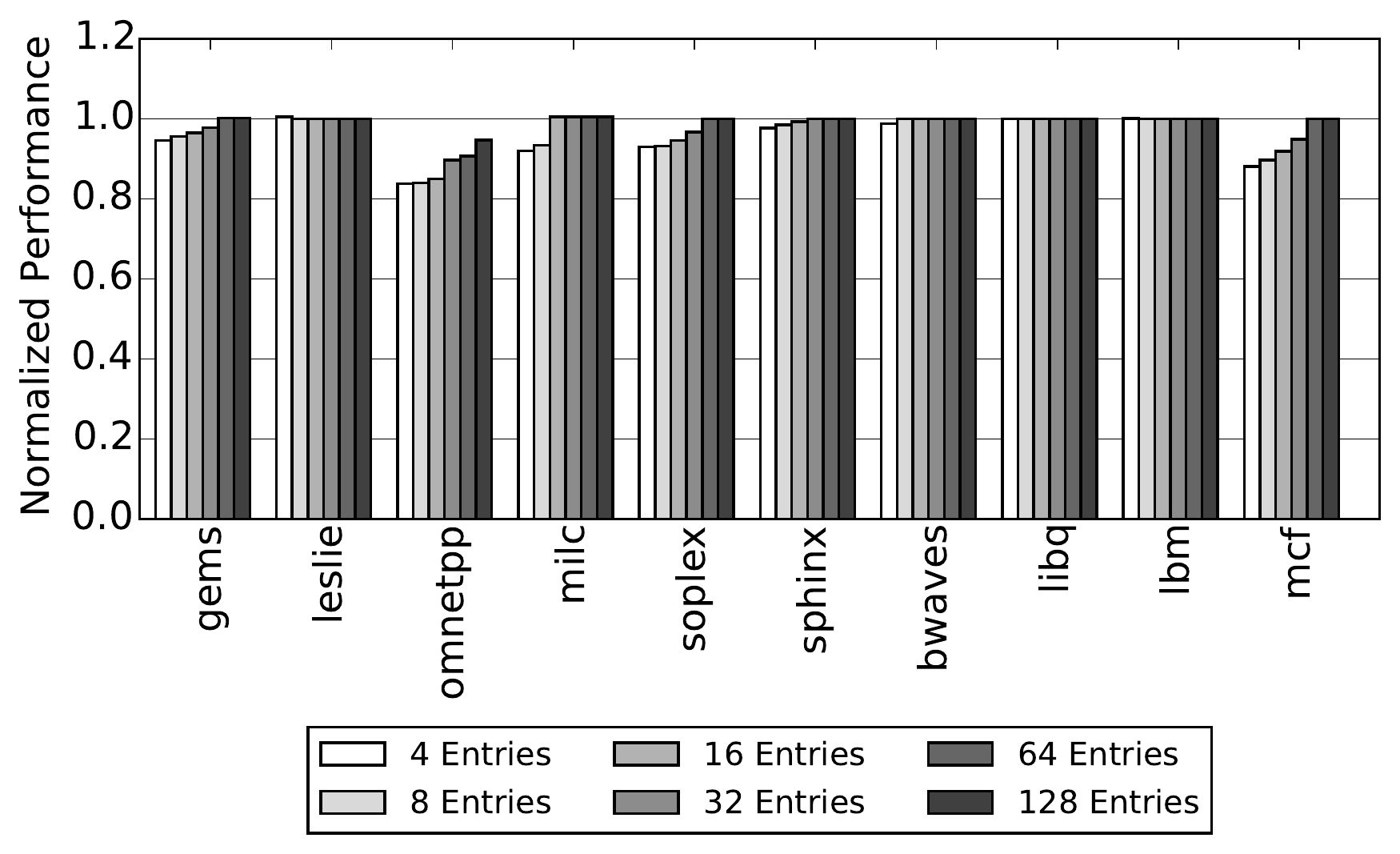}
	\caption{Sensitivity to the Number of Tracked PCs}
	\label{fig:trackPCSens}
\end{figure}

Figure \ref{fig:trackPCSens} shows performance for the stall oracle policy if the number of PC entries is varied from 4 to 128 (based on the 90\% percentile data in \ref{fig:trackPCSens}). The chart is normalized to a baseline with an unbounded amount of storage. Some applications maximize performance with a 4-entry cache, once again highlighting that the most recently used path is often advantageous for predicting future behavior. However, the 4-entry configuration results in a significant performance degradation on \textit{mcf} and \textit{omnetpp}. On average, the 32-entry configuration provides performance close to the unbounded baseline at lower cost (with the exception of \textit{omnetpp} which has the largest number of PCs that cause full-window stalls in Figure \ref{fig:missStallStack}). Therefore, I propose maintaining a 32-entry cache of PCs that tracks the operations that cause full-window stalls. If the processor is in a memory-intensive phase, the PC that has caused the highest number of full-window stalls is marked and used to generate a new dependence chain for use during runahead execution. To separate independent cache misses from dependent cache misses, the PC-Miss table is only updated if the operation blocking retirement has been determined to not be a dependent cache miss (Section \ref{sec:EMC:trace_gen}). This policy is described in Algorithm \ref{alg:emcIndChainMark}.

\begin{algorithm}
	\caption{Marking a Runahead Operation}
	\label{alg:emcIndChainMark}
	\begin{algorithmic}
	\IF{ROB full \textbf{AND} stall op is not dependent cache miss}
	\STATE Access PC-Miss Table with PC of op causing stall
	\IF{Hit}
	\STATE Increment miss counter for PC-entry
	\ELSE
	\STATE Allocate new PC-Miss Table entry
	\ENDIF
	\ENDIF
	\newline
	\IF{New Interval \textbf{AND} High MPKI}
	\STATE Mark PC with largest miss counter in PC-Miss Table
	\ENDIF
	\end{algorithmic}
\end{algorithm}

\begin{figure}
	\centering
	\includegraphics[width=5.7in, height=2.8in]{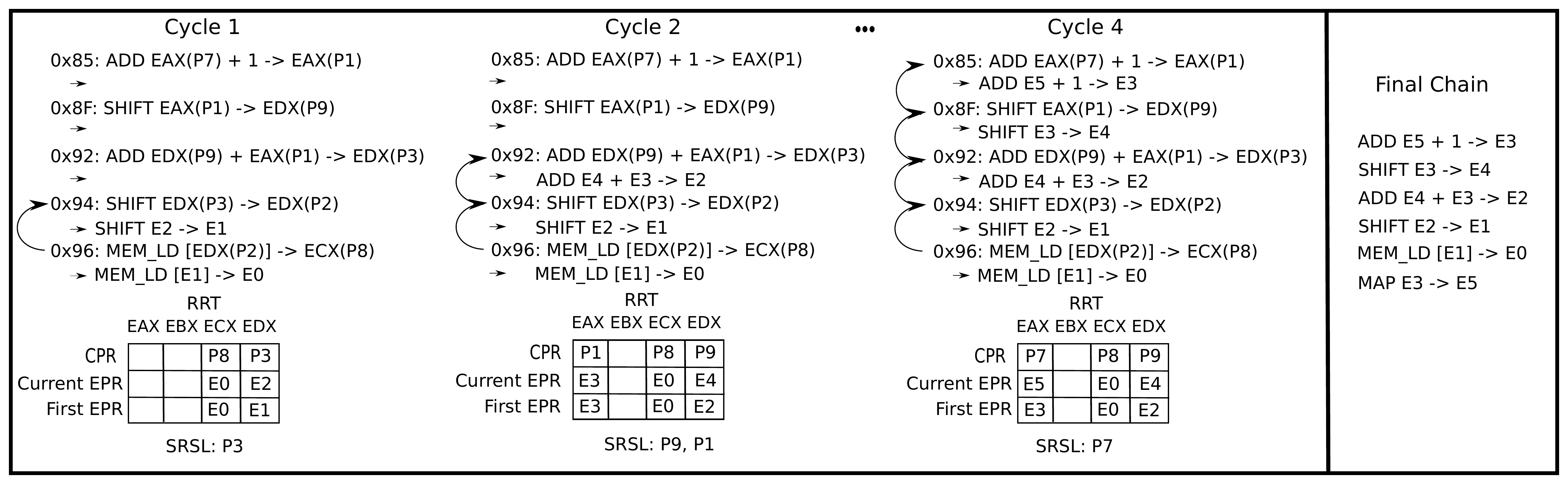}
	\caption{EMC Runahead Chain Generation}
	\label{fig:raemcDepChain}
\end{figure}

Once a PC is marked for dependence chain generation, the next time a matching PC is issued into the reservation stations the core begins the chain generation process. Dependence chain generation is a combination of the backwards data-flow walk in Section \ref{sec:raBuf:wakeup} and EMC dependence chain generation in Section \ref{sec:EMC:trace_gen}. A backwards data-flow walk is required to identify source operations, but operations must be renamed to execute at the EMC. Algorithm \ref{alg:emcIndChainGen} details the dependence chain generation process while Figure \ref{fig:raemcDepChain} provides an example chain of code adapted from \textit{mcf}. As in the dependent cache miss case (Section \ref{sec:EMC:trace_gen}), we leverage the register state available at the core to rename the chain for EMC execution. This is advantageous as the chain only has to be renamed once instead of every iteration at the EMC.

In Figure \ref{fig:raemcDepChain} the load at PC 0x96 has been marked for dependence chain generation.  In cycle 0 (Not shown in Figure \ref{fig:raemcDepChain}), the operation is identified and the destination register P8 is mapped to E0. Source register P2 is mapped to E1 and added to the source register search list (SRSL). These changes are recorded in the register remapping table (RRT). Note that the RRT from Section \ref{sec:EMC:trace_gen} has been slightly modified to include an additional row of EMC physical registers. This row is written once when an architectural register is mapped for the first time. It is then used to re-map all live-outs back to live-ins at the end of dependence chain generation and is required to allow the dependence chain to execute as if it was in a loop.

In cycle 1 the core searches all older destination registers for the producer of P2. From Chapter \ref{chap:raBuf}, the ROB has been modified to include a CAM on source and destination registers. These structures are used during the dependence chain generation process. If an operation is found, it is marked to be included in the dependence chain and read out of the ROB at retirement. The result of the search is found to be a SHIFT and the source register of the shift (P3) is remapped to E2 and enqueued in the SRSL. This process continues until the SRSL is empty. In cycle 2 P9 and P1 are remapped to E4 and E3 respectively. In cycle 3 the SHIFT at address 0x8F is remapped and in cycle 4 the ADD at address 0x85 is remapped and enqueues P7 into the SRSL. 

In cycle 5 P7 does not find any producers. This means that EAX (P7) is a live-in into the dependence chain. This result is recorded in the RRT. To be able to speculatively execute this dependence chain as if it was in a loop a new operation is inserted at the end of the final dependence chain. This ``MAP" operation moves the live-out for EAX (E3) into the live-in for EAX (E5) thereby propagating data from one dependence chain iteration to the next. Semantically, MAP also serves as a data-flow barrier and denotes the boundary between dependence chain iterations. MAP cannot issue at the EMC until all prior operations have issued. This restricts the out-of-order engine at the EMC but as the issue width of the EMC (2) is much smaller than dependence chain length MAP does not result in a negative performance impact. For wider EMC back-ends, future research directions could include unrolling dependence chains using live-in data to the maximum number of free EMC physical registers. A pure hardware dataflow implementation is also possible as the EMC does not retire any operations and is just prefetching.

\begin{figure}
	\centering
	\includegraphics[width=2.5in, height=2.5in]{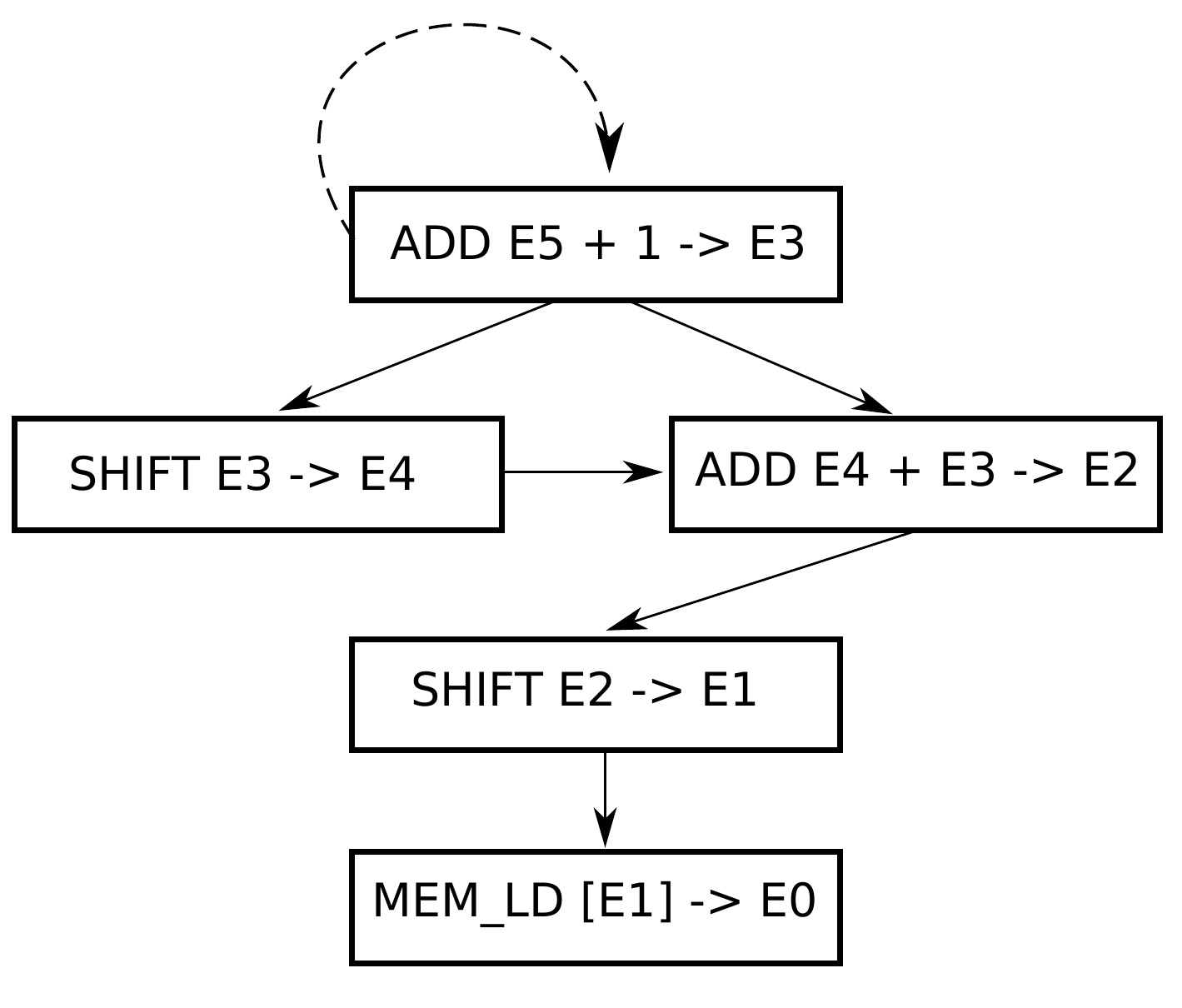}
	\caption{Dataflow Graph of Dependence Chain}
	\label{fig:traceDataFlow}
\end{figure}

The result of the MAP operation on the dataflow graph of the dependence chain in Figure \ref{fig:raemcDepChain} is shown in Figure \ref{fig:traceDataFlow}.  Solid arrows represent actual register dependencies while the dashed line shows the MAP operation feeding the destination register E5 back into the source register of the ADD. The final dependence chain including the MAP is shown to the right of Figure \ref{fig:raemcDepChain}. Once Algorithm \ref{alg:emcIndChainGen} has completed, the dependence chain is sent to the EMC along with a copy of the core physical registers used in the RRT to begin runahead execution.

\begin{algorithm}
	\caption{Dependence chain generation \newline MAXLENGTH: 32, SRSL: Source Register Search List, ROB: Reorder Buffer, DC: Dependence Chain, CAR: Core Architectural Register, CPR: Core Physical Register, EPR: EMC Physical Register}
	\label{alg:emcIndChainGen}
	\begin{algorithmic}
		\IF{PC Marked \textbf{AND} PC Issued}
		\STATE \textit{mapOperation}(missPCOp)
		\WHILE {SRSL != EMPTY \AND DC \textless MAXLENGTH}
		\STATE Dequeue CPR from SRSL
		\STATE Search ROB for older op that produces CPR
		\IF {Matching op with new PC found}
		\STATE \textit{mapOperation}(matchingOp)
		\IF{Matching op is load}
		\STATE Search store buffer for load address
		\IF{Store buffer match}
		\STATE Mark matching store for DC inclusion
		\STATE \textit{mapOperation}(matchingStore)
		\ENDIF
		\ENDIF
		\ELSE
		\STATE Mark CAR as live-in in RRT		
		\ENDIF
		\ENDWHILE
		\STATE MAP each live in register: First EPR -\textgreater Current EPR
		\ENDIF	
		\newline
		\STATE \textbf{function} \textit{mapOperation}(operation)
		\STATE \hspace{\algorithmicindent} Allocate First EPR for each new CAR in RRT
		\STATE \hspace{\algorithmicindent} Allocate Current EPR for destination CPR in RRT
		\STATE \hspace{\algorithmicindent} Allocate Current EPR for each new source CPR in RRT
		\STATE \hspace{\algorithmicindent} EPR = RRT[CAR] for each core source register
		\STATE \hspace{\algorithmicindent} Enqueue all new source CPR to SRSL
		\STATE \hspace{\algorithmicindent} Mark op for DC inclusion
		\STATE \textbf{end function}
	\end{algorithmic}
\end{algorithm}

\begin{figure}
	\centering
	\includegraphics[width=3.5in, height=2.2in]{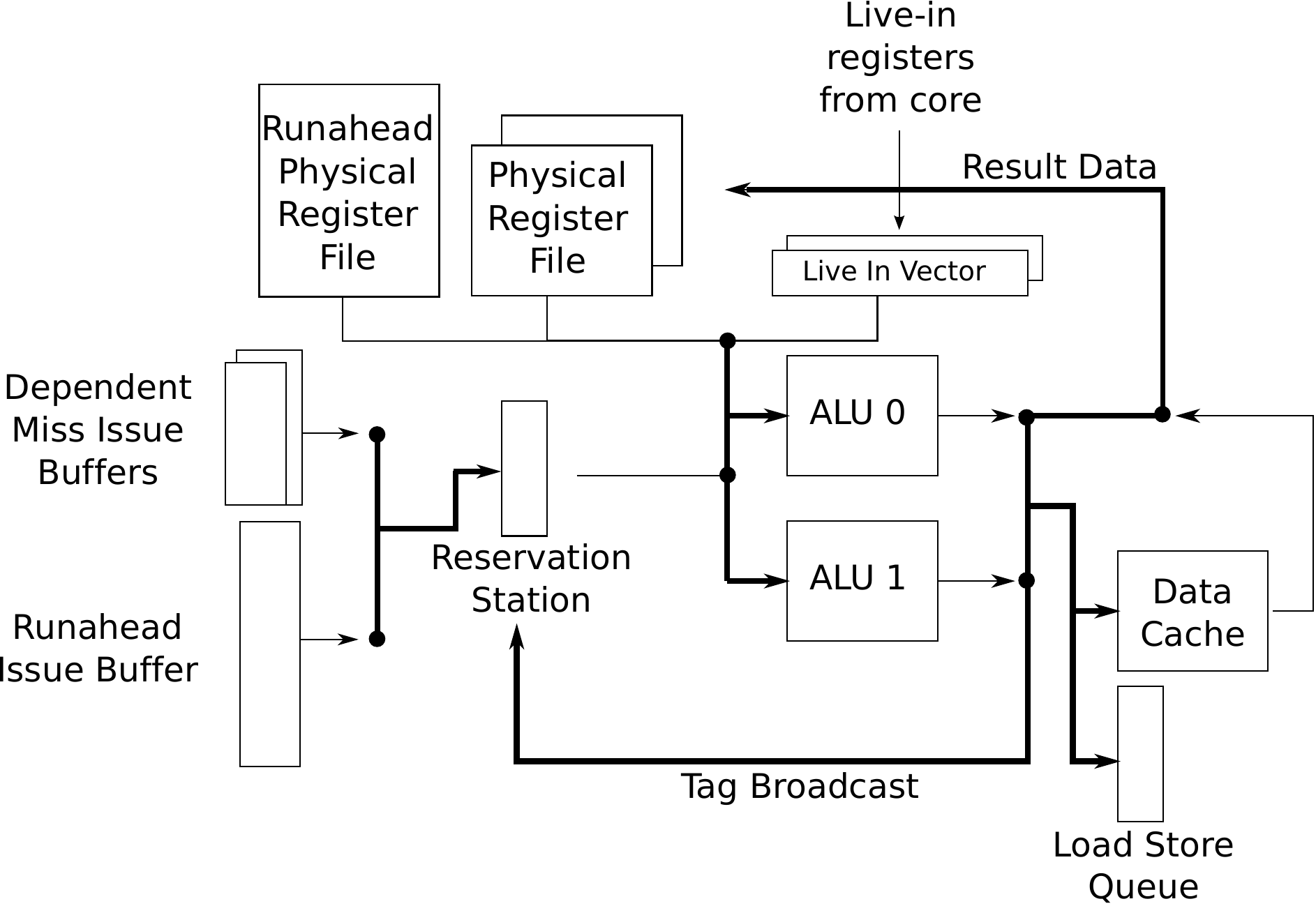}
	\caption{EMC Microarchitecture Runahead Modifications}
	\label{fig:raEMCDesign}
\end{figure}

Table \ref{tab:raEMCHW} lists the additional hardware storage overhead required to implement Algorithm \ref{alg:emcIndChainGen}. At the core, the PC-Miss table requires entires and counters to track operations that block retirement. The core must include additional storage for the dependence chain while it is being generated and a copy of all remapped physical registers to send to the EMC as input data. The EMC requires an additional hardware context to hold the runahead dependence chain and physical registers. These additions are shown in Figure \ref{fig:raEMCDesign}.

\begin{table*}[ht]
	\small
	\centering
	\begin{tabular}{|p{2.2in}|p{3.0in}|}
		\hline
		\textbf{Component} & \textbf{Bytes} \\
		\hline
		\hline
		\textbf{Core} & \\
		\hline
		PC-Miss Table Entries & 4 Bytes * 32 Entries = 128 Bytes \\
		\hline
		PC-Miss Table Counters & 4 Bytes * 32 Entries = 128 Bytes \\
		\hline
		RRT Capacity Increase & 4 Bytes * 32 Entries = 128 Bytes \\
		\hline
		Dependence Chain Buffer & 8 Bytes * 32 Entries = 256 Bytes \\
		\hline
		Register Copy Buffer & 4 Bytes * 32 Entries = 128 Bytes \\
		\hline
		Total New Core Storage & 768 Bytes \\
		\hline
		\hline
		\textbf{EMC} & \\
		\hline
		Runahead Physical Register File & 4 Bytes * 32 Entries = 128 Bytes \\
		\hline
		Runahead Chain Storage & 8 Bytes * 32 Entries = 256 Bytes\\
		\hline
		Total New EMC Storage & 384 Bytes\\
		\hline
	\end{tabular} 
	\caption{Additional RA-EMC Hardware Overhead}
	\label{tab:raEMCHW}
\end{table*}

\subsection{EMC Runahead Control}
\label{sec:scRaEMC:dist}

While Section \ref{sec:scRaEMC:hwDepPolicy} describes the hardware policy used to determine which dependence chain to execute at the EMC, it does not explore how to control runahead execution at the EMC. In this section I show that runahead request accuracy, and correspondingly EMC runahead performance, can be managed by the interval that the core uses to send dependence chain updates to the EMC. 

\begin{figure}
	\centering
	\includegraphics[width=\columnwidth]{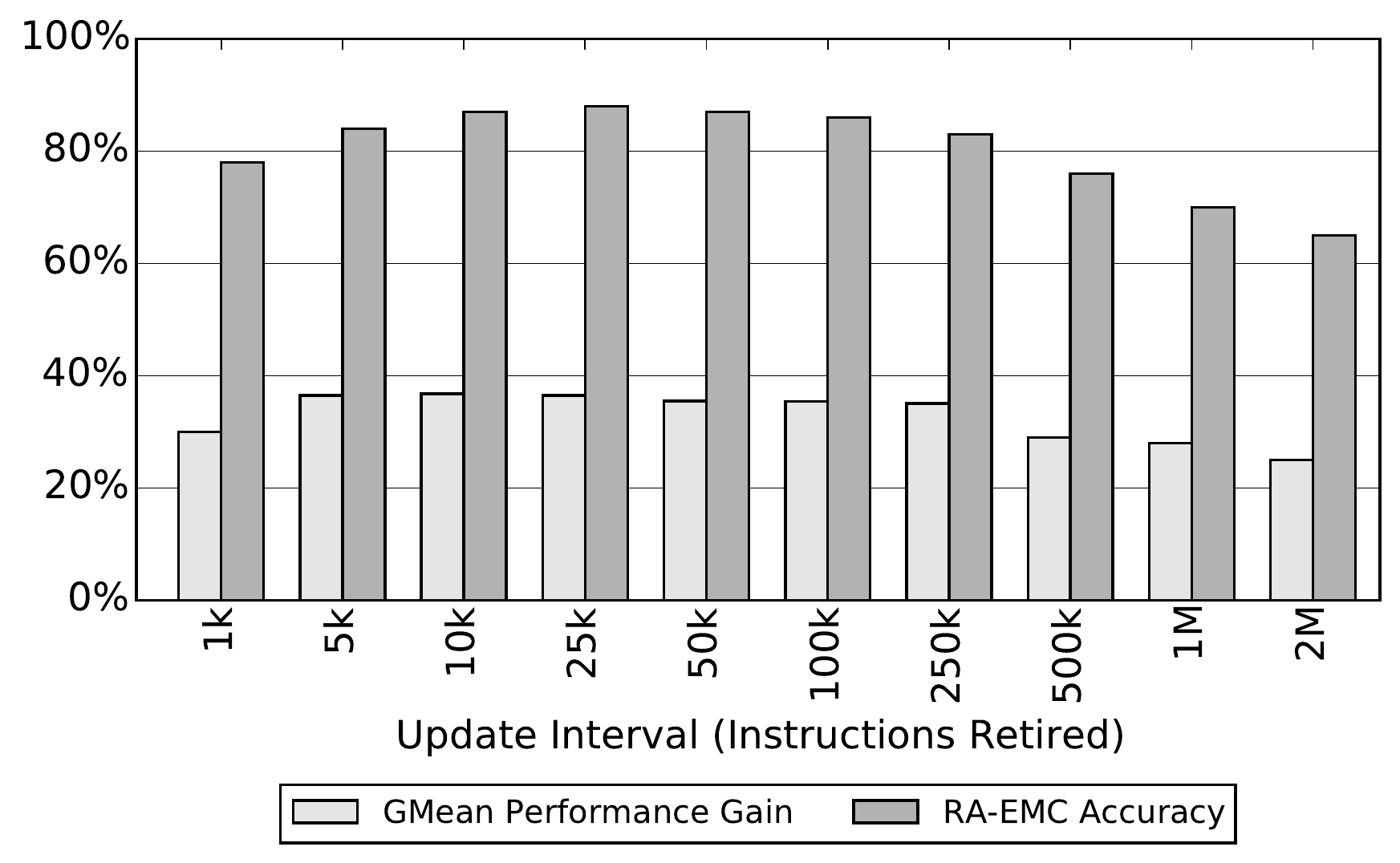}
	\caption{Sensitivity to Update Interval}
	\label{fig:chainDurSens}
\end{figure}

When a dependence chain is sent to the EMC for runahead execution, a copy of the core physical registers that are required to execute the first iteration of the dependence chain are also sent. This serves to reset runahead at the EMC. For example, if the core sends a new dependence chain to the EMC at every full window stall, the runahead interval length at the EMC for each dependence chain is simply the average time between full-window stalls. At every new full window stall, the processor will reset the state of the EMC to execute a new runahead dependence chain. This limits the distance that the EMC is allowed to run ahead of the main core. Therefore, Figure \ref{fig:chainDurSens} explores how modifying the dependence chain update interval impacts performance and runahead request accuracy. The x-axis varies the update interval based on the number of instructions retired at the core from one thousand instructions to 2 million instructions. There are two bars, the first bar is the average geometric mean performance gain of the memory intensive \textit{SPEC06} benchmarks. The second bar is the request accuracy defined as the percent of total lines fetched by runahead at the EMC (RA-EMC) that are touched by the core before eviction.

The main observation from Figure \ref{fig:chainDurSens} is that both average performance and runahead request accuracy plateaus from the 5k instruction update interval to the 100k update interval. From the 250k update interval to the 2M instruction interval both request accuracy and performance decrease. Average runahead request accuracy in the plateau of the chart is at about 85\%. This is about 10\% less than average runahead request accuracy with the runahead buffer (Figure \ref{fig:raTouch}). As both runahead accuracy and performance gain decrease above the 250k update interval, it is clear that allowing the EMC to runahead for too long without an update has a negative effect on performance as the application can move to a different phase. However, the 1k update interval also reduces performance without a large effect on runahead request accuracy. Table \ref{tab:raEMCIDiff} demonstrates why this occurs by listing the average number of instructions executed between when a runahead line is fetched by the EMC and when it is accessed by the core. On average, the 1k interval length has a much shorter runahead distance than the 5k or 10k interval. By frequently resetting the EMC, the core decreases EMC effectiveness in the 1k case. 

From the 5k interval length upwards, the values in Table \ref{tab:raEMCIDiff} stabilize to roughly three thousand instructions from when the value is prefetched into the last level cache by the EMC to when the core accesses the data on average. This number is controlled by the rate at which the EMC is able to issue memory requests into the cache hierarchy and is influenced by many factors including: EMC data cache hit rate, interconnect contention, LLC bank contention, memory controller queue contention, benchmark memory intensity, DRAM bank contention, and DRAM row buffer hit rate. If the entire memory system is abstractly viewed as a queue, Little's Law \cite{little:61} bounds the rate at which the EMC is able to issue independent cache misses at the queue size divided by the average time spent in the queue. Assuming a constant application phase where the memory intensity of the runahead dependence chain is constant, this is independent from the update interval of the EMC in the steady state. For the memory intensive \textit{SPEC06} applications, this plateau lasts from the 5k interval length until the 100k interval length on average. To reduce communication overhead, it is advantageous to control the EMC at the coarsest interval possible to maintain high performance. Therefore, based on Table \ref{tab:raEMCIDiff} and Figure \ref{fig:chainDurSens} I choose a 100k instruction update interval for runahead at the EMC. This interval length changes in Section \ref{sec:mcRaEMC:dynamicPolicy} when multi-core contention changes this argument and a dynamic interval length is advantageous.

\begin{table*}[ht]
	\small
	\centering
	\begin{tabular}{|p{.4in}|p{.4in}|p{.4in}|p{.4in}|p{.4in}|p{.4in}|p{.4in}|p{.4in}|p{.4in}|p{.4in}|}
		\hline
		1k & 5k & 10k & 25k & 50k & 100k & 250k & 500k & 1M & 2M \\
		\hline
		\hline
		732 & 2576 & 2864 & 2919 & 2898 & 2933 & 3002 & 3086 & 3074 & 3032 \\
		\hline
	\end{tabular} 
	\caption{Runahead Load to Use Distance (Instructions)}
	\label{tab:raEMCIDiff}
\end{table*}

\section{Methodology}
\label{sec:scRaEMC:meth}

For the single core evaluation, the simulator is set up as in Section \ref{sec:raBuf:method}. EMC parameters are as in Section \ref{sec:EMC:method}, except with only one dependent cache miss context and the addition of a runahead issue context as shown in Figure \ref{fig:raEMCDesign}. These settings are summarized in Table \ref{tab:systemConfig}. The system is evaluated on the medium and high memory intensive benchmarks from the \textit{SPEC06} benchmark suite.

\begin{table*}[ht!]
	\small
	\centering
	\begin{tabular}{|p{.75in}|p{4.5in}|}
		\hline Core & 4-Wide Issue, 256 Entry ROB, 92 Entry Reservation Station, Hybrid Branch Predictor, 3.2 GHz Clock Rate \\ 
		\hline L1 Caches & 32 KB I-Cache, 32 KB D-Cache, 64 Byte Lines, 2 Ports, 3 Cycle Latency, 8-way, Write-Through. \\ 
		\hline \hline L2 Cache &  1MB 8-way, 18-cycle latency, Write-Back. \\ 
		\hline \hline EMC \newline Compute & 2-wide issue. 8 Entry Reservation Stations. 32 Entry TLB. 64 Line Data Cache 4-way, 2-cycle access, 1-port. 1 Runahead dependence chain context with 32 entry uop buffer, 32 entry physical register file. 1 Dependent cache miss context with 16 entry uop buffer, 16 entry physical register file. Micro-op size: 8 bytes in addition to any live-in source data. \\
		\hline EMC \newline Instructions & Integer: add/subtract/move/load/store. \newline \hbox{Logical: and/or/xor/not/shift/sign-extend.} \\
		\hline Memory Controller & 64 Entry Memory Queue. \\ 
		\hline Prefetchers & Stream: 32 Streams, Distance 32. Markov: 1MB Correlation Table, 4 addresses per entry. GHB G/DC: 1k Entry Buffer, 12KB total size. All configurations: FDP \cite{fdp07}, Dynamic Degree: 1-32, prefetch into Last Level Cache. \\  
		\hline DRAM & DDR3\cite{dram:micron}, 1 Rank of 8 Banks/Channel, 8KB Row-Size, CAS 13.75ns. 800 MHz bus. 2 Channels. \\
		\hline
	\end{tabular} 
	\caption{System Configuration}
	\label{tab:systemConfig}
\end{table*}

\section{Results}
\label{sec:scRaEMC:results}

\subsection{Performance Results}
\label{sec:scRaEMC:results:perf}

\begin{figure}
	\centering
	\includegraphics[width=\columnwidth]{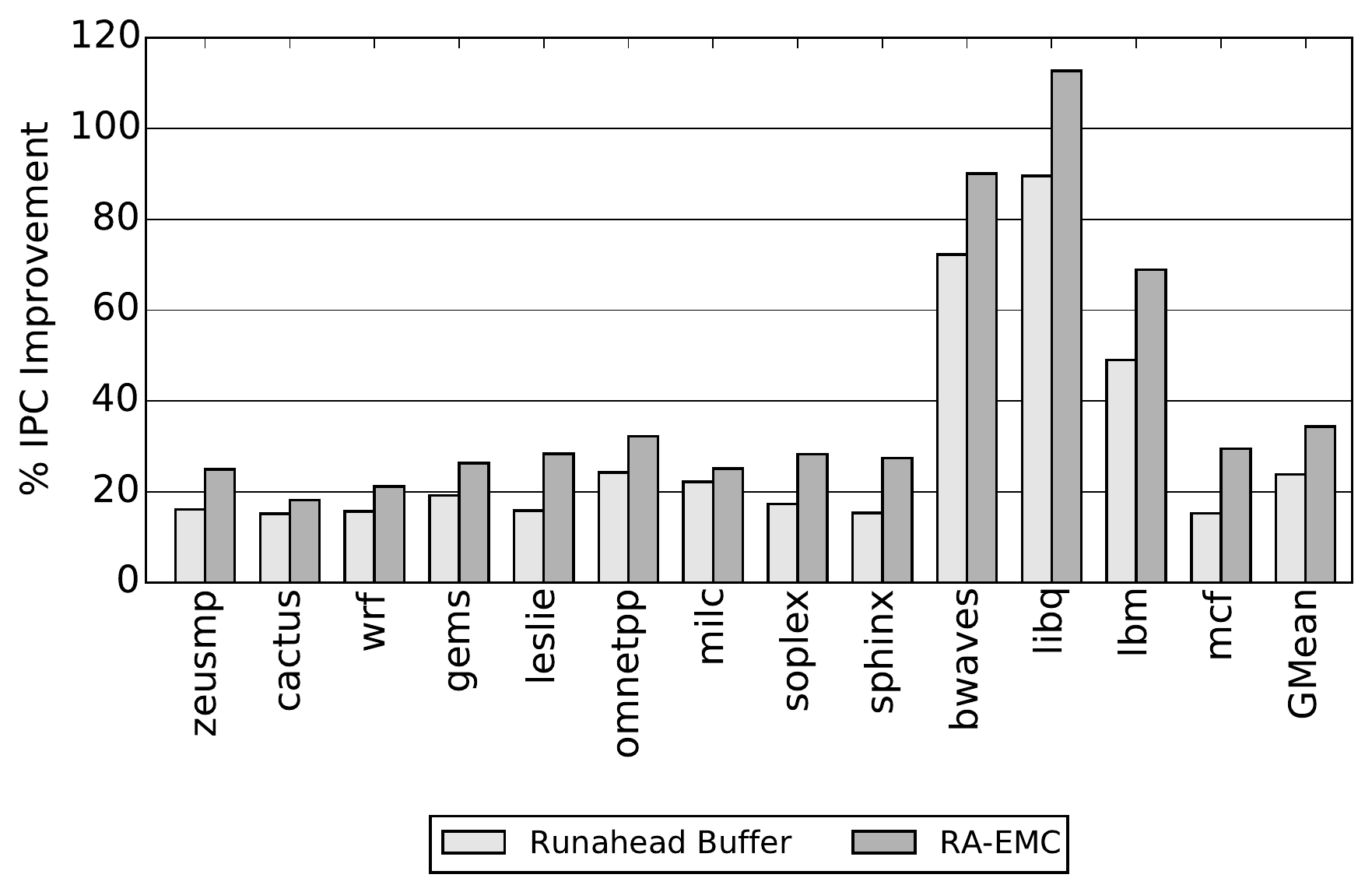}
	\caption{RA-EMC Performance}
	\label{fig:scRaEMC:perf}
\end{figure}

\begin{figure}
	\centering
	\includegraphics[width=\columnwidth]{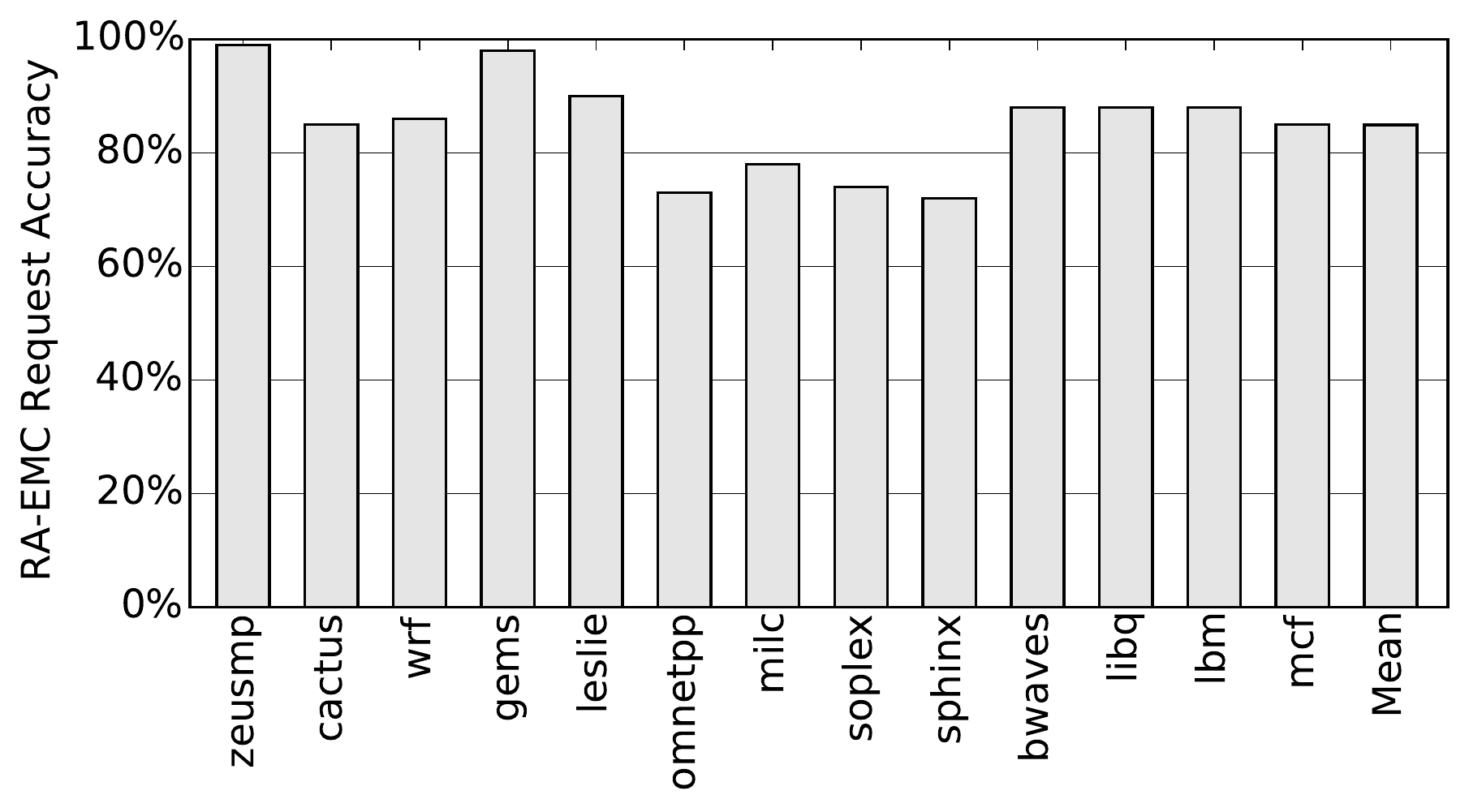}
	\caption{RA-EMC Accuracy}
	\label{fig:scRaEMC:acc}
\end{figure}

Figure \ref{fig:scRaEMC:perf} presents the performance results of runahead at the EMC (RA-EMC) as compared to the best runahead buffer algorithm (Section \ref{sec:scRaEMC:depChain}). RA-EMC improves performance over the runahead buffer for all of the high memory intensity \textit{SPEC06} benchmarks. The benchmarks with both very high memory intensity and high RA-EMC request accuracy in Figure \ref{fig:scRaEMC:acc} such as \textit{bwaves}, \textit{libquantum}, \textit{lbm}, and \textit{mcf} all show larger performance gains over the runahead buffer algorithm. On average, across the entire set of benchmarks the runahead buffer with the hardware stall policy improves performance by 23\% while RA-EMC improves performance by 35\%. The hardware stall policy results in a 7\% lower performance gain when compared to the stall oracle of Figure \ref{fig:threadPolicy}. Considering only the high memory intensity benchmarks the runahead buffer with the hardware stall policy improves performance by 27\% while RA-EMC increases performance by 41\%.

\subsection{RA-EMC Overhead}
\label{sec:scRaEMC:results:overhead}

While RA-EMC increases performance, it also leads to an increase in on-chip activity. With a 100k instruction interval length, the overhead of migrating dependence chains and register state to the EMC for runahead is very small as shown in Table \ref{tab:raEMCTrans}. The total number of data messages required to send register state and micro-operations to the EMC are listed under ``Data Messages". This is a .0006\% average increase in data ring activity. The length of the average dependence chain that is sent to the EMC is listed under ``Dependence Chain Length" for each benchmark.

\begin{table*}[ht]
	\centering
	\begin{tabular}{|c|ccccccc|}\hline
		\centering
		\multirow{4}{1.0in}{\centering Data Messages} & zeusmp & cactus & wrf & gems & leslie & omnetpp & milc \\
		& 992 & 1235 & 362 & 325 & 637 & 1402 & 972 \\\cline{2-8}
		& soplex & sphinx & bwaves & libq & lbm & mcf & \\
		& 668 & 1267 & 190 & 192 & 187 & 917 & \\\cline{2-8}
		\hline
		\multirow{4}{1.0in}{\centering Dependence Chain Length} & zeusmp & cactus & wrf & gems & leslie & omnetpp & milc \\
		& 10.7 & 12.9 & 3.9 & 3.5 & 4.8 & 15.1 & 10.5 \\\cline{2-8}
		& soplex & sphinx & bwaves & libq & lbm & mcf & \\
		& 5.2 & 13.3 & 2.1 & 2.0 & 2.0 & 9.8 & \\\cline{2-8}
		\hline
	\end{tabular}
	\caption{RA-EMC Communication Overhead}
	\label{tab:raEMCTrans}
\end{table*}

\begin{figure}
	\centering
	\includegraphics[width=\columnwidth]{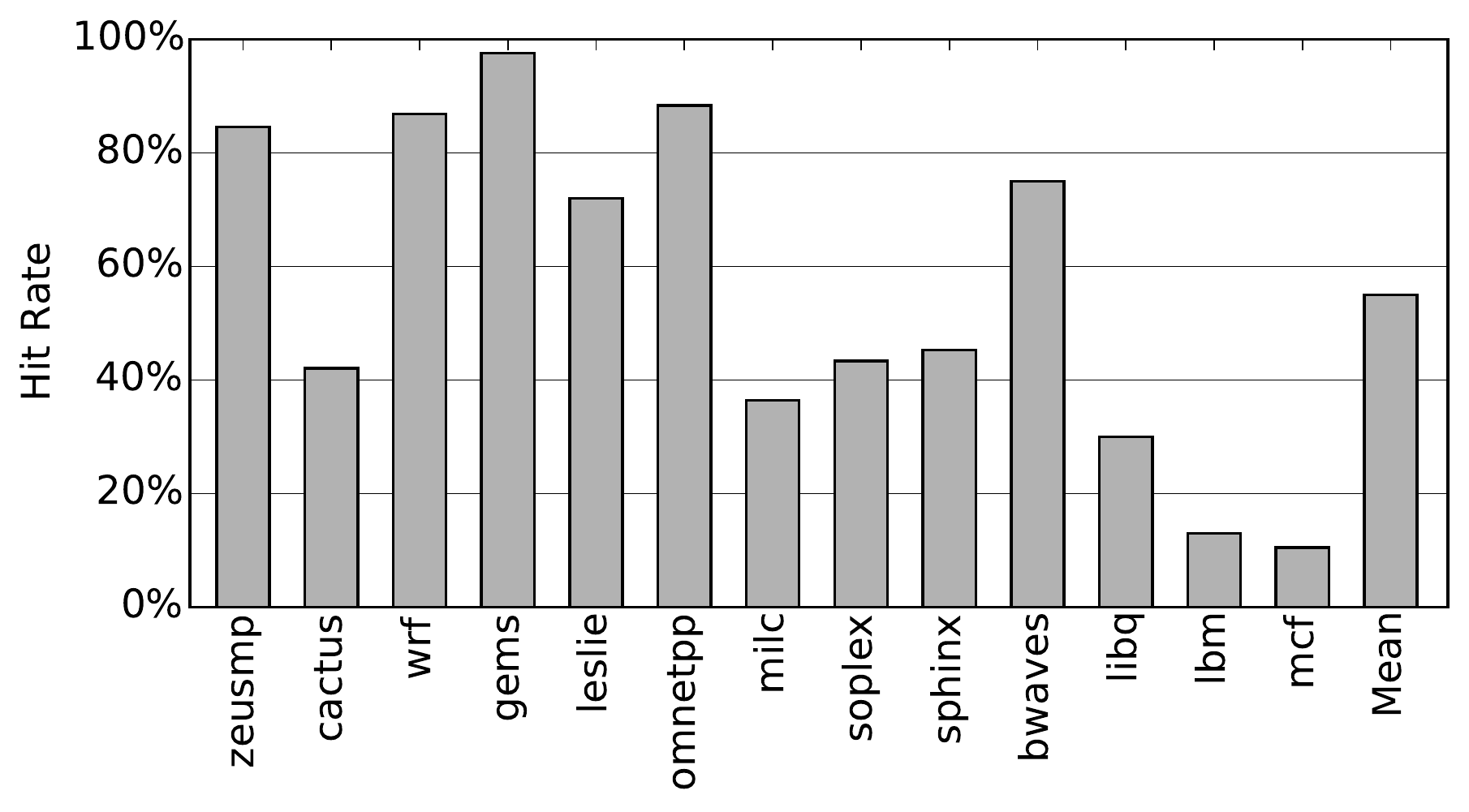}
	\caption{EMC Cache Hit Rate}
	\label{fig:scRaEMC:dcHR}
	\includegraphics[width=\columnwidth]{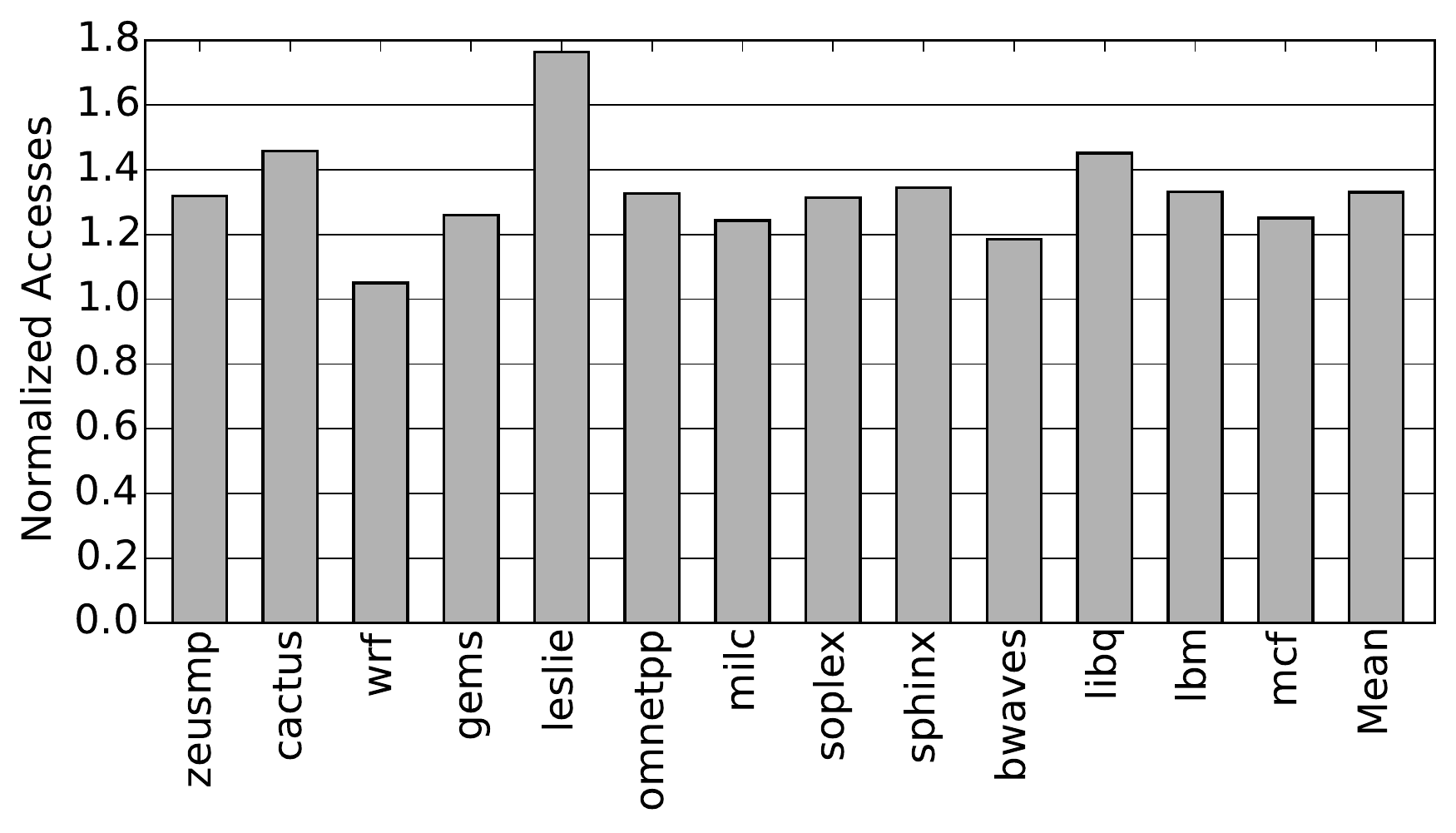}
	\caption{LLC Access Overhead}
	\label{fig:scRaEMC:L1Acc}
\end{figure}

While register state and the dependence chains do not constitute a large overhead, the EMC increases pressure on the on-chip cache hierarchy. EMC requests first query the EMC data-cache before accessing the LLC (Section \ref{sec:EMC:trace_exec}). The hit rate in the EMC data cache is shown in Figure \ref{fig:scRaEMC:dcHR} while the increase in LLC traffic is shown in Figure \ref{fig:scRaEMC:L1Acc}.

The benchmarks with very high hit rates in Figure \ref{fig:scRaEMC:dcHR} such as \textit{gems} or \textit{omnetpp} tend to have load accesses in their dependence chains that hit in the data cache of the core. These benchmarks require fast access to this shared data for high RA-EMC request accuracy. Figure \ref{fig:scRaEMC:dcHR} demonstrates that the 4kB EMC cache results in a 56\% cache hit rate. While this is a substantial fraction of operations, the trade-off to executing dependence chains at the EMC is the increased pressure that the remaining 44\% of loads place on the LLC. Figure \ref{fig:scRaEMC:L1Acc} shows that this results in a 35\% average increase in the number of LLC requests over a no-EMC baseline.

\subsection{RA-EMC + Prefetching}
\label{sec:scRaEMC:results:pref}

\begin{figure}
	\centering
	\includegraphics[width=\columnwidth]{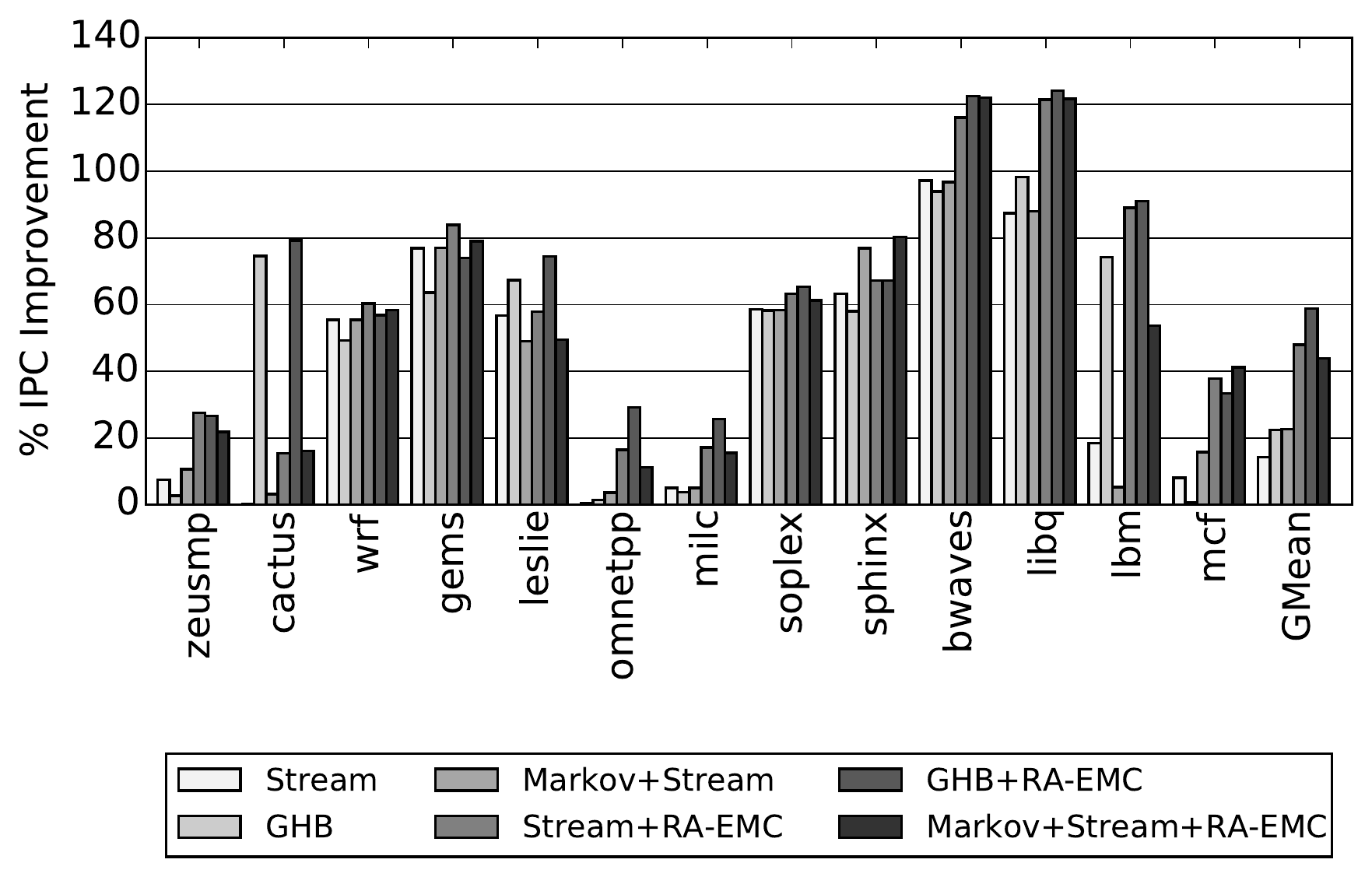}
	\caption{RA-EMC Performance with Prefetching}
	\label{fig:scRaEMC:perfpf}
	\centering
	\includegraphics[width=\columnwidth]{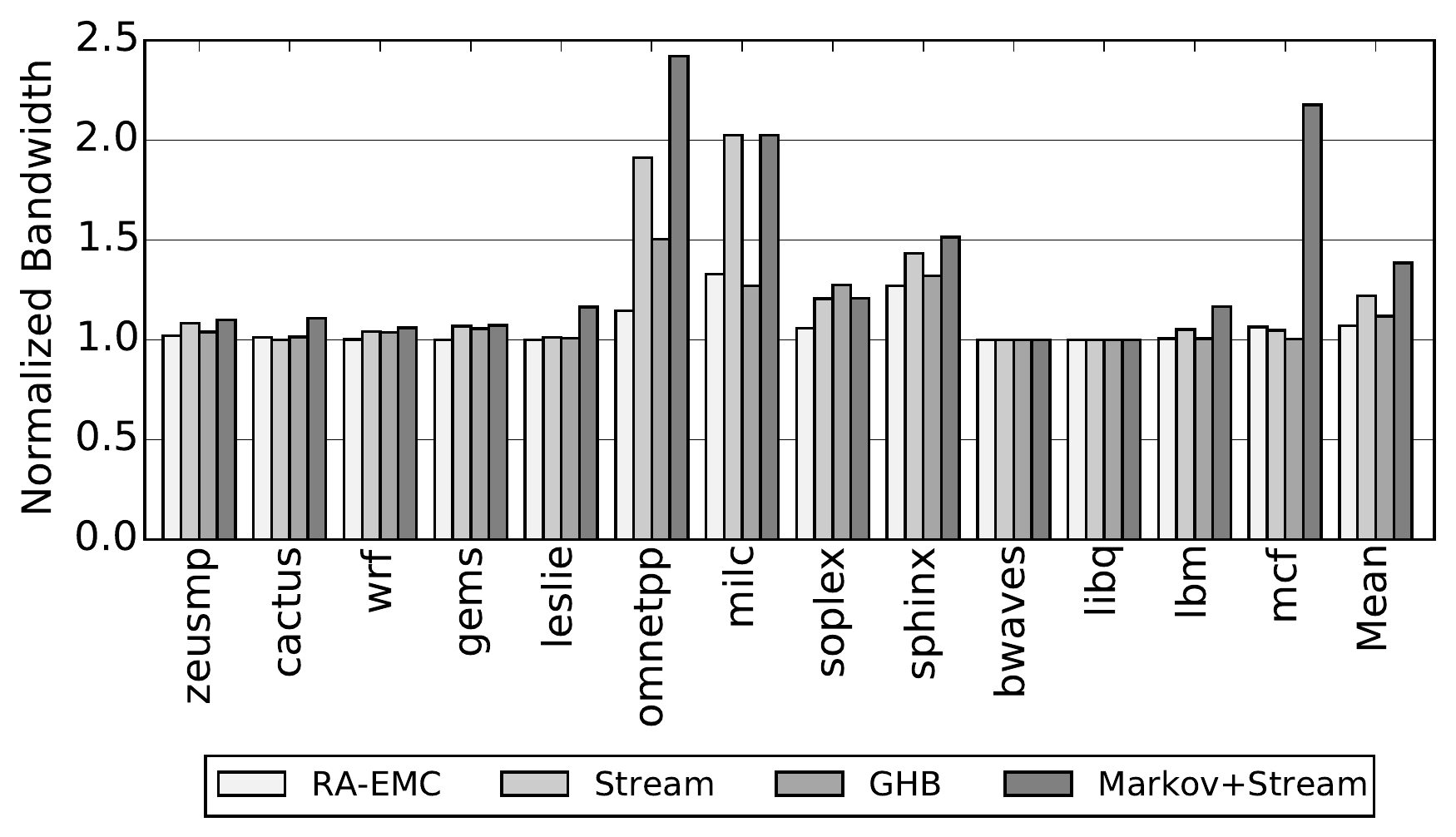}
	\caption{Normalized Bandwidth Overhead}
	\label{fig:scRaEMC:bw}
\end{figure}

Figure \ref{fig:scRaEMC:perfpf} demonstrates the performance impact when prefetchers are added to the system. Overall, the Stream, GHB, and Markov+Stream prefetchers increase performance by 14\%, 22\%, and 22\% on average respectively. As RA-EMC increases performance by 34\% across the medium/high memory intensity benchmarks and by 40\% across the high memory intensity benchmarks, RA-EMC out-performs all three prefetchers. Additionally, RA-EMC improves performance on average when combined with each prefetcher. As in both Sections \ref{sec:raBuf:Pref} and \ref{sec:raBuf:Pref} I observe that the highest performing system on average is the RA-EMC+GHB prefetcher.

Moreover, RA-EMC has the lowest bandwidth overhead of any of the evaluated prefetchers as shown in Figure \ref{fig:scRaEMC:bw}. I find that the Markov+Stream prefetcher uses the most additional bandwidth while the GHB prefetcher bandwidth consumption is comparable to RA-EMC. Applications with low RA-EMC request accuracy such as \textit{omnetpp}, \textit{milc}, \textit{soplex}, and \textit{sphinx} use more bandwidth than those with high accuracy such as \textit{lbm}.

The 34\% performance increase that RA-EMC provides is due to a decrease in the effective memory access latency visible to the core. The effective memory access latency is listed in Table \ref{tab:raEMCLat} for each of the medium/high memory intensity \textit{SPEC06} benchmarks. Effective memory access latency is defined as the number of cycles that it takes for a memory request to be satisfied (wake up dependent operations) after it misses in the first level data cache of the core. The GHB prefetcher results in a 30\% effective memory access latency reduction, the largest when considering the three prefetchers used in this evaluation. RA-EMC outperforms all three prefetchers by resulting in a 34\% reduction in effective memory access latency.

\begin{table*}[ht]
	\centering
	\begin{tabular}{|c|ccccccc|}\hline
		\centering
		\multirow{4}{1.0in}{\centering Baseline} & zeusmp & cactus & wrf & gems & leslie & omnetpp & milc \\
		& 98 & 243 & 89 & 135 & 101 & 131 & 151 \\\cline{2-8}
		& soplex & sphinx & bwaves & libq & lbm & mcf & Mean \\
		& 130 & 114 & 170 & 253 & 186 & 93 & 146 \\\cline{2-8}
		\hline
		\multirow{4}{1.0in}{\centering Stream PF} & zeusmp & cactus & wrf & gems & leslie & omnetpp & milc \\
		& 92 & 240 & 66 & 64 & 58 & 136 & 113 \\\cline{2-8}
		& soplex & sphinx & bwaves & libq & lbm & mcf & Mean \\
		& 98 & 65 & 152 & 200 & 94 & 90 & 113 \\\cline{2-8}
		\hline
		\multirow{4}{1.0in}{\centering GHB PF} & zeusmp & cactus & wrf & gems & leslie & omnetpp & milc \\
		& 96 & 175 & 62 & 62 & 58 & 131 & 116 \\\cline{2-8}
		& soplex & sphinx & bwaves & libq & lbm & mcf & Mean \\
		& 95 & 66 & 81 & 223 & 85 & 93 & 103 \\\cline{2-8}
		\hline
		\multirow{4}{1.0in}{\centering Markov + Stream PF} & zeusmp & cactus & wrf & gems & leslie & omnetpp & milc \\
		& 91 & 235 & 76 & 64 & 58 & 130 & 110 \\\cline{2-8}
		& soplex & sphinx & bwaves & libq & lbm & mcf & Mean \\
		& 99 & 63 & 153 & 225 & 103 & 90 & 115 \\\cline{2-8}
		\hline
		\multirow{4}{1.0in}{\centering RA-EMC} & zeusmp & cactus & wrf & gems & leslie & omnetpp & milc \\
		& 86 & 180 & 79 & 69 & 68 & 89 & 98 \\\cline{2-8}
		& soplex & sphinx & bwaves & libq & lbm & mcf & Mean \\
		& 69 & 55 & 84 & 183 & 90 & 78 & 95 \\\cline{2-8}
		\hline
	\end{tabular}
	\caption{Effective Memory Access Latency (Cycles)}
	\label{tab:raEMCLat}
\end{table*}

\subsection{Energy Results}
\label{sec:scRaEMC:results:energy}

\begin{figure}
	\centering
	\includegraphics[width=\columnwidth]{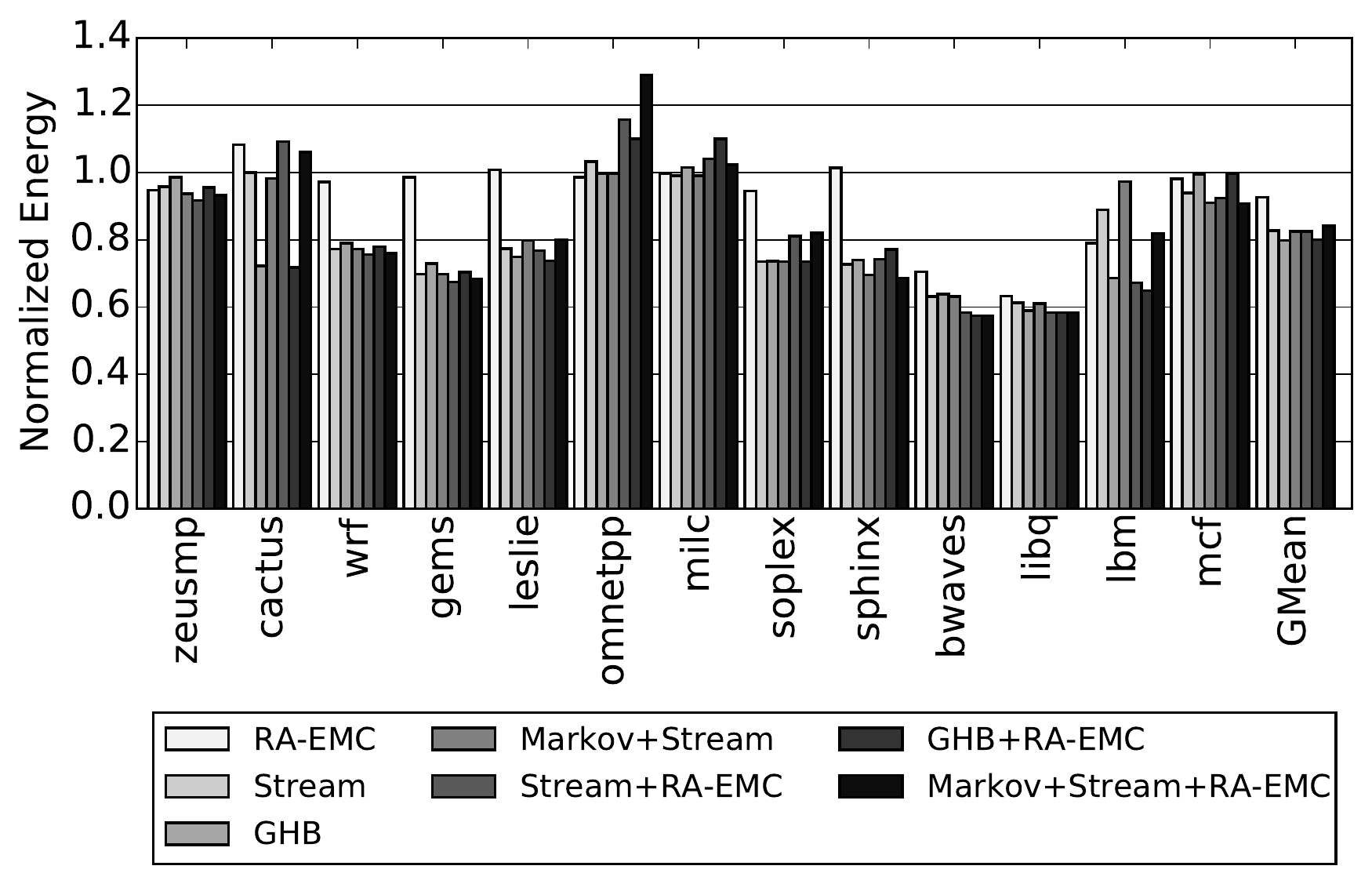}
	\caption{RA-EMC Energy Consumption}
	\label{fig:scRaEMC:energy}
\end{figure}

While RA-EMC may decrease energy consumption, it does so at the cost of additional on-chip computational hardware. Figure \ref{fig:scRaEMC:energy} demonstrates the effect of the RA-EMC on system (Chip+DRAM) energy consumption.

Overall, most of the benchmarks break even on energy consumption versus the baseline. The three benchmarks with high accuracy and low bandwidth overhead (\textit{bwaves}, \textit{libquantum}, and \textit{lbm}) show significant energy reductions, leading to a 10\% energy reduction over the no-prefetching baseline. As in the performance case, RA+EMC interacts favorably with the GHB prefetcher but increases energy consumption with the Markov+Stream prefetcher. As noted in Figure \ref{fig:scRaEMC:bw} the Markov+Stream prefetcher significantly increases bandwidth consumption, causing RA-EMC requests to be less effective.

RA-EMC relies on significantly cutting execution time to reduce static energy consumption since runahead causes an increase in dynamic energy consumption. In the single core case, this trade-off is more difficult to balance as the chip is smaller. However, sharing the RA-EMC in the multi-core case is evaluated in Section \ref{sec:mcRaEMC:sens} and results in a more significant reduction in energy consumption. Also note that the Chip + DRAM energy evaluation does not include other significant static power sources such as disk or hardware peripherals. Table \ref{tab:dynEnergy} breaks down the energy evaluation into static and dynamic components normalized to a no-prefetching baseline. The RA-EMC causes a 18\% reduction in static energy consumption but a 21\% increase in dynamic energy consumption on average.

\begin{table*}[ht]
	\centering
	\begin{tabular}{|c|ccccccc|}\hline
		\centering
		\multirow{4}{1.0in}{\centering Static Energy} & zeusmp & cactus & wrf & gems & leslie & omnetpp & milc \\
		& .90 & 1.1 & .94 & .84 & .88 & .90 & .95 \\\cline{2-8}
		& soplex & sphinx & bwaves & libq & lbm & mcf & Mean \\
		& .87 & .89 & .51 & .45 & .60 & .84 & .82 \\\cline{2-8}
		\hline
		\multirow{4}{1.0in}{\centering Dynamic Energy} & zeusmp & cactus & wrf & gems & leslie & omnetpp & milc \\
		& 1.01 & 1.01 & 1.02 & 1.3 & 1.20 & 1.21 & 1.10 \\\cline{2-8}
		& soplex & sphinx & bwaves & libq & lbm & mcf & Mean \\
		& 1.08 & 1.32 & 1.23 & 1.40 & 1.36 & 1.51 & 1.21 \\\cline{2-8}
		\hline
	\end{tabular}
	\caption{Normalized RA-EMC Static and Dynamic Energy}
	\label{tab:dynEnergy}
\end{table*}

\subsection{Sensitivity To System Parameters}
\label{sec:scRaEMC:results:sens}

In this Section I identify three key parameters to RA-EMC: LLC cache capacity, the number of memory banks and the threshold MPKI at which RA-EMC execution is marked to begin in Algorithm \ref{alg:emcIndChainMark}. RA-EMC performance and energy sensitivity to these parameters are listed in Table \ref{tab:raEMCSens}. The values used for these parameters in the evaluation (Section \ref{sec:scRaEMC:results:perf}) are bolded.

\begin{table}[htb*]
	\begin{minipage}[bht*]{1.00\columnwidth}
		\centering
		\footnotesize
		\begin{tabular}{|c|c||c|c||c|c||c|c|} \hline
			\multicolumn{8}{|c|}{{\bf LLC Cache Size} } \\ \hline 
			\multicolumn{2}{|c||}{512 KB} & \multicolumn{2}{c||}{\textbf{1 MB}} & \multicolumn{2}{c||}{2 MB} & \multicolumn{2}{c|}{4 MB} \\ \hline
			$\Delta$ Perf & $\Delta$ Energy & $\Delta$ Perf & $\Delta$ Energy & $\Delta$ Perf & $\Delta$ Energy & $\Delta$ Perf & $\Delta$ Energy \\ \hline
			22.3\% & 4.5\% & 34.3\% & -7.3\% & 36.1\% & -8.1\% & 35.8\% & -8.0\% \\ \hline
		\end{tabular}
	\end{minipage}
	
	\begin{minipage}[bht*]{1.00\columnwidth}
		\centering
		\footnotesize
		\begin{tabular}{|c|c||c|c||c|c||c|c|} \hline
			\multicolumn{8}{|c|}{{\bf Number of Memory Banks} } \\ \hline 
			\multicolumn{2}{|c||}{\textbf{8}} & \multicolumn{2}{c||}{16} & \multicolumn{2}{c||}{32} & \multicolumn{2}{c|}{64} \\ \hline
			$\Delta$ Perf & $\Delta$ Energy & $\Delta$ Perf & $\Delta$ Energy & $\Delta$ Perf & $\Delta$ Energy & $\Delta$ Perf & $\Delta$ Energy \\ \hline
			34.3\% & -7.3\% & 37.1\% & -7.9\% & 35.4\% & -7.7\% & 34.1\% & -7.7\% \\ \hline
		\end{tabular}
	\end{minipage}
	
	\begin{minipage}[bht*]{1.00\columnwidth}
		\centering
		\footnotesize
		\begin{tabular}{|c|c||c|c||c|c||c|c|} \hline
			\multicolumn{8}{|c|}{{\bf RA-EMC Threshold MPKI} } \\ \hline 
			\multicolumn{2}{|c||}{2} & \multicolumn{2}{c||}{\textbf{5}} & \multicolumn{2}{c||}{7} & \multicolumn{2}{c|}{10} \\ \hline
			$\Delta$ Perf & $\Delta$ Energy & $\Delta$ Perf & $\Delta$ Energy & $\Delta$ Perf & $\Delta$ Energy & $\Delta$ Perf & $\Delta$ Energy \\ \hline
			31.1\% & -3.1\% & 34.3\% & -7.3\% & 28.4\% & -1.5\% & 23.8\% & 3.3\% \\ \hline
		\end{tabular}
	\end{minipage}
	\begin{small}
		\caption{RA-EMC Performance and Energy Sensitivity}
		\label{tab:raEMCSens}
	\end{small}
\end{table}

RA-EMC shows some sensitivity to LLC capacity. If the LLC capacity is too small, as in the 512KB case, the runahead distance is limited by available cache capacity. Sensitivity to memory bandwidth is much smaller, as RA-EMC is able to be more aggressive as memory system bandwidth increases. The threshold MPKI to start runahead at the EMC also shows a large amount of performance sensitivity. If the threshold MPKI is too high, then RA-EMC is not able to prefetch effectively enough to amortize its static and dynamic energy overhead.

\vspace{-.2in}
\subsection{Dependent Miss Acceleration}
\label{sec:scRaEMC:results:perfDep}

\begin{figure}
	\centering
	\includegraphics[width=\columnwidth]{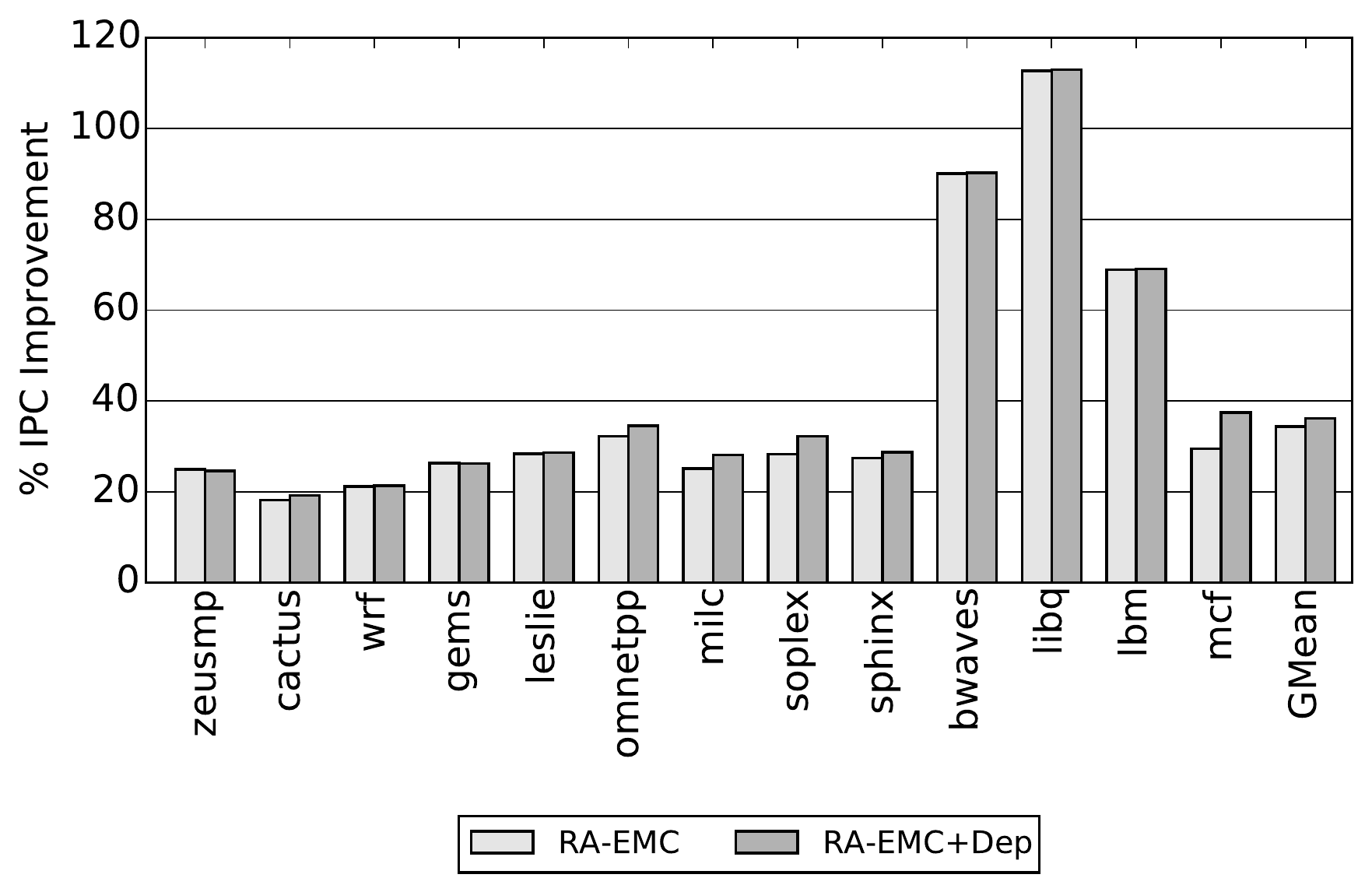}
	\caption{RA-EMC with Dependent Miss Acceleration}
	\label{fig:scRaEMC:depPerf}
\end{figure}

As demonstrated in Section \ref{sec:EMC:singleCore}, dependent miss acceleration does not have a large effect on single core performance since the small amount of on-chip contention does not have a large impact on effective memory access latency. However, Figure \ref{fig:scRaEMC:depPerf} shows the performance results of using both runahead and dependent miss acceleration at the EMC. Since dependent cache misses are critical to processor performance (they are currently stalling the home core pipeline) and RA-EMC requests are prefetches, dependent cache misses are given priority to issue if they are available. Figure \ref{fig:scRaEMC:depPerf} shows that benchmarks with high numbers of dependent cache misses (predominantly \textit{mcf}) increase further in performance when dependent miss acceleration is added to RA-EMC. This study is revisited in a multi-core context in Section \ref{sec:mcRaEMC:eval}.

\section{Conclusion}
\label{sec:scRaEMC:conc}

In this chapter I augmented the Enhanced Memory Controller (EMC) with the ability to continuously run ahead of the core without the interval length limitations of the runahead paradigm. The result is a 34\% average reduction in effective memory access latency and 37\% performance increase on the high memory intensity \textit{SPEC06} benchmarks. I show that a more intelligent decision to pick the dependence chain to use during runahead results in increased performance using both the runahead buffer and the EMC. In the next chapter, I evaluate the RA-EMC in a bandwidth constrained multi-core setting and demonstrate its impact as a shared resource that reduces effective memory access latency for both independent and dependent cache misses.

\chapter{Multi-core Enhanced Memory Controller Policies}
\label{chap:mcRaEMC}
\setlength{\epigraphwidth}{0.41\textwidth}

\section{Introduction}
\label{sec:mcRaEMC:Intro}

Chapter \ref{chap:scRaEMC} developed hardware techniques that allow a single core to use the EMC to continuously runahead during memory intensive phases, thereby reducing effective memory access latency for independent cache misses. In this Chapter, I expand the single-core system to a multi-core system. In the multi-core case, the EMC is a shared resource that all of the different cores contend over. Therefore, I will develop policies that allow the EMC to decide when it is best to runahead with a dependence chain from each core (Section \ref{sec:mcRaEMC:policies}). I will then combine the independent cache miss acceleration that runahead provides with the dependent miss acceleration mechanisms developed in Chapter \ref{chap:EMC} (Section \ref{sec:mcRaEMC:eval}). By combining these two mechanisms, I propose a complete mechanism that can reduce effective memory access latency for all cache misses. This is the first work that I am aware of that uses dependence chains to accelerate both independent and dependent cache misses in a multi-core context.

\section{Methodology}
\label{sec:mcRaEMC:meth}

As the policies in this chapter are experiment-driven, I first review the multi-core simulation model. As in Chapter \ref{chap:EMC} weighted speedup is used \cite{sna:tul00} as a performance metric.

\begin{equation}
\displaystyle Wspeedup = \sum\limits_{i=0}^{n-1} \frac{IPC^{shared}_{i}}{IPC^{alone}_{i}}
\label{eqn:wspeedup}
\end{equation}

The system configuration is shown in Table \ref{tab:mcRaEmc:systemConfig}, and is identical to the system in Chapter \ref{chap:EMC} with the exception of the single new runahead context (RA-EMC).  The workloads that I use for evaluation in this chapter are shown in Table \ref{tab:mcRaEmc:workloadChoices}. The ``High" workloads are labeled H1-H10 and consist of a random mix of high memory intensity benchmarks. The ``Mix" workloads are labeled ``M1-M5" and ``L16-L20" and consist of a random mix of 2 high intensity benchmarks/2 medium intensity benchmarks and 2 high intensity benchmarks/2 low intensity benchmarks respectively.  In addition to these combinations, I additionally show results for workloads that consist of four copies of each of the high and medium memory intensity benchmarks in Table \ref{tab:mcRaEmc:workloadClass}. I refer to these workloads as the ``Copy" workloads.

\begin{table*}[ht!]
	\small
	\centering
	\begin{tabular}{|p{.75in}|p{4.5in}|}
		\hline Core & 4-Wide Issue, 256 Entry ROB, 92 Entry Reservation Station, Hybrid Branch Predictor, 3.2 GHz Clock Rate \\ 
		\hline L1 Caches & 32 KB I-Cache, 32 KB D-Cache, 64 Byte Lines, 2 Ports, 3 Cycle Latency, 8-way, Write-Through. \\ 
		\hline \hline L2 Cache &  Distributed, Shared, 1MB 8-way slice per core, 18-cycle latency, Write-Back. 4 MB total. \\ 
		\hline Interconnect & 2 Bi-Directional rings, control (8 bytes) and data (64 bytes). 1 cycle core to LLC slice bypass. 1 cycle latency between ring stops. \\ 
		\hline \hline EMC \newline Compute & 2-wide issue. 8 Entry Reservation Stations. 4KB Data Cache 4-way, 2-cycle access, 1-port. 1 Runahead dependence chain context with 32 entry uop buffer, 32 entry physical register file. 2 Dependent cache miss contexts with 16 entry uop buffer, 16 entry physical register file. Micro-op size: 8 bytes in addition to any live-in source data.\\
		\hline EMC \newline Instructions & Integer: add/subtract/move/load/store. \newline Logical: and/or/xor/not/shift/sign-extend. \\
		\hline Memory Controller & Batch Scheduling \cite{mut:mos08}. 128 Entry Memory Queue. \\ 
		\hline Prefetchers & Stream: 32 Streams, Distance 32. Markov: 1MB Correlation Table, 4 addresses per entry. GHB G/DC: 1k Entry Buffer, 12KB total size. All configurations: FDP \cite{fdp07}, Dynamic Degree: 1-32, prefetch into Last Level Cache. \\  
		\hline DRAM & DDR3\cite{dram:micron}, 1 Rank of 8 Banks/Channel, 2 Channels, 8KB Row-Size, CAS 13.75ns.  CAS = $t_{RP}$ = $t_{RCD}$ = CL. Other modeled DDR3 constraints: BL, CWL, $t_{RC, RAS, RTP, CCD, RRD, FAW, WTR, WR}$. 800 MHz Bus, Width: 8 B.  \\
		\hline
	\end{tabular} 
	\caption{Multi-Core System Configuration}
	\label{tab:mcRaEmc:systemConfig}
\end{table*}

\begin{table}
	\small
	\centering
	\begin{tabular}{|p{.25in}|p{2.25in}||p{.25in}|p{2.25in}|}
		\hline H1 & bwaves+lbm+milc+omnetpp & M11 & soplex+gems+wrf+mcf \\ 
		\hline H2 & soplex+omnetpp+bwaves+libq & M12 &  milc+zeusmp+lbm+cactus  \\ 
		\hline H3 & sphinx3+mcf+omnetpp+milc & M13 & gems+wrf+mcf+omnetpp \\
		\hline H4 & mcf+sphinx3+soplex+libq  & M14 &  cactus+gems+soplex+sphinx3  \\ 
		\hline H5 & lbm+mcf+libq+bwaves & M15 &  libq+leslie3d+wrf+lbm \\ 
		\hline H6 & lbm+soplex+mcf+milc& L16 & h264ref+lbm+omnetpp+povray \\  
		\hline H7 & bwaves+libq+sphinx3+omnetpp & L17 & tonto+sphinx3+sjeng+mcf \\ 
		\hline H8 & omnetpp+soplex+mcf+bwaves & L18 & bzip2+namd+mcf+sphinx3 \\ 
		\hline H9 & lbm+mcf+libq+soplex & L19 & omnetpp+soplex+namd+xalanc \\
		\hline H10 & libq+bwaves+soplex+omentpp & L20 & soplex+bwaves+bzip2+perlbench \\ 
		\hline
	\end{tabular} 
	\caption{Multi-Core Workloads}
	\label{tab:mcRaEmc:workloadChoices}
\end{table}

\begin{table}
	\small
	\centering
	\begin{tabular}{|p{1.5in}|p{3.75in}|}
		\hline High Intensity \newline (MPKI \textgreater= 10) &  omnetpp, milc, soplex, sphinx3, bwaves, libquantum, lbm, mcf\\ 
		\hline Medium Intensity \newline (MPKI \textgreater=5) & zeusmp, cactusADM, wrf, GemsFDTD, leslie3d \\
		\hline Low Intensity \newline (MPKI \textless 5) & calculix, povray, namd, gamess, perlbench, tonto, gromacs, gobmk, dealII, sjeng, gcc, hmmer, h264ref, bzip2, astar, xalancbmk \\ 
		\hline
	\end{tabular} 
	\caption{SPEC06 Classification by Memory Intensity}
	\label{tab:mcRaEmc:workloadClass}
\end{table}

\section{Multi-core RA-EMC Policies}
\label{sec:mcRaEMC:policies}

In this Section I evaluate three different policies for determining which dependence chain to use during RA-EMC in a multi-core system. From Table \ref{tab:mcRaEmc:systemConfig}, note that the EMC is augmented with one runahead context. I show sensitivity to this number (Section \ref{sec:mcRaEMC:context}), but a single runahead context is optimal as it devotes all EMC resources in a given interval to accelerating a single benchmark, thereby maximizing runahead distance for that application.

\subsection{Policy Evaluation}
\label{sec:mcRaEMC:policyEval}

Three policies are evaluated in this section. All three policies are interval based. Initially the interval length is 100k instructions retired by the core that has provided a runahead dependence chain (as in Section \ref{sec:scRaEMC:dist}). At the end of each interval, the EMC selects a new dependence chain to use for runahead. Dependence chains are generated by each core (Section \ref{sec:scRaEMC:hwDepPolicy}).

The first policy is a round-robin policy. This policy picks a core from each eligible application in the workload in a round-robin fashion. An eligible application has an MPKI above the threshold (MPKI \textgreater 5, from Table \ref{tab:raEMCSens}). The chosen core then provides the EMC with a dependence chain to use during RA-EMC. This scheduling is repeated after the home core that generated the dependence chain notifies the EMC that it has retired the threshold number of instructions.

The second policy schedules a dependence chain for RA-EMC from an eligible application with the lowest IPC in the workload. By picking the benchmark with the lowest IPC, the EMC is able to accelerate the application that is performing the worst in the workload.

The third policy schedules a dependence chain from the eligible application with the highest score in the workload. Recall from Section \ref{sec:scRaEMC:hwDepPolicy} that the hardware stall policy assigns a score to each cache miss based on how often it blocks retirement. These scores are sent to the EMC in the third policy and the EMC notifies the core with the highest score to send a dependence chain for runahead execution. This policy prioritizes accelerating the dependence chain that is causing the workload to stall the most.

\begin{figure}
	\centering
	\includegraphics[width=\columnwidth]{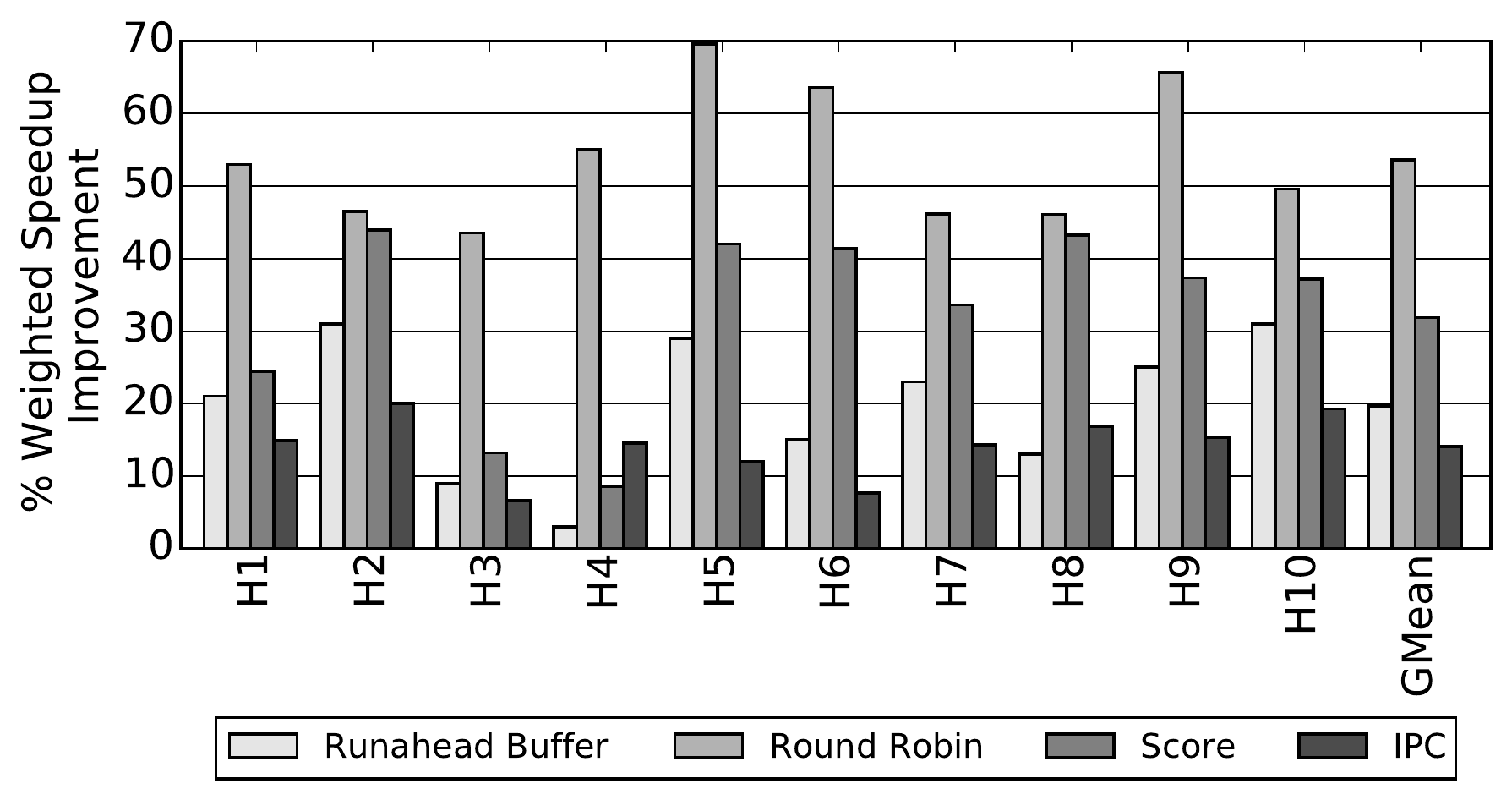}
	\caption{Multi-Core Policy on High Workloads}
	\label{fig:mcRaEMC:polHigh}
	\centering
	\includegraphics[width=\columnwidth]{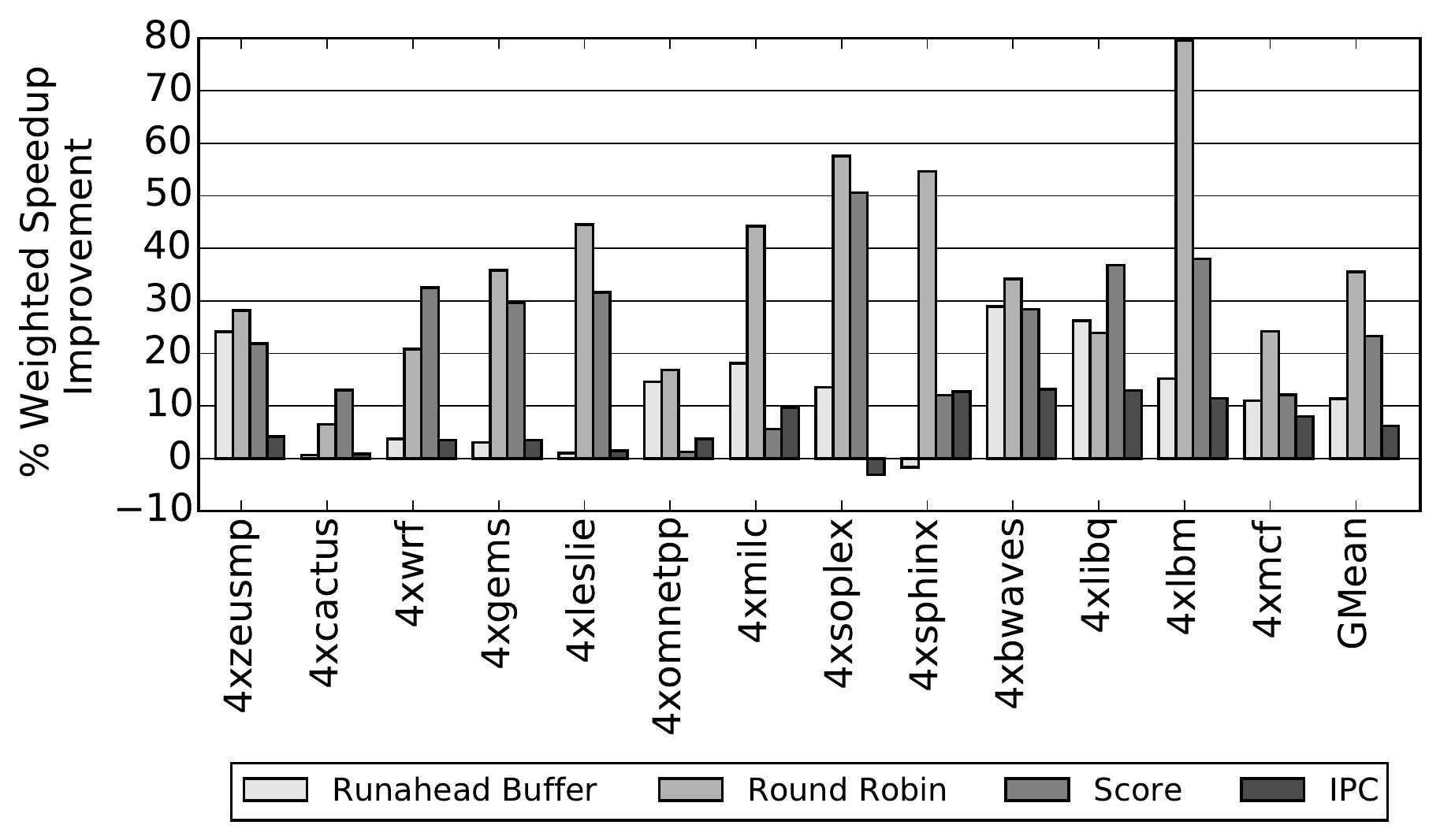}
	\caption{Multi-Core Policy on Copy Workloads}
	\label{fig:mcRaEMC:polCopy}
\end{figure}

Since the EMC is intended to accelerate high memory intensity workloads, I first concentrate on making policy decisions based on the results of the High and Copy workload sets. The performance results of these three policies are shown in Figure \ref{fig:mcRaEMC:polHigh} for the High workload set and in Figure \ref{fig:mcRaEMC:polCopy} for the Copy workloads. The first, second, and third policies are referred to as Round Robin, IPC, and Score respectively. Figures \ref{fig:mcRaEMC:polHigh} and \ref{fig:mcRaEMC:polCopy} also include a Runahead Buffer data point. In this configuration a runahead buffer is added to each core and allowed to runahead using the hardware stall policy (Section \ref{sec:scRaEMC:hwDepPolicy}).

From this study it is clear that the round-robin policy is the highest performing policy on average across both the High and Copy workloads. Examining the Copy workloads in more detail, the round-robin policy is the highest performing policy on all high intensity workloads except for 4x\textit{libq} where the Score policy performs best. The Score policy also comes close to matching round-robin performance on 4x\textit{bwaves}. Both \textit{libq} and \textit{bwaves} have a very small number of dependence chains that cause full-window stalls in Figure \ref{fig:missStallStack}. This indicates that the Score policy works best when there is a clear choice as to the dependence chain that is causing the workload to slow down the most. 

The runahead buffer results show that adding a runahead buffer to each core does not match the performance gains of the RA-EMC policies. The runahead buffer is not able to runahead for very long periods of time, reducing its performance impact (Section \ref{sec:scRaEMC:Intro}). The IPC policy performs very poorly. Table \ref{tab:mcRaEMC:raEMCIDiff} shows why this is the case for the Copy workloads where the IPC policy has a very small performance gain of only 6\%.

\begin{table*}[ht]
	\small
	\centering
	\begin{tabular}{|p{1in}|p{1in}|p{1in}|}
		\hline
		& Accuracy & Runahead \linebreak Distance \\
		\hline
		\hline
		Round-Robin & 85\% & 2343  \\
		\hline
		Score & 85\% & 2473  \\
		\hline
		IPC & 75\% & 3658  \\
		\hline
	\end{tabular} 
	\caption{RA-EMC Accuracy and Runahead Distance (Instructions)}
	\label{tab:mcRaEMC:raEMCIDiff}
\end{table*}

The IPC policy has both much lower memory request accuracy and a much larger runahead distance when compared to the round-robin and Score policies. Runahead distance is measured from the number of instructions that the core executes between when the EMC fetches a cache line and when the core accesses the line for the first time. The reason for this disparity is that by picking the benchmark with the smallest IPC every time, the IPC policy lengthens the number of cycles that the EMC executes a particular dependence chain.  This interval is initially statically set to 100k instructions. A benchmark with a very low IPC takes longer to reach this threshold relative to rest of the multi-core system. This means that the EMC runs ahead for more cycles than it would with a dependence chain from a different core, generating more runahead requests and hurting the cache locality of the other application. This observation motivates the need for a dynamic interval length in the multi-core RA-EMC system to control this effect. I explore a dynamic runahead interval in Section \ref{sec:mcRaEMC:dynamicPolicy}.

\subsection{Dynamically Adjusting Runahead Distance}
\label{sec:mcRaEMC:dynamicPolicy}

Table \ref{tab:mcRaEMC:raEMCIDiff} shows that a long RA-EMC update interval can lead to inaccurate runahead requests in a multi-core setting. Therefore, I propose a dynamic policy that tracks runahead request accuracy (similar to FDP \cite{fdp07}). Runahead requests set an extra-bit in the tag-store of each LLC cache line. Upon eviction, the EMC is notified if a runahead-fetched line was touched by the core. If this is the case, the EMC increments a useful counter. These counters are reset at the beginning of each runahead interval. Based on these counters, the EMC determines the length of each runahead interval as in Table \ref{tab:mcRaEMC:raInterval}.

\begin{table*}[ht]
	\small
	\centering
	\begin{tabular}{|p{1in}|p{.5in}|p{.5in}|p{.5in}|p{.5in}|}
		\hline
		& \textgreater95\% & \textgreater90\% & \textgreater85\% & \textless85\% \\
		\hline
		\hline
		Interval Length & 100k & 50k & 20k & 10k  \\
		\hline
	\end{tabular} 
	\caption{RA-EMC Accuracy and Interval Length (Retired Instructions)}
	\label{tab:mcRaEMC:raInterval}
\end{table*}

\begin{figure}
	\centering
	\includegraphics[width=\columnwidth]{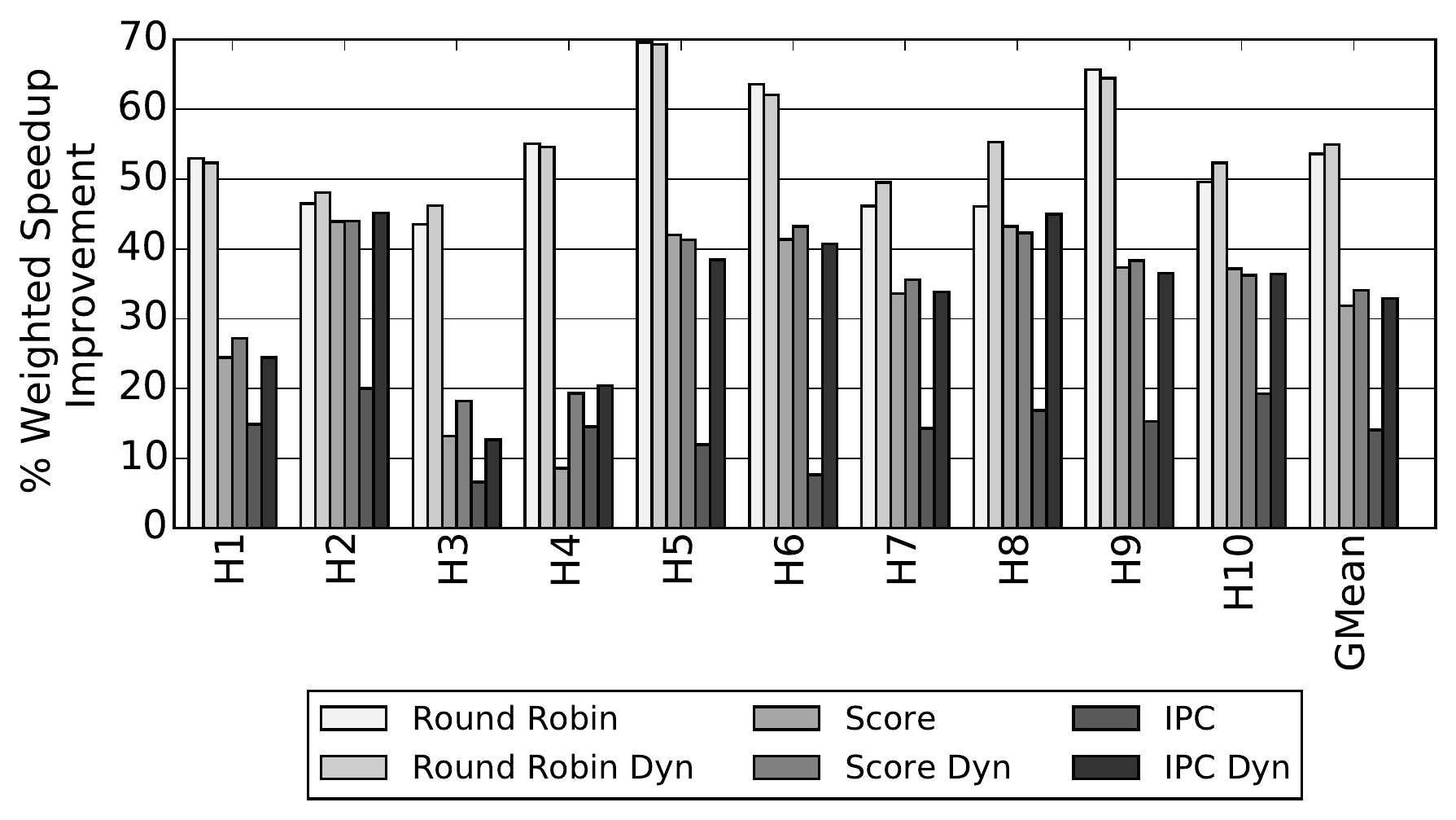}
	\caption{Dynamic Multi-core Policy on High Workloads}
	\label{fig:mcRaEMC:polHighT}
	\centering
	\includegraphics[width=\columnwidth]{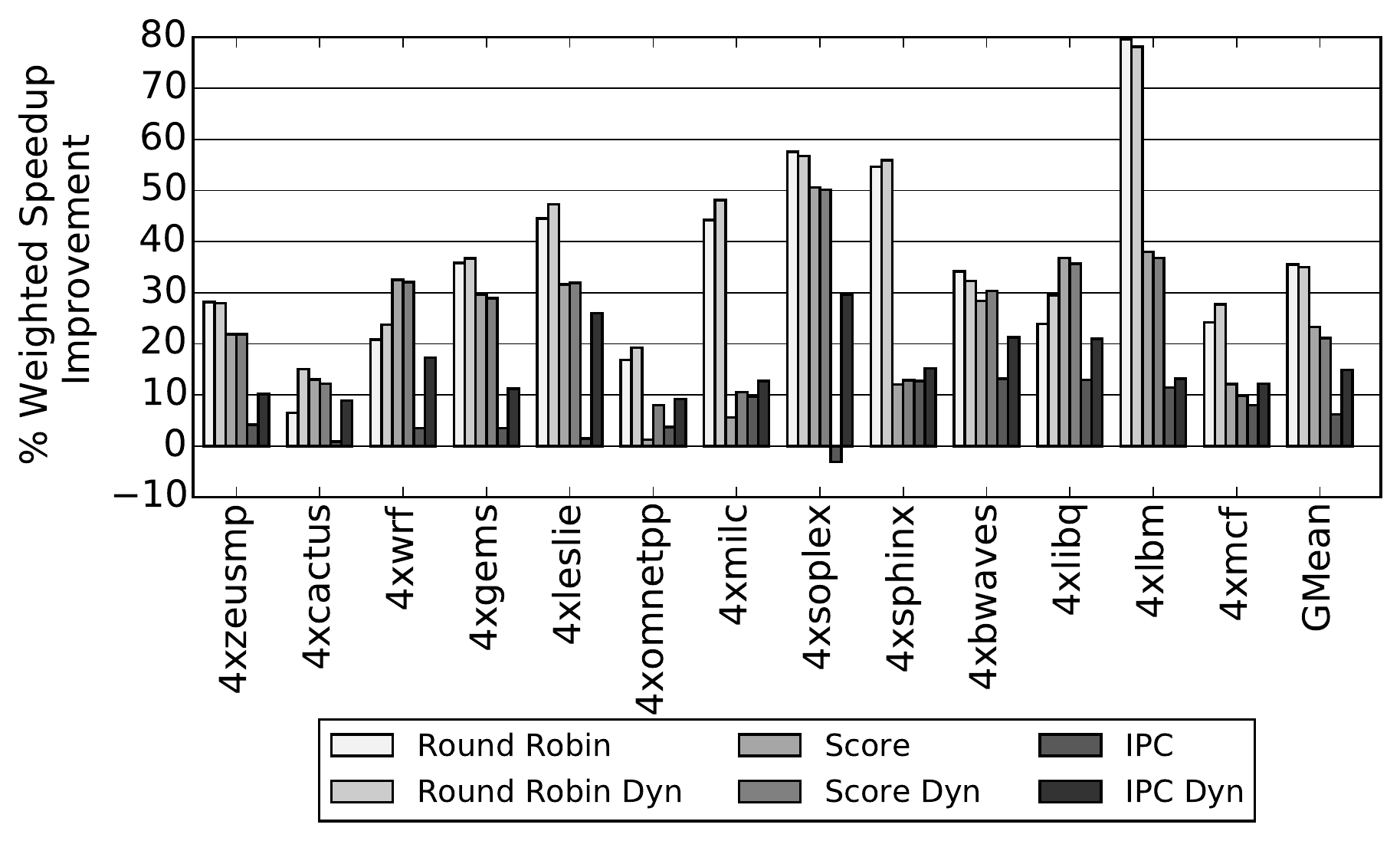}
	\caption{Dynamic Multi-core Policy on Copy Workloads}
	\label{fig:mcRaEMC:polCopyT}
\end{figure}

The performance results for the dynamic interval length policy are shown in Figure \ref{fig:mcRaEMC:polHighT} for the High workloads and Figure \ref{fig:mcRaEMC:polCopyT} for the Copy workloads. The runahead distance and accuracy for these dynamic polices are shown in Table \ref{tab:mcRaEMC:raEMCIDiffThrot}. Overall, all policies improve in runahead request accuracy with a dynamic interval length, but the result of the decrease in runahead distance has a much larger performance effect on low-performing workloads than high-performing workloads. The low-performing IPC policy shows the largest improvement, with a performance increase from 6\% on the Copy workloads to 15\%. On the High workloads IPC policy performance is increased from 14\% to 32\%. Yet, from this data, I conclude that the round robin policy is still the highest performing policy with a 55\% performance gain on the High workloads and a 37\% gain on the Copy workloads. This is roughly the same as the 53\% gain from Figure \ref{fig:mcRaEMC:polHigh} and the 37\% gain from Figure \ref{fig:mcRaEMC:polCopy}. This policy is used for the remainder of this evaluation.

\begin{table*}[ht]
	\small
	\centering
	\begin{tabular}{|p{1in}|p{1in}|p{1in}|}
		\hline
		& Accuracy & Runahead \linebreak Distance \\
		\hline
		\hline
		Round Robin & 91\% & 2040  \\
		\hline
		Score & 90\% & 2119  \\
		\hline
		IPC & 75\% & 2343  \\
		\hline
	\end{tabular} 
	\caption{Dynamic Runahead Accuracy and Distance}
	\label{tab:mcRaEMC:raEMCIDiffThrot}
\end{table*}

\subsection{Effect of Increasing RA-EMC Contexts}
\label{sec:mcRaEMC:context}

As shown in Table \ref{tab:mcRaEmc:systemConfig}, the EMC uses a single runahead dependence chain context for the policy analysis in this chapter. The EMC is designed to have the minimum capability to execute dependence chains (Section \ref{sec:EMC:Mechanism}). This results in a very lightweight hardware accelerator with a 2-wide issue capability, limited out-of-order, and a small data cache. If the EMC is multiplexed between runahead dependence chains every cycle on a very fine-granularity, overall performance gain degrades due to EMC resource contention. This is demonstrated in Table \ref{tab:mcRaEMC:raEMCCtx} where going from 1 to 2 runahead contexts reduces performance gain by half. While more aggressive EMC designs are possible, Section \ref{sec:mcRaEMC:dynamicPolicy} notes that even this lightweight design needs to be throttled down to maximize performance. I find that a single runahead context is sufficient to maximize runahead distance and this context can be multiplexed among high-memory intensity applications at coarse intervals.

\begin{table*}[ht]
	\small
	\centering
	\begin{tabular}{|p{2in}|p{1in}|p{1in}|}
		\hline
		& High & Copy \\
		\hline
		\hline
		1 Context &  55.3\% & 37.2\% \\
		\hline
		2 Contexts & 24.8\% & 20.8\% \\
		\hline
		4 Contexts & 10.7\% & 16.7\% \\
		\hline
	\end{tabular} 
	\caption{RA-EMC Context Performance Sensitivity}
	\label{tab:mcRaEMC:raEMCCtx}
\end{table*}

\section{Multi-core RA-EMC Evaluation}
\label{sec:mcRaEMC:eval}

To allow the EMC to accelerate both independent and dependent cache misses, in this Section I incorporate both dependent cache miss acceleration (Chapter \ref{chap:EMC}) and prefetching into the RA-EMC round robin policy.

\noindent\textbf{Dependent Miss Acceleration:}  To share the EMC between runahead operations and dependent cache miss chains I use a simple policy. Dependent cache misses are more critical than runahead requests since they are currently blocking retirement at the home core. Therefore, they are given priority at the EMC. If a dependent miss context has ready instructions it is given scheduling priority on the EMC. Otherwise, the EMC is allowed to execute runahead operations. 

\begin{figure}
	\centering
	\includegraphics[width=\columnwidth]{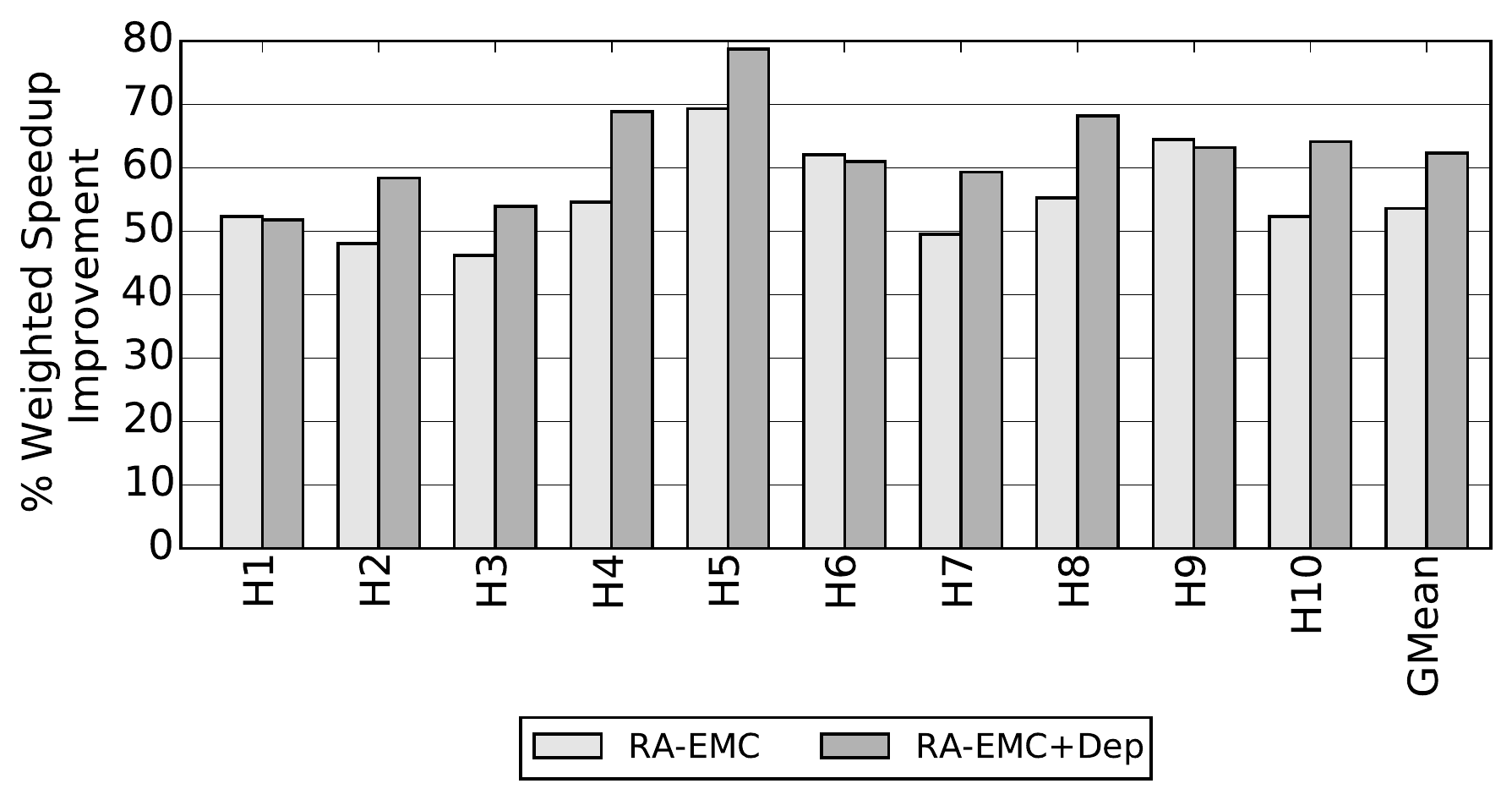}
	\caption{RA-EMC+Dep Performance on High Workloads}
	\label{fig:mcRaEMC:depHigh}
\end{figure}

\begin{figure}
	\centering
	\includegraphics[width=\columnwidth]{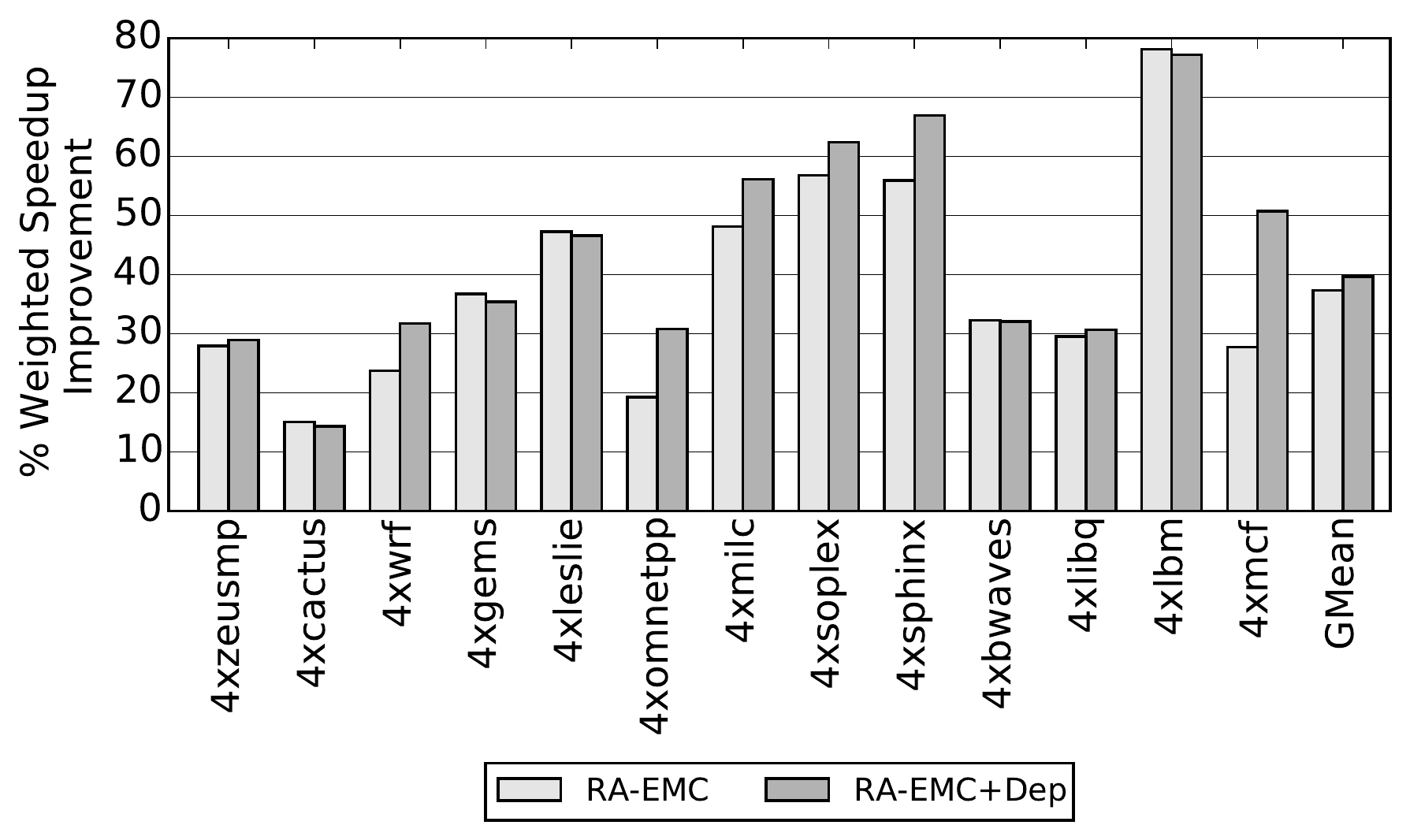}
	\caption{RA-EMC+Dep Performance on Copy Workloads}
	\label{fig:mcRaEMC:depCopy}
	\centering
	\includegraphics[width=\columnwidth]{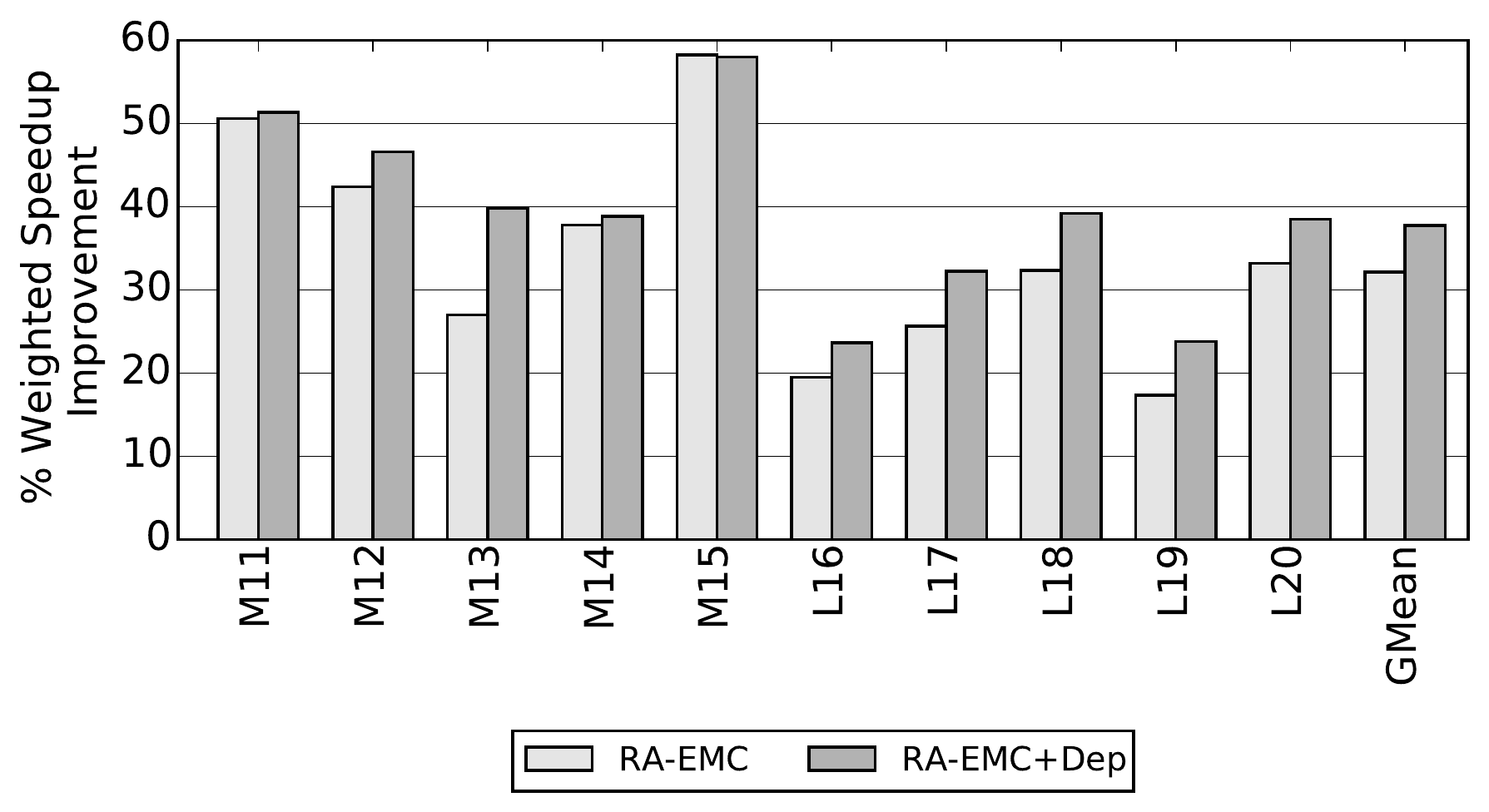}
	\caption{RA-EMC+Dep Performance on Mix Workloads}
	\label{fig:mcRaEMC:depMix}
\end{figure}

I evaluate RA-EMC+Dep (the combination of RA-EMC and dependent miss acceleration from Chapter \ref{chap:EMC}) on three sets of workloads. The High set and the Mix set in Table \ref{tab:mcRaEmc:workloadChoices} along with four copies of each of the high and medium intensity benchmarks in Table \ref{tab:mcRaEmc:workloadClass}. Results for the High/Copy/Mix workloads are shown in Figure \ref{fig:mcRaEMC:depHigh}/\ref{fig:mcRaEMC:depCopy}/\ref{fig:mcRaEMC:depMix} respectively.

Overall, the benefit of adding dependent miss acceleration is similar to the results of Section \ref{sec:EMC:result}. Workloads such as H1/H6/H9 that show small gains in Figure \ref{fig:EMC:highPerf} also show lower performance (Figure \ref{fig:mcRaEMC:depHigh}). Workloads such as H3, H4, H7, H8 all show performance gains over the RA-EMC policy. Adding dependent miss acceleration to the RA-EMC policy leads to a 8.7\% performance gain over the High workloads. The Copy workloads similarly show large performance gains on \textit{mcf} (22\%) and \textit{omnetpp} (12\%) while showing no gain on benchmarks with small numbers of dependent cache misses like \textit{bwaves} or \textit{libquantum}.

The Mix workloads show much smaller gains than the higher memory intensity workloads. The workloads with \textit{mcf} or \textit{omnetpp}, such as M13, perform well while RA-EMC+Dep does not improve performance over RA-EMC in the other cases.

Table \ref{tab:mcRaEMC:raEMCDepStats} lists the dynamic operation split between runahead chains and dependent cache miss chains at the EMC. Of all the operations executed at the EMC, only 3.2\% are operations in dependent cache miss chains for the High workload suite. This data supports the argument that available dependent cache miss chains need to be given priority over runahead operations at the EMC. Dependent cache misses are much more rare than runahead operations and gain high priority when they are available.  Table \ref{tab:mcRaEMC:raEMCDepStats} also lists the bandwidth overhead of the RA-EMC+Dep system. There is a small increase from the 7\% bandwidth overhead (Figure \ref{fig:scRaEMC:bw}) to a 11.2\% increase for RA-EMC+Dep. 

\begin{table*}[ht]
	\small
	\centering
	\begin{tabular}{|p{2in}|p{1in}|p{1in}|p{1in}|}
		\hline
		& High & Copy & Mix \\
		\hline
		\hline
		Dependent Ops Executed (\%) & 3.2\% & 2.4\% & 1.3\%  \\
		\hline
		Bandwidth Overhead & 11.2\% & 8.5\% & 5.4\%  \\
		\hline
	\end{tabular} 
	\caption{RA-EMC+Dep Statistics}
	\label{tab:mcRaEMC:raEMCDepStats}
\end{table*}

The effective memory access latency reduction for RA-EMC+Dep is listed in Table \ref{tab:mcRaEMC:raEMCDepLat}. Latencies are shown in cycles for each of the three evaluated workload sets. Effective memory access latency is measured from the time a memory access misses in the data-cache to the corresponding fill that wakes up dependent operations. This distribution is bimodal, with LLC hits taking fewer cycles than LLC misses. Therefore, higher intensity workloads have higher effective memory access latency, with the average latency of the High workload being the highest at 298 cycles. The RA-EMC+Dep reduces average effective memory access latency by 19\%/22\%/43\% for the High/Copy/Mix workloads respectively. The effective memory access latency improvement increases as workload memory intensity decreases. The reason for this is also shown in Table \ref{tab:mcRaEMC:raEMCDepLat} as the reduction in MPKI for each RA-EMC+Dep system is listed. The lower memory intensity applications have a higher relative reduction in MPKI.

\begin{table*}[ht]
	\small
	\centering
	\begin{tabular}{|p{1.75in}|p{2.5in}|p{1in}|}
		\hline
		& Effective Memory Access Latency & MPKI \\
		\hline
		\hline
		High Base & 258 & 23.9 \\
		\hline
		High RA-EMC+Dep & 210 & 19.9 \\
		\hline
		\hline
		Copy Base & 226 & 16.1 \\
		\hline
		Copy RA-EMC+Dep & 175 & 12.8 \\
		\hline
		\hline
		Mix Base & 159 & 11.3 \\
		\hline
		Mix RA-EMC+Dep & 90 & 8.5 \\
		\hline
	\end{tabular} 
	\caption{RA-EMC+Dep Effective Memory Access Latency Reduction}
	\label{tab:mcRaEMC:raEMCDepLat}
\end{table*}

\begin{figure}
	\centering
	\includegraphics[width=\columnwidth]{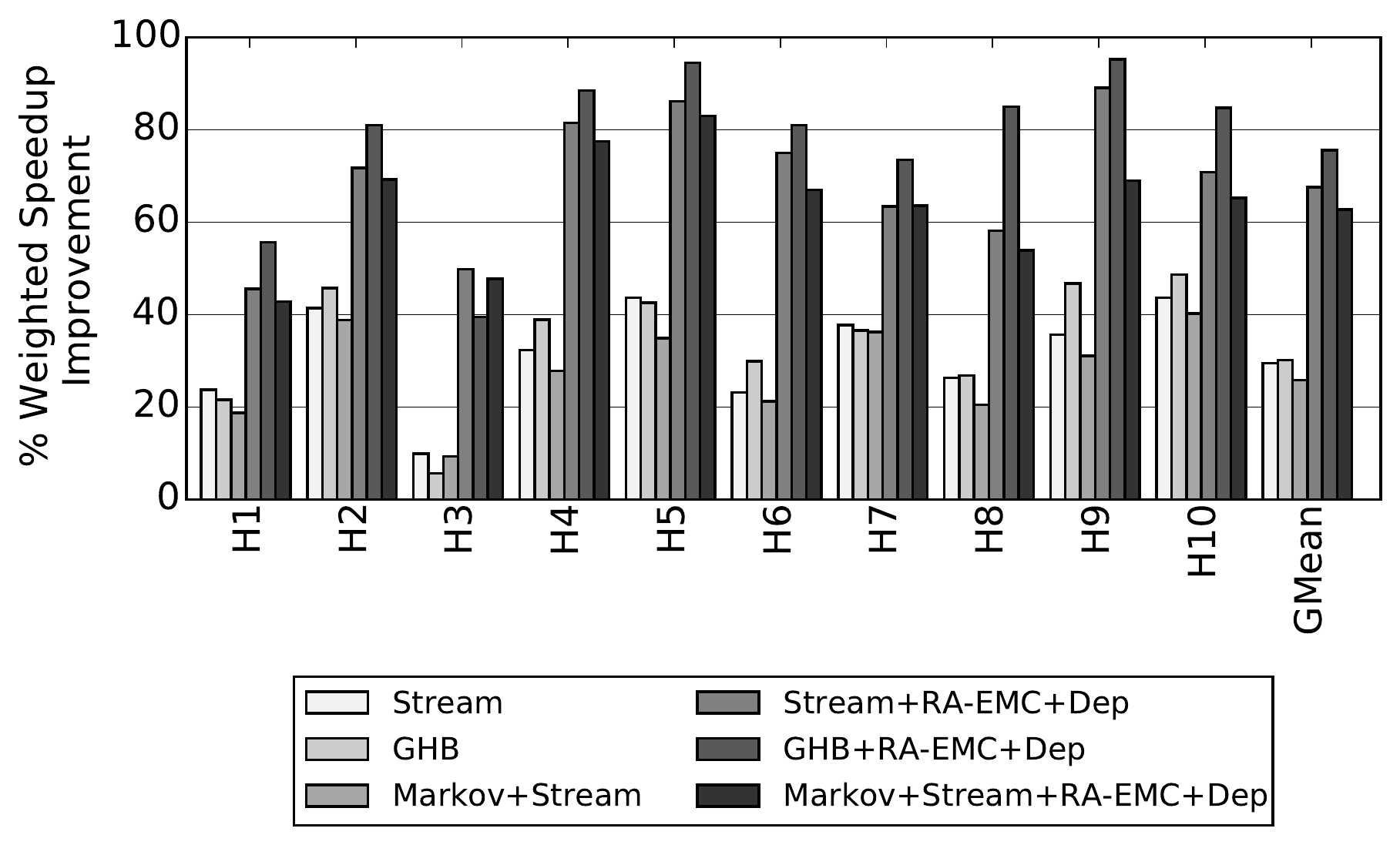}
	\caption{RA-EMC+Dep Performance with Prefetching on High Workloads}
	\label{fig:mcRaEMC:pfHigh}
	\centering
	\includegraphics[width=\columnwidth]{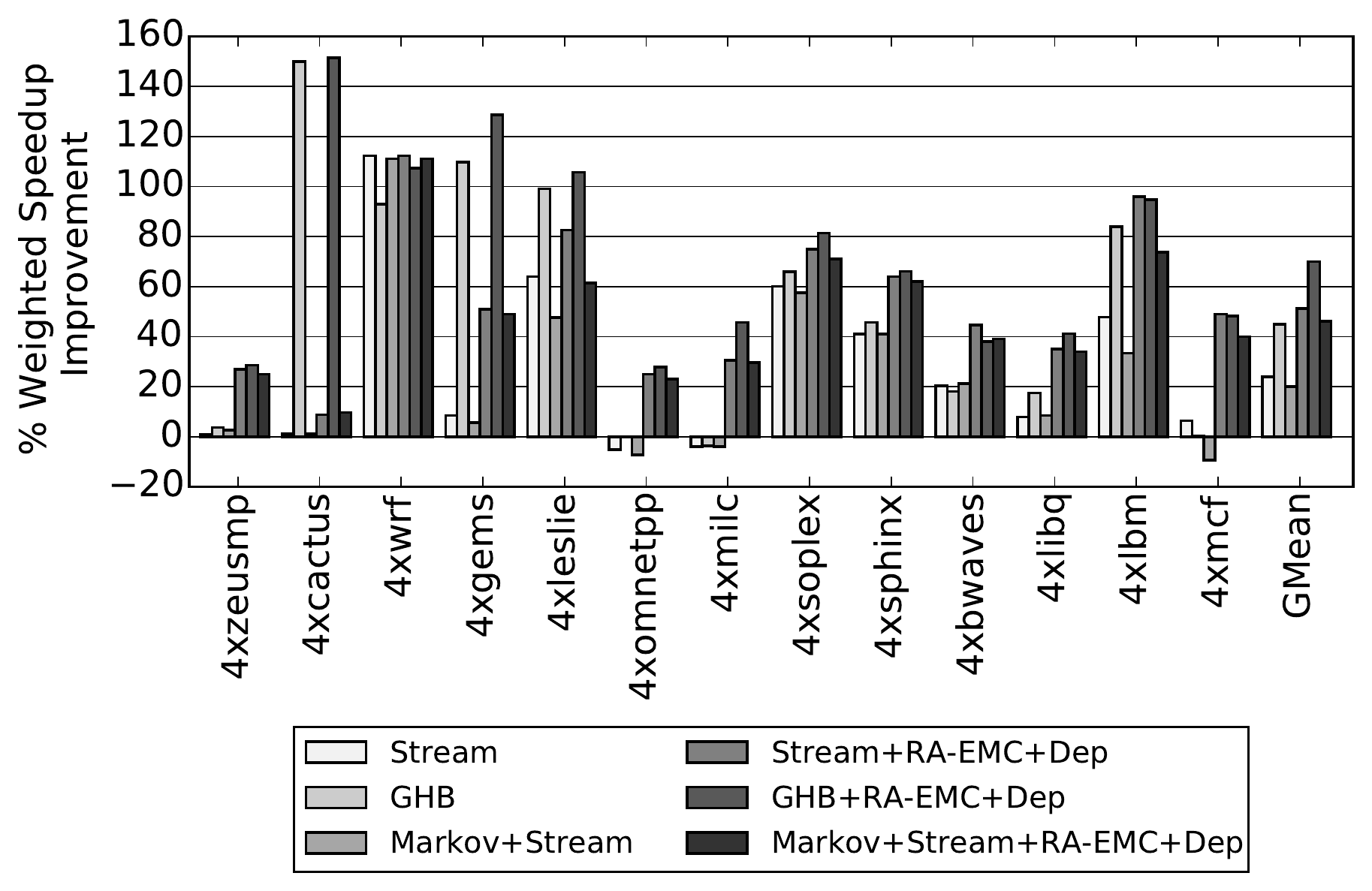}
	\caption{RA-EMC+Dep Performance with Prefetching on Copy Workloads}
	\label{fig:mcRaEMC:pfCopy}
\end{figure}

\begin{figure}
	\centering
	\includegraphics[width=\columnwidth]{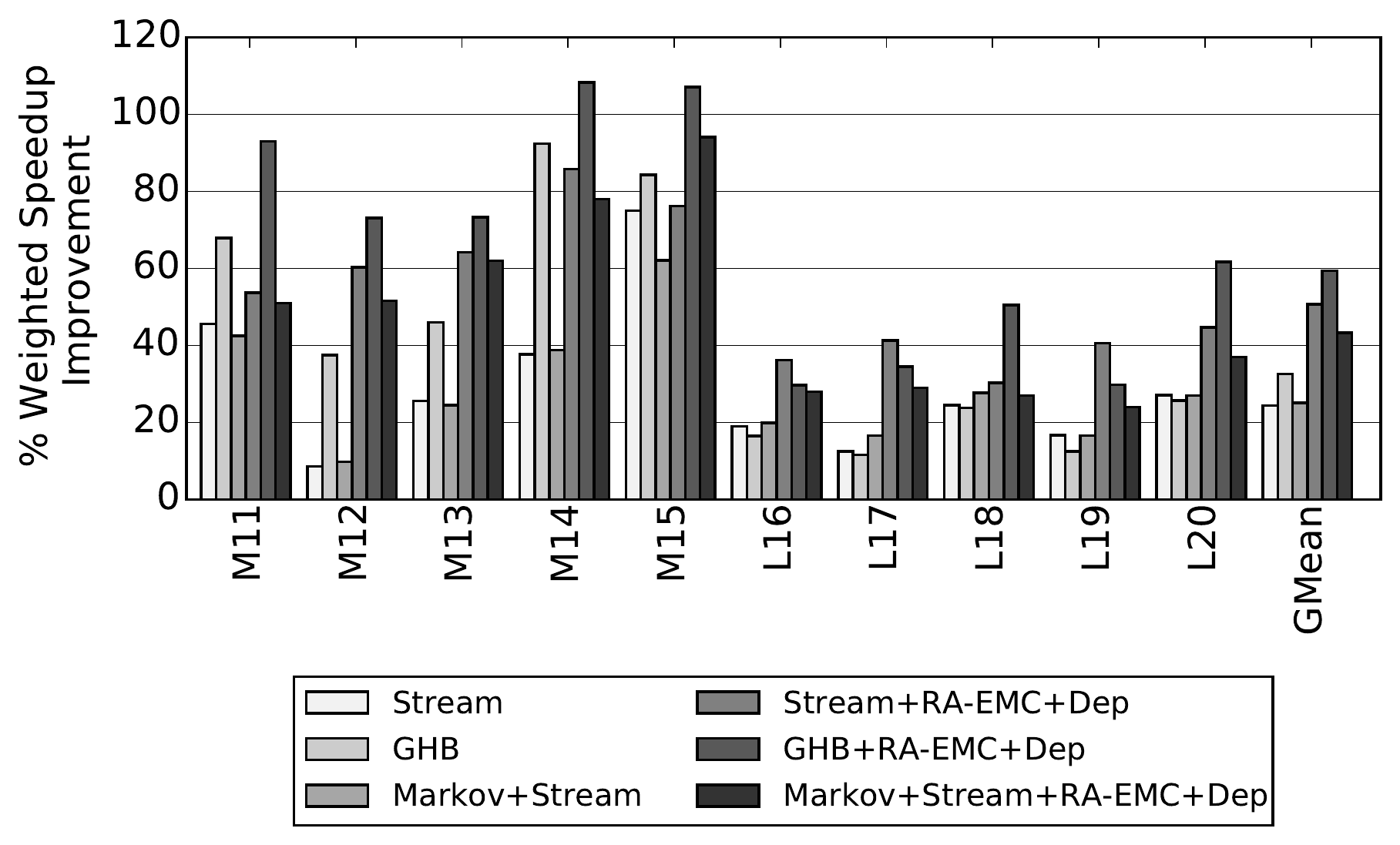}
	\caption{RA-EMC+Dep Performance with Prefetching on Mix Workloads}
	\label{fig:mcRaEMC:pfMix}
\end{figure}

\noindent\textbf{Prefetching:} Earlier chapters in this dissertation have demonstrated that prefetching increases performance when combined with the independent/dependent cache miss acceleration mechanisms that I have proposed. I find that this continues with RA-EMC+Dep. Figures-\ref{fig:mcRaEMC:pfHigh}/\ref{fig:mcRaEMC:pfCopy}/\ref{fig:mcRaEMC:pfMix} show performance for the High/Copy/Mix workload suites when combined with Stream/GHB/Markov+Stream prefetchers.

Once again, the results are an extension of those in Figure \ref{fig:EMC:highPerf} and Figure \ref{fig:scRaEMC:perfpf}. For the High workloads, the 62\% performance increase of RA-EMC+Dep in Figure \ref{fig:mcRaEMC:depHigh} is larger than any of the average performance increases of the evaluated prefetchers. On the lower memory intensity workloads, the GHB prefetcher alone performs as well as RA-EMC+Dep. On the Copy workloads, the GHB prefetcher results in a 45\% gain while RA-EMC+Dep results in a 40\% performance gain. On the Mix workloads the GHB prefetcher results in a 33\% gain while RA-EMC+Dep results in a 37\% gain. I conclude that the GHB prefetcher is the highest performing prefetcher among the evaluated on-chip prefetchers.

\begin{figure}
	\centering
	\includegraphics[width=\columnwidth]{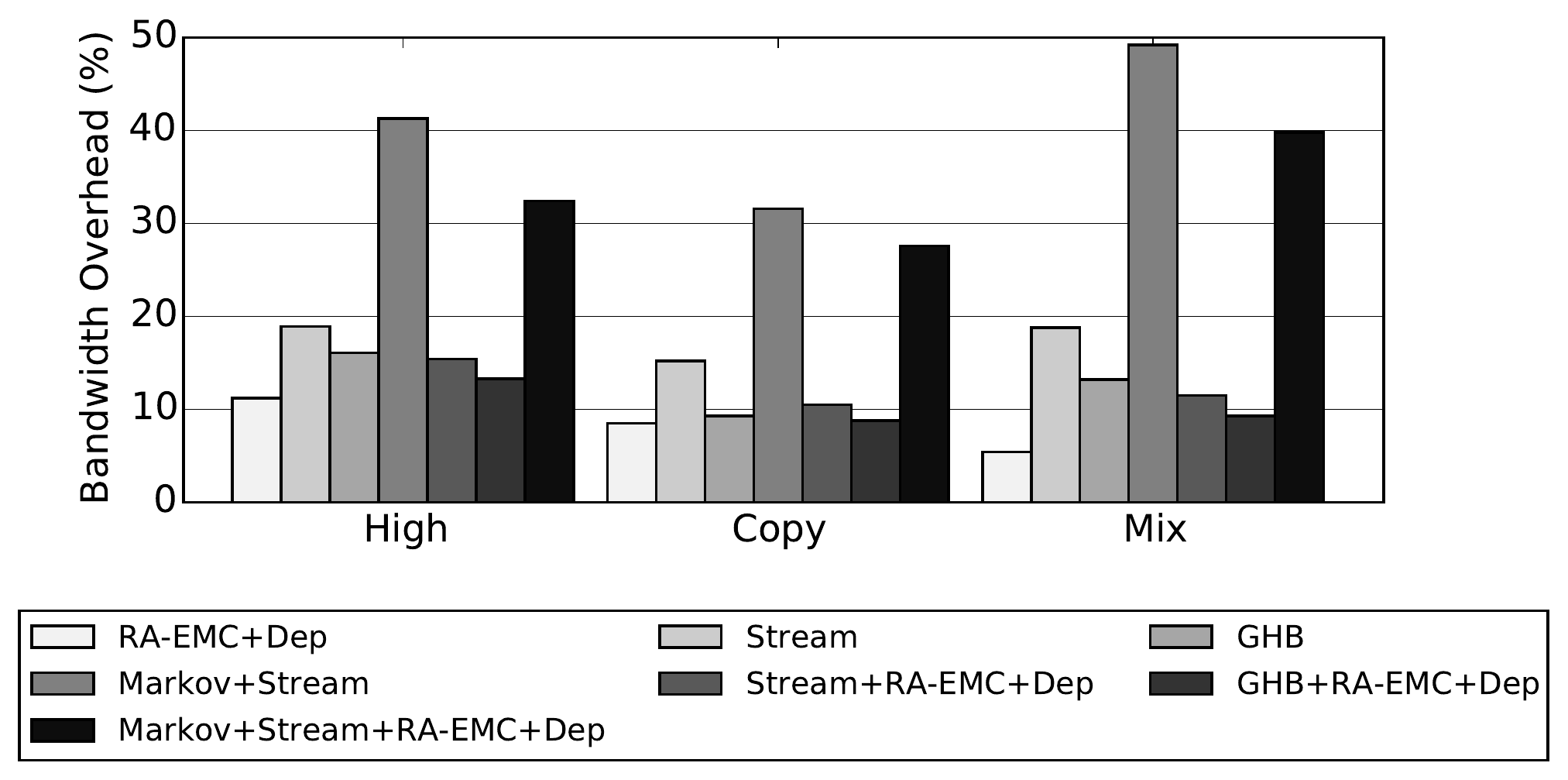}
	\caption{Average Bandwidth Overhead with Prefetching}
	\label{fig:mcRaEMC:bw}
	\centering
	\includegraphics[width=\columnwidth]{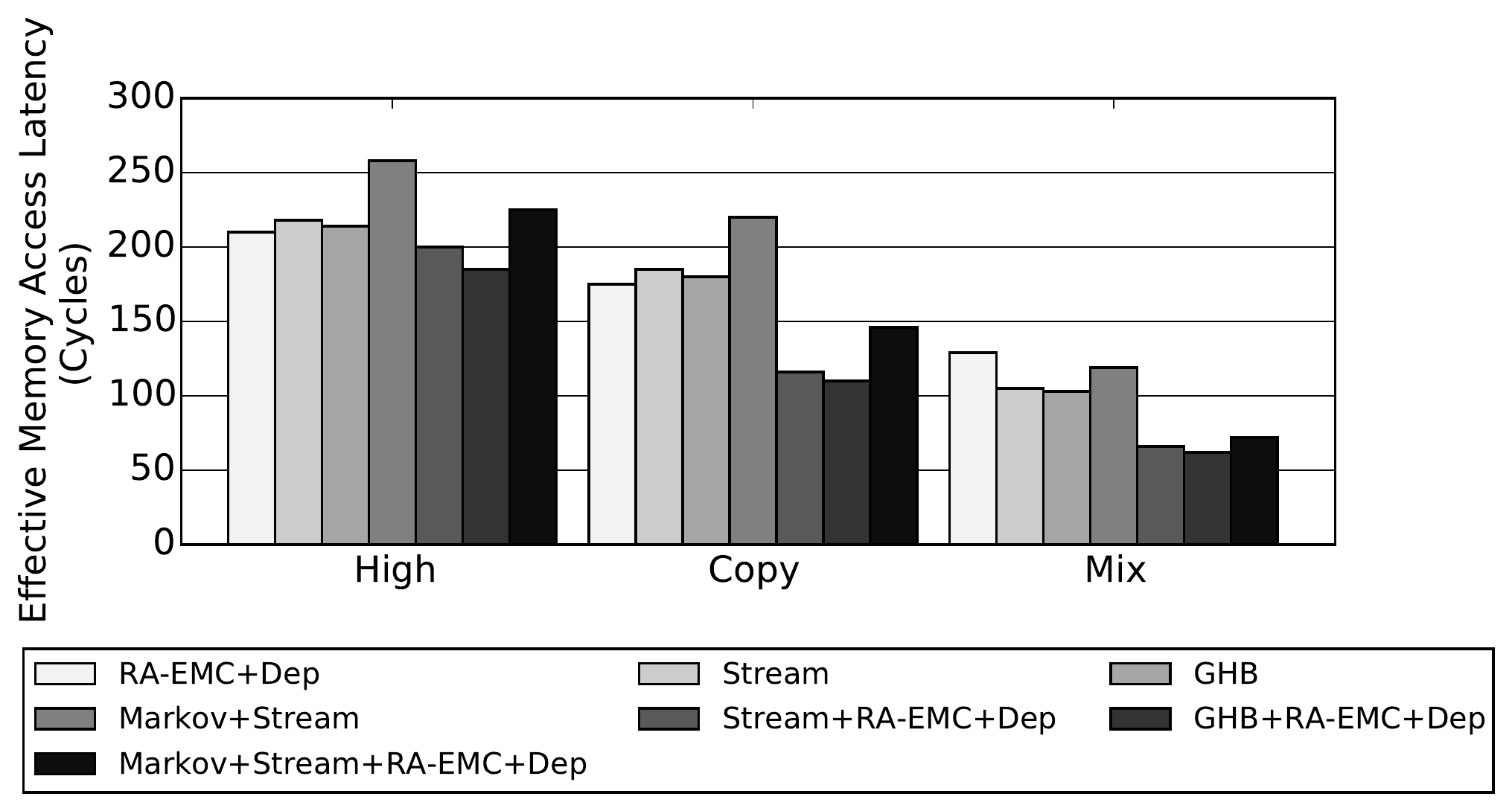}
	\caption{Average Effective Memory Access Latency with Prefetching}
	\label{fig:mcRaEMC:effLat}
\end{figure}

The highest performing system overall is the combination of GHB+RA-EMC+Dep. On the High/Copy/Mix workloads this system improves performance by 76\%/70\%/59\% over the no-prefetching baseline. The GHB prefetcher and RA-EMC+Dep complement each other well, due to the low bandwidth overhead of these two techniques. The highest bandwidth prefetcher, Markov+Stream, performs poorly with RA-EMC+Dep. The overall bandwidth consumption and effective memory access latency improvements of each of these systems are listed in Figures \ref{fig:mcRaEMC:bw}/\ref{fig:mcRaEMC:effLat} respectively.

\ignore{
\begin{table*}[ht]
	\small
	\centering
	\begin{tabular}{|p{1.75in}|p{1in}|p{1.7in}|}
		\hline
		& Bandwidth \newline Overhead (\%) & Effective Memory \newline Access Latency (Cycles) \\
		\hline
		\hline
		High RA-EMC+Dep & 11.2\% &  210 \\
		\hline
		High Stream & 18.9\% & 218 \\
		\hline
		High GHB & 16.1\% & 214\\
		\hline
		High Markov+Stream & 41.3\% & 258 \\
		\hline
		High \newline Stream+RA-EMC+Dep & 15.4\% & 200 \\
		\hline
		High \newline GHB+RA-EMC+Dep & 13.3\% & 185\\
		\hline
		High Markov+Stream+ \newline RA-EMC+Dep & 32.4\% & 225 \\
		\hline				
		\hline
		Copy RA-EMC+Dep & 8.5\% & 175 \\
		\hline
		Copy Stream & 15.2\% & 185 \\
		\hline
		Copy GHB & 9.3\% & 180 \\
		\hline
		Copy Markov+Stream & 31.6\% & 220\\
		\hline
		Copy \newline Stream+RA-EMC+Dep & 10.5\% & 116 \\
		\hline
		Copy \newline GHB+RA-EMC+Dep & 8.8\% & 110 \\
		\hline
		Copy Markov+Stream+ \newline RA-EMC+Dep & 27.6\% & 146 \\
		\hline
		\hline
		Mix RA-EMC+Dep & 5.4\% & 129 \\
		\hline
		Mix Stream & 18.8\% & 105 \\
		\hline
		Mix GHB & 13.2\% & 103 \\
		\hline
		Mix Markov+Stream & 49.2\% & 119 \\
		\hline
		Mix \newline Stream+RA-EMC+Dep & 11.5\% & 66 \\
		\hline
		Mix \newline GHB+RA-EMC+Dep & 9.3\% & 62 \\
		\hline
		Mix Markov+Stream+ \newline RA-EMC+Dep & 39.8\% & 72 \\
		\hline
	\end{tabular} 
	\caption{System Bandwidth Overhead with Prefetching}
	\label{tab:mcRaEMC:raEMCPFStats}
\end{table*}
}

\noindent \textbf{Throttling the EMC:} Since these RA-EMC+Dep and the GHB prefetcher complement each other particularly well, I extend the throttling policy (Section \ref{sec:mcRaEMC:dynamicPolicy}) to control both RA-EMC+Dep and the GHB prefetcher. By keeping track of the accuracy of each mechanism (defined by the percent of all prefetched lines that are accessed by the core prior to eviction) the EMC is able to throttle RA-EMC and the GHB prefetcher in a fashion similar to FDP \cite{fdp07}. If RA-EMC is more accurate than the GHB prefetcher then the GHB prefetcher is throttled down: the number of requests it is allowed to issue is reduced. If the GHB prefetcher is more accurate than RA-EMC, then the issue width of the EMC for runahead chains is reduced from 2 to 1. The performance effects of this throttling scheme are shown in Table \ref{tab:mcRaEMC:throtPerf}. This policy increases performance on workloads where the GHB prefetcher is more accurate than RA-EMC. It generally does not effect the performance of the high memory intensity workloads (High/Copy), but it does improve performance for the Mix workload set from 59\% to 65\%.

\begin{table*}[ht]
	\small
	\centering
	\begin{tabular}{|p{2.0in}|p{.75in}|p{.75in}|p{.75in}|}
		\hline
		& High & Copy & Mix\\
		\hline
		\hline
		Weighted Speedup Gain (\%) & 76.4\% & 71.0\%  & 65.3\% \\
		\hline
	\end{tabular} 
	\caption{RA-EMC+Dep+GHB Performance with Throttling}
	\label{tab:mcRaEMC:throtPerf}
\end{table*}

\subsection{Energy Evaluation}
\label{sec:mcRaEMC:energy}

\begin{figure}
	\centering
	\includegraphics[width=\columnwidth]{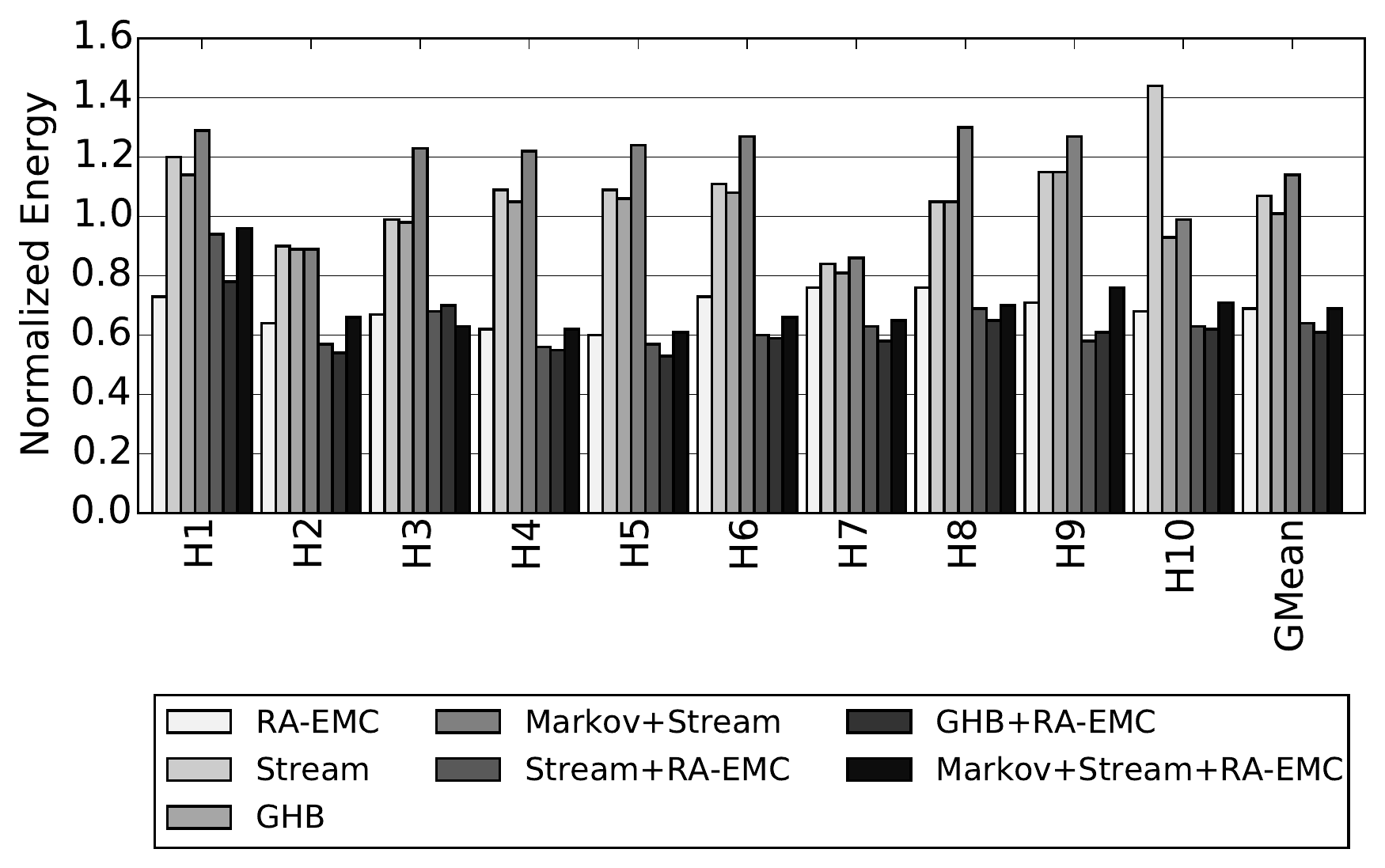}
	\caption{RA-EMC+Dep Energy Consumption on High Workloads}
	\label{fig:mcRaEMC:eHigh}
	\centering
	\includegraphics[width=\columnwidth]{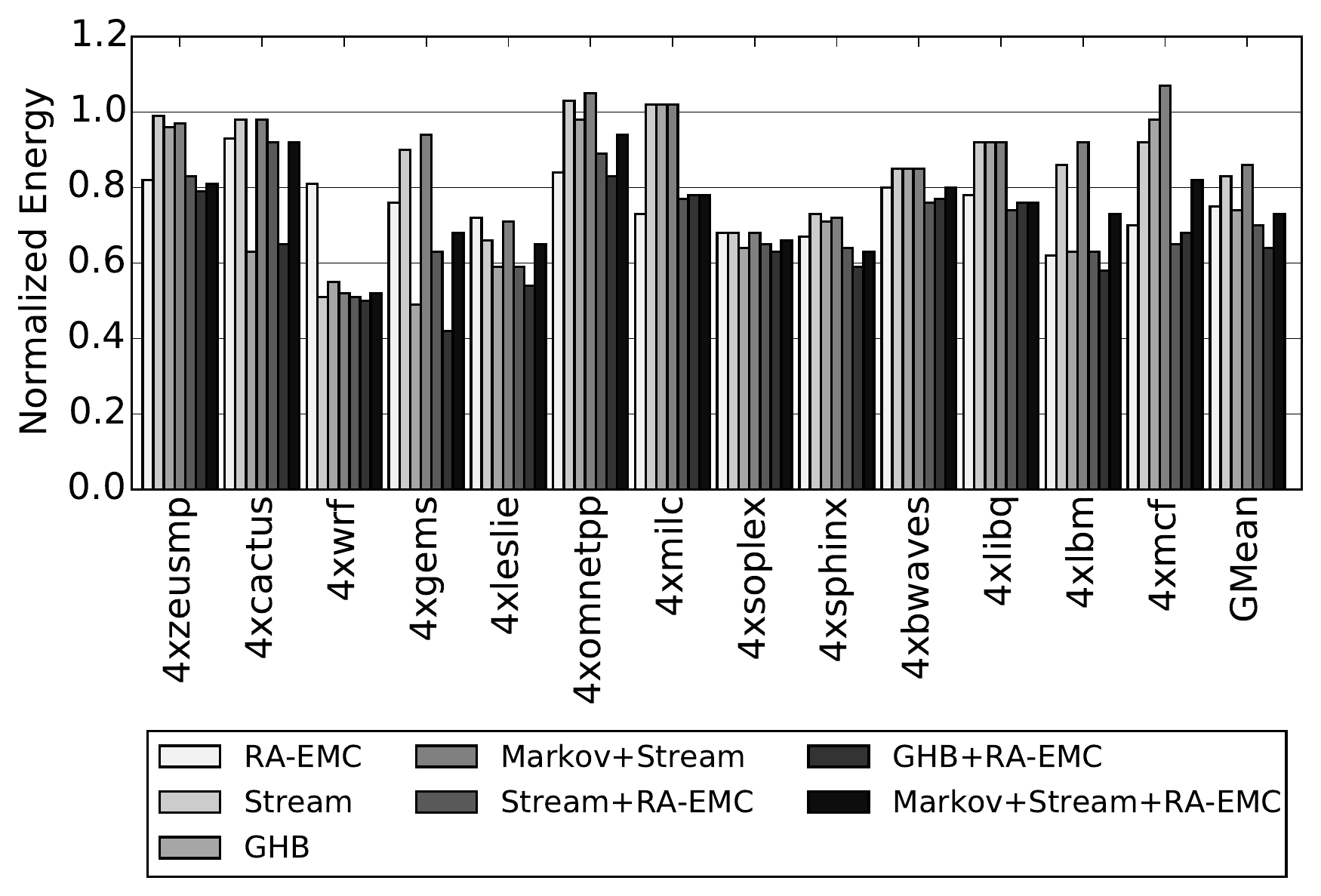}
	\caption{RA-EMC+Dep Energy Consumption on Copy Workloads}
	\label{fig:mcRaEMC:eCopy}
\end{figure}

\begin{figure}
	\centering
	\includegraphics[width=\columnwidth]{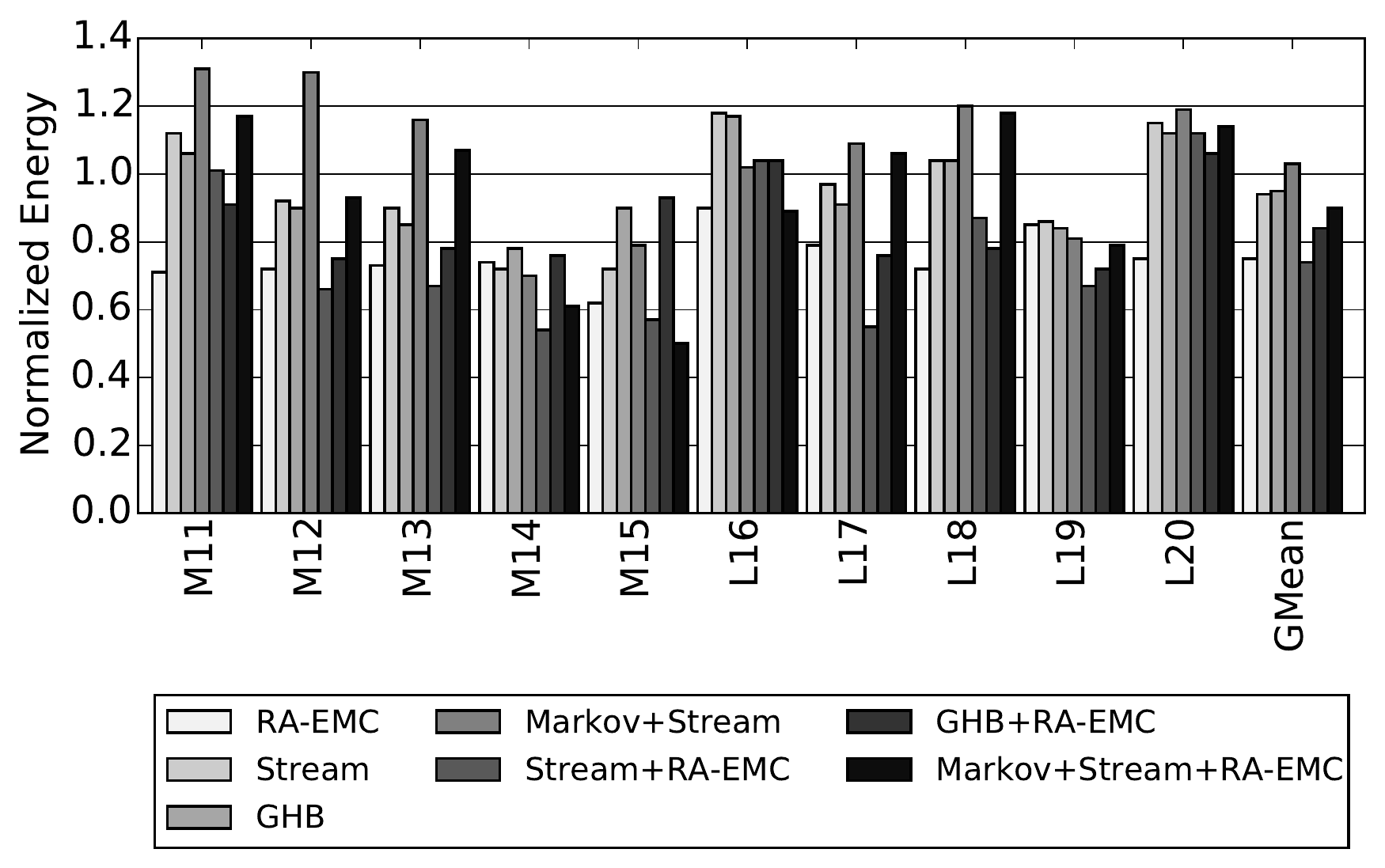}
	\caption{RA-EMC+Dep Energy Consumption on Mix Workloads}
	\label{fig:mcRaEMC:eMix}
\end{figure}

In contrast to the single core case in Chapter \ref{chap:scRaEMC} where the EMC led to a 7.8\% chip area overhead, the EMC is 2\% of total chip area in the multi-core case. This reduces EMC static energy impact. Moreover, the multi-core workloads run for longer than the single-core workloads due to multi-core contention. For example the multi-core run of 4x\textit{mcf} runs for 42\% more cycles than the single core run of \textit{mcf} in Chapter \ref{chap:scRaEMC}. Since these memory intensive applications already have low-activity factors, this leads to static energy dominating energy consumption. For 4x\textit{mcf}, static energy is 76.9\% of total system energy consumption. In contrast, static energy is 59.7\% of \textit{mcf} energy consumption in the single core case. These large static energy contributions relative to the small static energy cost of the EMC in the multi-core case mean that the large performance improvements from Section \ref{sec:mcRaEMC:eval} translate to large energy reductions. These reductions are shown in Figures-\ref{fig:mcRaEMC:eHigh}/\ref{fig:mcRaEMC:eCopy}/\ref{fig:mcRaEMC:eMix} for the High/Copy/Mix workloads.


I find that RA-EMC+Dep+GHB is the lowest energy consuming system in all three workload sets consuming 61\%/64\%/65\% of the energy in the no-prefetching baseline on average for the High/Copy/Mix workloads. RA-EMC+Dep is generally as energy efficient as the GHB prefetcher. However, in the High workloads RA-EMC+Dep has a large relative performance increase over GHB prefetching (Section \ref{sec:mcRaEMC:eval}). This leads to a 13\% energy reduction over GHB prefetcher in Figure \ref{fig:mcRaEMC:eHigh}.

While RA-EMC+Dep improves energy consumption in the multi-core case, the cost is an increase in on-chip contention. Table \ref{tab:mcRaEMC:raEMCEnergyStats} shows the increase in ring activity and L2 accesses for RA-EMC+Dep. The systems with prefetching but no RA-EMC+Dep do not effect these statistics and are excluded from the table. Overall, RA-EMC+Dep results in a roughly 30\% on-chip interconnect overhead and a 75\% increase in the number of LLC accesses. The LLC access overhead increases when prefetching is enabled, particularly when the Markov+Stream prefetcher is added to the system which generally causes a reduction in RA-EMC+Dep performance and accuracy.

\begin{table*}[ht]
	\small
	\centering
	\begin{tabular}{|p{1.75in}|p{.75in}|p{.8in}|p{.75in}|}
		\hline
		& Data Ring \newline Overhead & Control Ring \newline Overhead & LLC Access Overhead \\
		\hline
		\hline
		High RA-EMC+Dep & 26.0\% &  24.0\% & 75.1\% \\
		\hline
		High \newline Stream+RA-EMC+Dep & 40.5\% & 31.3\% & 81.3\% \\
		\hline
		High \newline GHB+RA-EMC+Dep & 38.5\% & 29.1\% & 78.3\%\\
		\hline
		High Markov+Stream+ \newline RA-EMC+Dep & 41.2\% & 34.5\% & 92.7\% \\
		\hline				
		\hline
		Copy RA-EMC+Dep & 22.9\% & 31.3\% & 63.5\%  \\
		\hline
		Copy \newline Stream+RA-EMC+Dep & 30.9\% & 38.1\% & 82.5\% \\
		\hline
		Copy \newline GHB+RA-EMC+Dep & 27.4\% & 33.2\% & 71.4\% \\
		\hline
		Copy Markov+Stream+ \newline RA-EMC+Dep & 35.9\% & 41.2\% & 87.8\%\\
		\hline
		\hline
		Mix RA-EMC+Dep & 29.5\% & 37.1\% & 65.3\%  \\
		\hline
		Mix \newline Stream+RA-EMC+Dep & 34.3\% & 41.2\% & 74.1\% \\
		\hline
		Mix \newline GHB+RA-EMC+Dep & 32.6\% & 38.5\% & 72.3\% \\
		\hline
		Mix Markov+Stream+ \newline RA-EMC+Dep & 38.1\% & 45.3\% & 86.7\% \\
		\hline
	\end{tabular} 
	\caption{System On-Chip Overhead}
	\label{tab:mcRaEMC:raEMCEnergyStats}
\end{table*}

\section{Sensitivity to System Parameters}
\label{sec:mcRaEMC:sens}

In this Section I identify three key parameters to the RA-EMC+Dep system: LLC cache capacity, the number of memory banks, and the number of cycles that it takes to access the LLC. RA-EMC performance and energy sensitivity to these parameters are listed in Table \ref{tab:mcRaEMCSens}. The values used for these parameters in the evaluation in Section \ref{sec:mcRaEMC:eval} are bold.

\begin{table}[htb*]
	\centering
	\begin{minipage}[bht*]{1.00\columnwidth}
		\centering
		\footnotesize
		\begin{tabular}{|c|c||c|c||c|c||c|c|} \hline
			\multicolumn{8}{|c|}{{\bf LLC Cache Size} } \\ \hline 
			\multicolumn{2}{|c||}{2 MB} & \multicolumn{2}{c||}{\textbf{4 MB}} & \multicolumn{2}{c||}{8 MB} & \multicolumn{2}{c|}{16 MB} \\ \hline
			$\Delta$ Perf & $\Delta$ Energy & $\Delta$ Perf & $\Delta$ Energy & $\Delta$ Perf & $\Delta$ Energy & $\Delta$ Perf & $\Delta$ Energy \\ \hline
			53.5\% & -29.6\% & 63.3\% & -31.3\% & 52.9\% & -33.3\% & 45.8\% & -34.2\% \\ \hline
		\end{tabular}
	\end{minipage}
	
	\begin{minipage}[bht*]{1.00\columnwidth}
		\centering
		\footnotesize
		\begin{tabular}{|c|c||c|c||c|c||c|c|} \hline
			\multicolumn{8}{|c|}{{\bf Number of Memory Banks} } \\ \hline 
			\multicolumn{2}{|c||}{\textbf{8}} & \multicolumn{2}{c||}{16} & \multicolumn{2}{c||}{32} & \multicolumn{2}{c|}{64} \\ \hline
			$\Delta$ Perf & $\Delta$ Energy & $\Delta$ Perf & $\Delta$ Energy & $\Delta$ Perf & $\Delta$ Energy & $\Delta$ Perf & $\Delta$ Energy \\ \hline
			61.0\% & -32.9\% & 63.3\% & -31.3\% & 51.0\% & -28.8\% & 50.1\% & -28.0\% \\ \hline
		\end{tabular}
	\end{minipage}
	
	\begin{minipage}[bht*]{1.00\columnwidth}
		\centering
		\footnotesize
		\begin{tabular}{|c|c||c|c||c|c||c|c|} \hline
			\multicolumn{8}{|c|}{{\bf LLC Latency (Cycles)} } \\ \hline 
			\multicolumn{2}{|c||}{13} & \multicolumn{2}{c||}{\textbf{18}} & \multicolumn{2}{c||}{23} & \multicolumn{2}{c|}{28} \\ \hline
			$\Delta$ Perf & $\Delta$ Energy & $\Delta$ Perf & $\Delta$ Energy & $\Delta$ Perf & $\Delta$ Energy & $\Delta$ Perf & $\Delta$ Energy \\ \hline
			67.1\% & -34.3\% & 63.3\% & -31.3\% & 43.9\% & -26.9\% & 38.5\% & -25.3\% \\ \hline
		\end{tabular}
	\end{minipage}
	\begin{small}
		\caption{Multi-Core RA-EMC Performance and Energy Sensitivity}
		\label{tab:mcRaEMCSens}
	\end{small}
\end{table}

RA-EMC+Dep shows significant performance sensitivity to very large LLC size (16MB), where the impact of runahead prefetching is diminished. RA-EMC+Dep also shows performance sensitivity to a large number of memory banks/channel (64 banks/channel) where the delay that the dependent miss acceleration at the EMC exploits is decreased. Overall, system energy reduction stays relatively constant in the LLC/memory bank sensitivity as chip size/DRAM bandwidth increases in both RA-EMC+Dep and the baseline. Increasing LLC access latency decreases RA-EMC+Dep benefit and reduces the energy consumption benefit over the baseline, while reducing LLC latency benefits RA-EMC+Dep. This is because all EMC cache misses result in LLC lookups and a low latency LLC is advantageous to EMC performance.

\section{Conclusion}
\label{sec:mcRaEMC:conclusion}

In this chapter I developed the mechanisms that are required to allow the Enhanced Memory Controller to accelerate both independent and dependent cache misses in a multi-core system. This proposal, RA-EMC+Dep is shown to outperform three on-chip prefetchers (Stream, GHB, and Stream+Markov). RA-EMC+Dep reduces effective memory access latency by 19\% on a suite of high memory intensity workloads. This is greater than the effective memory access latency reduction achieved by any of the three evaluated prefetchers. When combined with a GHB prefetcher, RA-EMC+Dep+GHB is the highest performing system, resulting in a 28.2\% reduction in effective memory access latency. I conclude that RA-EMC+Dep improves system performance by accelerating both independent and dependent cache misses.

\chapter{Conclusions and Future Work}
\label{chap:conclusion}
\setlength{\epigraphwidth}{0.41\textwidth}

The effective latency of accessing main memory is the largest impediment to high single thread performance. All of the applications in Figure \ref{fig:intro:stall} with an IPC of under one are high-memory intensity benchmarks. This dissertation studies the dynamic properties of the operations that lead to last level cache (LLC) misses. The key insight is that all LLC misses can be separated into two categories based on their dependence chains: independent cache misses and dependent cache misses. Using this insight, I develop hardware mechanisms to increase performance for high-memory intensity benchmarks. 

Independent cache misses have all of the source data that is required to generate the address of the memory access available on-chip. The reason that the processor pipeline stalls waiting for the data from an independent cache miss is that the back-end of the processor has limited resources and can not continuously fetch and execute operations. To mitigate this problem, this dissertation proposes an efficient version of runahead execution. By dynamically isolating and executing only the filtered dependence chains that lead to independent cache misses, the runahead buffer generates 57\% more memory level parallelism on average when compared to traditional runahead while consuming 17.5\% less energy. 

This dissertation then shows that the performance gain due to the runahead buffer is limited by the short length of each runahead interval. To solve this problem, I develop mechanisms to offload runahead dependence chains to a compute capable memory controller where they are speculatively executed continuously to prefetch data. This prefetching is shown to significantly reduce the effective memory access latency of subsequent demand requests and outperforms three on-chip prefetchers with a lower memory bandwidth overhead.

Dependent cache misses are difficult to accelerate as source data is not available on-chip and memory access addresses are data-dependent. This severely limits the ability of prefetchers to efficiently predict the address of a dependent cache miss far enough in advance to reduce the effective latency of accessing main memory. This dissertation shows that a predominant source of the effective memory access latency for dependent cache misses is a result of on-chip contention. This dissertation develops mechanisms to reduce the on-chip delay observed by a dependent cache miss by migrating the filtered dependence chain to a compute capable memory controller for execution when source data arrives from main memory. By executing dependent cache misses at the enhanced memory controller (EMC), these misses experience 20\% lower latency than if they were issued by the core. 

By combining dependent cache miss acceleration with continuous runahead execution at the EMC, the final mechanism in this dissertation, RA-EMC+Dep reduces effective memory access latency in a multi-core system by 19\% while increasing performance on a set of ten high-memory intensity workloads by 62\%. This is a greater performance increase and effective memory access latency reduction than any of the three on-chip prefetchers that are used in the evaluation. RA-EMC+Dep is the first combined mechanism that uses dependence chains to automatically accelerate both independent and dependent cache misses in a multi-core system.

RA-EMC+Dep requires additional compute hardware at the memory controller. This dissertation demonstrates that this proposed hardware has limited overhead and can be tailored to the task of executing the dependence chains that lead to cache misses. This enhanced memory controller (EMC) does not require large compute structures such as floating point or vector units. It does not require heavyweight out-of-order hardware structures such as register renaming or a wide super-scalar execution engine. The short dependence chains that the EMC executes do not require a large monolithic physical register file. The EMC does not need a front-end as it executes chains of decoded micro-ops that are provided by the main core. As dependent cache misses are rare when compared to independent cache misses, this dissertation demonstrates that these two acceleration mechanisms are easily combined at a lightweight EMC.

This dissertation makes a case for compute capable memory controllers and dynamically filtered code execution. These are two lightly explored areas that provide a route for hardware to reduce or eliminate the effects of effective memory access latency on memory intensive applications. There are several paths forward to improve the EMC that is proposed in this dissertation. The primary drawback to computation at the EMC is the increased pressure that is placed on the LLC. Yet, the necessity of this increased pressure is questionable. As the EMC executes speculative dependence chains it is not clear that EMC data cache needs to be placed inside the coherence domain of the multi-core processor. Opportunistic or lazy data updates to the EMC would reduce the remote-execution overhead on the multi-core chip. Taking this concept even further, the EMC is a prime location for value prediction. As the data values that the EMC uses do not need to be 100\% correct and up to date for prefetching, a value predictor may be good enough to avoid constant LLC requests.

Exploring the hardware structure of the EMC itself provides a different research direction. The proposed EMC still looks like the back-end of an out-of-order core. However, as dependence chains are short, a pure dataflow implementation could further reduce dynamic energy consumption and increase runahead distance. Exposing the EMC to the programmer could allow expert-programmers to hand code optimal dataflow threads that prefetch data in cohort with application phases.

Researchers can also build on the simple code filtering mechanisms that are developed in this dissertation to enable EMC-like engines to provide intelligent feedback and control throughout the cache hierarchy. For example, different flavors of pattern-matching prefetchers can be driven by feedback from how accurate runahead threads currently are and how useful each flavor of prefetcher is to future code. Cache replacement policies and cache partitioning algorithms can take into account reuse information from filtered dependence chains to guide policy decisions. Memory scheduling algorithms can use information about future memory accesses to determine which row buffers to hold open or to close. All current on-chip cache hierarchy policies guess at access patterns using local information. Intelligently controlling these policies globally based on future code segments is an exciting research direction.

Dynamically increasing single-thread performance is a challenging problem. This dissertation demonstrates that micro-architectural compute improvements are still capable of delivering performance increases. As main memory latencies remain roughly constant and do not improve relative to core frequency, architects must search for new and creative avenues to reduce effective memory access latency for applications that cannot hide long latency operations with parallelism.




%
%

\bibliographystyle{plain}
\bibliography{diss}



\end{document}